%% file: cars.tex
\documentclass{aa}
\usepackage{graphicx}
\graphicspath{{Images/}}
\usepackage{natbib}
\bibpunct{(}{)}{;}{a}{}{,}
\usepackage[Symbol]{upgreek}
\usepackage{hyperref}
\usepackage{url}
\usepackage{color}
\usepackage{txfonts}
\usepackage{stfloats}
\usepackage{amstext}

\usepackage{multirow}
\usepackage{multicol}
\usepackage{mathrsfs}

\begin{document} 
    \title{The Close AGN Reference Survey (CARS).}
    \subtitle{No obvious signature of AGN feedback on star formation, but subtle trends}
    \author{
    I. Smirnova-Pinchukova\inst{\ref{mpia}}
    \and B.~Husemann\inst{\ref{mpia}}
    \and T.~A.~Davis\inst{\ref{cardiff}}
    \and C.~M.~A.~Smith\inst{\ref{cardiff}}
    \and M.~Singha\inst{\ref{manitoba}}
    \and G.~R.~Tremblay\inst{\ref{harvard}}
    \and R.~S.~Klessen\inst{\ref{uni_heidelberg},\ref{uni_IZWR}}
    \and M.~Powell\inst{\ref{stanford}}
    \and T.~Connor\inst{\ref{JPL},\ref{michstate}}
    \and S.~A.~Baum\inst{\ref{manitoba}}
    \and F.~Combes\inst{\ref{paris}}
    \and S.~M.~Croom\inst{\ref{sydney}}    
    \and M.~Gaspari\inst{\ref{INAF},\ref{PU}}
    \and J.~Neumann\inst{\ref{portsmouth}}
    \and C.~P.~O'Dea\inst{\ref{manitoba}}
    \and M.~P\'erez-Torres\inst{\ref{granada},\ref{zaragoza}}
    \and D.~J.~Rosario \inst{\ref{Durham}}
    \and T.~Rose\inst{\ref{Durham}}
    \and J.~Scharw\"achter\inst{\ref{gemini}}
    \and N.~Winkel\inst{\ref{mpia}}
    }
    \institute{
    Max-Planck-Institut f\"ur Astronomie, K\"onigstuhl 17, 69117 Heidelberg, Germany \email{smirnova@mpia.de} \label{mpia}
    \and 
    School of Physics \& Astronomy, Cardiff University, Queens Buildings, The Parade, Cardiff, CF24 3AA, UK \label{cardiff}
    \and
    Department of Physics \& Astronomy, University of Manitoba, Winnipeg, MB R3T 2N2, Canada \label{manitoba}
    \and
    Center for Astrophysics $|$ Harvard \& Smithsonian, 60 Gardent St., Cambridge, MA 02138, USA \label{harvard}
    \and
    Universit{\"a}t Heidelberg, Zentrum für Astronomie, Institut für Theoretische Astrophysik, Albert-Ueberle-Str 2, D-69120 Heidelberg, Germany\label{uni_heidelberg}
    \and
    Universit{\"a}t Heidelberg, Interdisziplinäres Zentrum für Wissenschaftliches Rechnen, Im Neuenheimer Feld 205, D-69120 Heidelberg, Germany\label{uni_IZWR}
    \and
    Kavli Institute  of  Particle  Astrophysics  and  Cosmology,  Stanford University, 452 Lomita Mall, Stanford, CA 94305, USA\label{stanford}
    \and
    Jet Propulsion Laboratory, California Institute of Technology, 4800 Oak Grove Drive, Pasadena, CA 91109, USA \label{JPL}
    \and 
    Department of Physics \& Astronomy, Michigan State University, 567 Wilson Road, East Lansing, MI 48824, USA \label{michstate}
    \and
    Observatoire de Paris, LERMA, Coll\`ege de France, CNRS, PSL University, Sorbonne University, 75014, Paris, France \label{paris}
    \and 
    Sydney Institute for Astronomy, School of Physics, A28, The University of Sydney, NSW, 2006, Australia \label{sydney}
    \and
    INAF - Osservatorio di Astrofisica e Scienza dello Spazio, via P. Gobetti 93/3, I-40129 Bologna, Italy \label{INAF}
    \and 
    Department of Astrophysical Sciences, Princeton University, 4 Ivy Lane, Princeton, NJ 08544-1001, USA \label{PU}
    \and 
    Institute of Cosmology and Gravitation, University of Portsmouth, Burnaby Road, Portsmouth, PO1 3FX, UK \label{portsmouth}
    \and
    Instituto de Astrofísica de Andaluc\'{i}a, Glorieta de las Astronom\'{i}a s/n, 18008 Granada, Spain\label{granada}
    \and 
    Departamento de F\'{\i}sica Te\'orica, Facultad de Ciencias, Universidad de Zaragoza, E-50009 Zaragoza, Spain\label{zaragoza}
    \and 
    Centre for Extragalactic Astronomy, Durham University, DH1 3LE, UK \label{Durham}
    \and 
    Gemini Observatory/NSF's NOIRLab, 670 N. A'ohoku Place, Hilo, HI 96720, USA \label{gemini}
    }
   \date{}
   \abstract
   {Active Galactic Nuclei (AGN) are thought to be responsible for the suppression of star formation in massive $\sim10^{10}M_\sun$ galaxies. While this process is a key feature in numerical simulations of galaxy formation, it is not yet unambiguously confirmed in observational studies.}
   {Characterization of the star formation rate (SFR) in AGN host galaxies is challenging as AGN light contaminates most SFR tracers. Furthermore, the various SFR tracers are sensitive to different timescales of star formation from $\sim$a few to 100\,Myr. We aim to obtain and compare SFR estimates from different tracers for AGN host galaxies in the Close AGN Reference Survey (CARS) to provide new observational insights into the recent SFR history of those systems.}
   {We construct integrated panchromatic spectral energy distributions (SED) to measure the FIR luminosity as a tracer for the recent ($<$100\,Myr) SFR. In addition we use integral-field unit (IFU) observation of the CARS targets to employ the H$\alpha$ luminosity decontaminated by AGN excitation as a proxy for the current ($<$5Myr) SFR.}
   {We find that significant differences in specific SFR of the AGN host galaxies as compared with the larger galaxy population disappear once cold gas mass, in addition to stellar mass, is used to predict the SFR for a specific AGN host. Only a tentative trend with the inclination of the host galaxy is remaining such that SFR appears slightly lower than expected when the galaxies of unobscured AGN appear more edge-on along our line-of-sight, particular for dust-insensitive FIR-based SFRs. We identify individual galaxies with significant difference in their SFR which can be related to a recent enhancement or decline in their SFR history that might be related to various processes including interactions, gas consumption, outflows and AGN feedback.}
   {AGN can occur in various stages of galaxy evolution which makes it difficult to relate the SFR solely to the impact of the AGN. Our study shows that stellar mass alone is an insufficient parameter to estimate the expected SFR of an AGN host galaxy compared to the underlying non-AGN galaxy population. We do not find any strong evidence for a global positive or negative AGN feedback in the CARS sample. However, there is tentative evidence that 1) the relative orientation of the AGN engine with respect to the host galaxies might alter the efficiency of AGN feedback and 2) the recent SFH is an additional tool to identify rapid changes in galaxy growth driven by the AGN or other processes.}
   \maketitle

\section{Introduction}

The evolution of galaxies is driven by many different physical processes, which we can indirectly study by probing various galactic properties across cosmic time. One key parameter is the star-formation rate (SFR), which changes with redshift and discloses essential information about the evolution of galaxies \citep[e.g.][]{madau2014}. Furthermore, at lower redshifts the galaxy population shows a clear bi-modality in the colour-magnitude diagram \citep[e.g.][]{strateva2001,baldry2004,wyder2007,brammer2009} of which the specific SFR versus stellar mass (sSFR-M$_{\star}$) diagram \citep[e.g. ][]{brinchmann2004_sdssdr2, noeske2007, karim2011, renzini_peng2015_ms} is a more physical representation. Such diagrams separate star-forming from passive galaxies where the so-called \textit{star-forming main sequence} \citep[SFMS,][]{noeske2007} indicates the average location of star-forming galaxies as a function of stellar mass. Moreover, the SFMS tends to be present across a wide range of redshifts as a tight correlation \citep[e.g.][]{elbaz2007, speagle2014}. How this galaxy bi-modality is established is one of the open key questions in our current galaxy evolution model. One option is that accreting super-massive black holes (SMBH) release enough energy to suppress the star formation in their host galaxies as they pass quickly through from the blue cloud to the red sequence \citep[e.g.][]{faber2007,schawinski2014_quenching}, but also opposite tracks via rejuvenation of star forming have been proposed \citep[e.g.][]{hasinger2008,mancini2019,chauke2019}. Furthermore, environmental processes \citep[e.g.][]{balogh2004, peng2010, schaefer2019} and gas  consumption \citep[e.g.][]{colombo2020} may facilitate this transformative process.

The ability both to rapidly quench and maintain low-levels of star formation in red-sequence massive galaxies has often been attributed to the impact of active galactic nuclei (AGN). Radiatively-driven winds, mechanical energy and/or thermal pressure release from the AGN are physical process routinely incorporated into numerical simulations of galaxy evolution to solve the overcooling problem \citep[e.g.][]{vogelsberger2014a, schaye2015, pillepich2018, bassini2019, wittor2020}, so that the characteristics of the observed galaxy population can be reproduced \citep[e.g.][]{vogelsberger2014b,nelson2018}. Often two types of AGN feedback mode are being distinguished, which are called the quasar mode or radiative mode \citep[e.g.][]{diMatteo2005, hopkins2008, hopkins2010} and the radio mode also known as kinetic or maintenance mode \citep[e.g.][]{bower2006, croton2006, fabian2012, gaspari2020}. The quasar mode is thought to initially quench star formation through powerful radiatively driven winds that remove the gas from the galaxy  \citep[e.g.][]{nesvadba2008, feruglio2010, maiolino2012} and thereby suppress star formation. The radio mode is a heating mechanism affecting the gaseous halo that prevents continuous condensation of cold and warm gas, thereby quenching star formation over long time scales and maintaining a low level of star formation in red sequence galaxies \citep[e.g.][]{brueggen2002, mcnamara2005, gaspari2019, mcdonald2021}. Strong observational evidence for radio mode feedback has been collected from galaxy clusters hosting powerful radio galaxies \citep[e.g.][]{fabian2006, mcnamara2007}. The role of quasar mode feedback is much more controversial from an observational perspective. Various studies have reported \textit{negative} \citep{ho2005, nandra2007, schawinski2009, farrah2012, page2012, mullaney2015, shimizu2015, wylezalek2016, kakkad2017, catalan-torrecilla2017, bing2019, bluck2020, padilla2020, simcha2020, smith2020}, \textit{positive} \citep{kim2006, cresci2015, cresci2015b, bernhard2016, santoro2016, maiolino2017, koss2020} or \textit{no effect} from feedback \citep{elbaz2011, bongiorno2012, harrison2012, husemann2014_qdeblend3d, balmaverde2016, leung2017, woo2017, shangguan2018, scholtz2020} on the star formation rates of observed AGN host galaxies. 

One common difficulty in observational studies is the measurement of the basic AGN host galaxy parameters such as SFR and stellar mass, because the AGN light can dominate the total emission of the host galaxy across nearly the entire electro-magnetic spectrum from radio to X-rays. Hence, it can be challenging to measure the SFR or stellar mass from commonly used proxies and various strategies have been used in the literature to mitigate the issue. The stellar mass can be easily determined from optical spectra or NIR-optical photometry in obscured AGN \citep[e.g.][]{kauffmann2003_bpt,silverman2009,aird2012, mignoli2013} as the light from the central AGN is blocked by dust along our line of sight. However, the drawback is that AGN parameters such as black hole (BH) mass can usually not be directly inferred for obscured AGN, except for few very nearby systems using the narrow-line region (NLR) kinematics \citep[e.g.][]{barth2001,walsh2013}. 

The SFR of a galaxy is usually determined either from the UV radiation of young stars \citep[e.g.][]{salim2007}, the ionizing photons from \ion{H}{ii} regions using H$\alpha$ or [\ion{O}{ii}] \citep[e.g.][]{kennicutt1983,kewley2004}, the reheated warm and cold dust at MIR and FIR wavelengths \citep[e.g.][]{kennicutt1998}, or the radio emission produced in supernova explosions \citep[e.g.][]{condon1992}. All those tracers can be significantly contaminated by AGN emission even in obscured AGN. The UV or radio emission can be spatially separated through high-angular observations \citep{Fernandez2018,Rosario2021}. Alternatively, the overall spectral energy distribution (SED) is decomposed into various emitting components of a galaxy including AGN, star-formation heated dust and stars. Examples of such SED fitting codes are (X-)Cigale \citep{noll2009,yang2020}, Magphys \citep{daCunha2008}, AGNfitter \citep{agnfitter} or FortesFit \citep{rosario2019}, which can be used to infer stellar mass, IR-based SFR and AGN luminosity simultaneously even for unobscured type 1 AGN. One potential drawback of IR-based SFR tracers for AGN feedback studies is that it traces the SFR on relatively long timescales of $\sim$100\,Myr and therefore the SFR can be significantly over or underestimated for a strongly declining or rising SFR within that timescale \citep{hayward2014}. 

Considering that the quasar-mode feedback is expected to be a fast process, the H$\alpha$ line may be a more viable tracer for the feedback, considering that it probes the current SFR on $<5\textnormal{Myr}$ timescales. 
Unfortunately, the H$\alpha$ line can be strongly boosted by AGN ionization. In order to use H$\upalpha$ as a SFR tracer, optical emission lines diagnostic diagrams such as the Baldwin-Phillips-Terlevich diagrams \citep{baldwin1981_bpt} have been extensively used to distinguish between AGN and \ion{H}{ii} region excitation. A comprehensive classification scheme with such diagnostic diagrams has been developed \citet{kewley2006_dsf}. In particular with integral-field unit (IFU) spectroscopy, the usage of BPT diagrams has been extended to map the ionization conditions across galaxies to exclude regions from the SFR estimation dominated by AGN ionization using such BPT demarcation lines \citep[e.g.][]{husemann2014_qdeblend3d,nascimento2019_manga}. However, the superposition of different ionization conditions along the line-of-sight is a common scenario.  The observed AGN mixing sequence (MS) arises from such a superposition, and an analytic approach to separate the relative contribution was presented in \citet{davies2014_I, davies2014_II}, \citet{davies2016}, and \citet{davies2017}.

In this paper, we infer and compare the SFR estimated from H$\alpha$ and the IR for a sample of 41 galaxies as part of the Close AGN Reference Survey (CARS, \url{www.cars-survey.org}). CARS aims to shed light on the relationship between the central AGN and its host galaxy by following a spatially-resolved multi-wavelength approach. As presented in Husemann et al. (2021), the 41 CARS targets are a sub-sample of 99 luminous unobscured AGN at $z<0.06$ previously identified with the flux-limited Hamburg/ESO AGN survey \citep{wisotzki2000}. The AGN targeted by CARS therefore correspond to the most luminous unobscured AGN in the nearby Universe allowing a detailed investigation of AGN -- host galaxies interactions at relatively high spatial resolution.  

In this work we use both panchromatic SEDs (to estimate stellar masses and FIR-based SFRs), and analyze the CARS IFU data to infer H$\alpha$ based SFRs (adopting a new algorithm to separate the contribution to the H$\alpha$-line from other excitation mechanisms). In addition, we explore the use of different prescriptions to predict the expected SFR from the non-AGN galaxy population as control samples. This allows us to investigate links with AGN parameters depending on the actual SFR tracers and control sample properties being used as well as to distinguish between potentially declining and increasing star formation histories amongst the AGN host galaxies. 

Throughout the paper, we assume a flat cosmological model with $H_0 = 70$ km s$^{-1}$ Mpc$^{-1}$, $\Omega_\mathrm{M} = 0.3$, and $\Omega_{\Lambda} = 0.7$.

\section{Observational data set}

\begin{figure*}[t]
    \centering
    \includegraphics[width=1.\linewidth]{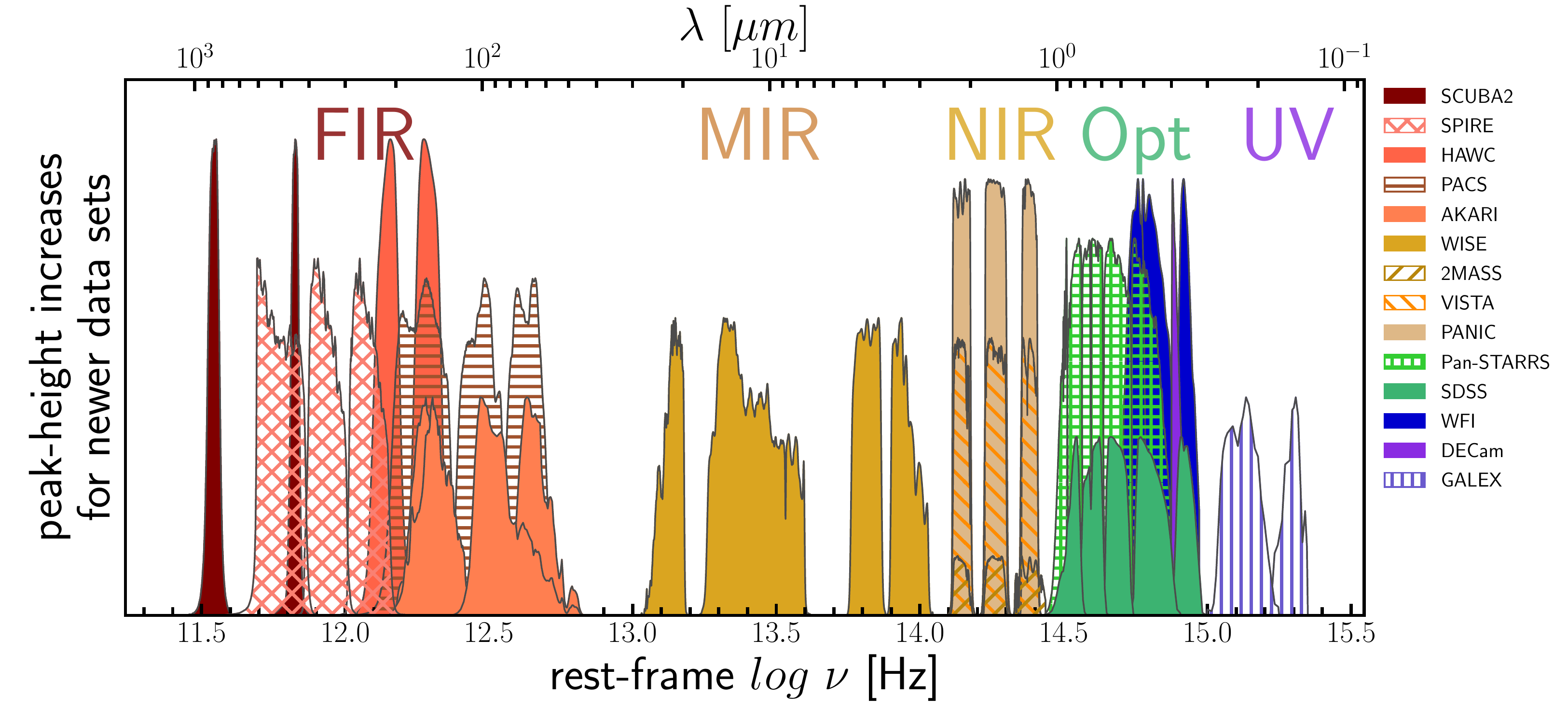}
    \caption{Photometric broad bands from a number of surveys and instruments used for the SED modelling. The $y$-axis is normalized roughly by the year of observational campaign to visually compare bands with similar coverage. When more than one option was available for certain bands, the preference was given to the band sets, which are temporally closer to each other (especially in near infrared and optical ranges) to minimize the impact of AGN variability.}
    \label{fig:sed_bands}
\end{figure*}

\subsection{Archival broad-band photometry}
We collect publicly available broad-band photometry from archival sources combined with dedicated observations for CARS targets to construct their panchromatic spectral energy distributions (SEDs). Where catalogs are not available or do not properly handle extended sources, we extract consistent aperture photometry from the public survey images. All photometric measurements are listed in Table~\ref{table:sed} and their origin is described below.

\subsubsection{Catalogue data}
We directly took data from the \textit{Herschel}/SPIRE point source catalogue \citep[250$\upmu$m, 350$\upmu$m, and 500$\upmu$m bands,][]{schulz2017_spire}, the AKARI/FIS bright source catalogue \citep[N60, N160, WIDE-L, and WIDE-S bands,][]{yamamura2009_akari_fis}, and the 2MASS extended source catalogue \citep[J, H, and K$_{\mathrm{S}}$ bands,][]{jarrett2000_2mass_xcs} without further processing. The GALEX source catalogue \citep[FUV and NUV bands,][]{bianchi2017_galex} were used only to compare the fluxes to our own aperture photometry measurements. 

\subsubsection{Aperture photometry}
For the following data sets we performed customized aperture measurements on the survey images: \textit{Herschel}/PACS 70, 100, and 160$\upmu$m images from the Herschel science archive\footnote{HSA, \url{http://archives.esac.esa.int/hsa/whsa/}}, 
WISE atlas images \citep{cutri2011_wise}, 2MASS J, H and K$_{\rm S}$ images from the interactive image service\footnote{\url{https://irsa.ipac.caltech.edu/applications/2MASS/IM/interactive.html}},
VISTA/Paranal J, H, and K$_{\rm S}$ band images from the Cambridge astronomical survey unit\footnote{CASU, \url{http://casu.ast.cam.ac.uk}},
optical \textit{grizy} photometric images\footnote{available at  \url{http://ps1images.stsci.edu}} from the Pan-STARRS DR1 \citep{chambers2016_pan-starrs} and \textit{ugriz} images\footnote{available at \url{https://dr12.sdss.org/}} from the SDSS-III DR12 \citep{alam2015_sdss}, and GALEX and Swift/UVOT ultraviolet images from the Barbara~A.~Mikulski Archive for Space Telescopes\footnote{MAST, \url{http://archive.stsci.edu/}}.\\
The elliptical apertures are defined for each instrument to cover the entire galaxy in the sequence of the bands. The aperture fluxes are computed using the AstroPy Python package \textsc{photutils} \citep[][ \url{photutils.readthedocs.io}]{photutils}. Since the images of CARS targets are extended at these wavelengths the apertures are not affected by beam smearing and PSF matching as done for higher redshift targets is not a concern for our sample.

\subsection{Galactic extinction and AGN broad-line correction}
The observed photometry for external galaxies is attenuated by the dust in the Milky Way depending on the specific line-of-sight. We apply an extinction correction to the NIR, optical, and UV photometry of the SED using the \citet{fitzpatrick1999} Milky Way attenuation curve. The normalization of the attenuation curve is anchored to the measured $V$ band extinction ($A_{V}$) along our line-of-sight towards the target is taken from \citet{schlafly2011} and retrieved from the NASA/IPAC Extragalactic Database (NED)\footnote{https://ned.ipac.caltech.edu/}. All corresponding photometric data in Table~\ref{table:sed} are provided after extinction correction.

The broad emission lines of unobscured AGN can significantly contribute to broad-band photometry. Because the SED model library used for the fitting assumes smooth power-law for the optical-UV accretion disc component we need to subtract the broad emission line contribution from the photometry. We used the available IFU spectroscopy of the CARS sample to measure H$\beta$ and H$\alpha$ fluxes, and subtract them from the optical bands (mainly $g$ and $r$).  The values presented in Table~\ref{table:sed} are presented after correction for the BLR contribution.

\subsection{Dedicated broad-band imaging}
Given the lack of available data in some bands for some of the CARS targets, we obtained dedicated observations  to fill the missing gaps in our photometry. Optical observations were mainly missing in the $u$ band which Pan-STARRS does not cover and HE\,0108$-$4743 and HE\,2211$-$3903 are not covered by SDSS or Pan-STARRS, given their low declination. Therefore, we obtained deep wide-field optical imaging with the Dark Energy Camera \citep[DECam,][]{
depoy2008_DECam} mounted to the 4m Blanco telescope of the Cerro Tololo Inter-American Observatory (CTIO) and with the Wide-Field Imager \citep[WFI,][]{baade1999_WFI} mounted to the 2.2m telescope of the La Silla observatory. Missing NIR imaging were obtained with the PAnoramic Near-Infrared Camera \citep[PANIC,][]{cardenas2018_PANIC} mounted to the 2.2m telescope at the Calar Alto Observatory. Finally, deeper FIR observations were obtained for most of the CARS targets undetected with AKARI using the High-resolution Airborne Wideband Camera Plus \citep[HAWC+,][]{harper2018_HAWC} aboard the Stratospheric Observatory for Infrared Astronomy (SOFIA) and the Submillimetre Common-User Bolometer Array 2 \citep[SCUBA-2,][]{holland2013_SCUBA2} camera mounted to the James Clark Maxwell Telescope (JCMT). The data characteristics and data reduction of all those new observations are described in more detail below. 

\subsubsection{DECam optical observations}
DECam imaging in the $u$ band was obtained for 14 CARS targets during the nights of 1--3 April 2017 as part of the program 2017A-0914 (PI: Grant Tremblay). DECam covers almost 3\,sq. degree using a mosaic of 62 2\,k$\times$4\,k CCDs with a pixel size of 0.27\arcsec. The total exposure times for each target ranged from 1200\,s to 4800\,s. The data were automatically reduced by the DECam community pipeline \citep{valdes2014_DECam} and retrieved from the NOAO archive. Absolute flux calibration for individual images was achieved by comparing detected stars in the field with the photometrically-calibrated SkyMapper DR1 \citep{wolf2018_SkyMapper} catalogue magnitudes.

\subsubsection{WFI optical observations}
WFI imaging in the $UBV$ bands was obtained for HE\,0108$-$4743 and HE\,2211$-$3903 to provide optical photometry given the missing Pan-STARRS and SDSS coverage. The observations were taken during a larger observing run from 18-28 October under program 0100.A-9003 (PI: Bernd Husemann). WFI covers a 34\arcmin$\times$33\arcmin\ FoV using a mosaic of 8 2\,k$\times$4\,k CCDs. At the time of observations 2 of the CCDs were broken so that a two-pointing offset scheme plus dithering was used to cover the same field. A total exposure time of 3600s was obtained for the $U$ band. 

The images were fully reduced and combined with the latest version of the THELI pipeline\footnote{https://github.com/schirmermischa/THELI} (version 3) \citep{schirmer2013_THELI}. The absolute magnitude zero-point for each combined image was obtained by comparing stellar counts with catalogue values from the SkyMapper DR1 and APASS DR1 \citep{henden2009_APASS} for the $UBV$ bands.

\begin{sidewaystable*}
\small
\caption{Multi-band aperture photometry measurements}\label{table:sed}
\input{Tables/photometry_SED/photometry_SED_optical.tex}
\end{sidewaystable*}

\setcounter{table}{\value{table}-1}
\begin{sidewaystable*}
\small
\caption{continued.}
\input{Tables/photometry_SED/photometry_SED_IR.tex}
\end{sidewaystable*}

\setcounter{table}{\value{table}-1}
\begin{sidewaystable*}
\small
\caption{continued.}
\input{Tables/photometry_SED/photometry_SED_FIR.tex}
\end{sidewaystable*}

\subsubsection{PANIC NIR observations}
Additional NIR imaging was also obtained with the Panoramic Near-Infrared Camera \citep[PANIC,][]{cardenas2018_PANIC} mounted to the 2.2m telescope at the Calar Alto observatory. Although PANIC consists of 4 Hawaii-2RG detectors only 2 were operational at the time of observations of which only one had nominal performance. 20 CARS targets were observed as part of programmes F17-2.2-014 and  H17-2.2-008 (PI: Bernd Husemann). For the first program only the best performing chip was used covering a FoV of 15\arcmin$\times$15\arcmin. For the second program a mosaic strategy was used to cover the nominal 30\arcmin$\times$30\arcmin with the two operational detectors. The sampling of PANIC is 0\arcsec45 per pixel. For the analysis in this paper we only use a small part of the images covering the AGN host galaxies so that the full FoV is not relevant. 

Total exposure times ranged from 1-- 2\,hours for the J, H, Ks bands, but not all NIR bands could be obtained for all the targets. A spiral dither pattern was used along all observations for background subtraction, excluding satellite tracks and image cosmetics. The PANIC exposures were reduced and combined with the dedicated PANIC pipeline\footnote{https://github.com/ppmim/PAPI}. The astrometric registration and absolute photometry was further optimized with the PhotometryPipline \citep[PP,][]{mommert2017_PP} software package using 2MASS as photometric reference.

\subsubsection{SCUBA-2 FIR observations}
SCUBA-2 is a 10\,000 pixel bolometer camera simultaneously operating at 450\,$\upmu$m and 850\,$\upmu$m and covering an area of 41\,arcmin$^2$ on the sky \citep{holland2013_SCUBA2}. Effective beam sizes are 10\arcsec\ and 15\arcsec\ at 450\,$\upmu$m and 850\,$\upmu$m, respectively.

Given the angular size of our AGN hosts on the sky, we observe using the `Constant Velocity (CV) Daisy' mapping mode which provides uniform sensitivity in the central 3\arcmin\ of the observation. The `CV Daisy' is a circular scanning pattern designed so that the target is always within the field-of-view of the array throughout the integration while moving at a constant 155\arcsec/s. The observation provides usable coverage out to $\sim$6.0\arcmin\ in radius, but beyond a radius of 1.5\arcmin\ the map sensitivity decreases rapidly. 

The observations were performed in service mode between February and December 2019 during grade 3 weather conditions ($0.08<\tau_{225}<0.12$) as part of project M19AP019 (PI: Timothy Davis). On-source exposure times ranged from 1\,h to 7.5\,h. The resulting maps were reduced using the Dynamic Iterative Map Maker (DIMM) within the sub-mm user reduction facility \citep[SMURF,][]{2013Chapin}. For a full overview of the procedure see \citet{2013Chapin}.

The data were reduced using the blank field reduction in order to detect point sources within the map. The maps were then calibrated using the standard flux calibration factors of FCF$_{450}$= 491\,Jy beam$^{-1}$pW$^{-1}$ and FCF$_{850}$= 537\,Jy beam$^{-1}$pW$^{-1}$ \citep{2013MNRAS.430.2534D}. To improve the point source detection, a matched filter was applied to the maps. This matched filtering caused a $\sim$10\% loss in flux (determined by inserting artificial sources into the map and comparing the flux before and after), so an additional 10\% was applied to the FCFs to account for this.

\begin{figure*}
\centering
  \includegraphics[width=.45\linewidth]{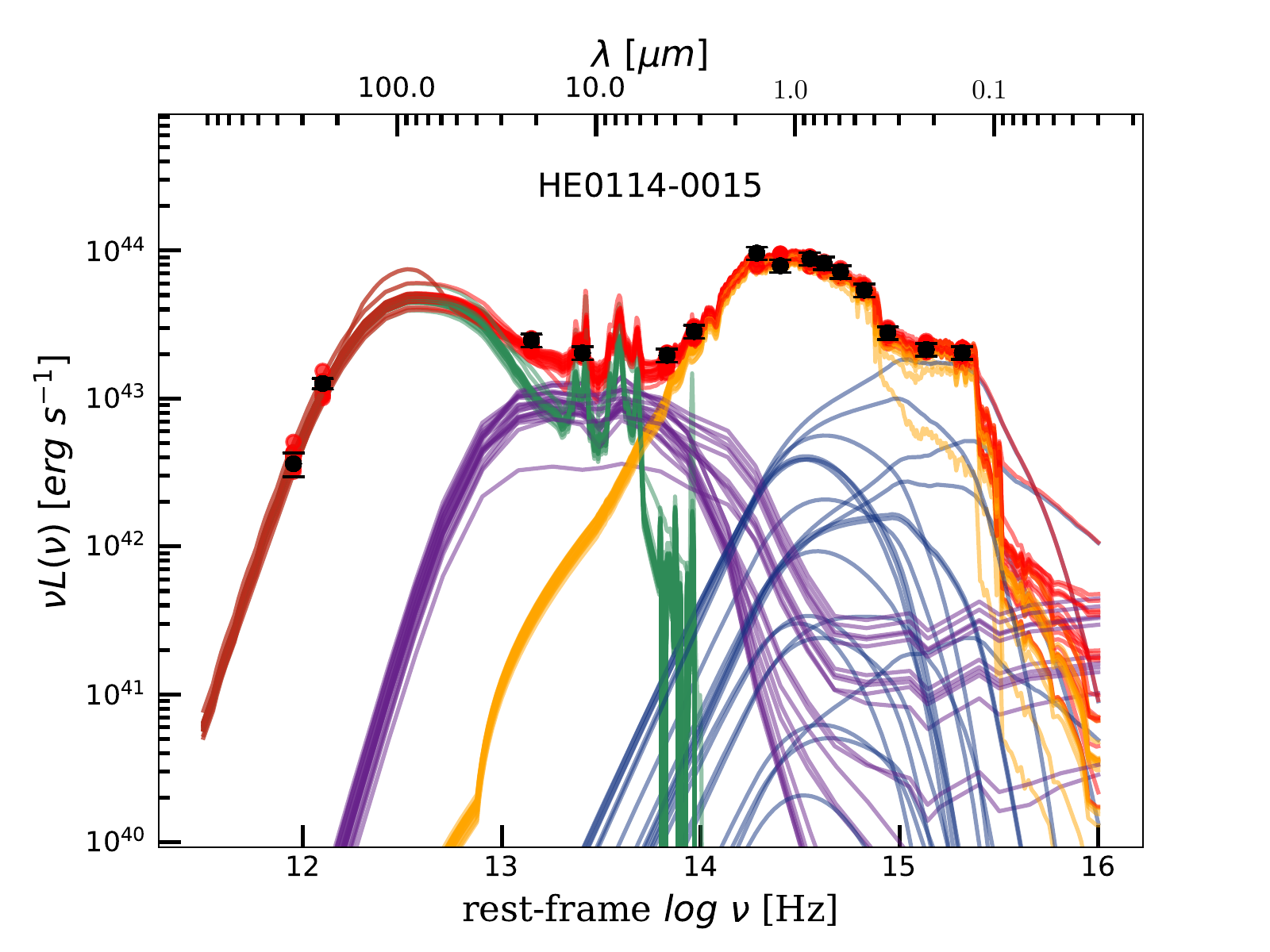}
  \includegraphics[width=.45\linewidth]{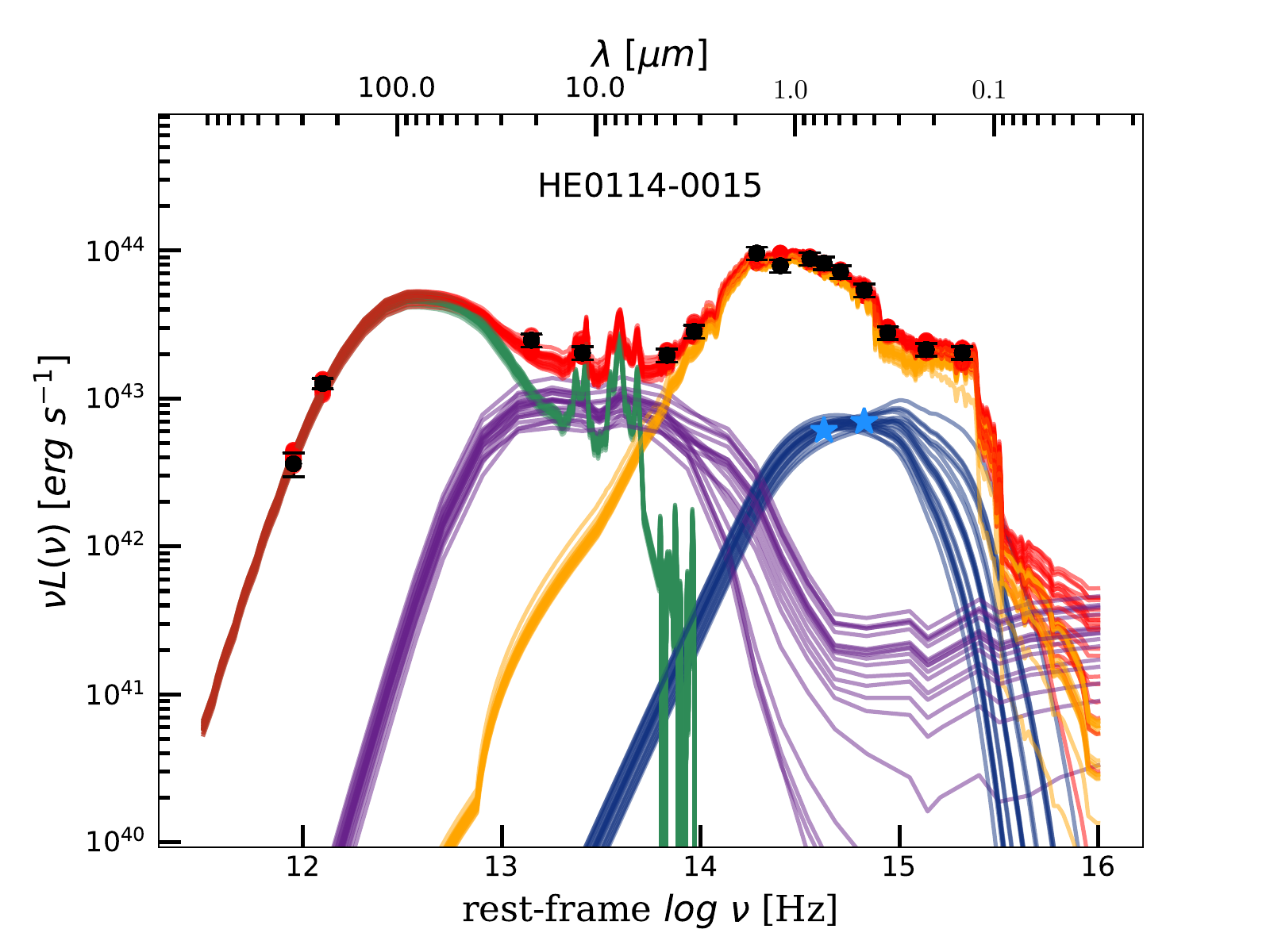}
    \caption{SED for HE\,0114$-$0015 and the best-fit model determined by \texttt{AGNfitter} \citep{agnfitter}. 20 MCMC realizations of the SED model are shown for the broadband photometric data (black points with error bars) where the red points are the corresponding predictions from the model. The red lines represent the total model, which consists of four components: the cold and warm dust in star-forming regions (green lines), the torus of AGN-heated dust (purple lines), the stellar continuum (yellow lines), and the AGN accretion disk (blue lines). 
    On the right panel is an additional constraint to the accretion disc model (e.g. \textit{g} and \textit{i} AGN photometry is included).}
    \label{fig:sed_example}
\end{figure*}

\subsubsection{SOFIA HAWC+ FIR observations}
The HAWC+ FIR camera \citep[][]{harper2018_HAWC} aboard \textit{SOFIA} was used to obtain FIR photometry for CARS targets that could not be detected with all-sky \textit{AKARI} FIR survey and that were not targeted with \textit{Herschel}. Observations were taken for 8 out of 11 proposed targets as part of a SOFIA survey program (Plan ID: 07\_193, PI: Bernd Husemann) with exposure times ranging from 15\,min to 60\,min. HE\,0227$-$0913 was observed in the $E$ band  (214\,$\mu$m) while all other targets were observed in the $D$ band (154\,$\mu$m). The FWHM of the HAWC+ beam is 18\farcs2 and 13\farcs5 for the $E$ and $D$ band, respectively. All data were reduced by the HAWC+ SOFIA team using a dedicated instrument pipeline. The pipeline was run specifically with the faint object flag to account for the faint signal in our targets.

\subsection{Integral-field spectroscopy observations}
IFU observations for the CARS sample were mainly obtained with the MUSE instrument at the Very Large Telescope (VLT) \citep{bacon2010_muse} for 37 targets under programs 094.B-0345(A), 095.B-0015(A), 099.B-0242(A), and 099.B-0249(B), with the VIMOS instrument at the VLT \citep{LeFevre:2003} for 2 targets under program 083.B-0801(A), and with the PMAS instrument at the 3.5m Calar Alto telescope \citep{roth2005_PMAS} for 2 targets under program H18-3.5-010. MUSE covers a $1\arcmin\times1\arcmin$ FoV at a sampling of $0\farcs2$, VIMOS covers a $27\arcsec\times27\arcsec$ FoV at a sampling of $0\farcs67$ and PMAS covers a $16\arcsec\times16\arcsec$ FoV with a sampling of 1\arcsec. All details of the IFU observations and their data reduction are presented in Husemann et al. (2021) to which the interested reader is referred to. 

For the present paper, we are working with emission line maps that were created after the AGN point-like emission was removed from the reconstructed IFU datacube as described in Husemann et al. (2021). Emission-line fluxes are extracted by fitting Gaussian profiles after modelling and subtracting the stellar continuum with \textsc{PyParadise} \citep[see][Husemann et al. 2021, for details]{walcher2015_pyparadise,weaver2018_pyparadise} from the AGN-subtracted data. All the details of this process and its application to the CARS IFU data are presented in Husemann et al. (2021). For the analysis presented in this paper we use the 2D emission line maps of [\ion{O}{iii}] $\lambda5007$, H$\upbeta$, [\ion{N}{ii}] $\lambda6583$, H$\upalpha$, [\ion{S}{ii}] $\lambda\lambda$6716,6731 and their associated errors. All those data are also accessible from the CARS data release 1 at \url{http://cars.aip.de}.

\section{Analysis}
\subsection{Modelling the panchromatic spectral energy distribution}
\label{SEDmodel}
In order to infer primary physical parameters such as the stellar mass $M_\star$ or the SFR, we need to model the panchromatic SED constructed for all the sources. We decided to perform the SED fitting with the publicly available \texttt{AGNfitter} package \citep{agnfitter}. \texttt{AGNfitter} models the observed photometry as a super-position of various template libraries for the stellar continuum, the AGN accretion disc emission, AGN-heated dust emission from the torus and the cold dust emission excited by star formation. 

The stellar library is taken from \citet{Bruzual2003} and parametrized by a stellar age of the galaxy and an exponential decay time of the starburst $\tau$ which can be further modified by reddening and normalized in absolute flux. The accretion disc emission is represented by an empirical composite spectrum generated from 259 type-1 AGN from the Sloan Digital Sky Survey \citep[SDSS,][]{York2000}, which can be reddened using the Small Magellanic Cloud reddening law of \citet{Prevot1984}. This leaves the flux normalization and the reddening as the only two free parameters for this component. The hot dust component of the AGN is reconstructed from the empirical SED library collected by \citet{Silva2004} based on photometric observations of AGN that were modelled with the radiative transfer code GRASIL \citet{Silva1998} to obtain the full NIR SEDs. The library is divided in a range of absorbing neutral hydrogen column densities $N_H$ through the torus along our line-of-sight as the main free parameter in addition to the normalization. Semi-empirical starburst template spectra from \citet{CharyElbaz2001} and \citet{DaleHalou2002} are used to model the cold dust component from star formation which are denoted by a luminosity parameter and can be normalized. 

\begin{table*}
\caption{AGN host galaxy properties inferred from SED modelling}\label{table:host_parameter}
\centering
\input{Tables/host_properties/host_properties.tex}
\end{table*}

\texttt{AGNfitter} uses the \texttt{emcee} package \citep{emcee} for an affine-invariant Markov-Chain Monte Carlo (MCMC) ensemble sampling to explore the 10-dimensional parameter space associated with the underlying template library of spectra. The advantage of \texttt{AGNfitter} is that it natively accepts upper limits in the photometry, allows to set priors on parameters and additional constraints can be incorporated easily to break degeneracies. Particularly, one big issue in the SED modelling of luminous type 1 AGN is that the accretion disc model is competing with the stellar population model in the UV--optical rest-frame wavelength regime. This leads to a degeneracy between the accretion disc and stellar continuum model when fitting the observed integrated broad-band photometry as shown in Fig.~\ref{fig:sed_example} (left panel). 

This degeneracy could be solved by linking the UV radiation field with the corresponding re-emission in the IR as an additional constraint implemented for example in the \texttt{CIGALE} or \texttt{MAGPHYS} SED fitting codes. However, this method requires an assumption of the UV escape fractions and dust content of the galaxies, which can be very different for starburst galaxies and AGN-dominated galaxies. Due to the low redshift of the CARS sample, it is possible to spatially separate the AGN and host galaxy light using 2D image synthesis modelling even in ground-based images. This allows us to obtain direct measurements for the AGN brightness in the optical bands. The AGN magnitudes in the \textit{g} and \textit{i} bands are obtained by modelling the ground-based images as a super-position of two Sersic profiles and a point source for the AGN with the \texttt{galfit} package \citep{Peng2002,peng2010} as described in Husemann et al. (2021). Hence, we adjusted the likelihood function within the \texttt{AGNfitter} code to consider these additional photometric data as constraints solely for the AGN accretion disc component model. The effectiveness of this approach is shown in Fig.~\ref{fig:sed_example} in comparison to the modelling without those additional AGN constraints.

With the additional photometric constraints  we reduced the uncertainties of the various physical quantities that are directly computed by \texttt{AGNfitter} from the posterior parameter distribution. For our work we are most interested in the stellar mass $M_\star$ that are based on  the stellar population model and its associated luminosity normalization. In addition we report the integrated luminosity of the torus model ($L_\mathrm{tor}$), and the luminosities of the cold dust component integrated between entire IR 8--1000\,$\mu\mathrm{m}$
 ($L_\mathrm{8-1000\,\mu\mathrm{m}}$) wavelength range and a restricted range of 42.5--122.5\,$\mu\mathrm{m}$  ($L_\mathrm{42.5-122.5}\,\mu\mathrm{m}$) for literature comparisons. \texttt{AGNfitter} estimates a SFR from the integrated IR luminosity using the calibration established by \citet{murphy2011}:
\begin{equation}
 \left(\frac{\mathrm{SFR}_\mathrm{IR}}{[\mathrm{M}_\sun\,\mathrm{yr}^{-1}]}\right) = 3.88\times10^{-44}\left(\frac{L_\mathrm{8-1000\,\mu\mathrm{m}}}{[\mathrm{erg\,s}^{-1}]}\right)\label{eq:SFR_IR}
\end{equation}
In cases where the cold dust SED is not well constrained due to upper limits in the FIR photometry, we determine $5\sigma$ upper limits from the posterior distribution function of the integrated luminosities and SFR. We list all the inferred parameters from SED fitting as described above in Table~\ref{table:host_parameter} and all the corresponding SED models with \texttt{AGNfitter} are shown in Fig.~\ref{fig:SED_apx} of the appendix.

\subsection{Measuring H$\upalpha$-based star formation rates}
\label{halphasfr}
The IFU observations for the CARS sample as presented in Husemann et al. (2021) are well suited to measure the current star formation rate of the host galaxies in comparison to the FIR-based SFR. Various calibrations have been determined to convert the extinction-corrected H$\upalpha$ luminosity into a SFR from which we use the calibration of \citep{calzetti2007}
\begin{equation}
 \left(\frac{\mathrm{SFR}_{\mathrm{H}\upalpha}}{[M_\sun\,\mathrm{yr}^{-1}]}\right) = 5.3 \times 10^{-42} \left(\frac{\mathrm{L}_{\mathrm{H}\upalpha}}{[\mathrm{erg\,s}^{-1}]}\right)\label{eq:SFR_Ha}
\end{equation}
in this work to estimate the integrated SFR based on the integrated H$\upalpha$. However, not all of the H$\upalpha$ emission is associated with star-forming \ion{H}{ii} regions in AGN host galaxies. A detailed spatially-resolved emission-line diagnostic analysis is therefore essential for AGN host galaxies to separate the contribution from AGN photoionization and star-forming $\ion{H}{II}$ regions to the excitation of H$\upalpha$.

\begin{figure*}
    \centering
        \includegraphics[width=1.\linewidth]{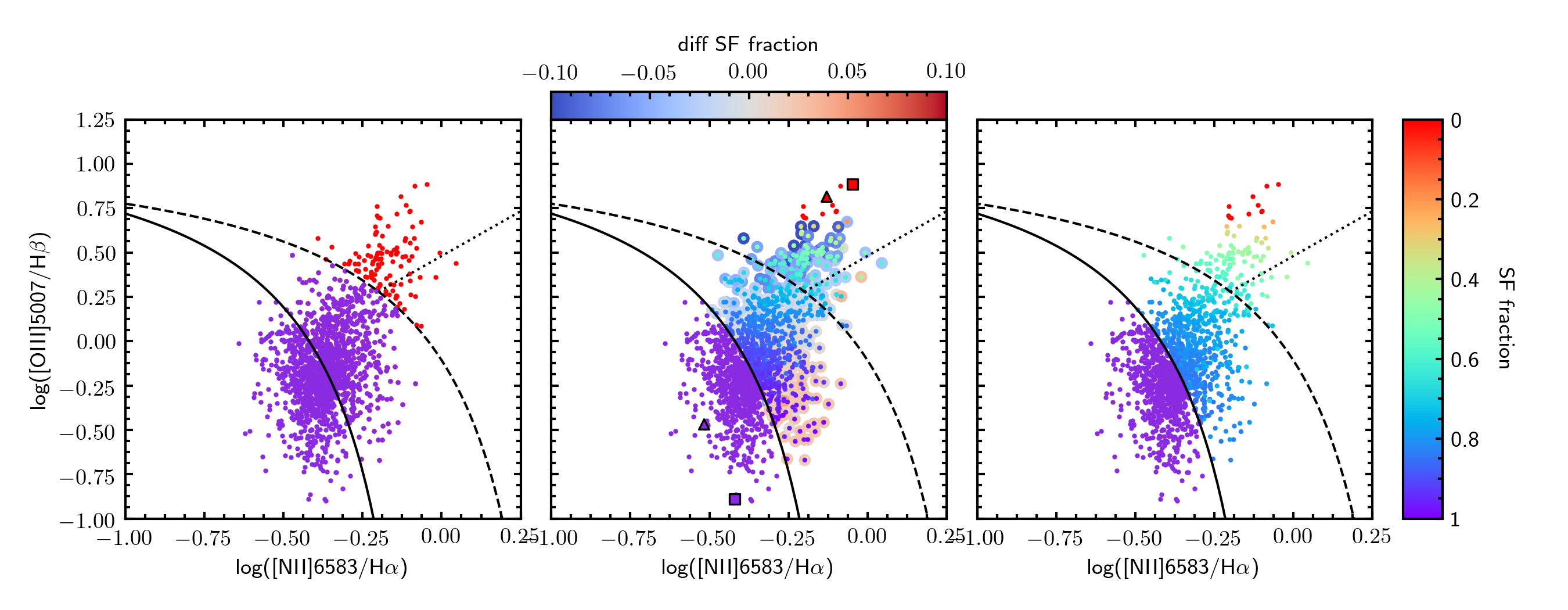}
    \caption{BPT diagram of HE\,0853$+$0102 as an example for a mixing sequence between \ion{H}{ii} regions and AGN photoionization across the host galaxy. We highlight three ways of determining the SF fraction from the  BPT diagram. \textit{Left panel:} A simple demarcation line approach where everything above the \citet{kewley2001_bpt} line is classified as pure AGN and below as pure star formation. \textit{Middle panel:} Simple linear combination with a $\chi^2$ fit using a fixed line vector as a reference point for the characteristic AGN and \ion{H}{ii} line ratios. The colored halos around the points represent the difference between the calculated SF fraction adopting  two different different reference pairs (triangle and square symbols) to highlight the systematic uncertainties introduced by a certain choice. \textit{Right panel:} MCMC fitting results  of the star formation fraction adopting a large cloud of potential reference points for the AGN (red points) and \ion{H}{ii} region (magenta points) which are explored during the MCMC sampling. }
    \label{fig:bpt_ms}
\end{figure*}

In order to provide reliable H$\upalpha$-based SFR estimates, we use a combination of line maps from pixel-by-pixel and binned IFU cubes after subtracting the AGN point-source emission as described in Husemann et al. (2021) in more detail. While the pixel-by-pixel maps provide high-spatial resolution information for bright emission line regions, the binned maps provide higher S/N information on the more extended ionised gas. We only consider the diagnostic lines H$\upalpha$, H$\upbeta$, [$\ion{N}{II}$] and [$\ion{O}{III}$] as detected above a 3$\upsigma$ level. For the maps we also include the weaker spaxels where only H$\upalpha$ S/N is required to be more than 3.

Here, we consider several ways to correct the H$\upalpha$ flux by the AGN contribution as described in more detail in the following subsections. All those methods leverage the potential of the IFU data to perform flexible spatially-resolved emission-line diagnostics that open possibilities for de-contamination compared to traditional narrow-band imaging or long-slit spectroscopy.

\subsubsection{Demarcation lines}
The simplest approach to correct for the AGN affected H$\upalpha$ emission is to ignore those spaxels displaying AGN ionization and integrate only the H$\upalpha$ flux from regions clearly associated with \ion{H}{ii} regions. Specific BPT demarcation lines are often used as hard boundaries to distinguish between different ISM ionization processes in galaxies even though physically (due to line-of-sight blending and other physical processes) those demarcations cannot be hard boundaries. 

The theoretical "maximum starburst" demarcation line proposed by \citet{kewley2001_bpt} divides the region where ionization could still be explained by pure star-formation from the region where AGN ionization is required to contribute to the emission. Similarly, the empirical `pure star-forming' BPT demarcation line proposed by \citet{kauffmann2003_bpt} defines a part in the BPT assumed to be powered purely by \ion{H}{ii} regions whereas other mechanisms can play a role in the remaining part. The gap in between the stricter "pure star-forming" and the "maximum starburst" demarcation lines is often referred to as the mixed or composite BPT region.

The AGN contamination to H$\upalpha$ across the BPT and, in particular, through the composite region can be tackled in different ways. \citet{nascimento2019_manga} considers the pure star-forming region and the composite region to be entirely associated with star formation assuming that the AGN contribution in the composite region compensates for the `hidden' star formation. \citet{wylezalek2018_manga} assigns a fixed fraction of 80\% of AGN ionization contribution to H$\upalpha$ in the BPT region above the "maximum starburst'' demarcation line and only a 20\% contribution to the composite region. We use the former approach (illustrated in Fig.~\ref{fig:bpt_ms}, left panel) to be able to consistently compare our results (Table~\ref{tab:ionized_gas}) with previous works and the more sophisticated methods as described below.

\subsubsection{BPT mixing sequence}
In contrast to the fixed contribution assumed before, there is likely a smooth transition of ionization contributions across galaxies. AGN host galaxies therefore often show a mixing sequence which appears as an elongated structure in the BPT diagram. This mixing sequence (MS) usually exhibits a prominent dependency on the distance from the AGN. As proposed by \citet{davies2014_II, davies2014_I, davies2016, davies2017}, one can pick basis vectors of emission line ratios to characterize pure AGN and star-forming ionization and treat the composite data points of the MS as a linear combination of the basis vectors:
\begin{equation}
    % \left\{
    % \begin{split}
        \left(
        \begin{matrix} [\ion{O}{III}] / \mathrm{H}\upbeta \\ \\ [\ion{N}{II}] / \mathrm{H}\upalpha \\ \\ [\ion{S}{II}] / \mathrm{H}\upalpha \end{matrix}
        \right)_\mathrm{MS} 
        = f_\mathrm{AGN} \times 
        \left(
        \begin{matrix} [\ion{O}{III}] / \mathrm{H}\upbeta \\ \\ [\ion{N}{II}] / \mathrm{H}\upalpha \\ \\ [\ion{S}{II}] / \mathrm{H}\upalpha \end{matrix}
        \right)_\mathrm {AGN}
        + f_\mathrm{SF} \times 
        \left(
        \begin{matrix} [\ion{O}{III}] / \mathrm{H}\upbeta \\ \\ [\ion{N}{II}] / \mathrm{H}\upalpha \\ \\ [\ion{S}{II}] / \mathrm{H}\upalpha \end{matrix}
        \right)_\mathrm{SF} \;. 
\label{eq:model}
\end{equation}
Here, the parameters $f_\mathrm{AGN}$ and $f_\mathrm{SF}$ represent the non-negative linear coefficient for the AGN and star-forming ionization fractions, correspondingly.  An important  additional basic constraint of $f_\mathrm{AGN} + f_\mathrm{SF} = 1$ ensures that flux is preserved. In principle more excitation mechanisms could be added to this equation, but this will lead to degeneracies in the basis vector as they are not fully orthogonal.

The emission line columns are normalized to either H$\upalpha$ or H$\upbeta$, which minimizes the impact of internal extinction within the host on the results. Here, we include the $[\ion{S}{II}] / $H$\alpha$ line ratio from the other classical BPT diagram and more line ratios could in principle be considered if S/N ratios are high enough to be informative. 

The idea behind the method is to pick emission-line basis vectors at the extreme ends of the mixing sequence, which should best reflect the assumption of two completely independent excitation mechanisms. The process of picking such emission-line ratio basis vectors allows considerable freedom and uncertainty. For example, the selected pair of basis vectors may not fully capture the different ISM conditions across the galaxies such as metallicity or ionization parameter. In the Fig.~\ref{fig:bpt_ms}, middle panel, the example of two different bases pairs - triangle and square symbols - leads to a 10\% difference in the resulting SF fraction.

For  more meaningful  results, the uncertainties caused by the choice of the basis vector need to be estimated and included into the final SFR uncertainty. With Monte Carlo Markov Chain (MCMC) algorithms it is possible to use the Bayesian approach and also take into account the uncertainty of the underlying model. So, our solution to the basis-picking problem is to define a large number of basis ratio sets and let the MCMC sampling to choose basis vectors for a certain spaxel and, more importantly, take the uncertainty of this choice into account. We use the MCMC python package \textsc{emcee} to develop a python package called \texttt{rainbow}\footnote{Publicly available open-source at \url{https://gitlab.com/SPIrina/rainbow}} designed to handle the mixing sequence fitting. 
For each given spaxel it tries various combinations of basis vectors and determines the posterior probability distribution of the parameters through maximising the likelihood function: 
\begin{equation}
    \begin{matrix}
    \ln p(\boldsymbol{y}|\boldsymbol{x}_\mathrm{AGN},\boldsymbol{x}_\mathrm{SF}, f_\mathrm{SF}, \boldsymbol{\sigma}_y, \mathscr{F}) = -\frac{1}{2} \sum \left[ \frac{(\boldsymbol{y} - \boldsymbol{y}_\mathrm{model})^2}{\boldsymbol{\sigma}^2} + \ln(2 \pi \boldsymbol{\sigma}^2) \right], \\ \\
    \boldsymbol{\sigma}^2 = \boldsymbol{\sigma}_y^2 + \boldsymbol{y}_\mathrm{model}^2 \times \mathscr{F}^2;
    \end{matrix}
\end{equation} 
where $\boldsymbol{y}$ represents the vector of emission line ratios of the fitted spaxel, $\boldsymbol{\sigma}_y$ is the corresponding error vector, $\boldsymbol{x}_\mathrm{AGN}$ and $\boldsymbol{x}_\mathrm{SF}$ represent basis vectors, and the model is $\boldsymbol{y}_\mathrm{model}(\boldsymbol{x}_\mathrm{AGN},\boldsymbol{x}_\mathrm{SF},f_\mathrm{SF}) = (1-f_\mathrm{SF}) \boldsymbol{x}_\mathrm{AGN} + f_\mathrm{SF} \boldsymbol{x}_\mathrm{SF}$ such as the vector equation~\ref{eq:model}.
The likelihood function here is a Gaussian where the variance is underestimated by the fractional amount parameter $\mathscr{F}$, that represents an uncertainty of the model.
The star-forming ionization fraction $f_\mathrm {SF}$ and its uncertainty then is inferred from the probability distribution.
An example of the MCMC fitting result from \texttt{rainbow} is shown in the right panel of Fig.~\ref{fig:bpt_ms}.

% bases picking criteria
Our criteria of picking SF emission-line ratio basis vectors from the mixing sequence are:
\begin{enumerate}
    \item SF bases should be below the "pure star-forming line" \citep{kauffmann2003_bpt};
    \item the signal-to-noise ratio of all the involved lines should be 3$\upsigma$ or more ($> 3\upsigma$ threshold is used to reduce the number of the basis vectors which speeds-up the fitting process);
    \item a certain radius may be specified to bound the basis vectors closer to the center and represent a narrower range of metallicities and ionization parameters.
\end{enumerate}
The criteria of picking AGN (or shock/evolved star, as described in section~\ref{sec:bpt_morph}) photoionization basis vector from the mixing sequence are the following:
\begin{enumerate}
    \item $[\ion{O}{III}] / \mathrm{H}\beta$ is above a certain threshold, this threshold cannot be same for all galaxies, as it depends on the strength of the AGN, and, therefore, is set to a value between 0.5 --- 0.8 (0.25 --- 0.5 for LINERs) depending on a galaxy, so that only the top spaxels are selected;
    \item a certain distance from the AGN may be specified to separate the AGN basis in the central spatial region and extended AGN cloud.
\end{enumerate}
The adopted selection will be shown on a galaxy-by-galaxy case in Fig.~\ref{fig:bpt_morphology_objects} for the entire sample in the appendix.

\begin{figure*}
    \centering
       \includegraphics[width=1.\linewidth]{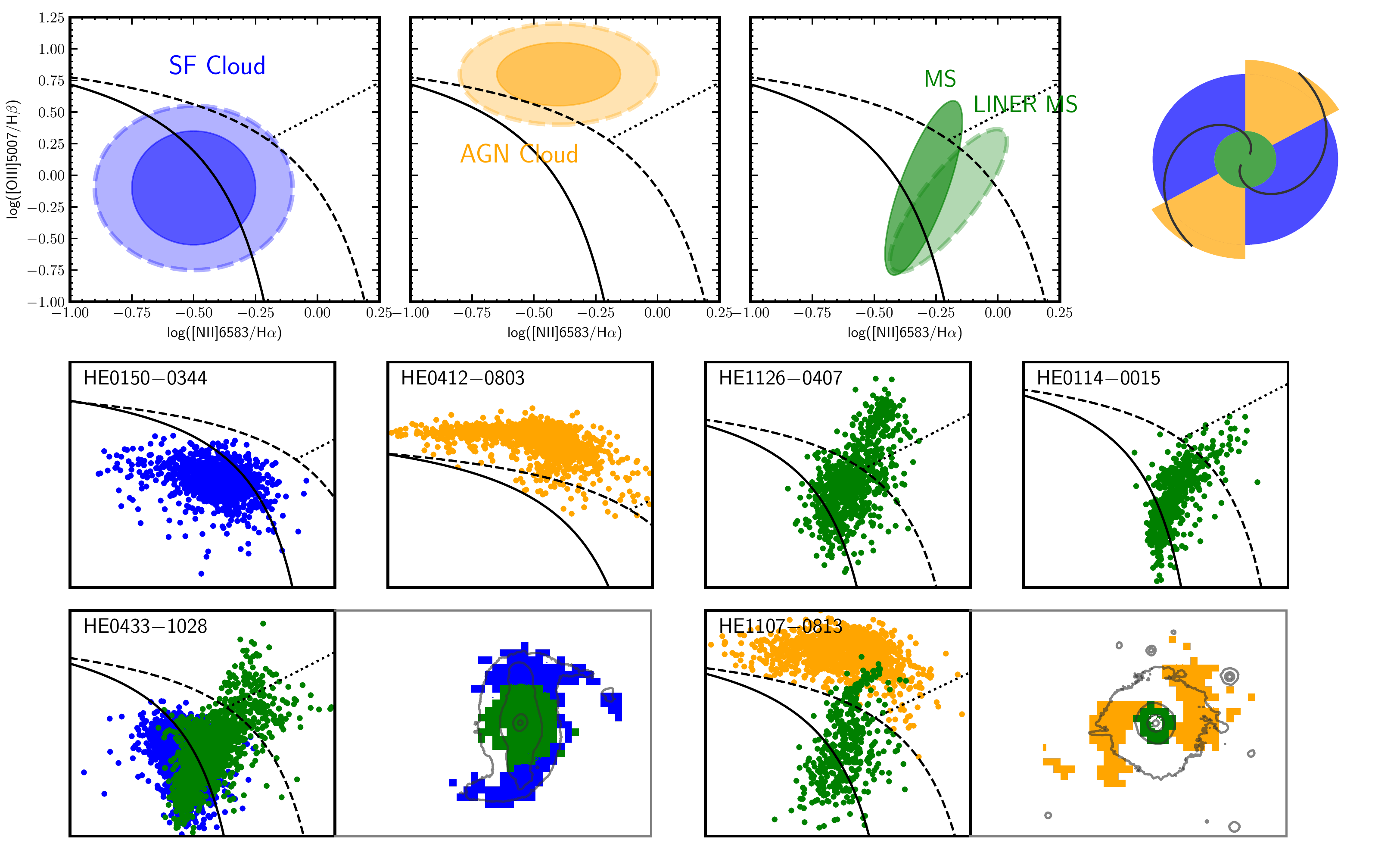}
    \caption{Schematic sketch of the BPT morphology populations. \textit{Upper row:} On the right there is a representation of a galaxy with a central region in green, the rest of the galactic body in blue, and an AGN ionization cone in yellow. These regions populate specific areas on the BPT diagram: \textit{left diagram:} star-forming cloud with two different sizes; \textit{middle diagram:} AGN-ionised cloud with two sizes; \textit{right diagram:} mixing sequence elongated towards the AGN area and a mixing sequence elongated towards the LINER area. \textit{Middle row:} BPT diagram schematic examples for the SF cloud, AGN cloud, mixing sequence and LINER mixing sequence dominated galaxies. \textit{Lower row:} BPT diagram and spatial map examples for SF cloud and AGN cloud combined with a mixing sequence in one galaxy.}
    \label{fig:bpt_morphology}
\end{figure*}

\subsubsection{BPT morphology}
\label{sec:bpt_morph}
The idea of treating the mixing sequence as a signature of physical mixing of $\ion{H}{II}$ regions and AGN ionised regions suggests that different populations of spaxels on a BPT diagram can be differentiated based on their properties. We name this approach \textit{BPT morphology}, as the idea is to distinguish different shapes on the diagram (see Fig.~\ref{fig:bpt_morphology}):\\
1) The \textit{star-forming cloud} is an extended structure located under the "maximum starburst line" of \citet{kewley2001_bpt} on a BPT diagram. This emission arises from across the star forming body of a galaxy and, therefore, contains emission from regions with different metallicities, densities, and ionization parameters. The variance of these ISM parameters results in the extensive shape that covers large area in the BPT diagram. Although it spreads out to the composite area between the "pure star-forming" and the "maximum starburst" demarcation lines on the BPT diagram, we assume that there is no need to clean this group of spaxels from the AGN contamination. See HE~0150$-$0344 (Fig.~\ref{fig:bpt_morphology} middle row, left) as an example of a galaxy without AGN but with the spaxels spreading to the composite BPT area.\\
2) The \textit{AGN-ionised cloud} is a structure which spans the upper region of a BPT diagram and spreads to the left side towards lower [\ion{N}{ii}]/H$\upalpha$ line ratios. This corresponds to the extended narrow-line region or AGN-illuminated gas. Spatially, this region can reach from the AGN across the entire galaxy and even outside of the galaxy's main body. This group of spaxels does not contribute to the galaxy's SFR, although we can infer the upper limits considering hidden star formation as highlighted below.\\
3) The \textit{mixing sequence} is already described in the section above; it is an elongated structure on a BPT diagram which tends to be located in the central region of a galaxy. 
However, not all mixing sequences end in the AGN ionised area of the BPT diagram, where some are extending from the $\ion{H}{II}$ regions area to the LINER area likely due to different excitation mechanisms for the LINER emission. Nevertheless, the LINER mixing sequence can be treated in the same way as the AGN mixing sequence, following the same assumption of the two main excitation sources --- star formation and LINER excitation. The model (Eq.~\ref{eq:model}) will then have $f_{\mathrm{LINER}}$ instead of $f_{\mathrm{AGN}}$ with the same flux preservation constrain $f_\mathrm{LINER} + f_\mathrm{SF} = 1$.

An AGN host galaxy can have one of the described populations or even a combination of those (see the galaxy-by-galaxy cases in  Appendix~\ref{app:bpt_morphology}). 
We separate star-forming clouds and AGN-ionised clouds with a simple yet efficient method: manually introducing fixed spatial radii for each individual galaxy (listed on the Table~\ref{tab:spatial_radii} in Appendix~\ref{app:bpt_morphology}) and separating the regions with the circles as on the bottom examples of Fig.~\ref{fig:bpt_morphology} and Fig.~\ref{fig:bpt_morphology_objects}.
After separating different populations we assume that the SF cloud and AGN-ionised cloud have 100\% and 0\% contribution to the star-formation rate, correspondingly. The mixing sequence is modelled, out to a certain radius, with \textsc{rainbow} as described in the previous section to compute SF fractions along the sequence.\\
However, we attempt to correct all populations for the AGN contribution (similarly as with the mixing sequence). Given that some small contribution to the H$\alpha$ emission from star formation may be hidden in galaxies which display only an AGN cloud morphology, we estimate an upper limit for the  SF contribution adopting AGN bases or SF bases from the other galaxies with a prominent mixing sequence and apply our \texttt{rainbow} analysis.

\begin{table*}
\caption{H$\alpha$-based SFRs and metallicity estimated from the IFU emission lines}\label{tab:ionized_gas}
\small
\centering
\input{Tables/gas_properties/gas_properties.tex}
\end{table*}

\subsubsection{Gas-phase metallicity}
Another important characteristic of the ionised gas is the metallicity. The gas-phase metallicity tracks the immediate enrichment history of the ISM due to the evolution of stars and their metal yields across the galaxy. Undisturbed disc galaxies typically have negative metallicity gradients \citep{Sanchez:2014} and their central metallicity correlates with the stellar mass of a galaxy \citep{Tremonti:2004,Kewley:2008}. As stellar evolution is a long-term process, outflows and inflows of gas on short timescales can significantly affect the observed metallicity distribution and therefore can be used as a key diagnostic to understand the origin and motions of gas on galactic scales. For example a flattening and dilution of gas-phase metallicity have been observed during galaxy mergers and interactions \citep[e.g.][]{Ellison:2008,Kewley:2010,Thorp:2019}, and in barred galaxies \citep{Martin1994,Sanchez:2014} where low-metallicity gas from the outskirts is efficiently transported towards the centre. On the other hand, gas outflows can enrich the circum-galactic medium with metals from the galaxy center which has been observationally confirmed from absorption line studies \citep[e.g.][]{bordoloi2011,tumlinson2011,bouche2012,nielsen2015,schroetter2019}.

\begin{figure}
  \includegraphics[width=0.5\textwidth]{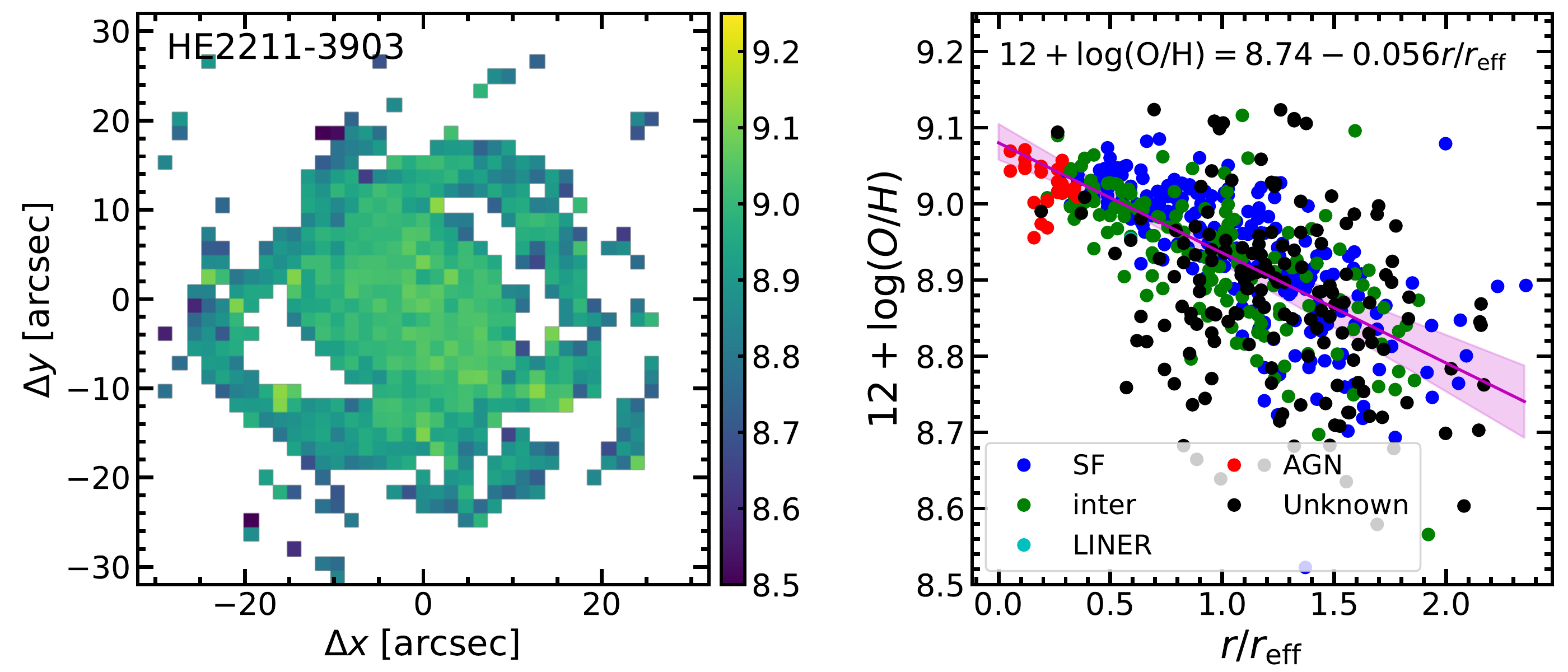}
  \caption{Example of the 2D and radial metallicity distribution for HE2211$-$3903. The oxygen abundance was measured using the N2S2 index which is almost insensitive to the excitation. On the left panel the full 2D distribution is shown and on the right panel a radial projection in units of the effective radius taken from the single Sersic model.  A linear fit is performed and shown as the red line.}\label{fig:metal_dist_example}
 \end{figure}

Measuring gas-phase metallicities across AGN host galaxies is more complicated because all strong-line metallicity diagnostics are calibrated for \ion{H}{ii} regions. Their application to regions photoionised by the AGN is invalid in the majority of cases, but specific calibrations have been developed \citep[e.g.][]{storchi-Bergmann1998}.  The metallicity measurements can in principle be restricted to \ion{H}{ii} regions ionised by star formation (as identified from BPT diagnostics) in AGN host galaxies  \citep[e.g.][]{Husemann:2014}, but this approach greatly limits suitable targets and radial coverage. The [\ion{N}{ii}]/H$\alpha$ (N2 index) line ratio is one of the prominent strong-line metallicity calibrators \citep{Pettini:2004,Marino:2013} for \ion{H}{ii} regions. Photoionization models show that the N2 index should also trace the metallicity in AGN photoionised regions \citep[e.g.][]{Groves:2006} with a different scale. That the situation is more complex was shown by \citet{Stern:2013} as they found a secondary dependence of the N2 index with AGN luminosity. At the same time they discovered that the [\ion{N}{ii}]/[\ion{S}{ii}] (N2S2 index) is well recovering the mass-metallicity relation of galaxies for AGN hosts independent of AGN luminosity. Indeed, the N2S2 index is generally a good metallicity calibrator \citep{Dopita:2016} also for SF ionised \ion{H}{ii} regions, but not as widely used given that the diagnostic lines are significantly fainter. 

In \citet{Husemann:2019a}, we established a N2S2 index calibration based on the SDSS galaxy sample, and used it to map the gas-phase metallicity across the CARS AGN galaxy HE~1353-1917. The metallicity pattern revealed that the ENLR of the edge-on galaxy follows a very similar radial metallicity than the star forming disc. This let us conclude that the diffuse extra-planar gas was expelled by SN-driven winds rather than an outflow from the central AGN. Here, we apply the N2S2 index calibration established in \citet{Husemann:2019a} for all CARS targets to infer the metallicity gradient and absolute scale of the central metallicity. In order to directly use the metallicity determined for SDSS based on \citet{Tremonti:2004} we determined an $\mathrm{N2S2}=\log([\ion{N}{ii}]\lambda\,6583/[\ion{S}{ii}]\lambda\lambda\,6716,6731$) calibration for this SDSS metallicity scale in the similar way as described in \citet{Husemann:2019a}:
\begin{equation}
 12+\log(\mathrm{O/H}) = 8.875+0.827\times\mathrm{N2S2}-0.288\times\mathrm{N2S2}^2\label{eq:N2S2}
\end{equation}

Based on this calibration we estimated the oxygen abundance across the galaxies and reconstruct the radial metallicity gradient, as shown for the face-on disc galaxy HE2211$-$3903 in Fig.~\ref{fig:metal_dist_example}. For this galaxy a negative radial metallicity gradient is clearly recovered with a slope of $\alpha_\mathrm{O/H} = -0.055$\,dex\,$r_\mathrm{eff}^{-1}$ with a zero-point central metallicity of $12+\log(\mathrm{O/H})=8.732$. As expected from the work of \citet{Stern:2013}, the N2S2 index indeed recovers a  matching absolute metallicites even in the case of strong AGN ionization without significant offset in metallicity as shown in Fig.~\ref{fig:metal_dist_example}.   Hence, the 2D map of the estimated oxygen abundance does not show any features related to the varying ionization conditions throughout its disc. While the [\ion{N}{ii}] and [\ion{S}{ii}] emission lines may not exactly originate from the same location within \ion{H}{ii} region, due to the ionization structure of the nebulae on 100\,pc scales \citep[e.g.][]{Sanders:2020,Mannucci:2021}, our calibration based on SDSS spectra (3\arcsec\ apertures) consistently include also part of the surrounding diffuse gas similar to our MUSE observation with typical resolutions of 0\farcs7-1\farcs0 covering 300--1000\,pc depending on redshift. Because we are unable to isolate individual \ion{H}{ii} regions with CARS, any calibration based on isolated \ion{H}{ii} would be invalid.

This methodology therefore allows us an initial characterization of the metallicity for the entire sample, as listed in Table~\ref{tab:ionized_gas} and maps shown in Fig.~\ref{fig:metals_apx}. Notably, we exclude a close star-forming companion in case of HE~0203$-$0031 and HE~1017$-$0305 from the radial fitting and fixed the center  to the AGN position in all cases. As the absolute metallicity scale is strongly dependent on the underlying calibration we also transform our oxygen abundance estimate from the original O3N2 calibration scale as used in \citep{Husemann:2019a} to the one of \citet{Tremonti:2004} so that we can more easily apply calibrations from the non-AGN SDSS sample as discussed below. Uncertainties are determined by bootstrapping; re-fitting the linear relation after randomly sub-selecting 80\% of the data points.  

\begin{figure*}
    \centering
        \includegraphics[width=1.\linewidth]{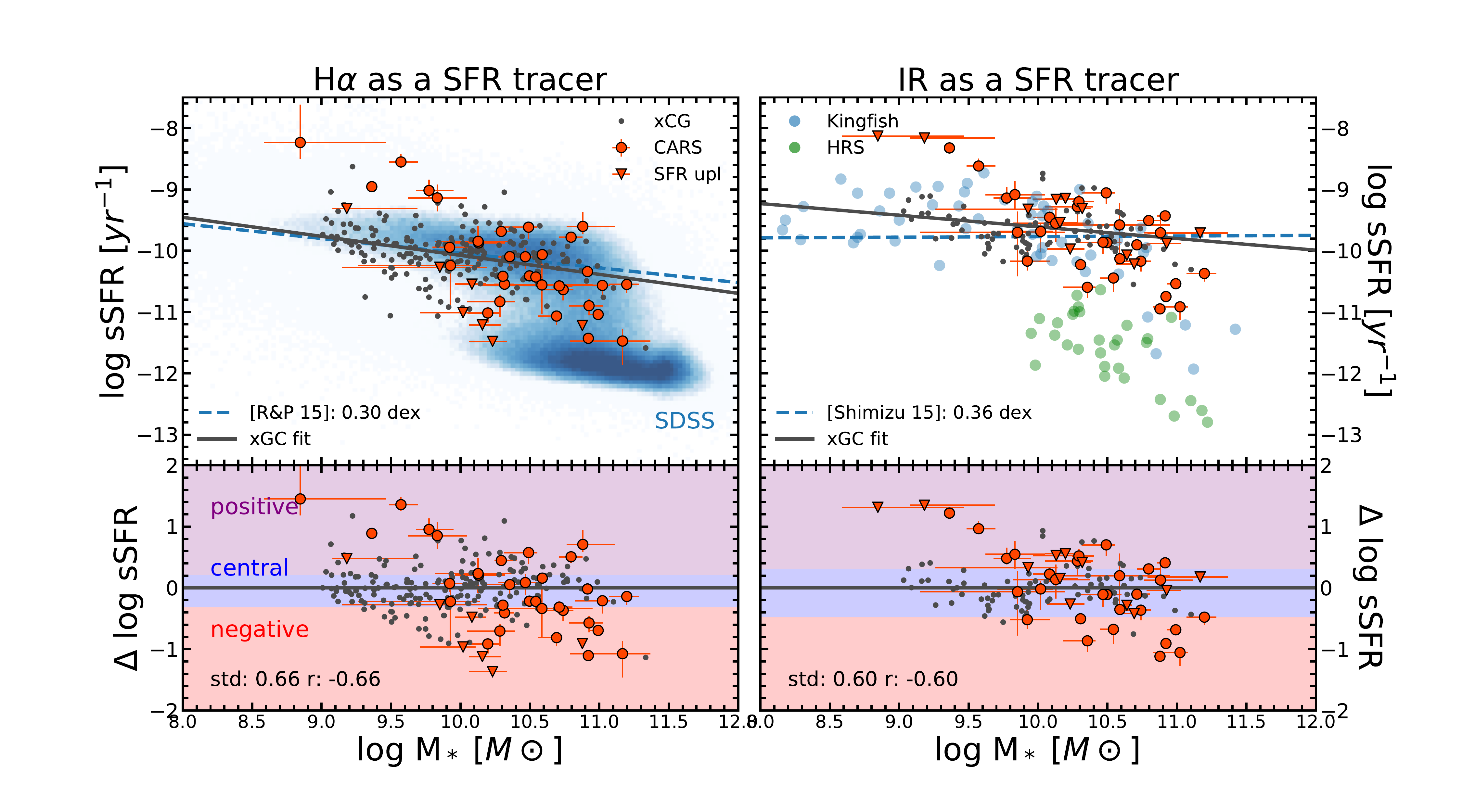}
    \caption{Comparison of the sSFR  against stellar mass for H$\alpha$- and IR-based SFR tracers. The CARS objects are shown as red points with error bars, while xCOLD GASS (xCG) are in small gray dots, KINGFISH are in blue and Herschel Reference Survey (HRS) in green. Specific SFR is calculated from H$\upalpha$ luminosity \textit{(left panel)} derived after BPT morphology analysis and \textsc{rainbow} fitting using the SFR formula $\mathrm{SFR}_{\mathrm{H}\upalpha} = 5.3 \cdot 10^{-42} \times \mathrm{L}_{\mathrm{H}\upalpha}$ \citep{calzetti2007}. IR specific SFR \textit{(right panel)} derived from the \textsc{AGNfitter} modelling using the \citet{murphy2011} SFR formula. The stellar mass is taken from the \textsc{AGNfitter} modeling parameters. The background on the left plot is the density map of SDSS galaxies \citep{brinchmann2004_sdssdr2}. Linear models to the star-forming main sequence are shown as black solid lines from \citet{renzini_peng2015_ms} and \citet{shimizu2015}, respectively. The best-fit linear relation from the xCG sample is shown as the blue dashed line in both cases. In the lower panel we show the residuals around the adopted star-forming main sequence where the colored bands (red, blue, and purple) emphasize three bins used further in Fig.~\ref{fig:bar_plot}}.
    \label{fig:sfr_mstar}
\end{figure*}

\section{Results and discussions}
\subsection{Does star formation rate depend on AGN parameters?}
In order to investigate the potential impact of AGN on the total SFR of their host galaxies we collect the derived CARS host galaxies parameters such as AGN-corrected integrated H$\upalpha$ and IR luminosity and associated SFRs together with the stellar mass from our analysis presented above. 
From the SFRs we compute the specific SFR (sSFR) by dividing with the stellar mass, i.e. $\mathrm{sSFR} = \mathrm{SFR}/M_\star$ and compare it against the stellar mass as shown in the Fig.~\ref{fig:sfr_mstar} for the H$\upalpha$ and IR-based SFRs, respectively.

We compare the CARS host galaxies with the non-AGN galaxy population at low redshifts using the data from SDSS \citep{brinchmann2004_sdssdr2, abazajian2009_sdssdr7}, xCOLD GASS \citep{saintonge2017}, KINGFISH \citep{skibba2011_kingfish}, and the Herschel Reference Survey \citep[HRS, ][]{smith2012_hrs, boselli2015_hrs}. Those reference samples highlight the position of the so-called star-forming main-sequence (SFMS) as determined by various authors and is indicated on Fig.~\ref{fig:sfr_mstar}, with blue dashed lines. These were derived from SDSS galaxies in the H$\upalpha$ diagram by \citet{renzini_peng2015_ms} and \cite{ shimizu2015} for the IR diagram. 

Those SFMS relations are derived from different data sets, and as our goal is to achieve consistency between the two diagrams, we consistently determine the SFMS from the xCOLD GASS (xCG) sample. The xCG is a subsample from the SDSS sample, therefore H$\upalpha$, IR, stellar mass, and also CO(1$-$0) and metallicity data are available from the literature sources listed above. As there is also a BPT classification of the objects, we exclude AGN and LINER galaxies to define a clean star-forming subsample of galaxies (hereafter \textit{training sample}) resulting in 197 objects for the H$\upalpha$ diagram and 86 objects for the IR diagram. The linear SFMS fits to the training sample data and the corresponding $R^2$ scores (defined below in equation~\ref{eq:r2}) are:

\begin{equation}
    \begin{matrix}
    \log L_{{\mathrm{H}}\upalpha} = (0.69 \pm 0.03) \log M_{\star} + (34.3 \pm 0.3); R^2 = 0.50 \\ \\
    \log L_\mathrm{IR} = (0.81 \pm 0.03) \log M_{\star} + (35.7 \pm 0.3); R^2 = 0.62
    \end{matrix}
\end{equation}
Our SFMS linear relations are consistent with the literature determinations shown in Fig.~\ref{fig:sfr_mstar} within the reported scatter.

In order to assess the differences between the CARS hosts and the training sample we compute the residuals along the SFMS relation (Fig.~\ref{fig:sfr_mstar} bottom panel). The CARS objects exhibit a scatter around the SFMS of $0.69$~dex based on H$\upalpha$ and $0.60$~dex based on IR SFR, respectively, ignoring the upper limits. Moreover, the residuals reveal a clear negative correlation with a Pearson correlation coefficient of $r=-0.69$ and $r=-0.61$ for the H$\upalpha$ and IR SFR, respectively.  This negative trend might be na{\"i}vely interpreted as a result of star-formation quenching where higher mass galaxies appear to be more passive potentially due to the effects of the AGN \citep[e.g.][]{cicone2014, saintonge2017, lacerda2020} considering our AGN selection.

However, the stellar mass is not the only parameter which controls the SFR of a galaxy. The cold gas content was shown to be another fundamental parameter linked to the SFR and explaining part of the scatter in SFR perpendicular to the SFMS \citep{tacconi2018, colombo2020, piotrowska2020, ellison2020}. Furthermore, the gas-phase metallicity is being discussed to be linked to the SFR  and stellar mass, the so-called fundamental mass-metallicity relation \citep{lequeux1979, mannucci2010, yates2012, curti2020}, but see also \citet{sanchez2017,sanchez2019}.  In particular, the metallicity could be related to the metallicity-dependent conversion of $L_\mathrm{CO}$ to the total cold gas mass \citep[e.g.][]{genzel2012, bolatto2013, carleton2017, utomo2017} or the general ability of the gas to cool to form star-forming gas clouds.

In order to quantitatively compare the previous model with the ones introduced further we use the coefficient of determination or the $R^2$ score \citep{draper1998_r2}. It provides an indication of how well the fit is for the given dataset with the best possible score of $1.0$. Unsuitable model can result in negative $R^2$ scores and an $R^2$ score of $0.0$ would result from a constant model that always predicts the expected value. For a dataset $y_1 ... y_n$ with fitted values $y_{1 \,\, \mathrm{model}} ... y_{n \,\, \mathrm{model}}$ and a mean value $\overline{y} = \frac{1}{n}\sum_{i=1}^{n} y_i$ the $R^2$ score is calculated as following:
\begin{equation}
    R^2 = 1 - \frac{\sum_i(y_i - y_{i \,\, \mathrm{model}})^2}{\sum_i(y_i - \overline{y})^2}.
    \label{eq:r2}
\end{equation}

\begin{figure*}
    \centering
        \includegraphics[width=1.\linewidth]{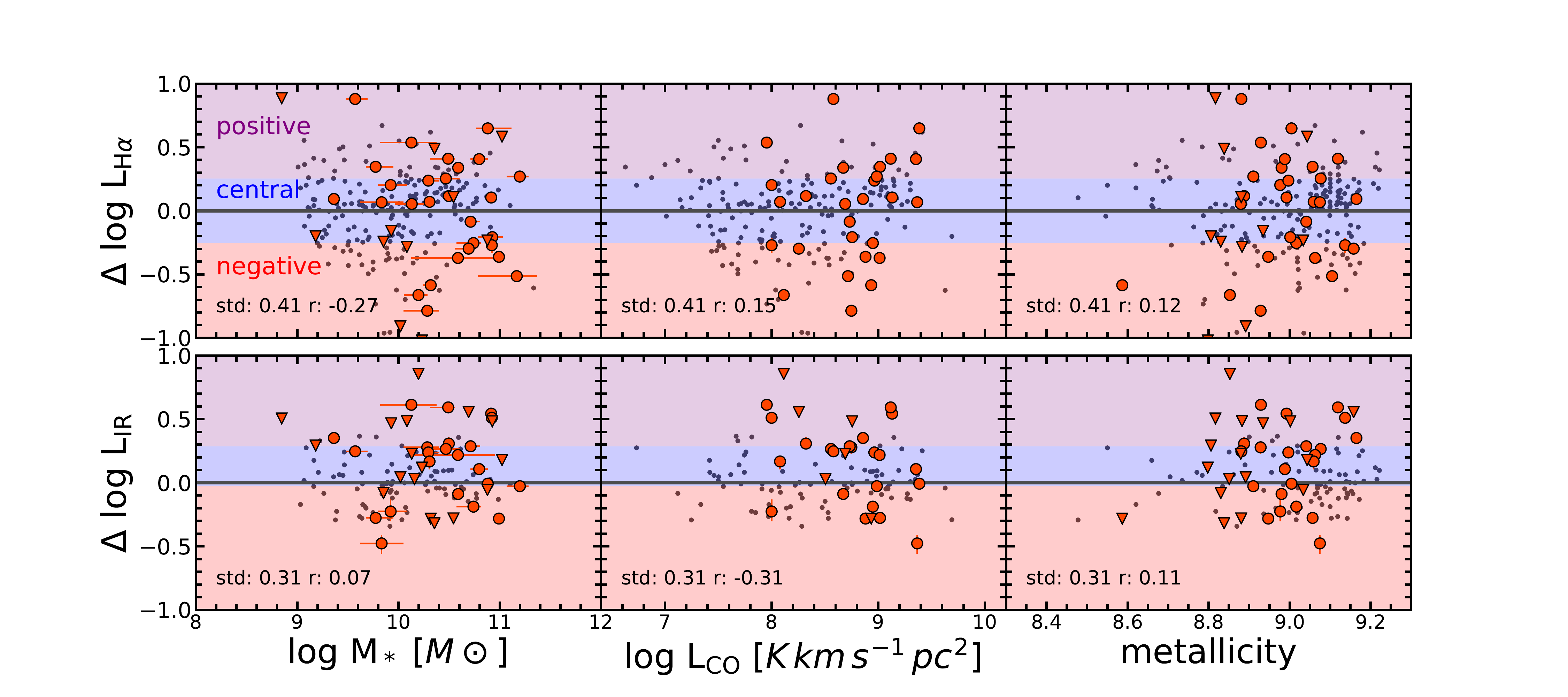}
    \caption{Difference between measured and expected H$\upalpha$ (upper panels) and FIR luminosity (lower panels) as a function of stellar mass, CO(1-0) luminosity and 12+$\log(\mathrm{O/H})$ ionised gas metallicity. Here we used our more complex model (Eq.~\ref{eq:SFR_IR} and  Eq.~\ref{eq:SFR_Ha}) to predict the SFR-sensitive luminosities. The black dots are the data from our xCG training sample and the red symbols are the CARS data with triangles indicating upper limits. The colored bands (red, blue and purple) define three bins in residual luminosity as discussed in the text and used for Fig.~\ref{fig:bar_plot}. }
    \label{fig:residuals}
\end{figure*}

As our model above does not take into account the cold gas content nor the gas-phase metallicity of the host galaxy, we expand our one-dimensional linear model into a multi-dimensional linear model for the SFR (hereafter \textit{SFMS+gas} model) with stellar mass ($M_\star$), CO(1-0) luminosity ($L_\mathrm{CO}$) and metallicity ($12+\log(\mathrm{O/H})$) as the independent parameters.  Here, we use the CO$(1-0)$ luminosity from \citet{saintonge2017} as the main cold gas mass proxy and the oxygen abundance $12+\log(\mathrm{O/H})$ determined from the ionised gas emission lines in the SDSS spectra as determined by \citet{Tremonti:2004} as the main gas-phase metallicity parameter.  This leads to the following formulas for the linear multi-dimensional model

\begin{align}
    \log(L_\mathrm{H\upalpha}/[\mathrm{erg\,s}^{-1}]) = & (-0.01 \pm 0.06) \log(M_{\star} / [M_\sun]) \nonumber \\
    & + (0.62 \pm 0.04) \log(L_\mathrm{CO}/[\mathrm{K\,km\,s}^{-1}\mathrm{pc}^{2}]) \nonumber \\ 
    & + (-0.01 \pm 0.07) (12+\log (\mathrm{O/H})) \nonumber \\
    & + (36.2 \pm 0.5)
\end{align}

\begin{align}
    \log(L_\mathrm{IR}/[\mathrm{erg\,s}^{-1}]) = & (-0.13 \pm 0.06) \log(M_{\star}/ [M_\sun]) \nonumber \\
    &+ (0.98 \pm 0.04) \log(L_\mathrm{CO}/[\mathrm{K\,km\,s}^{-1}\mathrm{pc}^{2}])\nonumber \\
    &+  (- 1.13 \pm 0.09) (12+\log(\mathrm{O/H}))\nonumber \\
    &+ (47.0 \pm 0.7) 
\end{align}
with $R^2$ scores of $R^2 = 0.64$ and $R^2 = 0.89$, respectively.
 
 As gas content is by far the most dominant driver compared to metallicity  and metallicity may not be always measurable, we also determine relations that only rely on stellar mass and gas content:
 \begin{align}
    \log L_\mathrm{H\upalpha} = & (-0.01\pm0.06) \log(M_{\star}/[M_\sun]) \nonumber\\
                         & + (0.62\pm 0.04) \log(L_\mathrm{CO}/[\mathrm{K\,km\,s}^{-1}\mathrm{pc}^{2}]) \nonumber\\
                         & + (36.2 \pm 0.3) \\
    \log L_\mathrm{IR} = & (-0.19\pm 0.05) \log(M_{\star}/[M_\sun]) \nonumber\\
                  & + (0.82\pm 0.04)  \log(L_\mathrm{CO}/[\mathrm{K\,km\,s}^{-1}\mathrm{pc}^{2}]) \nonumber\\
                  & + (38.7\pm0.3)
  \end{align}
The quality of these models can be represented by $R^2$ scores of $R^2 = 0.64$ and $R^2 = 0.84$, respectively. As expected the quality of the model does not become significantly worse and can be securely used for all samples where metallicity is not available. Here, we will use the full model including metallicity because this parameter is available to us and so we can use it in our analysis. In any case the choice of the models does not alter our final results given the similarity of both models and low impact of metallicity.

\begin{figure*}
    \centering
        \includegraphics[width=1.\linewidth]{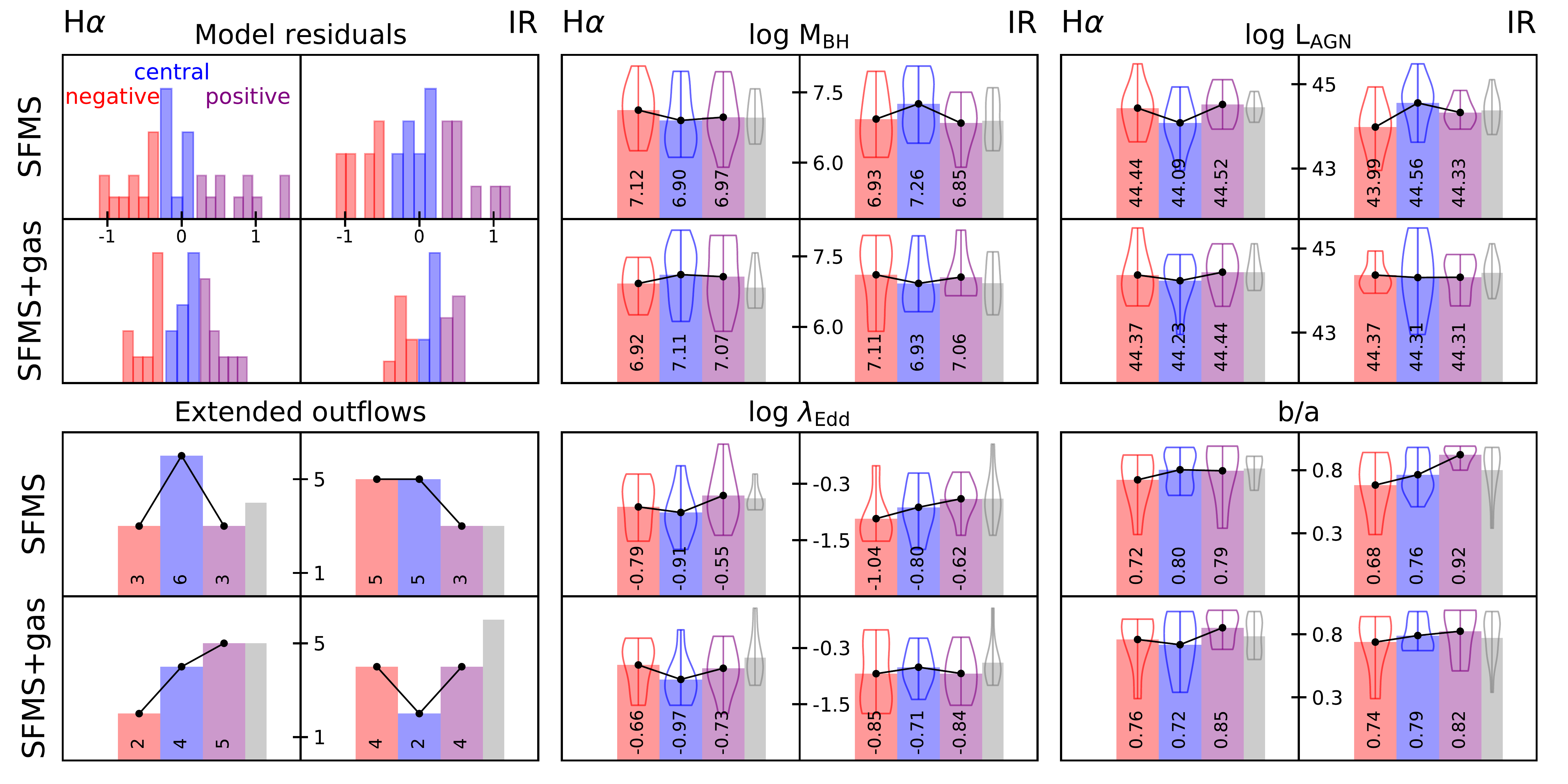}
    \caption{Comparison of SFR residuals from the SFMS and SFMS+gas model against various AGN parameters. The \textit{upper left} panel shows the histograms of the residuals for SFMS and SFMS+gas for both H$\upalpha$ and IR models, with the red, blue, and purple colors corresponding to the previous plots (Fig.~\ref{fig:sfr_mstar} and Fig.~\ref{fig:residuals}, background colors). The other panels show the comparison of the AGN parameters taken from Husemann et al.(2021) and Singha et al. (2021): logarithm of BH mass (\textit{upper center}), logarithm of AGN bolometric luminosity (\textit{upper right}), number of the objects, containing extended outflows (\textit{lower left}), logarithm of Eddington ratio (\textit{lower center}), and b/a morphological parameter (\textit{lower right}). Each of the four models is represented with three bins (from left to right: \textit{negative} red, \textit{central} blue, \textit{positive} purple) and the forth bin with the upper limits (grey). The error bars of the bins also represent the distribution of the parameters within the bins. The black dots and lines highlight the trend of the mean values that are written over the bins.}
    \label{fig:bar_plot}
\end{figure*}

After building the new model to predict SFR using the xCG training sample, we apply it to the CARS objects with the additional information of L$_{\mathrm{CO}(1-0)}$ taken from \citet{bertram2007} and $12+\log(\mathrm{O/H})$ inferred from the CARS IFU observations as described above. With the new \textit{SFMS+gas} model we obtain new predictions for the expected SFR of individual CARS objects and present the corresponding residuals in Fig.~\ref{fig:residuals}. In comparison with the previous residuals (Fig.~\ref{fig:sfr_mstar} bottom panel), the residuals are significantly smaller. The scatter is reduced by a factor of two as the trend with the gas content is taken into account. The initially observed correlation with stellar mass is significantly flattened, with a correlation coefficient consistent with no remaining mass dependence. One potential issue may arise for AGN hosts as the abundance of cold gas has been suspected to be linked with the AGN either by expelling or heating the cold gas \citep[e.g.,][]{carniani2017, kakkad2017,perna2018,bischetti2021,circosta2021}. Considering that the residual sSFR scatter is significantly reduced when incorporating cold gas mass in the SFR prediction indicates that this potential effect is negligible for our sample.   

In order to understand if there are any remaining link between the residuals in sSFR and various AGN parameters we divide the residual space into three bins (\textit{negative}, \textit{central}, and \textit{positive} residual bins) and compare the mean values of the AGN parameters for each bin. The bins are created such, that the number of objects in the bins is approximately equal (with a small excess of objects in the central bin). Objects with upper limits are treated separately as we cannot necessarily associate them to a specific bin. The central residual bin for IR luminosity is shifted slightly towards higher values because of the larger number of upper limits that are too high to be informative. Those objects with upper limits are missing to populate the central and negative residuals bins, but are mostly included in the case of H$\alpha$. As we are only interested in the relative changes along the residuals we focussed to have sufficient objects in each bin for statistical reasons. In Fig.~\ref{fig:bar_plot},  we compare  bins in residual SFR for the SFMS and the new \textit{SFMS+gas} models for both H$\upalpha$ and IR SFR tracers  allowing us to explore trends with BH mass, AGN luminosity, Eddington ratio and $b/a$ host galaxy axis ratio as reported by Husemann et al. (2021) and Singha et al. (2021) for both proxies.

We find no trends of the residuals to systematically change with the $\log \mathrm{M}_{\mathrm{BH}}$ and $\log \mathrm{L}_{\mathrm{AGN}}$ given the uncertainties. While a putative correlation is seen with the fraction of objects containing extended outflows in the \textit{SFMS+gas} model using the $H\upalpha$ proxy, it is not reflected in the IR tracer and therefore not robust. 
While the  logarithm of the Eddington ratio  ($\log \uplambda_{\mathrm{Edd}}$) has a slight increasing trend (e.g. higher AGN Eddington ratio in SFR excess systems) for the SFMS model, such a trend vanishes for the SFMS+gas model. It is important to note that most of the previous studies that investigated the impact of AGN on the SFR  only used stellar mass as a proxy for the expected SFR \citep[e.g.][]{page2012, husemann2014_qdeblend3d, shimizu2015, balmaverde2016,bernhard2016, catalan-torrecilla2017, circosta2018, scholtz2020}. The inclusion of the cold gas mass to predict the expected SFR in addition to the stellar mass can significantly alter the derived conclusions about potential positive, negative or no AGN feedback on their host galaxies. It is therefore crucial for future studies to take more parameters into account than just stellar mass to properly characterize the parent population of non-AGN galaxies.

In Fig.~\ref{fig:bar_plot} the geometrical parameter $b/a$ (tracing the galaxy inclination) shows a clear trend for both H$\upalpha$ and IR SFRs for the SFR model. When comparing the b/a distribution for the negative and positive residual bins with Kolmogorov-Smirnov and Anderson-Darling tests we find a highly significant difference between them for the IR SFR with a $p$-value of 0.003, but less significant for H$\upalpha$.  This suggests that dust extinction or hidden star formation is not the driver for this trend. The trend significantly flattens for the SFMS+gas model considering that the overall scatter in the residuals is greatly reduced, so that the difference become insignificant for our sample size.  A potential explanation for the tentative trend might be the orientation of the AGN. The CARS objects are unobscured AGN hosts, which means that the ionisation cone of the AGN is pointed to the observer. Hence, the cross-section of the AGN ionization cone which can interact with the ISM is increased when the host galaxies are closer to an edge-on geometry. Such misalignments of the galaxies rotation axis and the central AGN engine can have an important influence on the impact of the AGN on the host galaxy, as has already been proposed for various individual galaxies \citep[e.g.][]{garcia_burillo2014_ngc1068, gallimore2016, cresci2015_ngc5643, alonso_herrero2018_ngc5643, husemann2019_he1353, smirnova_pinchukova2019}. With the CARS sample we start to see a potential systematic trend across the population where the potential of the AGN to suppress SFR might also dependent on the relative orientation of the central engine with respect to the galaxy. However, our CARS sample is strongly limited by the low number of more edge-on systems and a bigger sample in needed to gain more insight into this process. Our results highlight that the AGN luminosity may not be the only factor determining the ability of the AGN to impact star formation. The efficiency with which the released energy can couple to the host galaxy adds complexity to the picture, where geometry is certainly just one of many additional parameters to be considered.

\subsection{Comparison of SFR tracers for AGN host galaxies}
Another complexity in investigating the impact of AGN on star formation is caused by the fact that different SFR indicators trace different timescales of star formation. H$\upalpha$ emission is related to the most recent star formation ($\sim$5\,Myr) as excited necessarily by hot O stars, whereas IR traces the dust heated by a wider range of stars, and, therefore traces SF over longer timescales ($\sim$100 Myr; e.g. \citealt{kennicutt1998_schmidt_law}, \citealt{hayward2014}, \citealt{flores2021_fire}). Both SFR tracers are expected to be implicitly linked to the CO luminosity as a molecular gas mass tracer, because the cold gas is the necessary seed reservoir of which new stars can be formed. 

\begin{figure*}
    \centering
        \includegraphics[width=.45\linewidth]{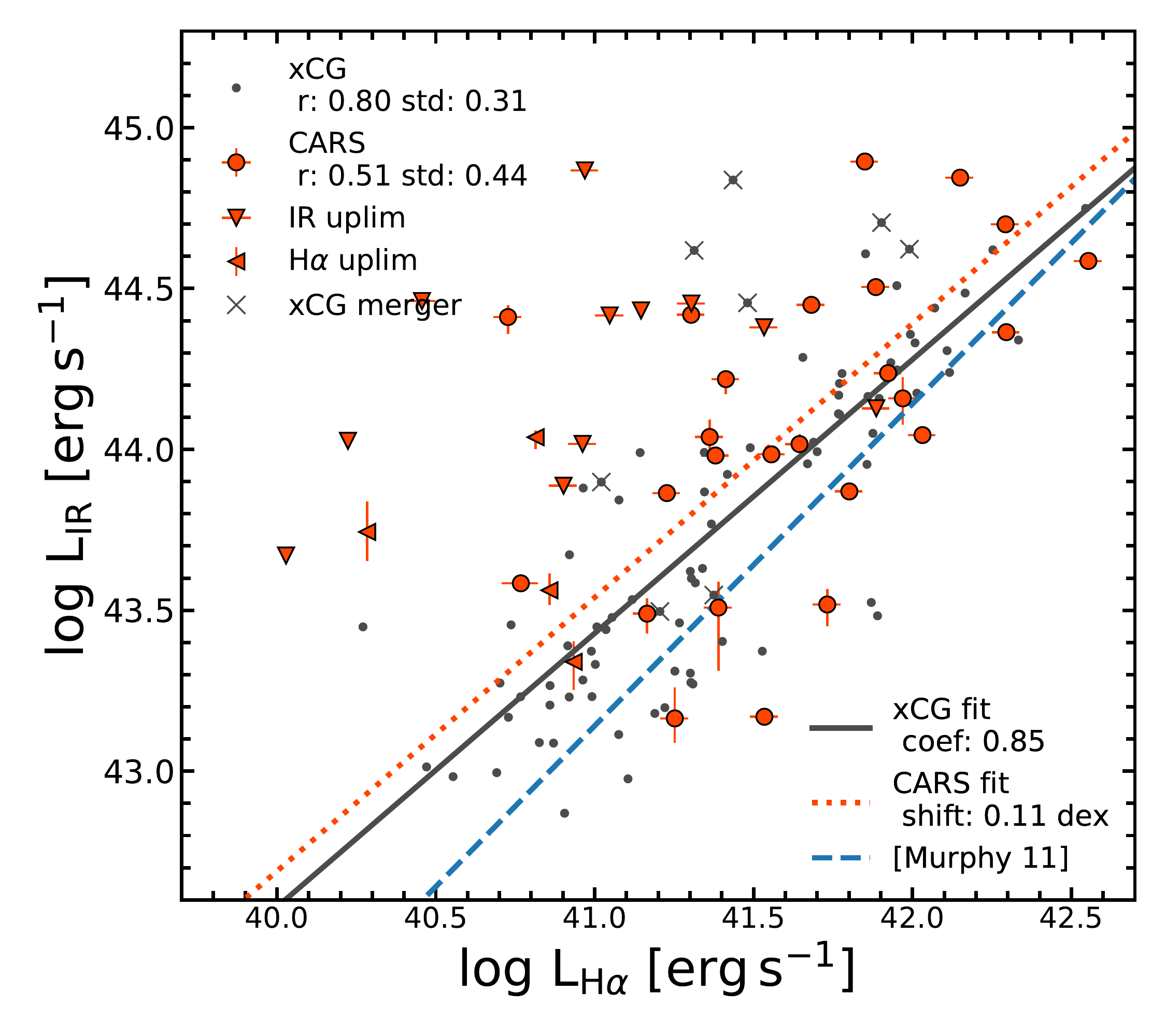}
        \includegraphics[width=.45\linewidth]{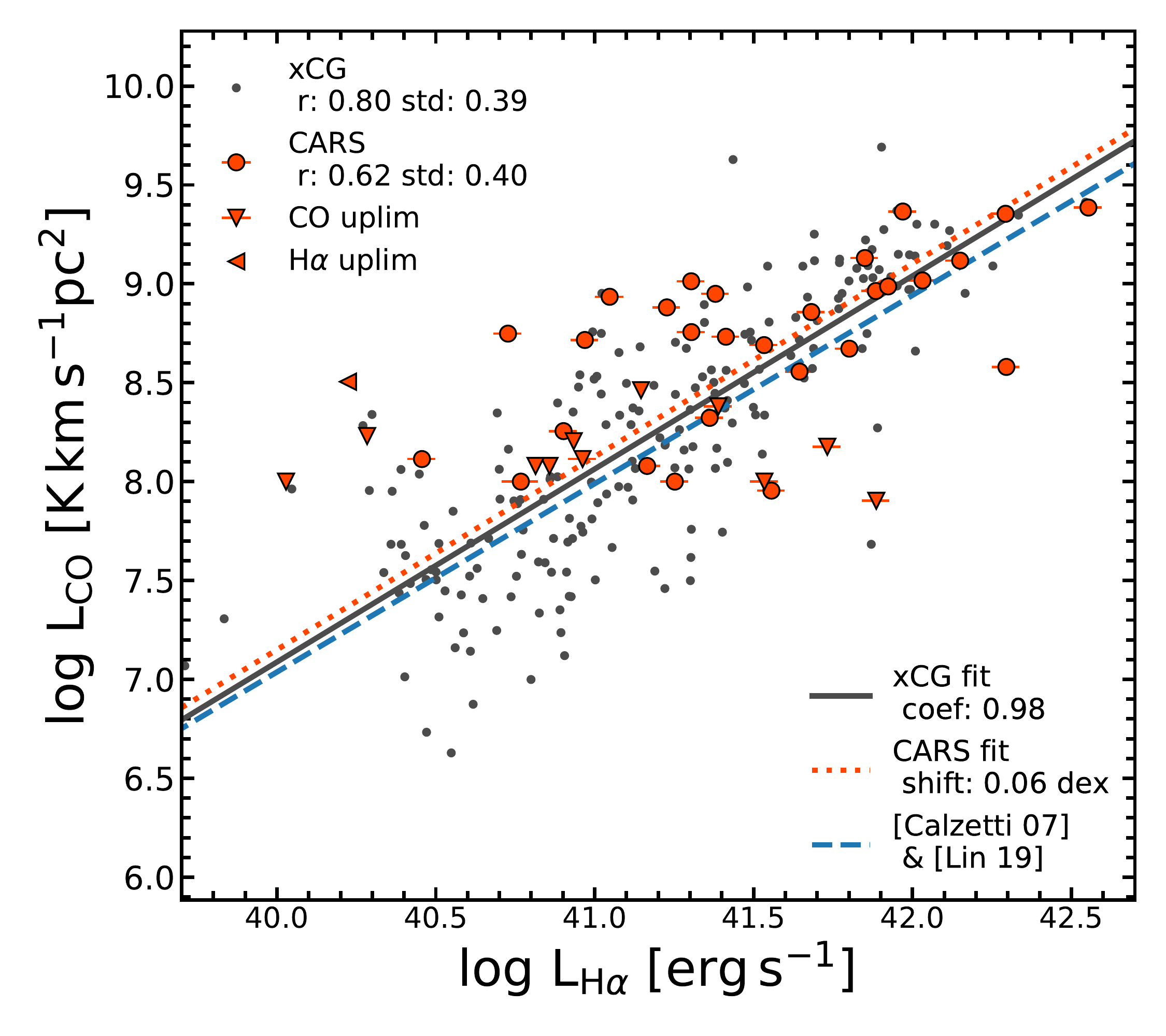}
    \caption{
    The 8--1000\,$\upmu$m IR luminosity $L_\mathrm{IR}$ \textit{(left panel)} and the CO~(1$-$0) luminosity $L_\mathrm{CO}$ \textit{(right panel)} plotted against H$\upalpha$ luminosity. The training sample from the xCG (grey dots) is compared to the CARS objects (red circles and triangles for the 5$\upsigma$ upper limits). Galaxy mergers in the xCG sample as classified by GalaxyZoo2 are additionally marked (black crosses) on the left panel. The blue dashed line corresponds to Eq.~\ref{eq:IR_Ha} and Eq.~\ref{eq:CO_Ha} are derived by the combination of literature calibrations as described in the main text. The black lines represent a linear relation fitted to the XCOLD GASS reference sample, see Eq.~(\ref{eq:IR_Ha_xCold}) and Eq.~(\ref{eq:CO_Ha_xCold}). The red dotted line is fitted to the CARS objects with the same slope as the black line, therefore highlight the shift in the normalization.}
    \label{fig:ir_co_halpha}
\end{figure*}

By combining the SFR calibration for the IR and H$\upalpha$ luminosity (Eq.~\ref{eq:SFR_IR} and Eq.~\ref{eq:SFR_Ha}) we can make a prediction for their expected relation. Furthermore, we can link the SFR to the CO luminosity given the calibration of \citet{lin2019_co_sfr}. This leads to the following relation between the H$\upalpha$ luminosity and the IR and CO luminosity, respectively.
\begin{eqnarray}
    \log\left(\frac{L_\mathrm{IR}}{[\mathrm{erg\,s}^{-1}]}\right) &=& \log\left(\frac{L_\mathrm{H\upalpha}}{[\mathrm{erg\,s}^{-1}]}\right) + 2.14\label{eq:IR_Ha}\\
    \log\left(\frac{L_\mathrm{CO}}{[\mathrm{K\,km\,s}^{-1}\mathrm{pc}^{2}]}\right) &=& 0.95 \log\left(\frac{L_\mathrm{H\upalpha}}{[\mathrm{erg\,s}^{-1}]}\right) - 31.05\label{eq:CO_Ha}
\end{eqnarray}
Comparing the three luminosities with each other in Fig.~\ref{fig:ir_co_halpha} we find that the CARS sample and the training sample from the xCG are significantly offset with respect to the predicted relation between H$\upalpha$ and IR luminosity. Despite this, the relation between $L_\mathrm{CO}$ and $L_\mathrm{H\upalpha}$ as predicted from the combined  \citet{calzetti2007} and \citet{lin2019_co_sfr} calibrations agree well with our data. This points to the notion that either the theoretical assumption made by \citet{murphy2011} is inaccurate or that some of the ongoing star formation is completely obscured in H$\upalpha$ which cannot be recovered with an extinction correction. Nevertheless, the xCG non-AGN sample can be used as a reference because the impact of extinction on the FIR/H$\upalpha$ ratio should be comparable to our AGN host galaxies. We therefore performed a linear fit directly on the xCG data and obtained the following relation which are shown as the black lines in Fig.~\ref{fig:ir_co_halpha}:
\begin{eqnarray}
    \log\left(\frac{L_\mathrm{IR}}{[\mathrm{erg\,s}^{-1}]}\right) &=& 0.85 \log\left(\frac{L_\mathrm{H\upalpha}}{[\mathrm{erg\,s}^{-1}]}\right) + 8.55 \label{eq:IR_Ha_xCold}\\
    \log\left(\frac{L_\mathrm{CO}}{[\mathrm{K\,km\,s}^{-1}\mathrm{pc}^{2}]}\right) &=& 0.97 \log\left(\frac{L_\mathrm{H\upalpha}}{[\mathrm{erg\,s}^{-1}]}\right) - 31.95\label{eq:CO_Ha_xCold}
\end{eqnarray}

Even though the CARS sample scatter is close to the comparison sample scatter, the CARS galaxies are on average located $\sim$0.2\,dex above the fitted L$_{\mathrm{IR}}$ to L$_{\mathrm{H}\upalpha}$ trend. The question here is whether this offset has a physical meaning or is just caused by low-number statistics and unknown biases of the CARS sample. Here, we claim that this offset is physical, based on two sets of evidence: 1) the CARS data are in a good agreement with the non-AGN reference sample on the L$_{\mathrm{CO}}$ to L$_{\mathrm{H}\upalpha}$ plot; 2) the training sample also has a few galaxies, located at similarly high L$_{\mathrm{IR}}$/L$_{\mathrm{H}\upalpha}$ ratios which drive the offset in the CARS sample. The xCG objects in this specific area of the L$_{\mathrm{IR}}$/L$_{\mathrm{H}\upalpha}$ diagram are almost exclusively populated with galaxy mergers according to the GalaxyZoo2 classification \citep{willett2013,hart2016}. Those mergers seem to have a lower instantaneous SFR$_{\mathrm{H}\upalpha}$ compared to the slightly longer timescale IR SFR, which can be expected from the rapid star formation history evolution and bursts of star formation on 100\,Myr timescale that are observed and predicted by detailed galaxy simulation \citep[e.g.][]{mihos1992,barnes2004,springel2005,diMatteo2008,hopkins2013}. Considering that the offset for the CARS sample is mainly caused by individual galaxies with similarly high L$_{\mathrm{IR}}$/L$_{\mathrm{H}\upalpha}$ ratio, as in the non-AGN merger sample, some AGN host galaxies potentially had higher SFR in the past (over a $\sim$100\,Myr timescale).

Whether the potential difference in the star formation history is caused by a recent excess of star formation, which might actually be linked to a delayed BH growth \citep[e.g.][]{wild2010}, or by ongoing suppression of star formation as expected from AGN feedback remains unclear. To shed light on these different possibilities we discuss the difference between current and recent star formation as probed by H$\alpha$ and FIR emission in the following for individual objects where more information is available and can be interpreted.

\begin{figure}
    \centering
        \includegraphics[width=1.\linewidth]{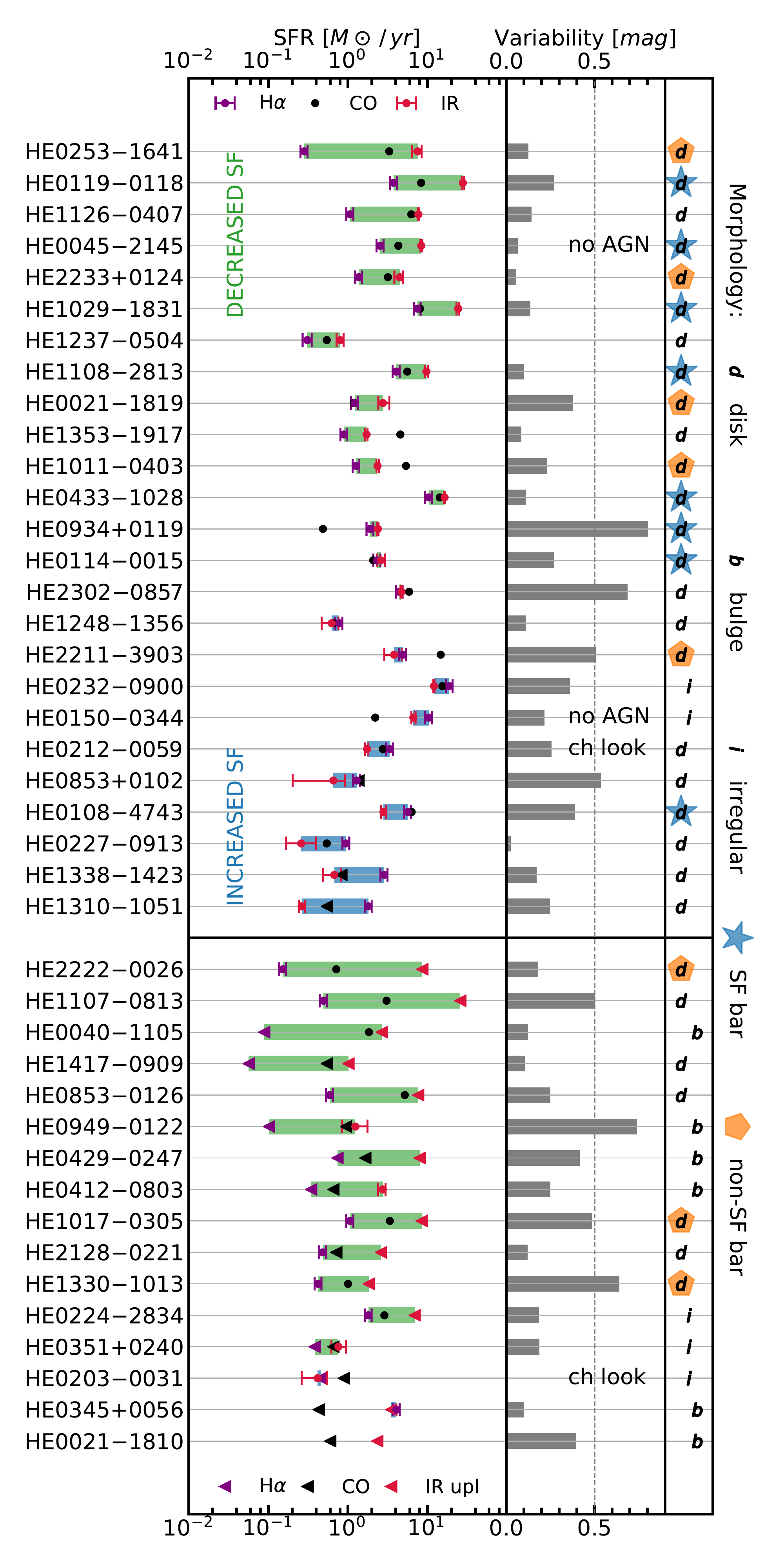}
    \caption{Comparison of different SFR measurements for individual objects. \textit{Left column:} H$\upalpha$-, IR-, and CO(1$-$0)-based SFRs for the CARS objects sorted by the $\mathrm{SFR}_{\mathrm{IR}} - \mathrm{SFR}_{\mathrm{H}\upalpha}$ difference. \textit{Central column:} near infrared variability; the objects with no AGN or known changing look AGN are also marked here. \textit{Right column:} morphology of the host galaxies. \textit{Lower panel} contains the objects with either H$\upalpha$ and/or IR SFR uncertainties (upper limits or missing data).}
    \label{fig:long_plot}
\end{figure}

\subsection{Individual comparison of current and recent SFRs}
In the previous subsection we compared different SFR tracers for the CARS sample with respect to the non-AGN comparison sample. Now we look at the SFR difference derived from the different SFR proxies for individual CARS galaxies and try to discuss the potential origins. We visualize the SFR results together with host galaxy morphologies and an indication for the AGN variability in Fig.~\ref{fig:long_plot}. The AGN variability is set here by the brightness difference in the NIR photometry between old 2MASS and more recent VISTA or PANIC observations separated by a few years. The H$\upalpha$-, IR- and CO-based SFR are again defined according to Eq.~(\ref{eq:SFR_Ha}) and Eq.~(\ref{eq:SFR_IR}), respectively. We plot objects with upper limits in either of the SFR traces separately as any difference between the SFR cannot be inferred on an individual basis. 

The sign of the relative difference of H$\upalpha$- and IR-based SFR is highlighted in green and blue color when the IR-based SFR is higher or lower compared to H$\upalpha$, respectively.  Assuming that the SFR timescale is the prime physical explanation for the difference \citep[e.g.][]{hayward2014,davies2015}, where H$\upalpha$ probes more recent star formation than the IR, we can in principle differentiate between declining or enhanced star formation. In Fig.~\ref{fig:long_plot} we order the objects from the strongest decline to the highest enhancement in SFR. The objects HE0253$-$1641, HE0119$-$0118, HE1126$-$0407, HE0045$-$2145, HE2233$+$0124 and HE1029$-$1831 belong to the declining SFR group and HE0150$-$0340, HE0212$-$0059, HE0853$+$0102, HE0108$-$4743, HE0227$-$0913, HE1338$-$1423, and HE1310$-$1051 belong to the galaxies with potentially enhanced star formation. Below we discuss different scenarios for those individual cases that may provide physical explanations, but certainly require further tests and observations to verify. 

All of these galaxies are disc-dominated and none are bulge-dominated or strongly interacting systems. Strong interactions therefore seem to play a minor role, but we still identify cases where this might be important but not obvious. In addition, bars might suppress the star formation dynamically due to bar quenching \citep[e.g.][]{Khoperskov2018,Fraser-McKelvie2020}, but we identified non-star forming and star forming bars in the decreasing SF group, based on the analysis of \citet{neumann2019}. While the majority of the galaxies with declining SFR are indeed barred galaxies, most of them appear star-forming contrary to expectation if they would suppress star formation. AGN or starburst-driven winds may play a role in individual cases of HE~1126$-$0407 and HE~0045$-$2145, which have both declining SFR. HE~1126$-$0407 (aka PG~1126$-$041) is known to have a powerful ultra-fast outflow driven by the AGN \citep{wang1999,giustini2011} which may couple more efficiently on kpc scales \citep{marasco2020} as the galaxy is significantly inclined with respect to our line-of-sight and central AGN engine orientation. Furthermore, HE~0045$-$2145 was misclassified in the Hamburg/ESO survey because of the broader lines caused by a starburst-driven outflow (Nevin et al. in prep.). In both cases the outflow might be related to the anticipated decline in SFR.  The targets HE~0253$-$1641 and HE~0119$-$0118 show high gas dispersion in the AGN-ionised region on kpc scales (see Fig.~B.1 in Husemann et al. (2021)) which also points to a past or ongoing galactic outflow event in these systems. Such outflow signatures are much weaker in HE2233$+$0124, but the galaxy is also more edge on with a misaligned ionization cone similar to HE~1126$-$0407 so that the impact of the AGN may be amplified. The cause of the potential decline in the SFR of HE~1029$-$1831 is less clear, but the stellar population modelling in high-angular NIR IFU spectroscopy with SINFONI revealed a recent circumnuclear starburst about 100\,Myr ago that is rapidly declining \citep{busch2015}. Although the cause of this decline is hard to directly link to AGN feedback or simple gas consumption, it clearly supports the timescale interpretation of the difference in the H$\alpha$ and FIR-based SFR.

For the objects with increased SFR, we identified HE1310$-$1051 to exhibit a strong interaction with a minor companion \citep{husemann2014_qdeblend3d} and HE0150$-$0344 to be a strongly-interacting non-AGN system. In those two cases, the interactions are most likely responsible for a recent burst of star formation. We cannot pin-point any obvious reasons why the star formation may be enhanced in HE~0212$-$0059, HE~0853$+$0102, HE~0108$-$4743, HE~0227$-$0913 and HE~1338$-$1423, but we note that the bolometric luminosity is close to the Eddington luminosity for the last three sources as reported in Husemann et al. (2021) due to their rather low BH mass with respect to the AGN luminosity.  This is consistent with the scenario discussed in Husemann et al. (2021) that low BH mass AGN are likely to be observed in earlier phase of the AGN cycle. In this case, the circumnuclear starburst would be observed much closer than 100\,Myr in time with respect to its peak activity. Indeed, narrow-line Seyfert 1 galaxies as high-Eddington ratio AGN show systematically higher SFR based on PAH emission detected with Spitzer \citep{sani2010} compared to the broad-line Seyfert 1 counterparts with lower Eddington ratios. This highlights the intriguing connection between the AGN and the circumnuclear starburst, where AGN with increasing SFR are potentially young AGN in a fueling mode powered by the starburst.

Lastly, we obtained some information on AGN variability as a side-product of our analysis when we compared old 2MASS NIR photometry with the more recent VISTA or PANIC observations. Interestingly, we see  that the targets with declining SFR show systematically less variability on a few tens of years timescale. It is still speculative whether this points to a much more stable energy output of the AGN on longer time-scales and thereby enhancing the impact as a cumulative effect of energy release over time. However, it shows that the process of AGN feedback is complex and the time evolution of the AGN phase with respect to the galaxy needs to be considered as well to get a comprehensive picture. Indeed, in current self-regulated feedback models (\citealt{gaspari2020} for a review) the AGN is expected to flicker on-off with rapid variability that increases toward low-mass systems often due to the chaotic cold accretion feeding the SMBH (e.g., \citealt{gaspari2015,tremblay2018,rose2019}).

\section{Summary and Conclusions}
In this paper, we presented a complete census of the integrated SFR properties across the entire CARS sample using the entire multi-wavelength data set available for this local AGN host galaxy sample. In particular, we inferred robust stellar masses and IR-based SFR from panchromatic SED modelling with \textsc{AGNfitter} and H$\upalpha$-based SFR from a careful analysis of spatially-resolved optical emission-line diagnostics with our new \texttt{rainbow} algorithm. Using the large xCG sample of local non-AGN galaxies as a control and training set we built different models to predict the SFR for our AGN host galaxies to investigate the role of the AGN in terms of star formation feedback. Our main conclusions from this analysis can be summarised as follows:
\begin{itemize}
    \item We find that stellar mass alone is an insufficient proxy for the expected SFR in AGN host galaxies.  The cold gas content and possibly the metallicity are crucial to consider in order to avoid artificial trends with AGN parameters that could mimic expected AGN feedback trends. 
    \item No systematic suppression of SFR could be detected with respect to the non-AGN galaxy reference sample and there is also no trend with AGN luminosity. 
    \item A potential link between lower than expected SFRs and the axis ratio $b/a$ of the AGN host galaxies was identified. As our sample contains only unobscured AGN, the central engine (and thus the ionization cones) of low $b/a$ systems must be mis-aligned with the galaxy rotation axis for inclined discs, leading to a higher cross-section of the AGN radiation field which can interact with the galaxy disc. This tentative trend is much weaker and currently insignificant for the SFR+gas model given the low-number of strongly inclined systems, but it may imply that such mis-alignments could amplify the coupling of the released AGN energy with the cold gas disc of the galaxy and thereby impact star formation more efficiently. This should be confirmed with bigger AGN host galaxies samples. 
    \item Interpreting the IR and H$\upalpha$ SFR tracers as proxies for the recent star formation history on $\sim$100\,Myr and $\sim$5\,Myr timescales respectively, we identified systems with decreasing or increasing SFR. The declining SFR cases might often be associated with galactic outflows while the increasing SFR cases can be associated with interaction or potentially with a young AGN phase. 
\end{itemize}

That we cannot find any strong evidence for a global positive or negative AGN feedback on the SFR across the entire CARS sample is in agreement with various recent studies reporting no immediate impact of AGN on the star formation \citep[e.g.][]{scholtz2020}. We emphasize that predicting the expected SFR of galaxies is difficult, and using stellar mass alone may not be sufficient. The cold gas mass is fundamental, and its use in such relations allows one to capture the significant scatter in the star-forming main sequence (as already demonstrated in several works for the non-AGN population; e.g.  \citealt{colombo2020,popesso2020}). Despite the lack of a global and obvious impact of AGN on star formation, we discover subtle effects that should be investigated in the future. Most importantly, the relative orientation of the AGN central engine and associated ionization cones may be relevant for the cross-section of release AGN energy and the cold gas of the galaxy. This is most prominent in disc galaxies and previously studied in several individual cases \citep[e.g.][]{cecil2001,morganti2015,mahony2016,mukherjee2018,husemann2019_he1353}, but CARS reveals a potential systematic trend that should be explored with larger samples in the future. 

The non-detection of a relation of AGN luminosity, BH mass and Eddington ratio with the global SFR may be related to the different time scales of the AGN phase and star formation in galaxies. In case the AGN phase is short, there would not be enough time passed to see the impact on the global star formation when selecting AGN samples rather than post-starburst system as discussed in the review of \citet{alexander2012} and in \citep{hickox2014}. Indeed, the CARS sample suggest a potential correlation with the duration of a luminous AGN phase as a function of BH mass (Husemann et al. 2021), which is further corroborated by current models of AGN feedback self-regulated via chaotic cold accretion \citep[e.g.,][]{gaspari2020}. The predicted durations are of the order of 1~Myr for a single AGN phase which would be clearly too short to suppress the star formation in the entire host galaxy and can explain our observations. Still, the circumnuclear SFR could be affected on these timescales. We observed interesting patterns of increasing and decreasing SFR by comparing the IR and H$\upalpha$ SFR tracers among the sample that we can partially attribute to outflows, circumnuclear star formation and galaxy interaction. This highlights the potential of this approach, and also the complexity in the galaxy properties to be considered. We plan to expand the diagnostics of the star formation history determination by inferring radio-based SFR for the CARS sample which probe intermediate SFR timescale of a few tens of Myr, filling the gap in the H$\upalpha$ and IR-based SFRs. Furthermore, we will expand our SFR investigation by zooming into the circumnuclear region of the galaxies in the CARS sample in the future. This requires the construction of an appropriate control sample with similar resolution which was beyond the scope of this paper. 

Overall, we identify cold gas content, relative AGN engine orientation with respect to the host galaxy, as well as the time domain variability as potential key parameters that need to be explored in the future to understand the impact of AGN on their galaxy on a population wide basis. This leads to obvious challenges in the sample selection, sample size and parameter space to be measured to gain more insights into the putative AGN feedback process. 

\begin{acknowledgements}
We thank the anonymous referee for helpful comments that improved the quality of the manuscript.
We thank Mischa Schirmer for his support with the usage of the updated Theli pipeline v3. 
ISP greatly appreciates financial support from the DLR through grant 50OR2006. 
BH is financially supported through DFG grant GE625/17-1. ISP and BH also acknowledge travel support from the DAAD via grant 57509925.
TAD acknowledges support from the UK Science and Technology Facilities Council through grant ST/S00033X/1.
MG acknowledges partial support by NASA Chandra GO8-19104X/GO9-20114X and HST GO-15890.020/023-A, and the \textit{BlackHoleWeather} program.
SB, CO, and MS acknowledge support from the Natural Sciences and Engineering Research Council (NSERC) of Canada.
MPT acknowledges financial support from the State Agency for Research of the Spanish MCIU through the
"Center of Excellence Severo Ochoa" award to the Instituto de Astrofísica de Andalucía (SEV-2017-0709)
and through grants PGC2018-098915-B-C21 and PID2020-117404GB-C21 (MCI/AEI/FEDER, UE).
The work of TC was carried out at the Jet Propulsion Laboratory, California Institute of Technology, under a contract with NASA.

Based on observations collected at the European Organization for Astronomical Research in the Southern Hemisphere under ESO programme 094.B-0345(A) and 095.B-0015(A).\\
Based on observations collected at the Centro Astronómico Hispano-Alemán (CAHA) at Calar Alto, operated jointly by Junta de Andalucía and Consejo Superior de Investigaciones Científicas (IAA-CSIC).\\

This project used data obtained with the Dark Energy Camera (DECam), which was constructed by the Dark Energy Survey (DES) collaboration. Funding for the DES Projects has been provided by the US Department of Energy, the US National Science Foundation, the Ministry of Science and Education of Spain, the Science and Technology Facilities Council of the United Kingdom, the Higher Education Funding Council for England, the National Center for Supercomputing Applications at the University of Illinois at Urbana-Champaign, the Kavli Institute for Cosmological Physics at the University of Chicago, Center for Cosmology and Astro-Particle Physics at the Ohio State University, the Mitchell Institute for Fundamental Physics and Astronomy at Texas A\&M University, Financiadora de Estudos e Projetos, Fundação Carlos Chagas Filho de Amparo à Pesquisa do Estado do Rio de Janeiro, Conselho Nacional de Desenvolvimento Científico e Tecnológico and the Ministério da Ciência, Tecnologia e Inovação, the Deutsche Forschungsgemeinschaft and the Collaborating Institutions in the Dark Energy Survey.

The Collaborating Institutions are Argonne National Laboratory, the University of California at Santa Cruz, the University of Cambridge, Centro de Investigaciones Enérgeticas, Medioambientales y Tecnológicas–Madrid, the University of Chicago, University College London, the DES-Brazil Consortium, the University of Edinburgh, the Eidgenössische Technische Hochschule (ETH) Zürich, Fermi National Accelerator Laboratory, the University of Illinois at Urbana-Champaign, the Institut de Ciències de l’Espai (IEEC/CSIC), the Institut de Física d’Altes Energies, Lawrence Berkeley National Laboratory, the Ludwig-Maximilians Universität München and the associated Excellence Cluster Universe, the University of Michigan, NSF’s NOIRLab, the University of Nottingham, the Ohio State University, the OzDES Membership Consortium, the University of Pennsylvania, the University of Portsmouth, SLAC National Accelerator Laboratory, Stanford University, the University of Sussex, and Texas A\&M University.

Based on observations at Cerro Tololo Inter-American Observatory, NSF’s NOIRLab (NOIRLab Prop. ID 2017A-0914; PI: Grant Tremblay), which is managed by the Association of Universities for Research in Astronomy (AURA) under a cooperative agreement with the National Science Foundation.
The Pan-STARRS1 Surveys (PS1) and the PS1 public science archive have been made possible through contributions by the Institute for Astronomy, the University of Hawaii, the Pan-STARRS Project Office, the Max-Planck Society and its participating institutes, the Max Planck Institute for Astronomy, Heidelberg and the Max Planck Institute for Extraterrestrial Physics, Garching, The Johns Hopkins University, Durham University, the University of Edinburgh, the Queen's University Belfast, the Harvard-Smithsonian Center for Astrophysics, the Las Cumbres Observatory Global Telescope Network Incorporated, the National Central University of Taiwan, the Space Telescope Science Institute, the National Aeronautics and Space Administration under Grant No. NNX08AR22G issued through the Planetary Science Division of the NASA Science Mission Directorate, the National Science Foundation Grant No. AST-1238877, the University of Maryland, Eotvos Lorand University (ELTE), the Los Alamos National Laboratory, and the Gordon and Betty Moore Foundation.\\
The James Clerk Maxwell Telescope is operated by the East Asian Observatory on behalf of The National Astronomical Observatory of Japan; Academia Sinica Institute of Astronomy and Astrophysics; the Korea Astronomy and Space Science Institute; Center for Astronomical Mega-Science (as well as the National Key R\&D Program of China with No. 2017YFA0402700). Additional funding support is provided by the Science and Technology Facilities Council of the United Kingdom and participating universities and organizations in the United Kingdom and Canada.\\
This publication makes use of data products from the Wide-field Infrared Survey Explorer, which is a joint project of the University of California, Los Angeles, and the Jet Propulsion Laboratory/California Institute of Technology, funded by the National Aeronautics and Space Administration.\\
\textit{Herschel} is an ESA space observatory with science instruments provided by European-led Principal Investigator consortia and with important participation from NASA.\\
This publication makes use of data products from the Two Micron All Sky Survey, which is a joint project of the University of Massachusetts and the Infrared Processing and Analysis Center/California Institute of Technology, funded by the National Aeronautics and Space Administration and the National Science Foundation.\\
This research has made use of the NASA/IPAC Infrared Science Archive, which is funded by the National Aeronautics and Space Administration and operated by the California Institute of Technology.\\
This research has made use of the APASS database, located at the AAVSO web site. Funding for APASS has been provided by the Robert Martin Ayers Sciences Fund.\\
Supported by the international Gemini Observatory, a program of NSF’s NOIRLab, which is managed by the Association of Universities for Research in Astronomy (AURA) under a cooperative agreement with the National Science Foundation, on behalf of the Gemini partnership of Argentina, Brazil, Canada, Chile, the Republic of Korea, and the United States of America.\\
The Science, Technology and Facilities Council is acknowledged by JN for support through the Consolidated Grant Cosmology and Astrophysics at Portsmouth, ST/S000550/1. 
\end{acknowledgements}

\bibliographystyle{aa}
\bibliography{references}

\clearpage
 \begin{appendix}

 \section{SED models}\label{sec:sed}
 \begin{figure*}
       \includegraphics[width=0.33\textwidth]{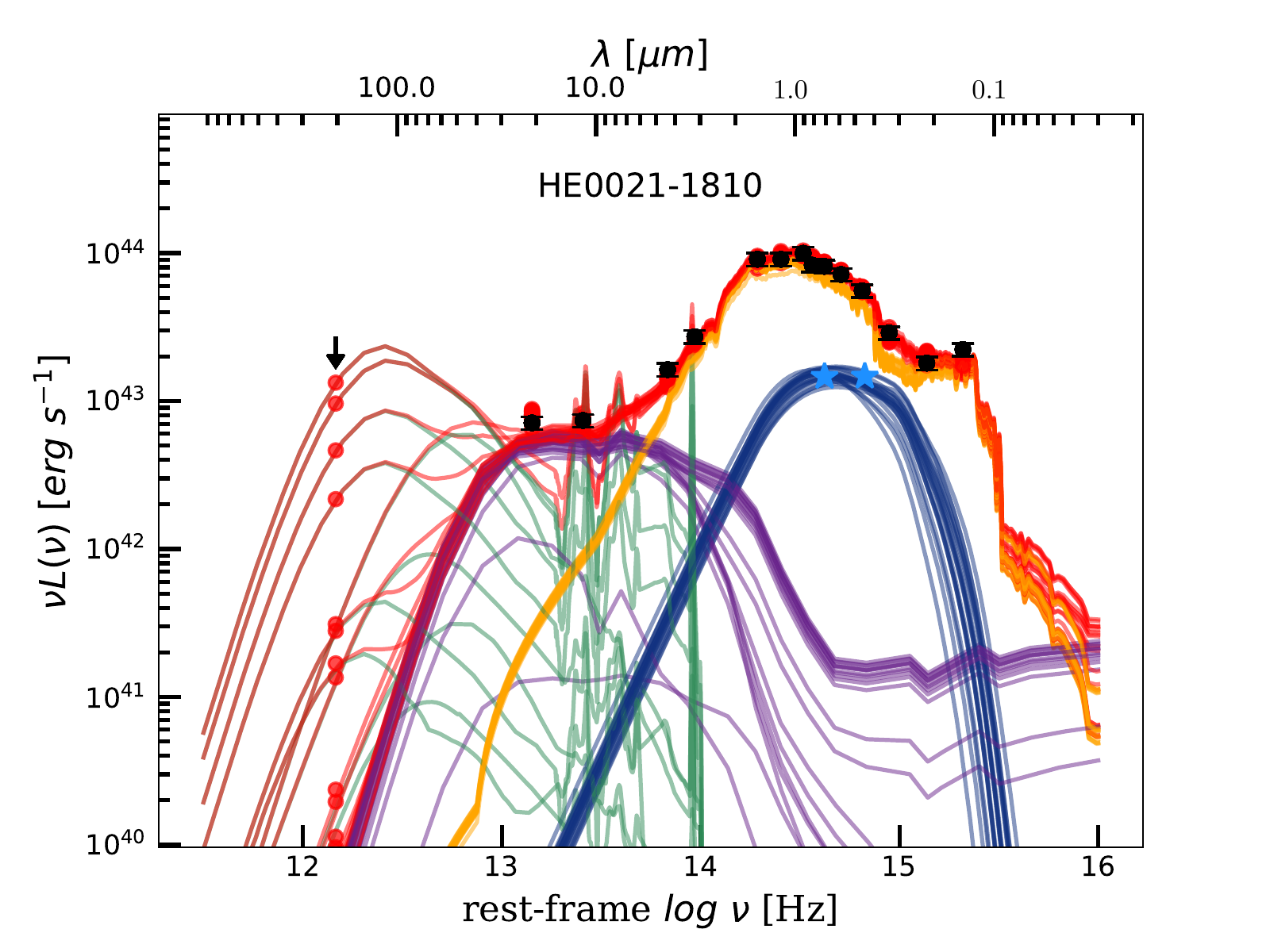}
       \includegraphics[width=0.33\textwidth]{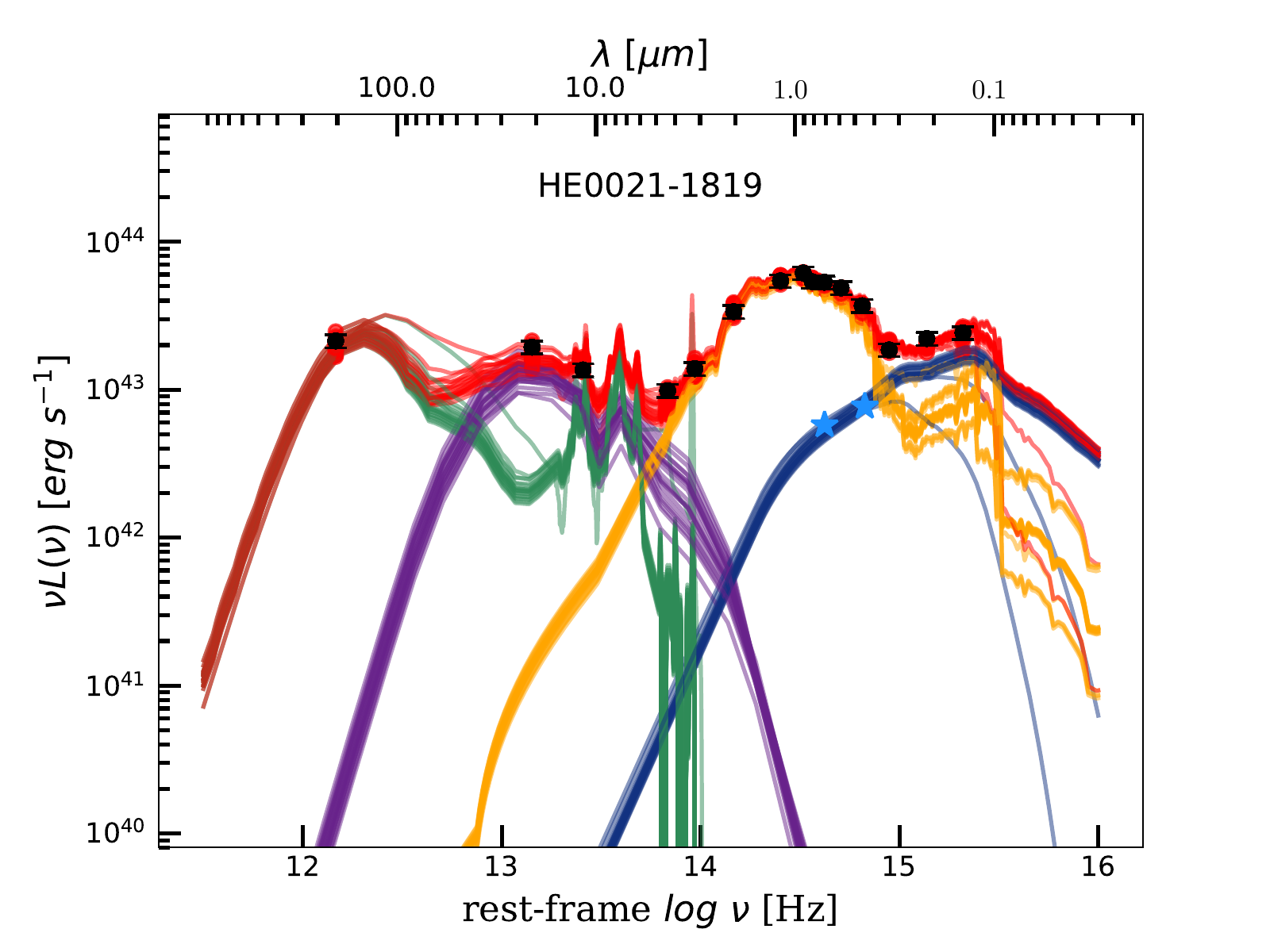}
        \includegraphics[width=0.33\textwidth]{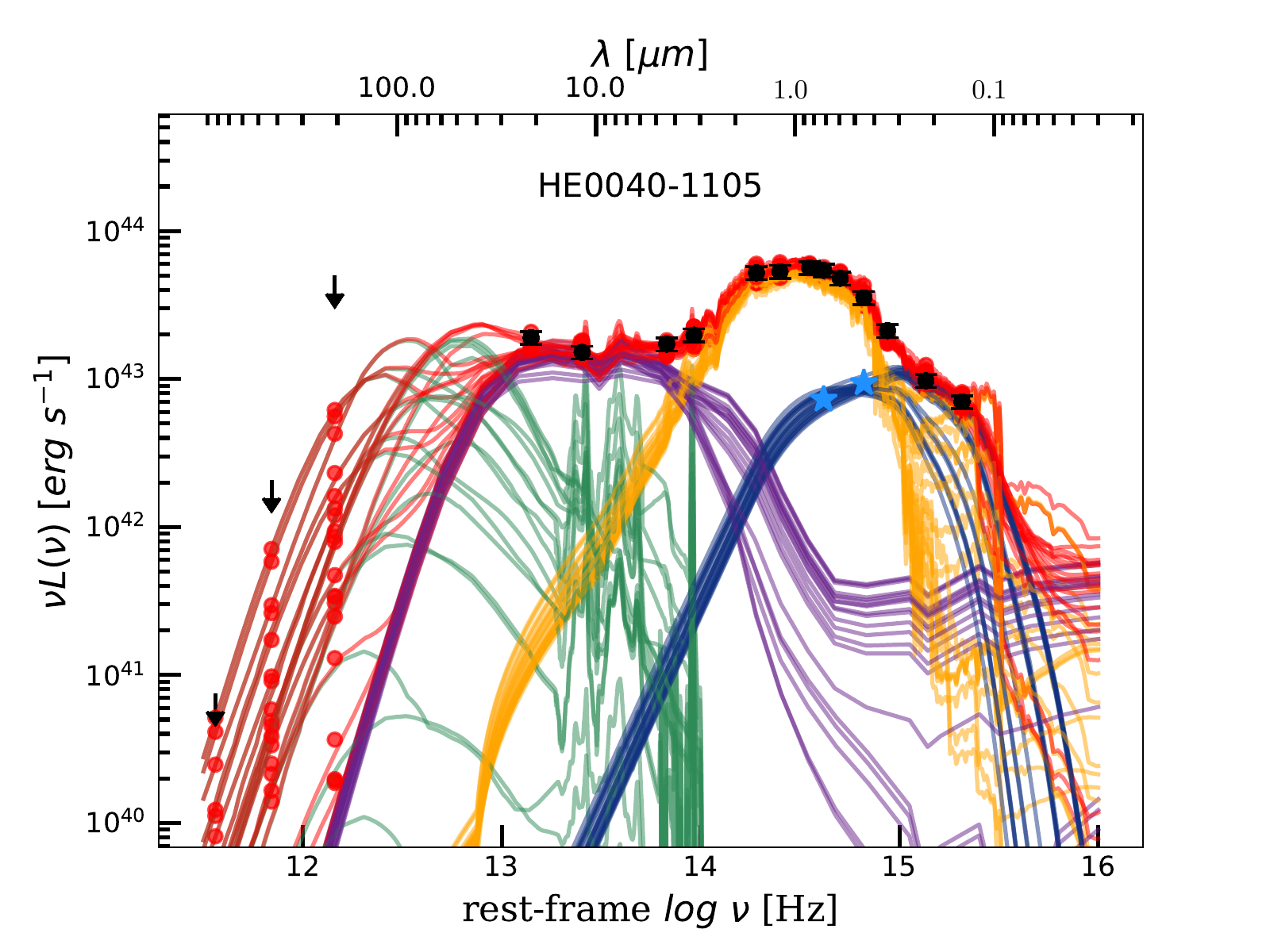}\\
        \includegraphics[width=0.33\textwidth]{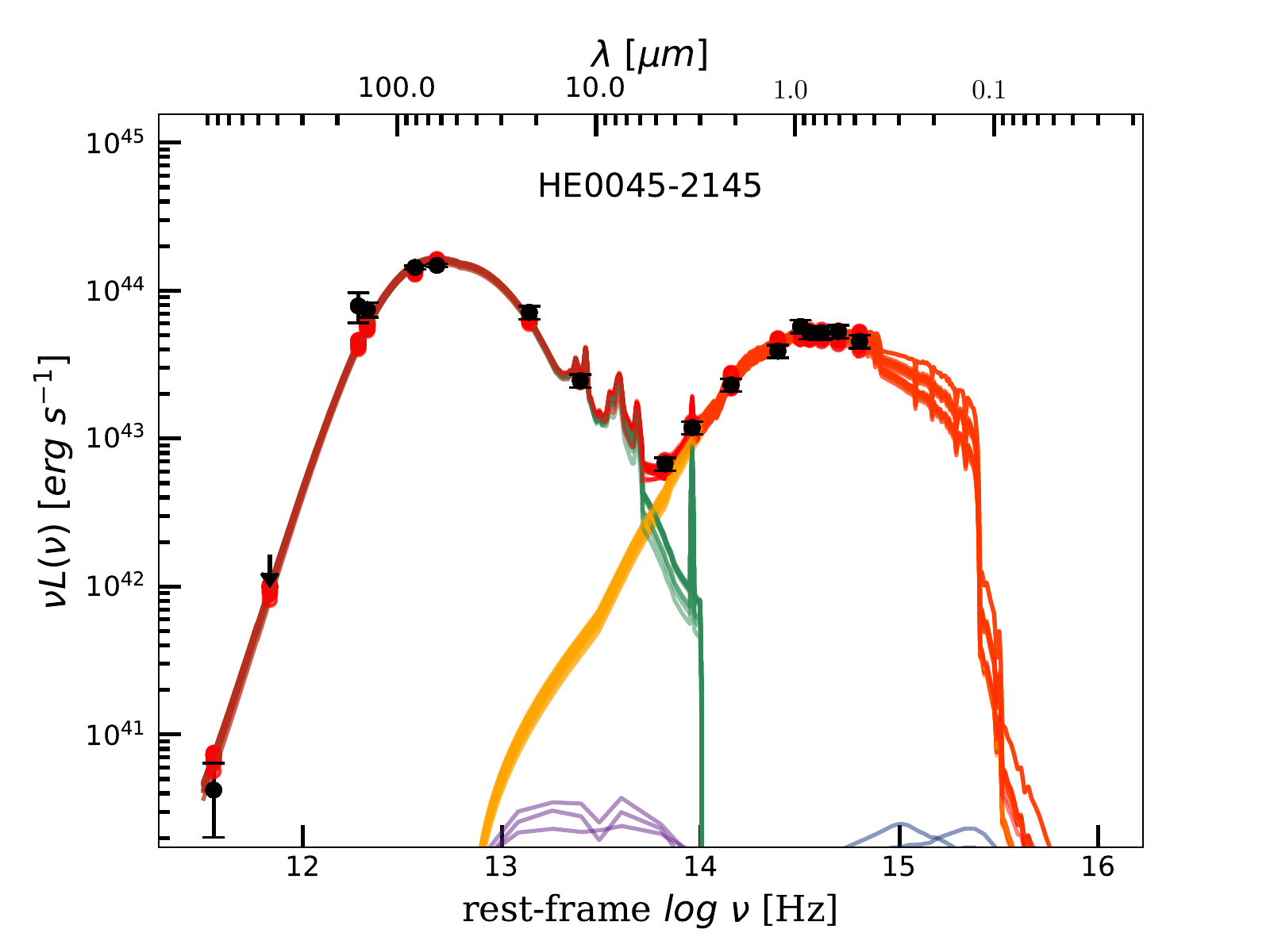}
        \includegraphics[width=0.33\textwidth]{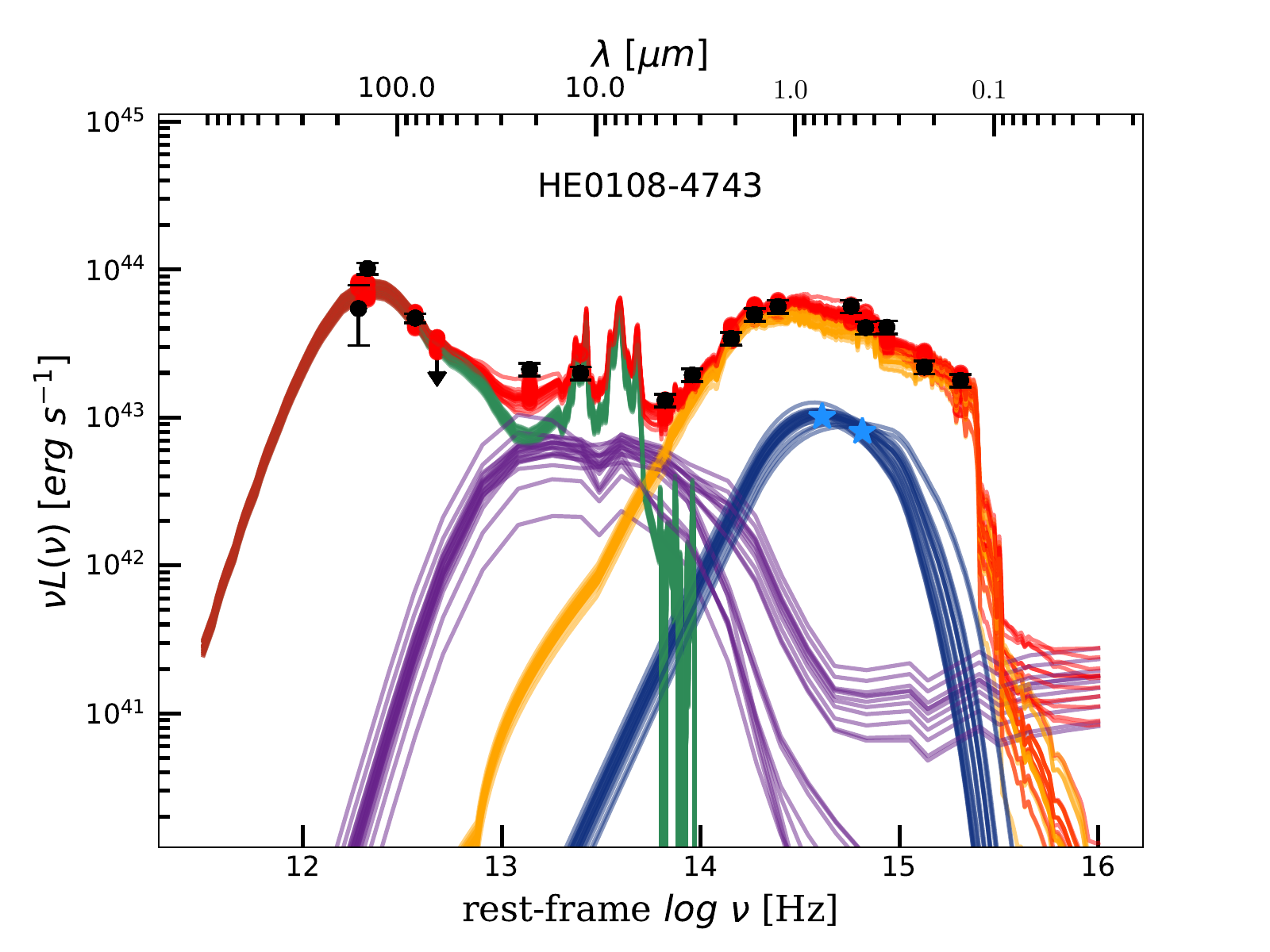}
        \includegraphics[width=0.33\textwidth]{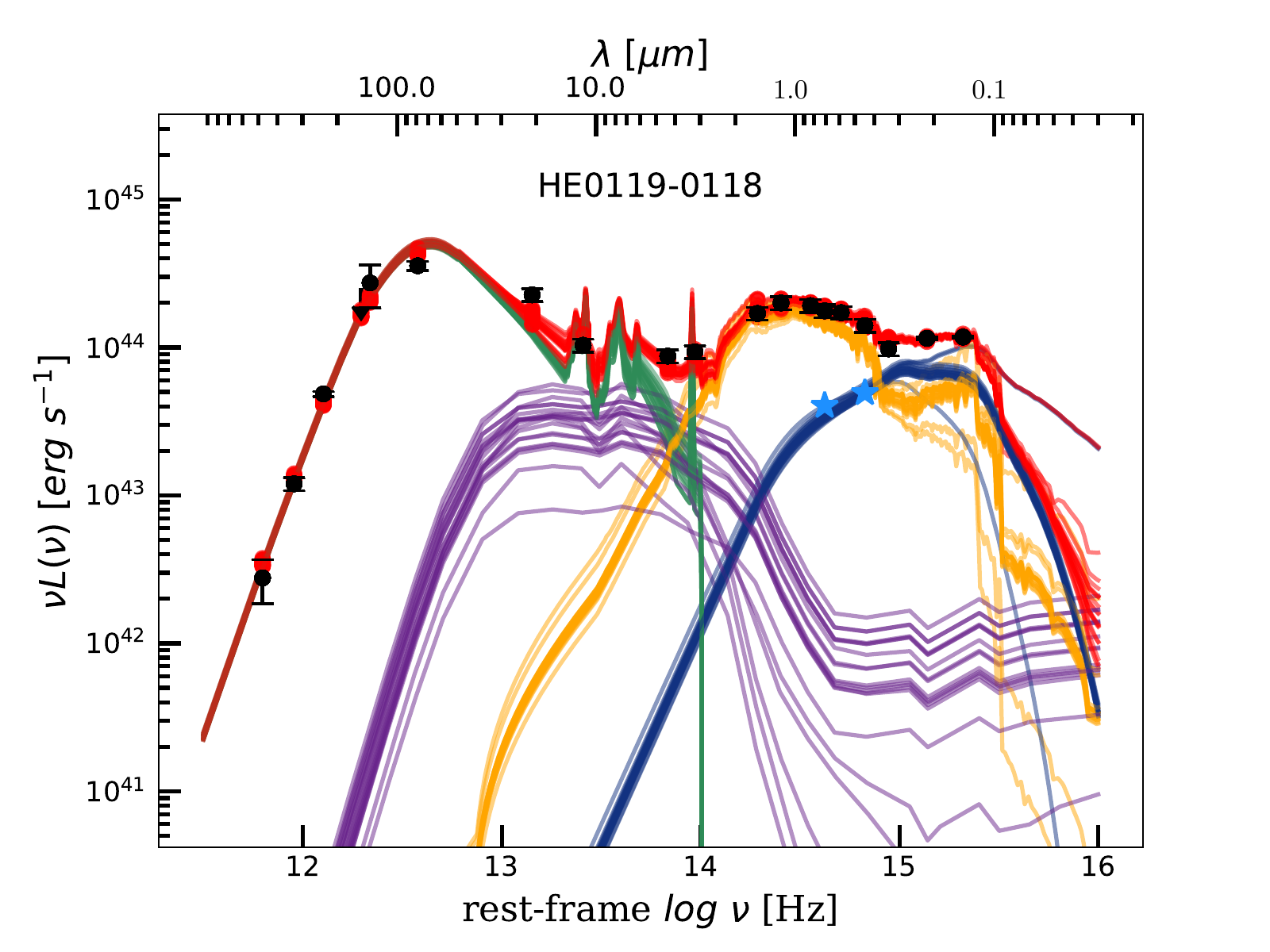}\\
        \includegraphics[width=0.33\textwidth]{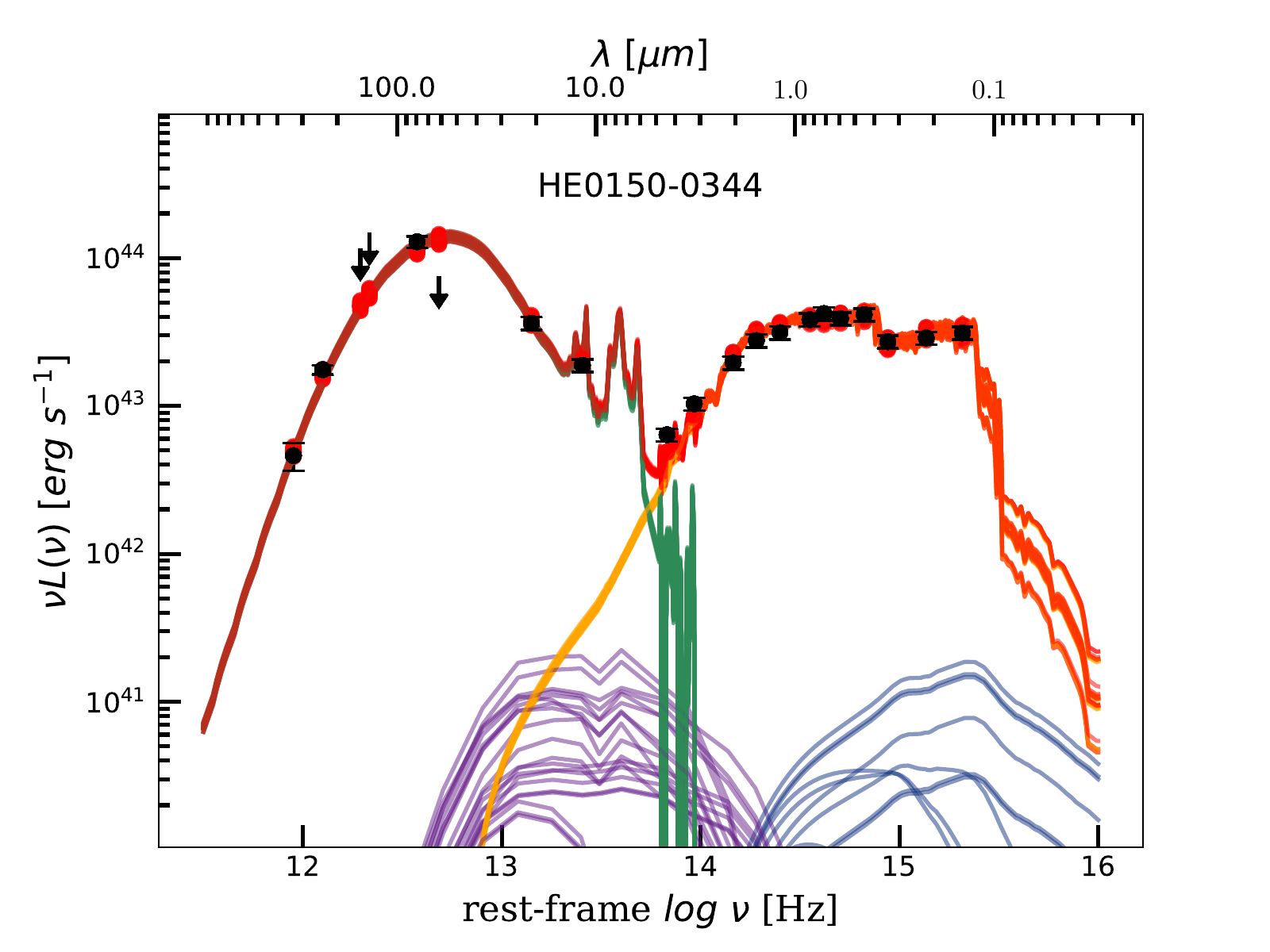}
        \includegraphics[width=0.33\textwidth]{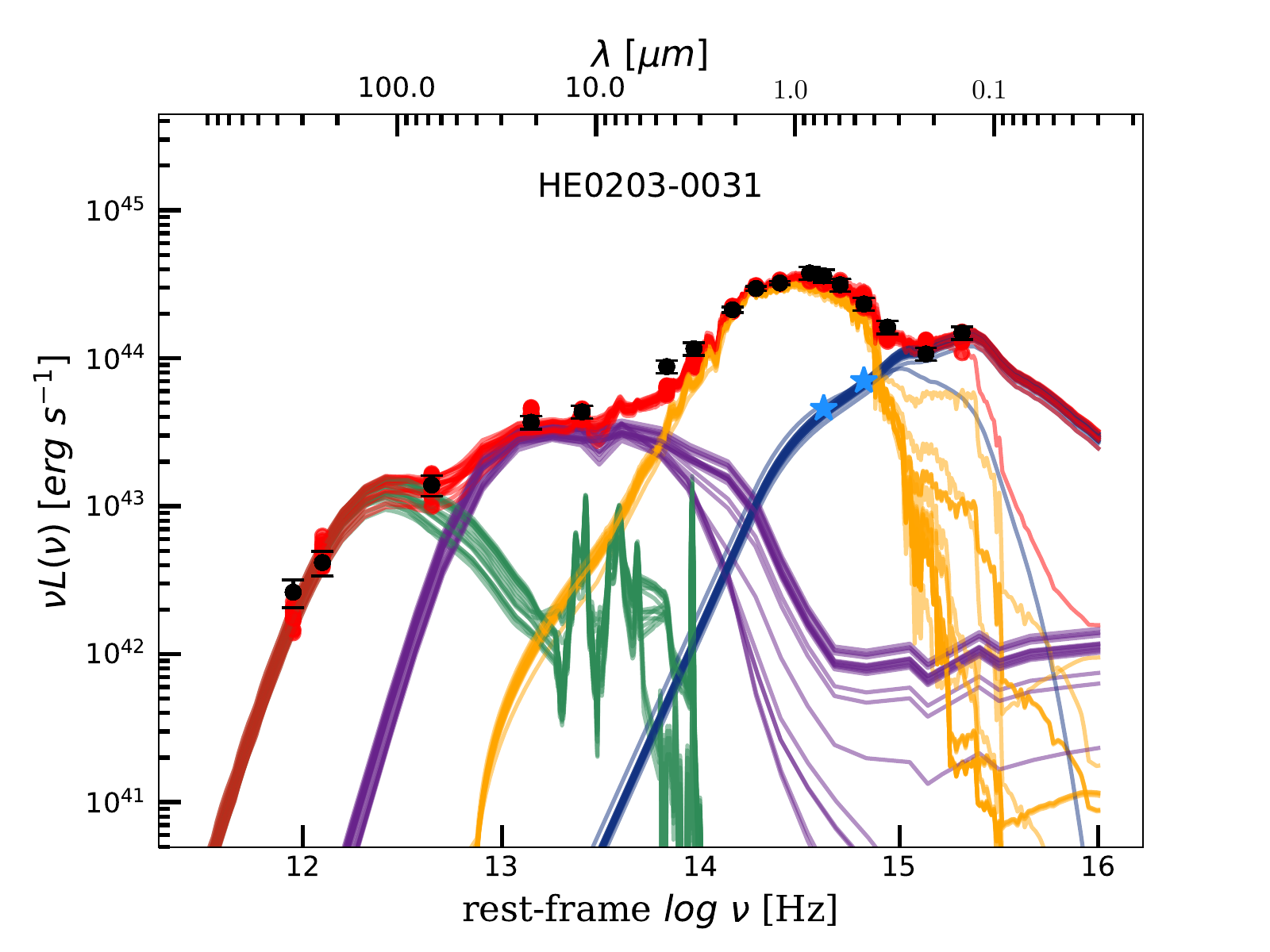}
        \includegraphics[width=0.33\textwidth]{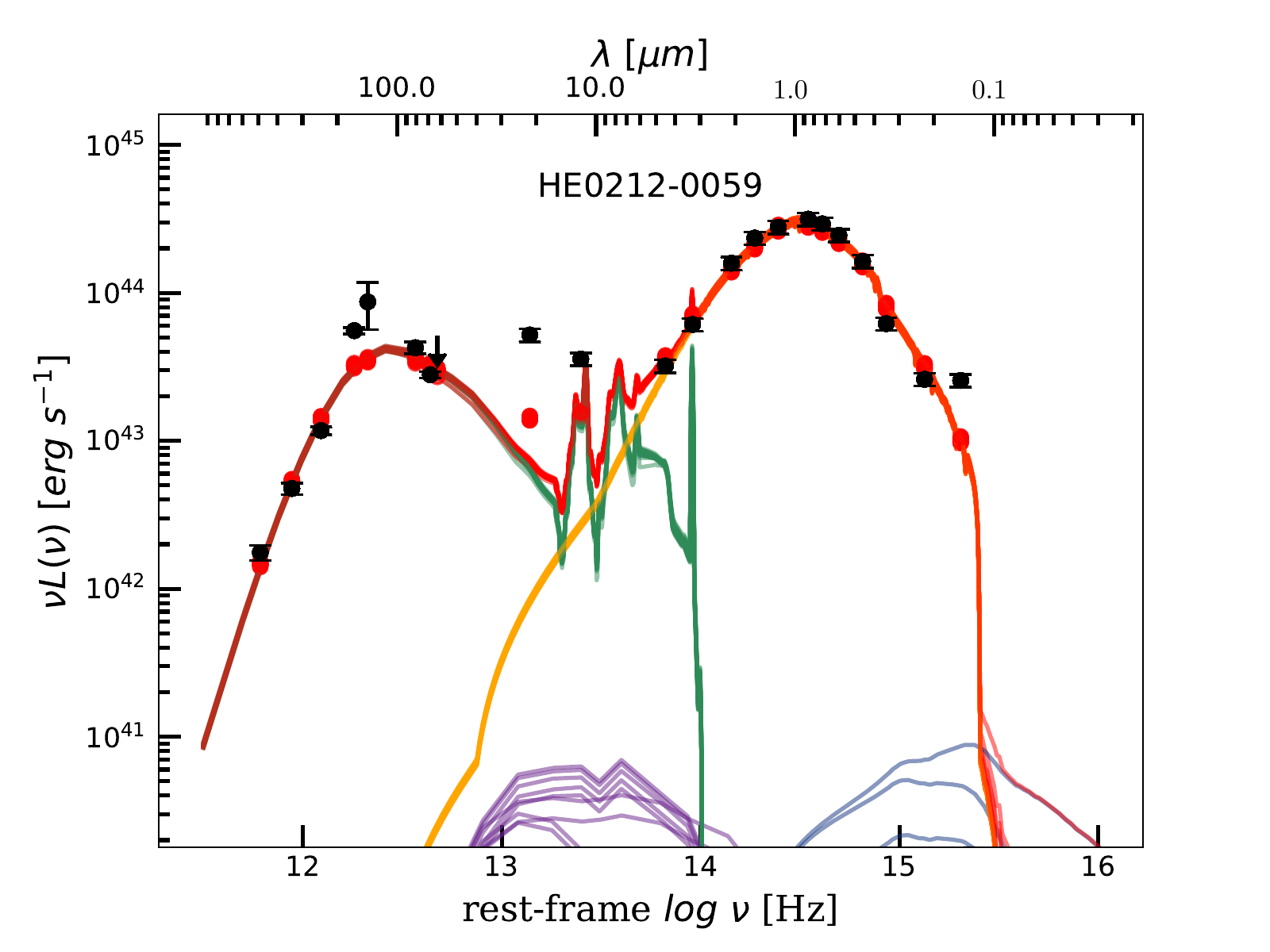}\\
        \includegraphics[width=0.33\textwidth]{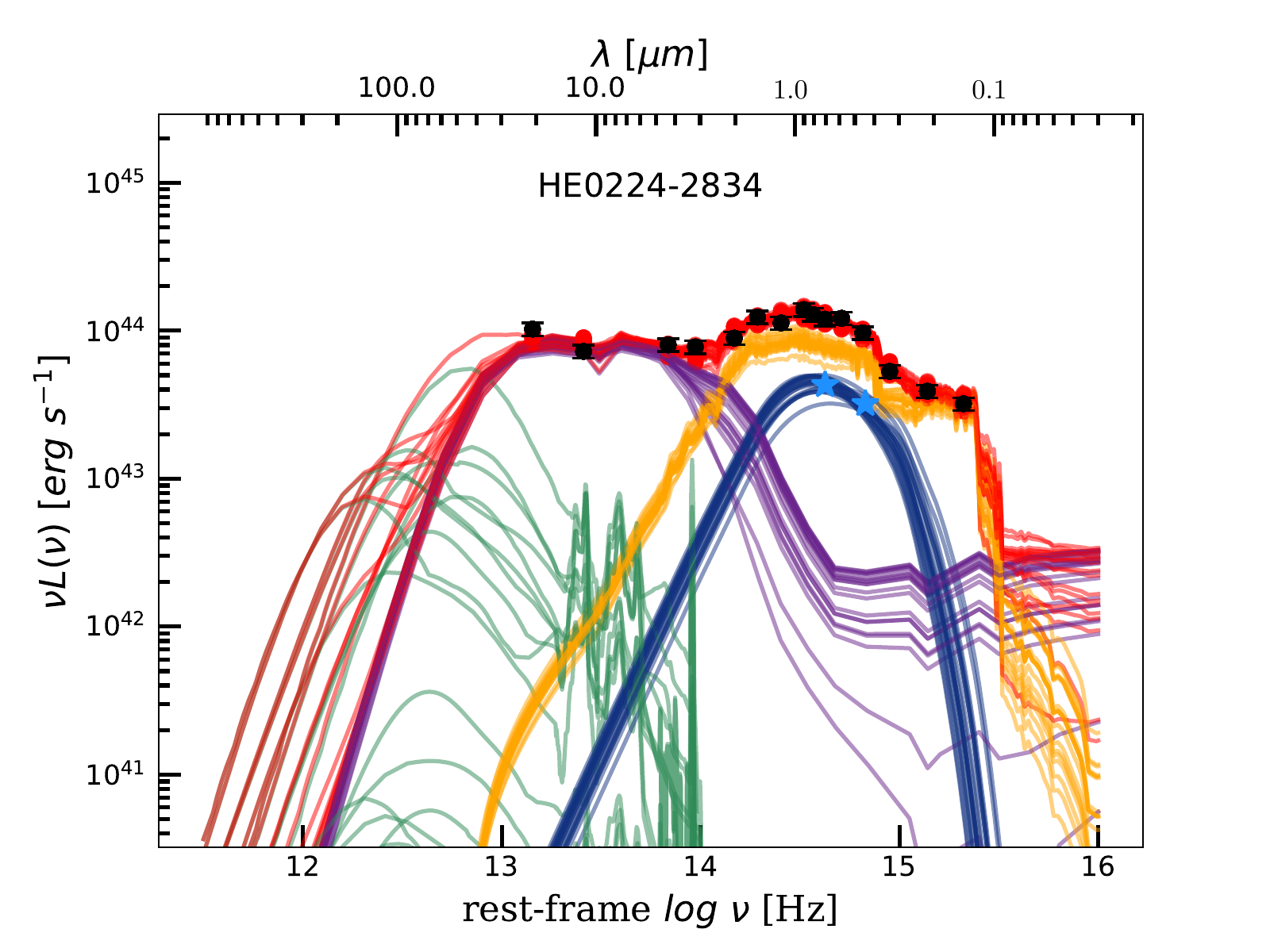}
        \includegraphics[width=0.33\textwidth]{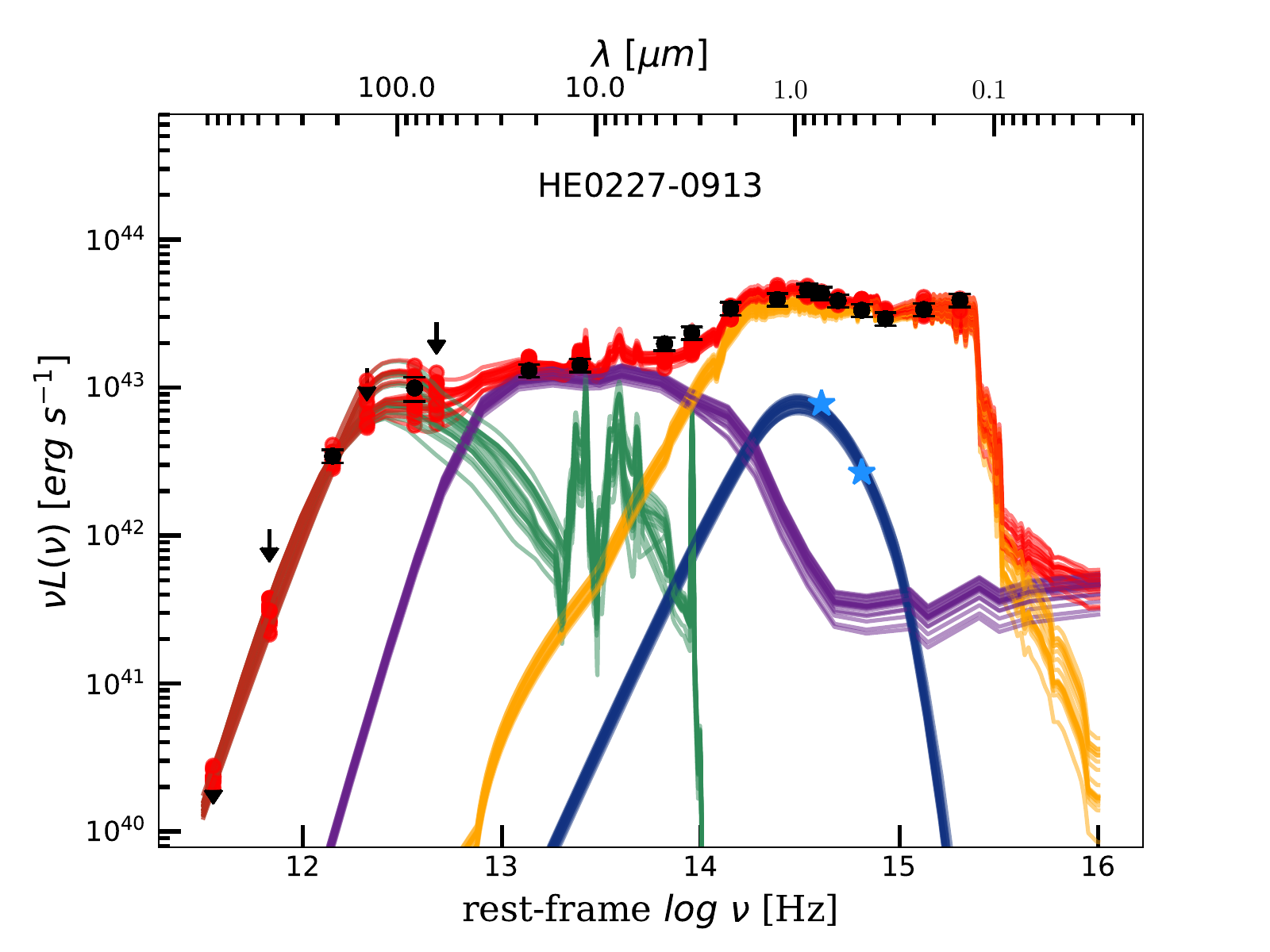}
        \includegraphics[width=0.33\textwidth]{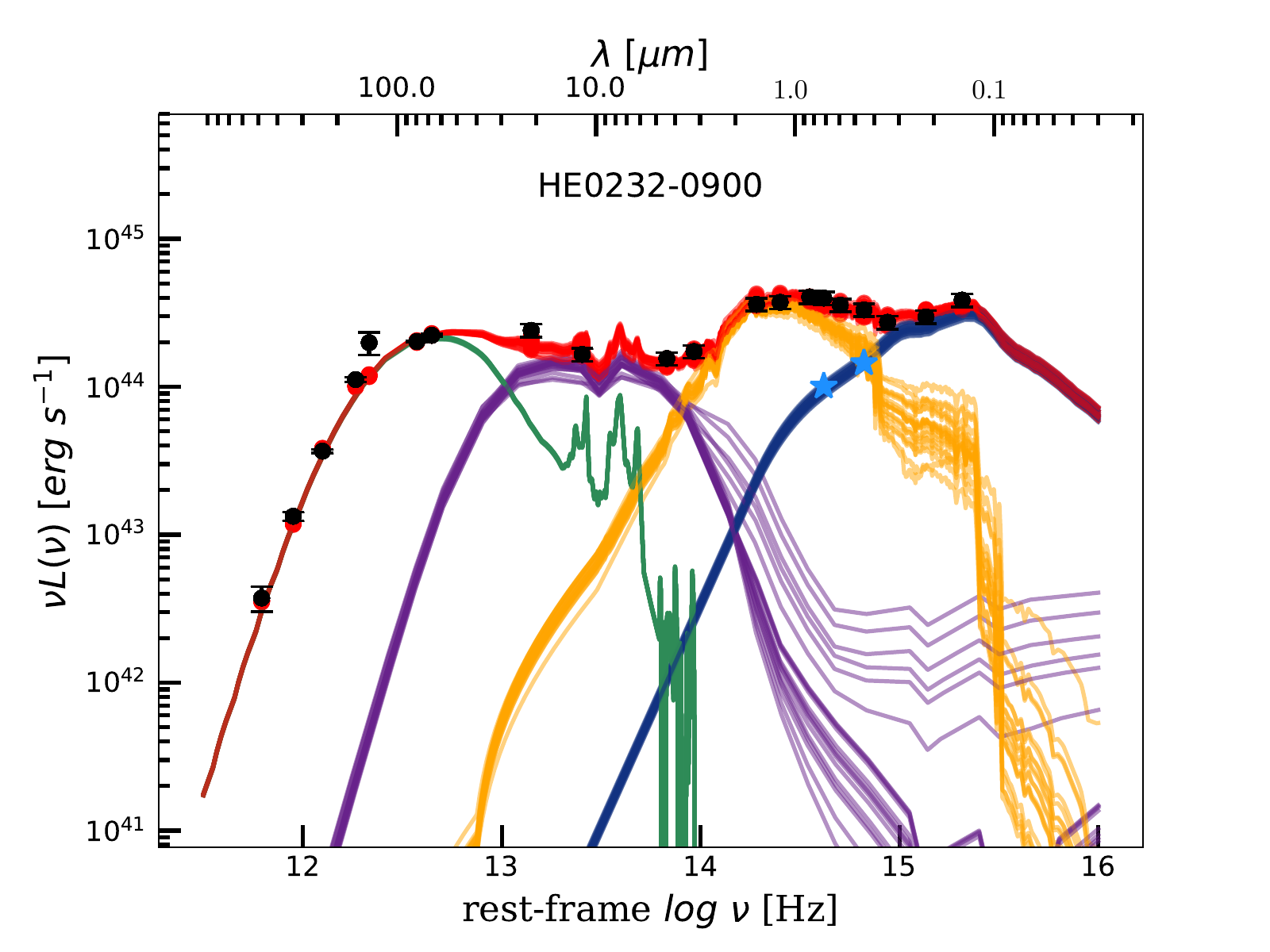}\\
        \includegraphics[width=0.33\textwidth]{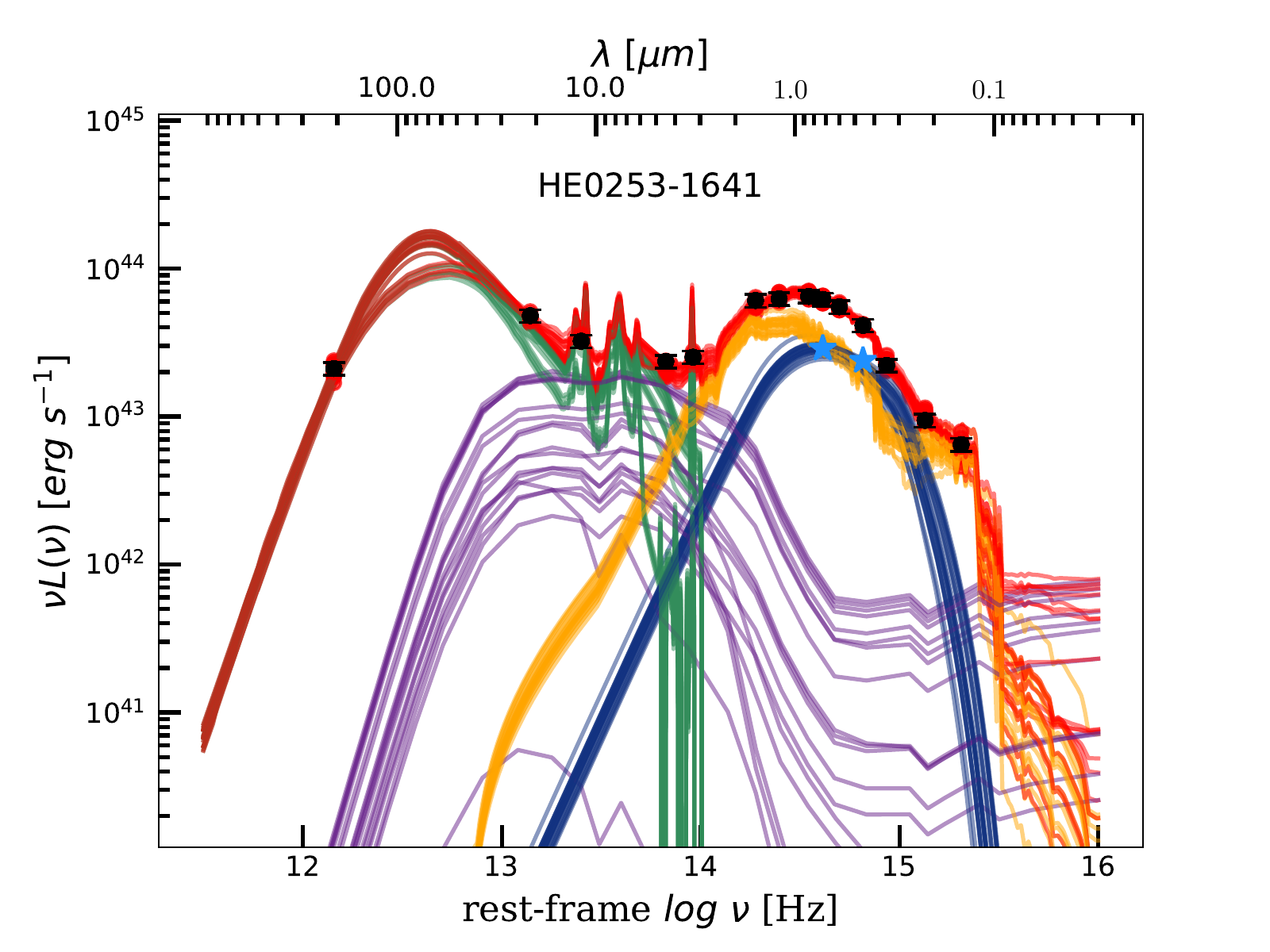}
        \includegraphics[width=0.33\textwidth]{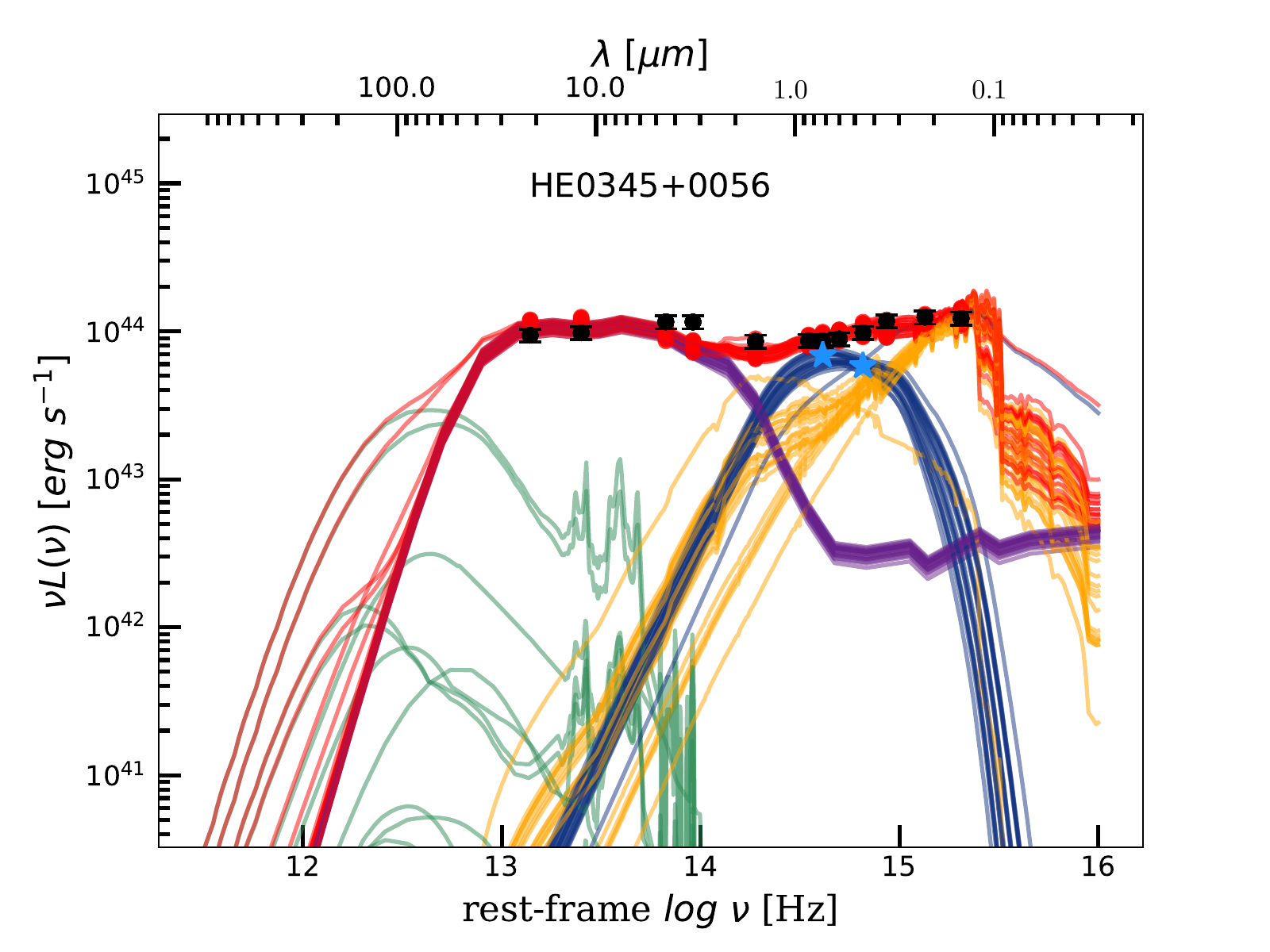}
        \includegraphics[width=0.33\textwidth]{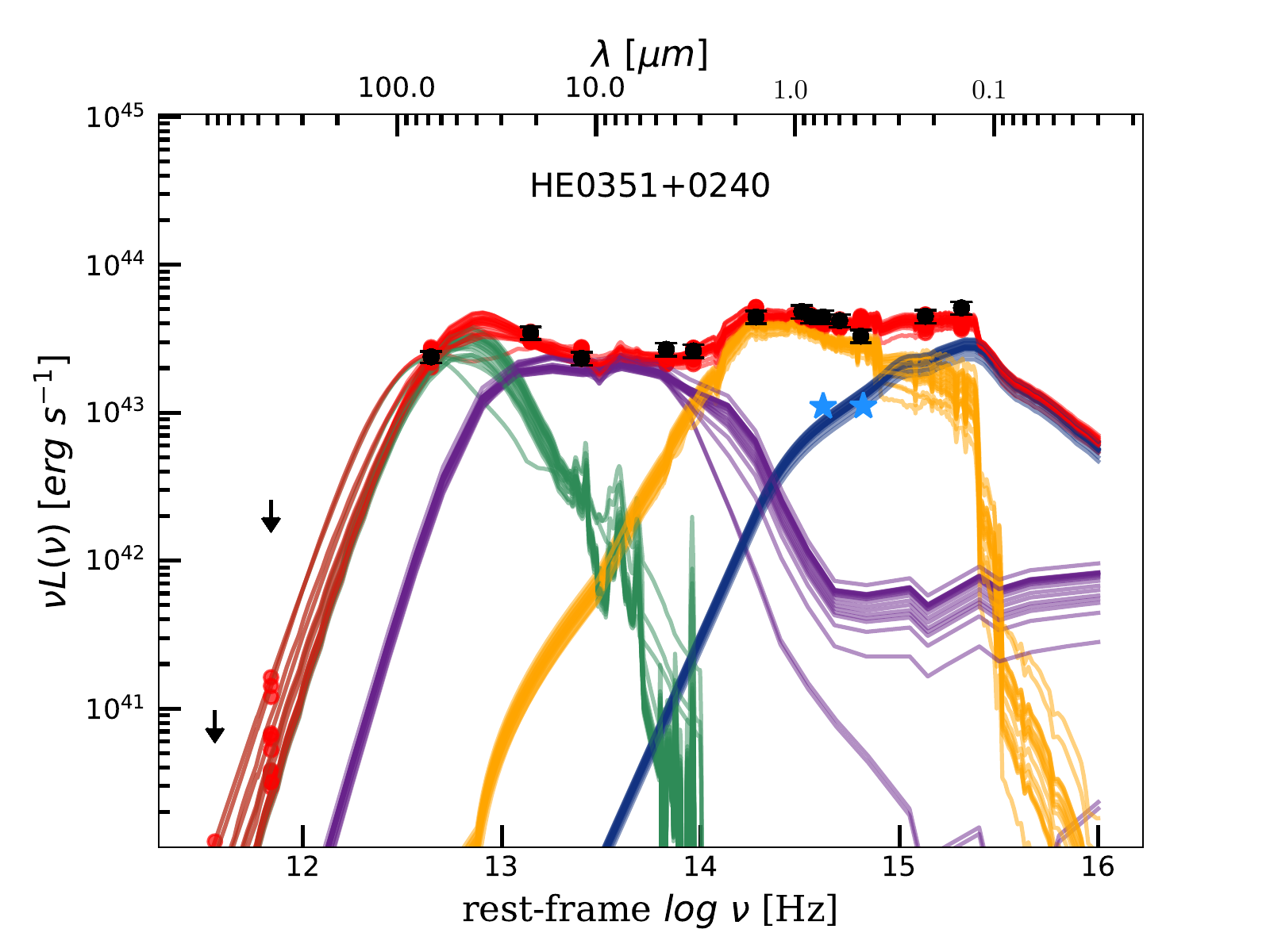}
         \caption{Overview of the \texttt{AGNfitter} SED modelling results for the entire CARS sample. Black data points represent the measured panchromatic photometry collected for each object. Upper limits are highlighted as black arrows. The red dots are the predicted photometry for XX MCMC relations from the superposition of individual SED components. The red, yellow, blue, purple and green lines correspond to the total SED, the stellar component, the AGN component, the hot dust component and the cold dust componet, respectively. }\label{fig:SED_apx}
 \end{figure*}
\addtocounter{figure}{-1}
\begin{figure*}        
        \includegraphics[width=0.33\textwidth]{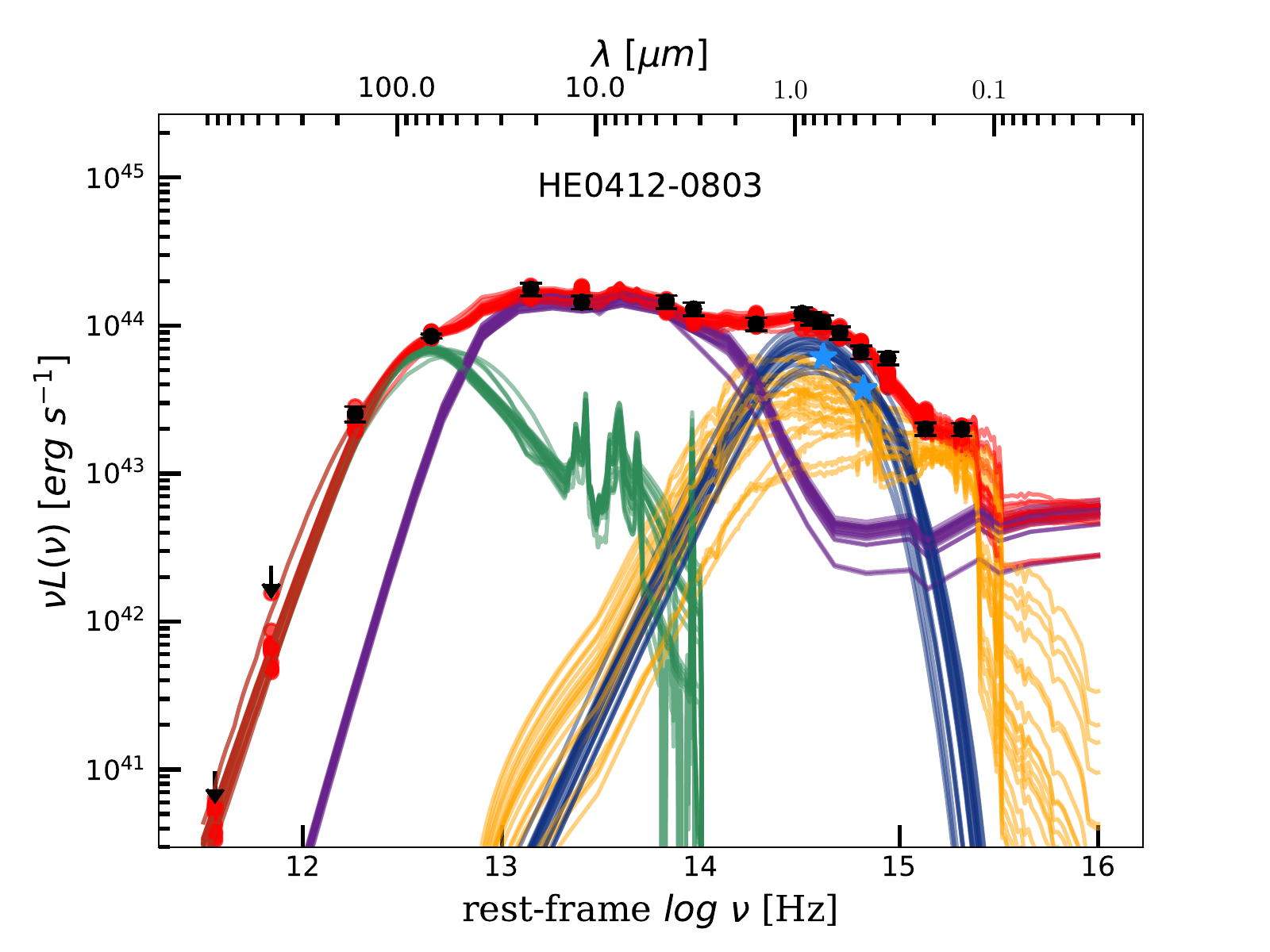}
        \includegraphics[width=0.33\textwidth]{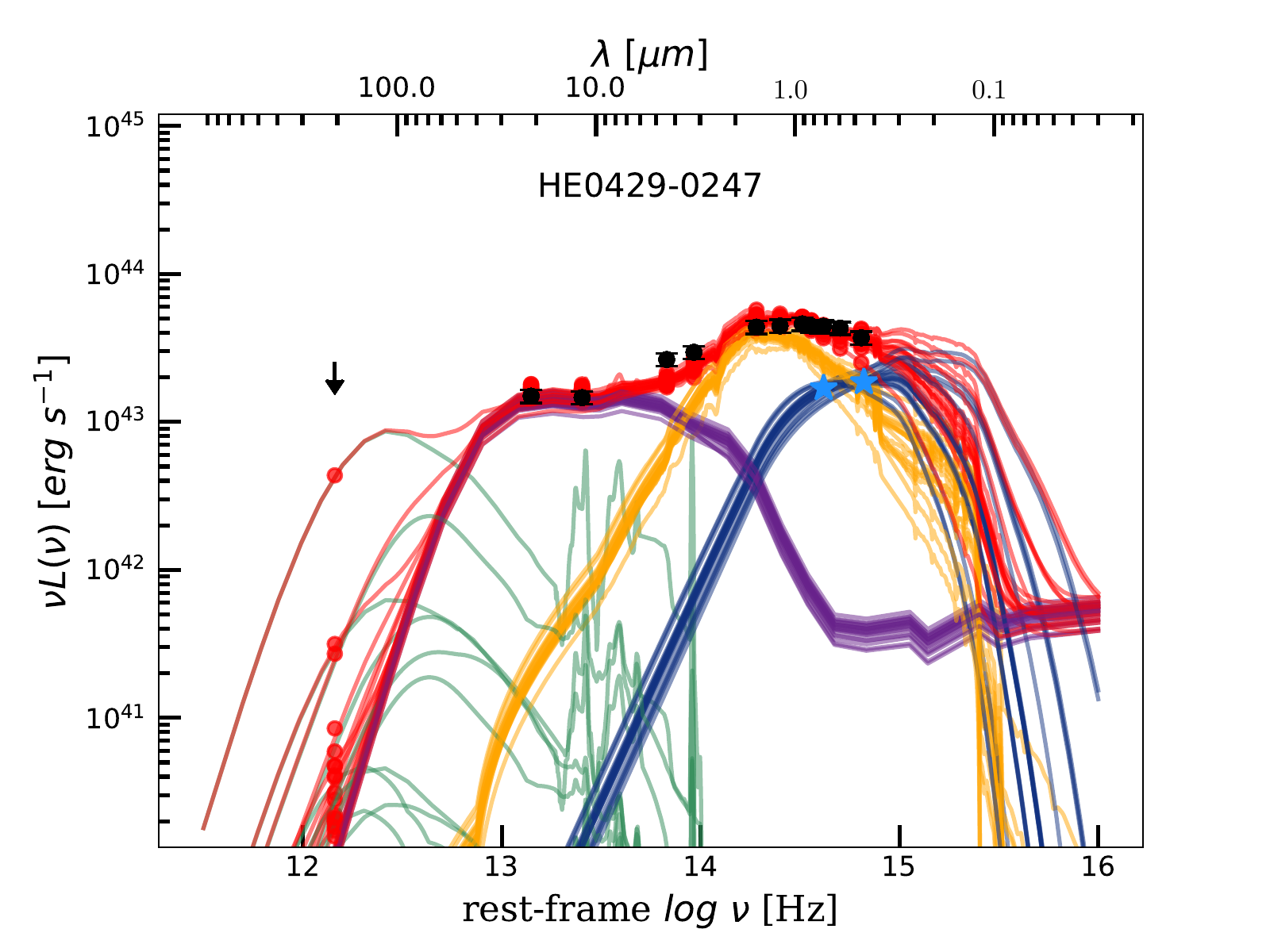}
        \includegraphics[width=0.33\textwidth]{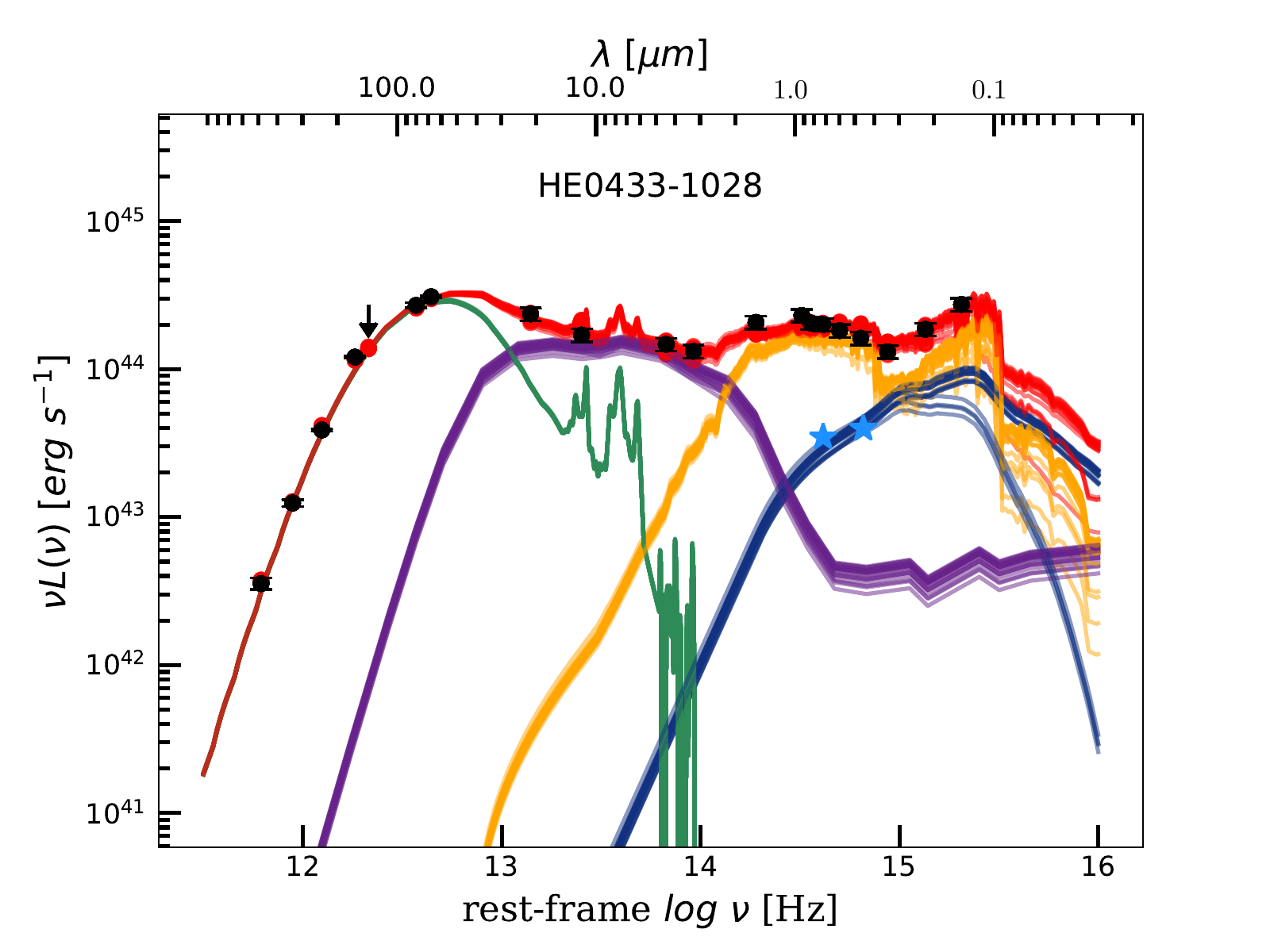}\\
        \includegraphics[width=0.33\textwidth]{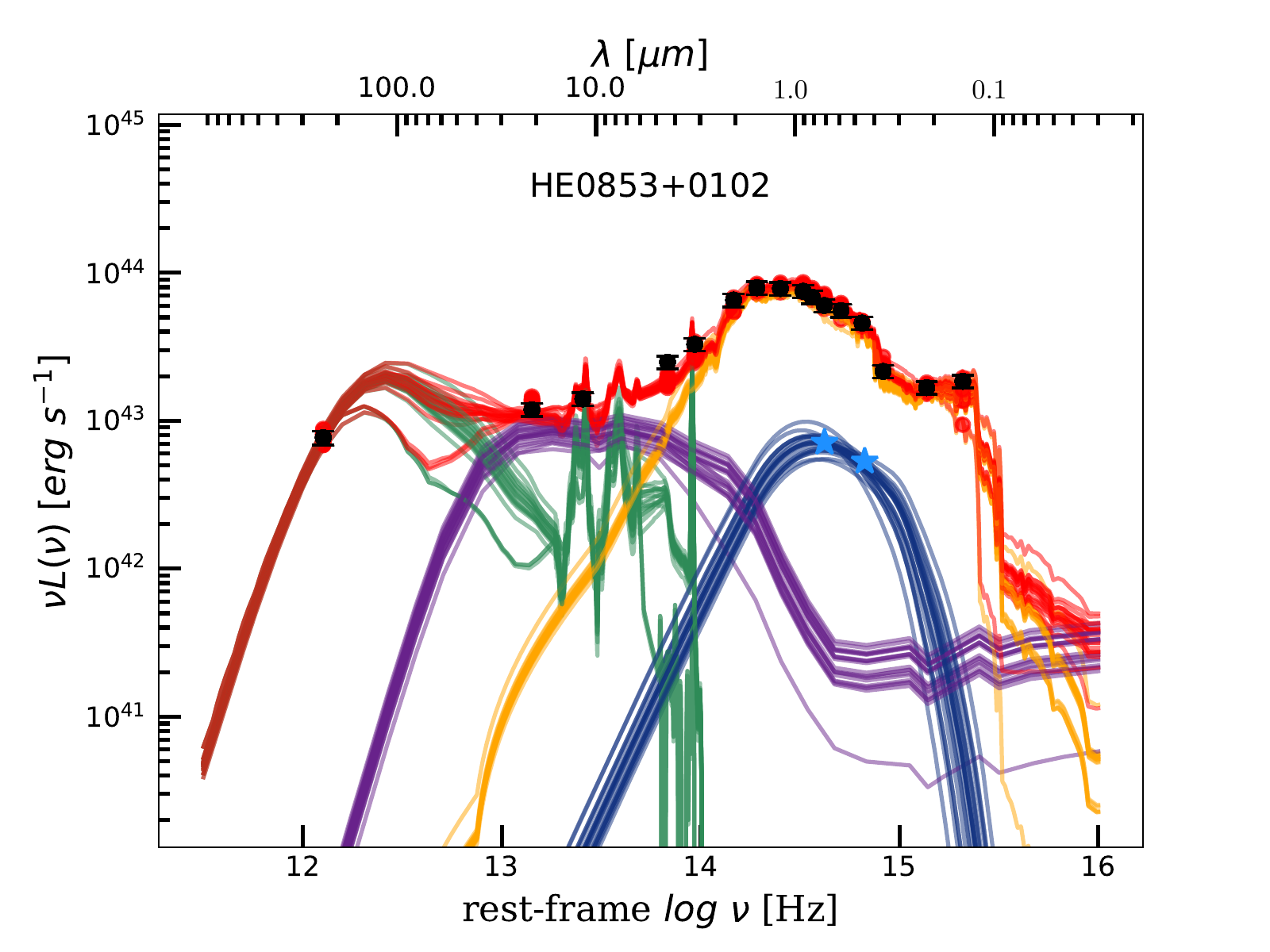}
        \includegraphics[width=0.33\textwidth]{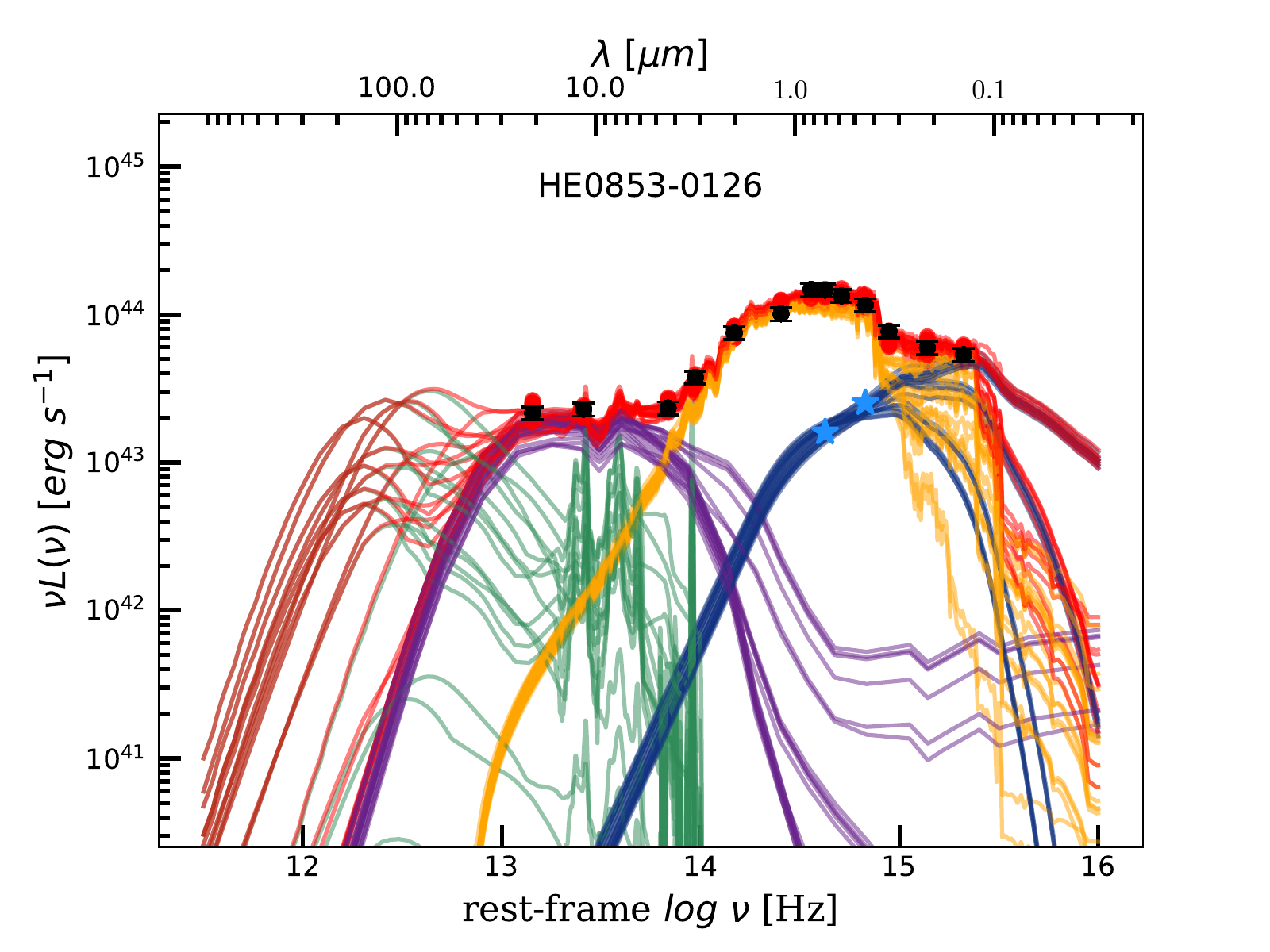}
        \includegraphics[width=0.33\textwidth]{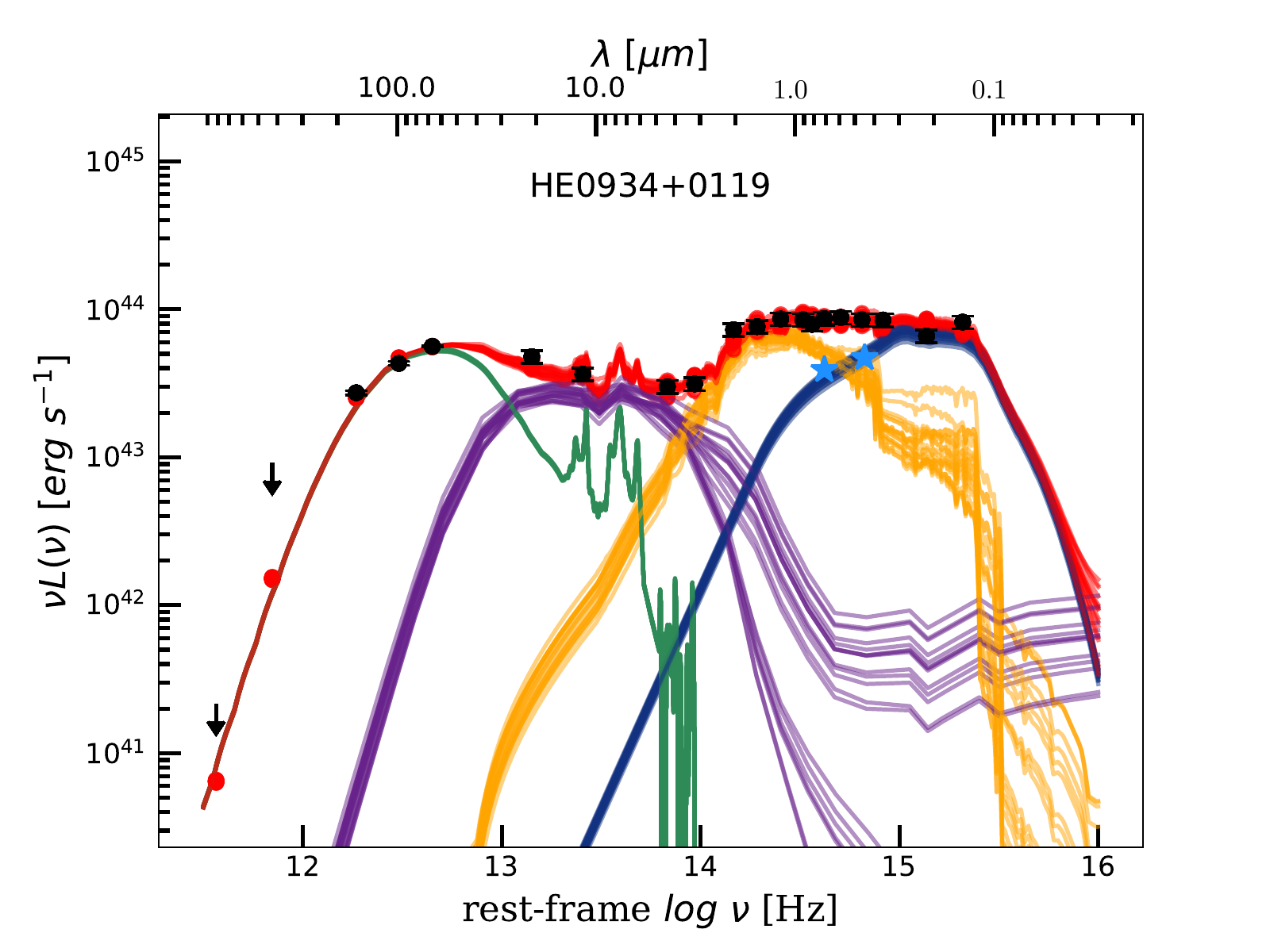}\\
        \includegraphics[width=0.33\textwidth]{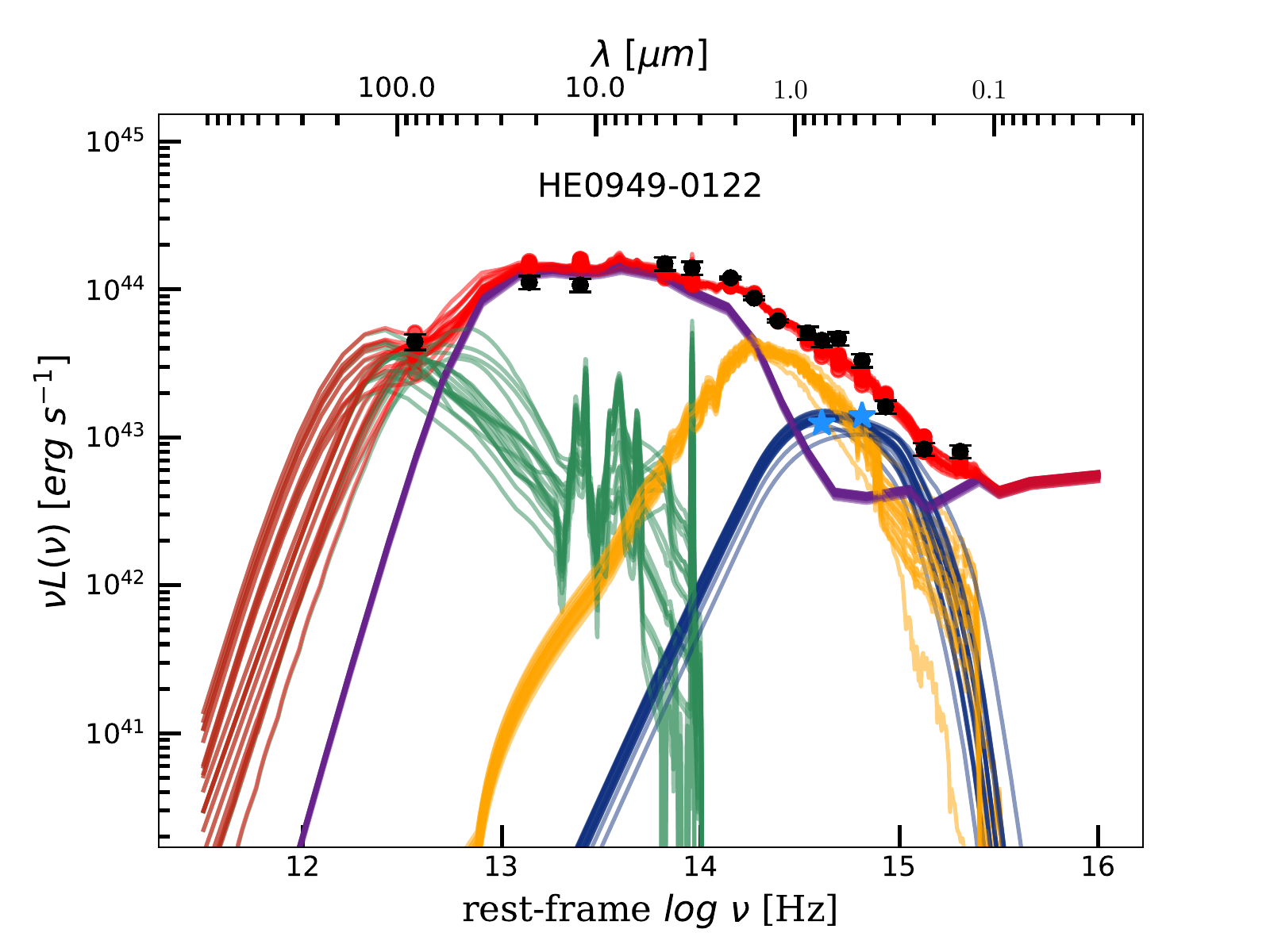}
        \includegraphics[width=0.33\textwidth]{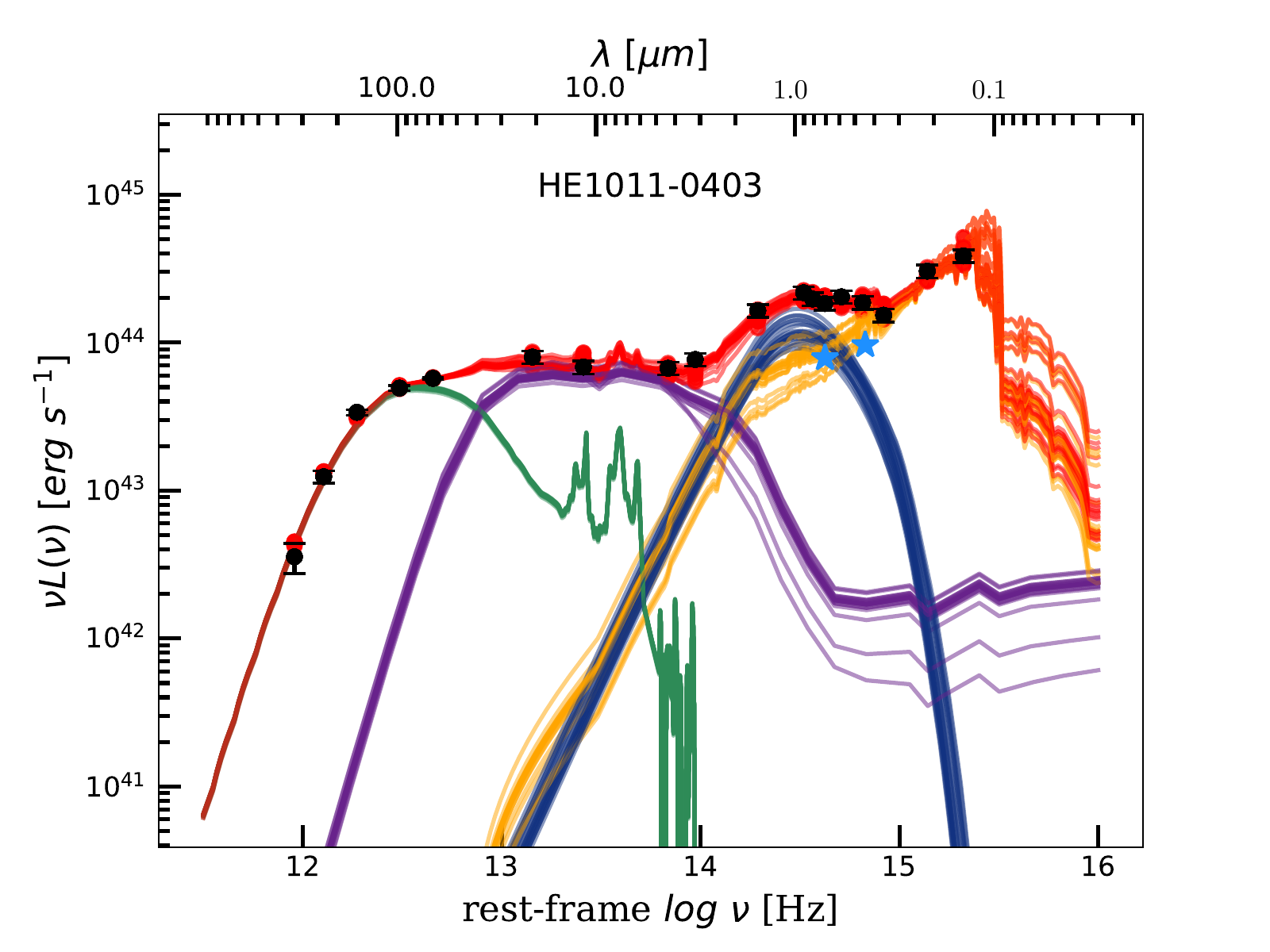}
        \includegraphics[width=0.33\textwidth]{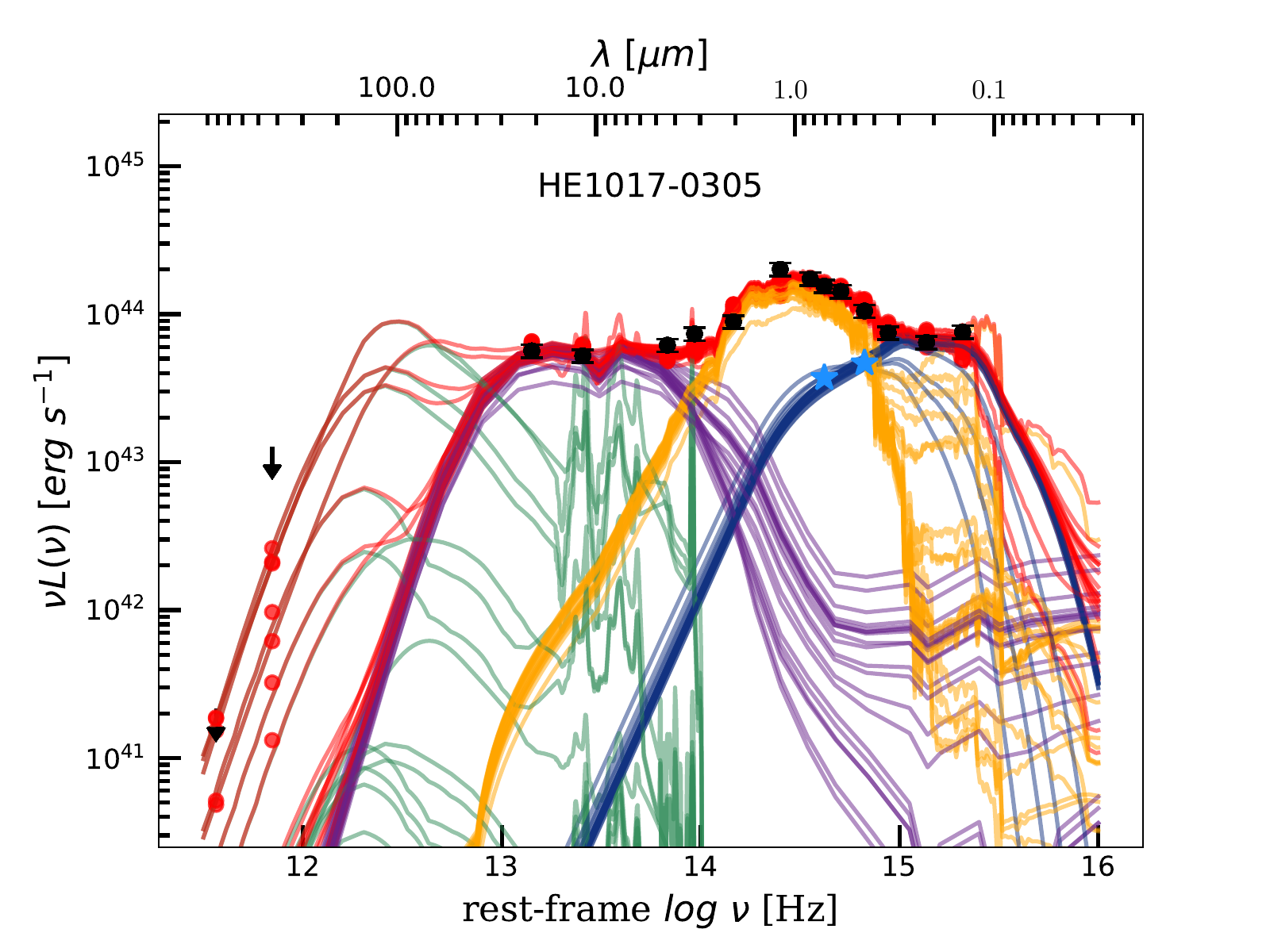}
        \includegraphics[width=0.33\textwidth]{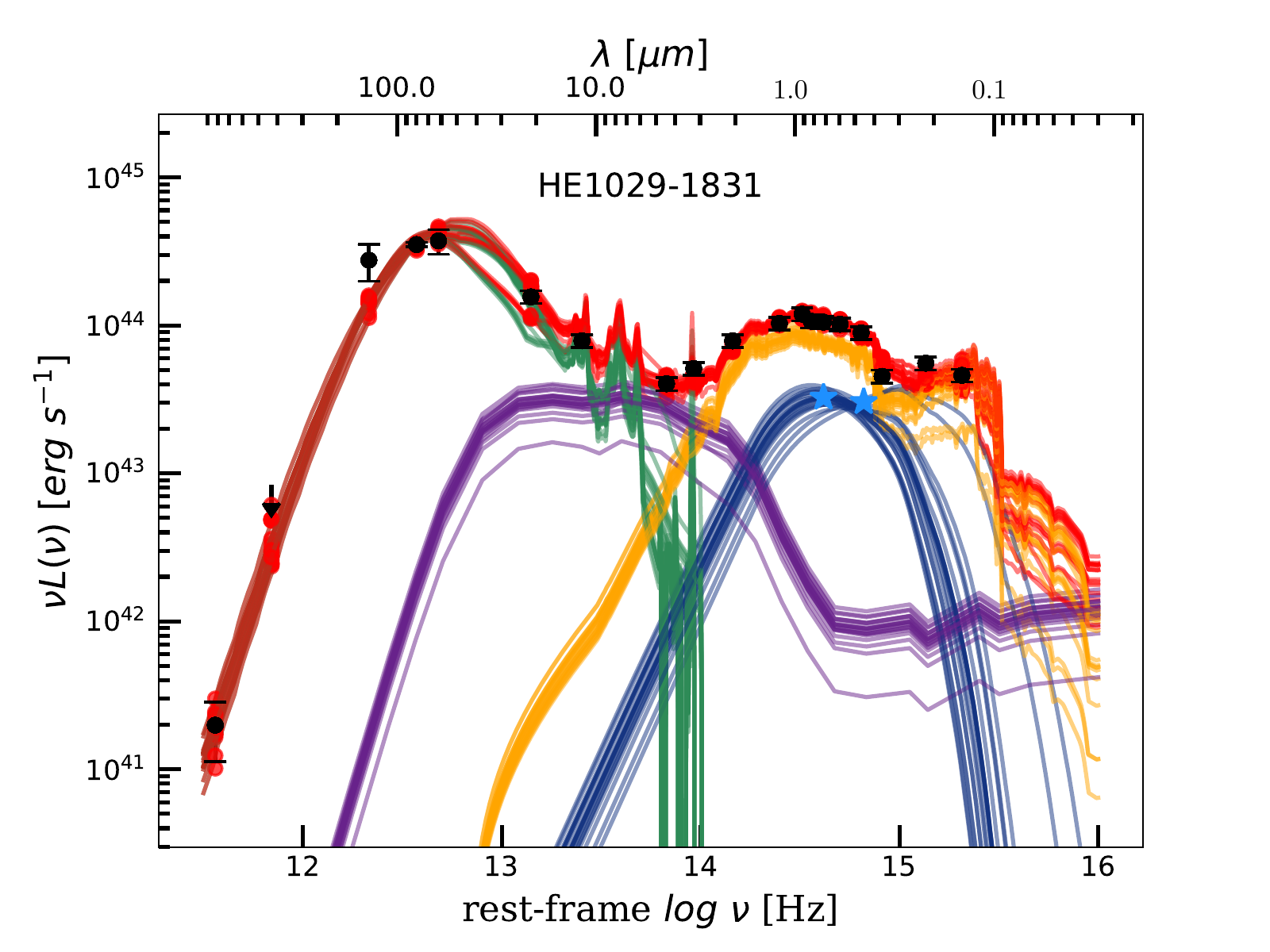}
        \includegraphics[width=0.33\textwidth]{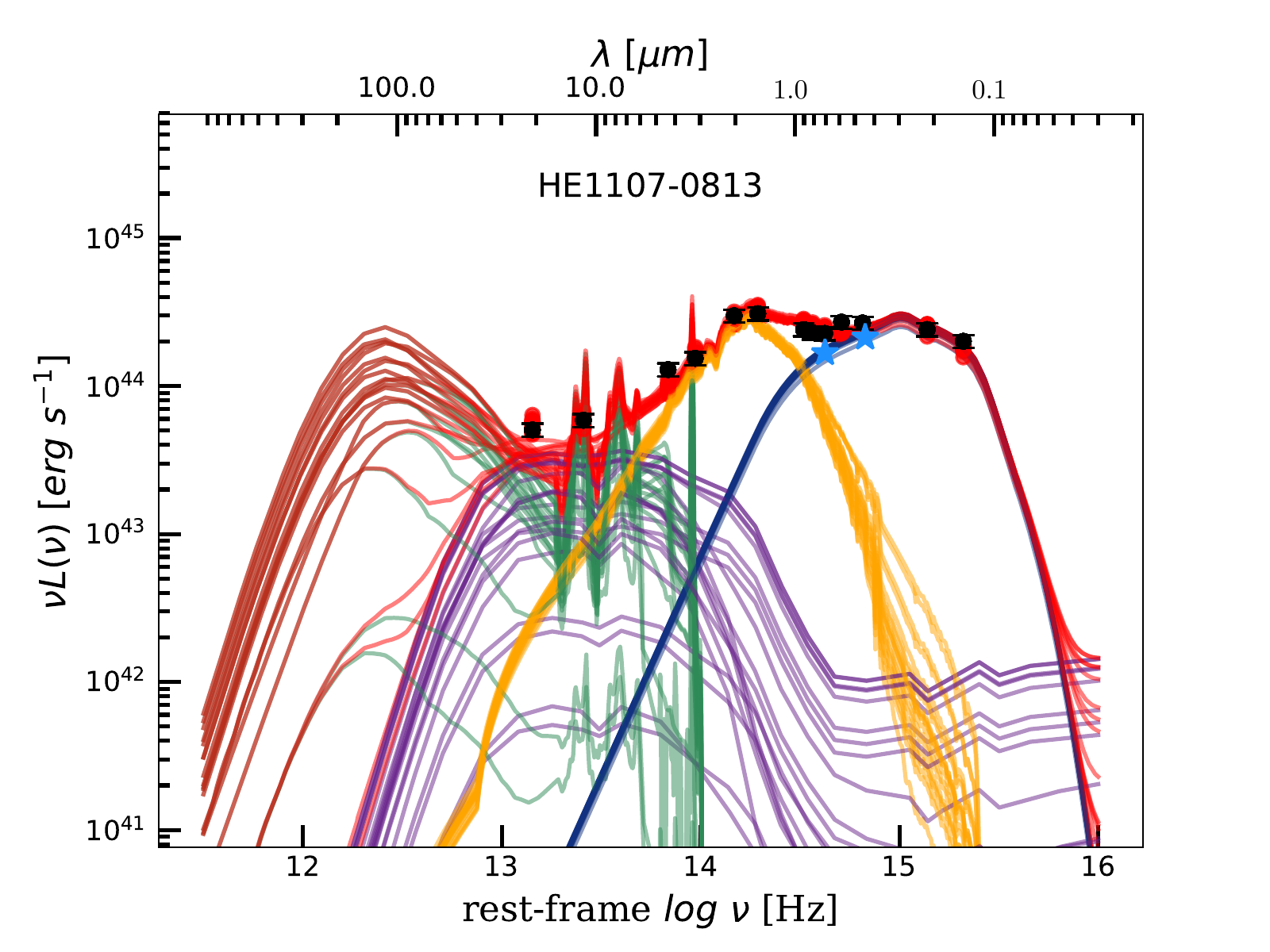}
        \includegraphics[width=0.33\textwidth]{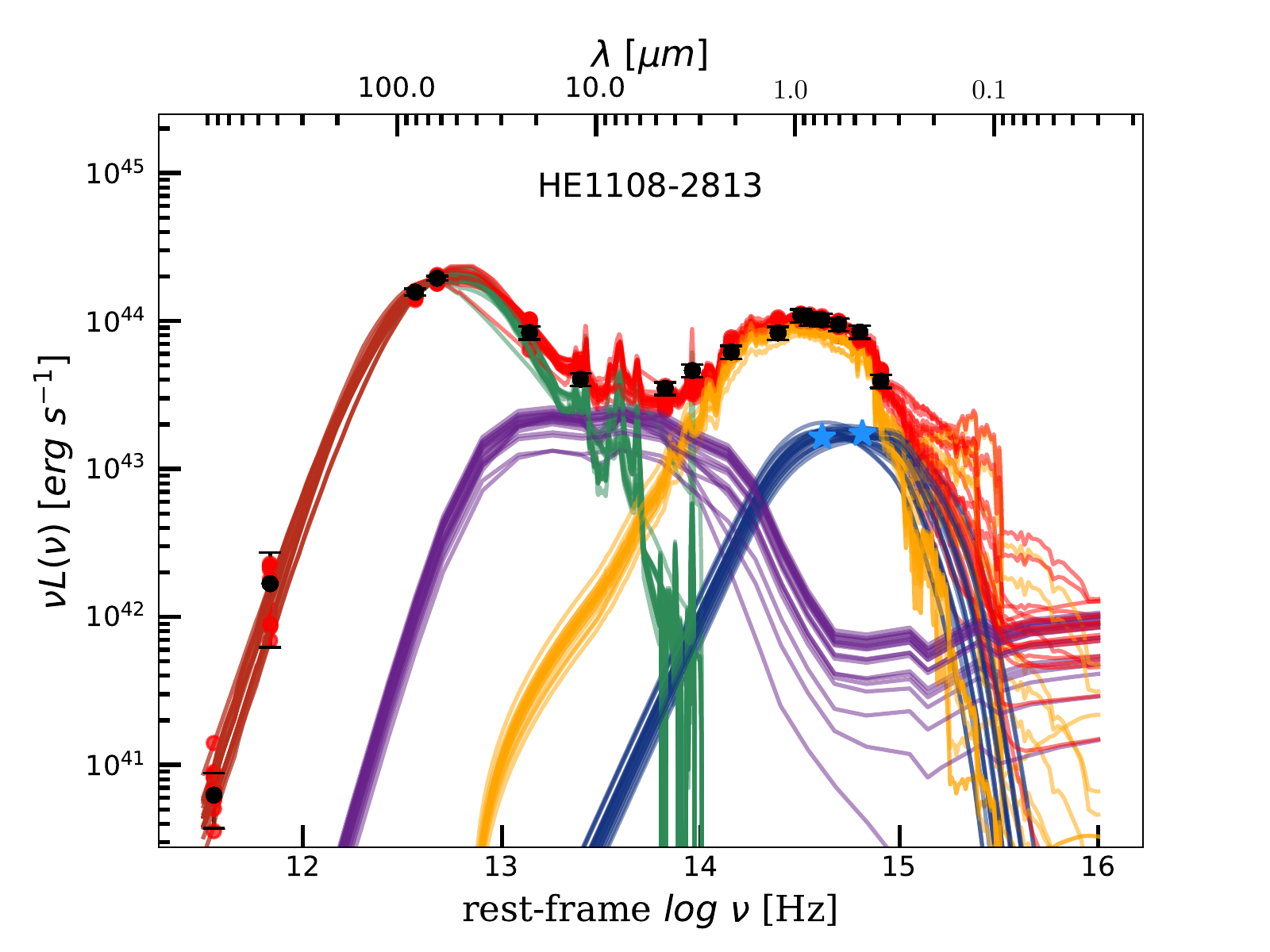}\\
        \includegraphics[width=0.33\textwidth]{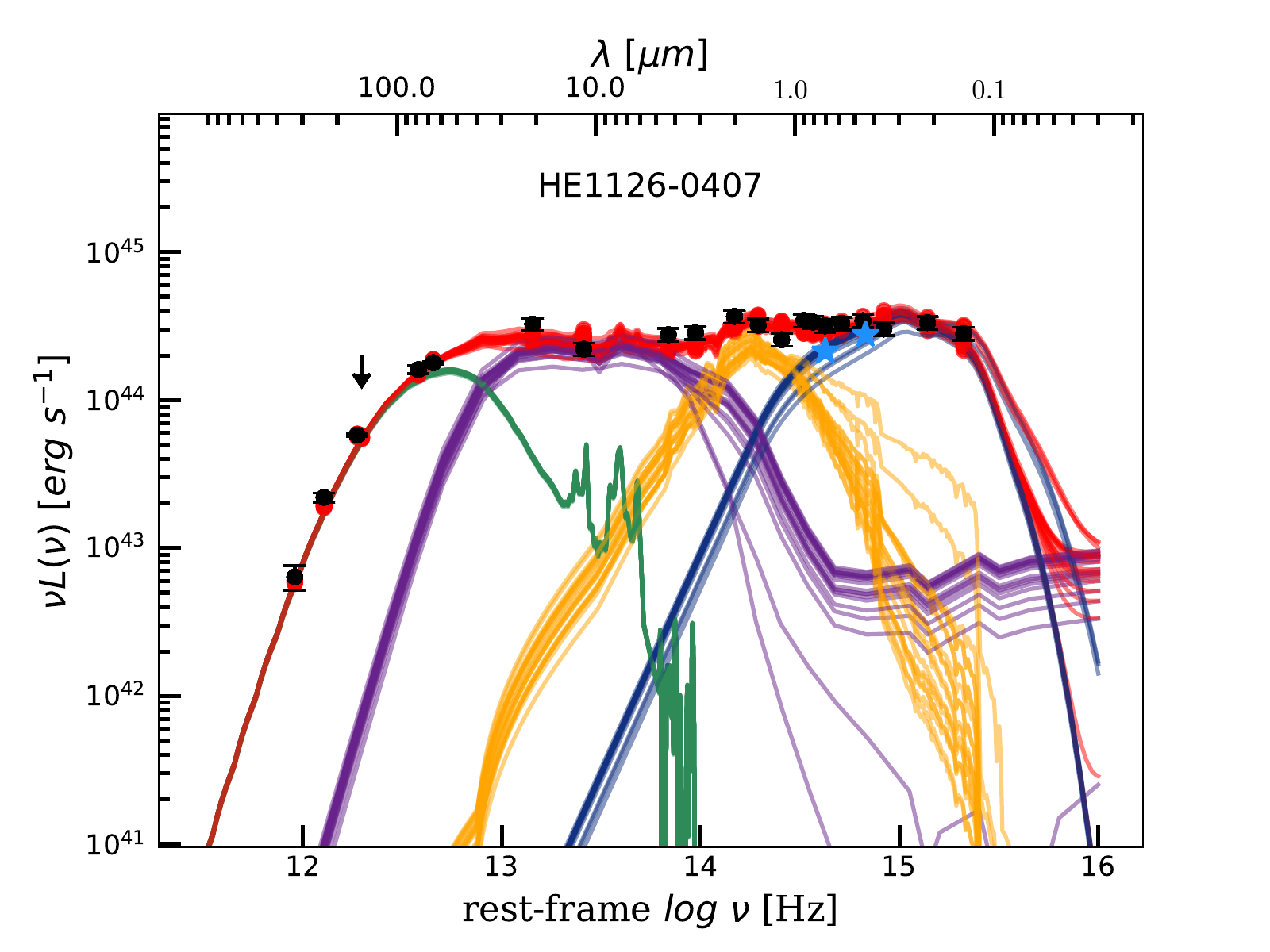}
        \includegraphics[width=0.33\textwidth]{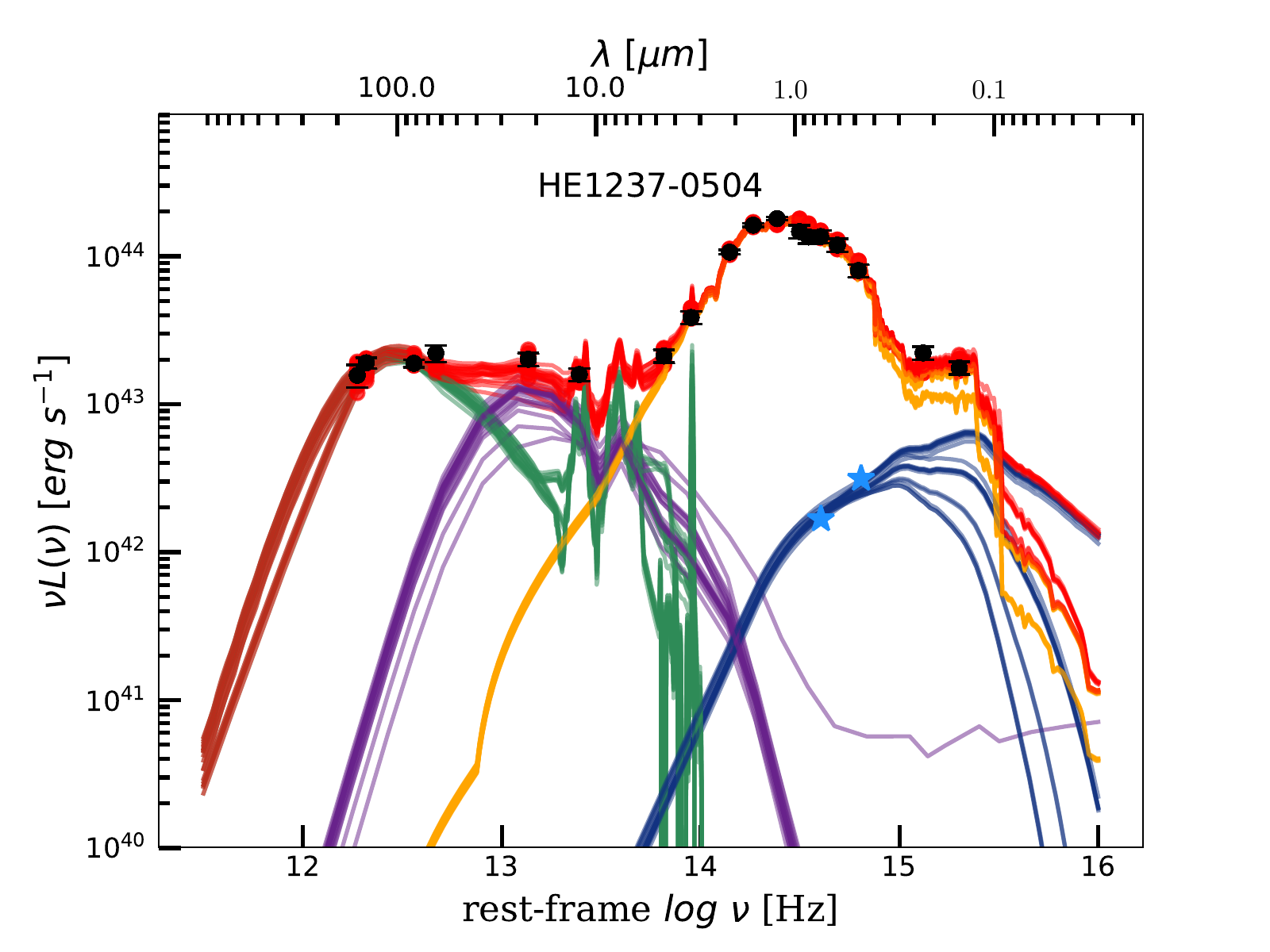}
        \includegraphics[width=0.33\textwidth]{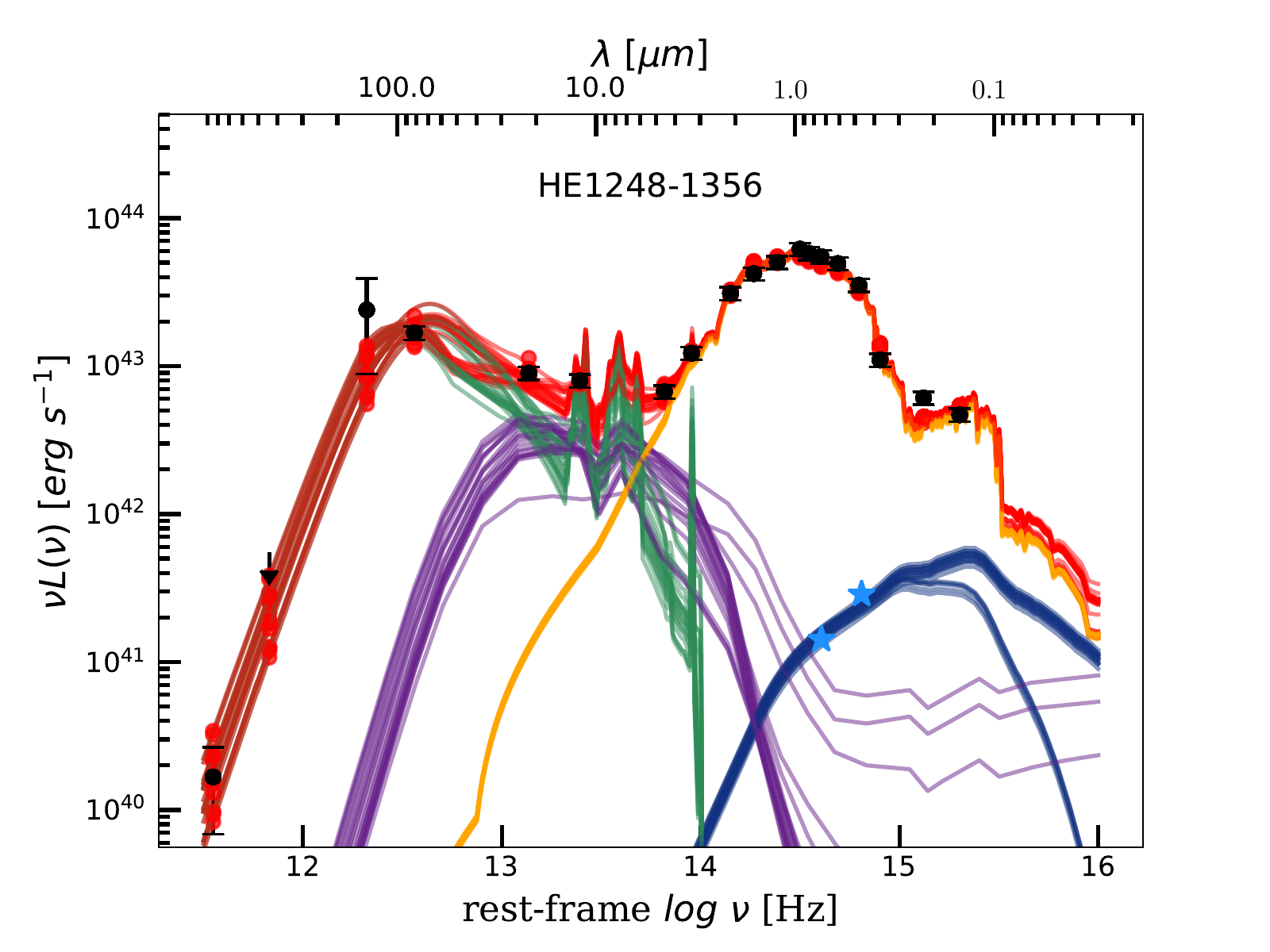}
        \caption{Continued.}
\end{figure*}
\addtocounter{figure}{-1}
\begin{figure*}
        \includegraphics[width=0.33\textwidth]{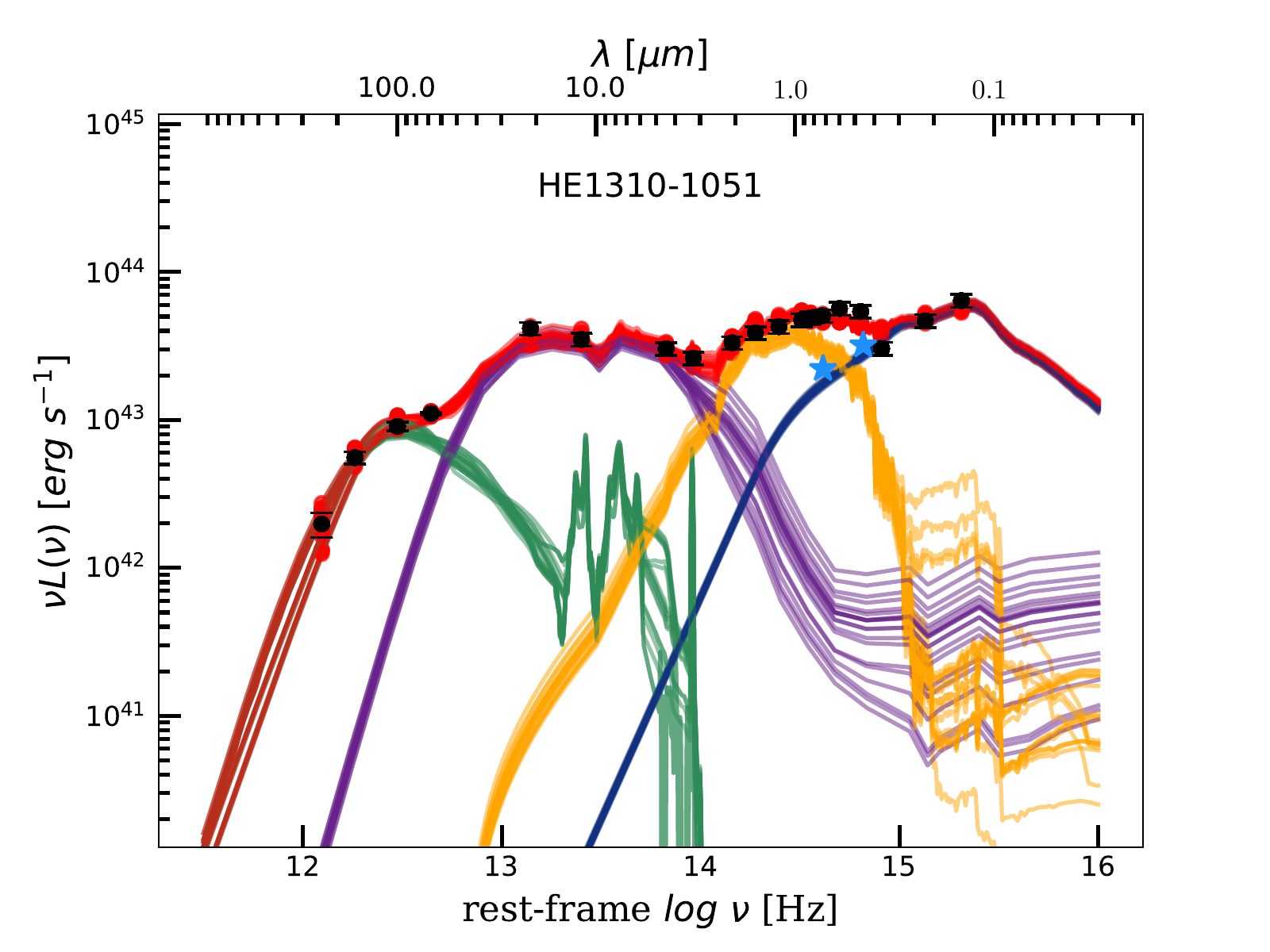}
        \includegraphics[width=0.33\textwidth]{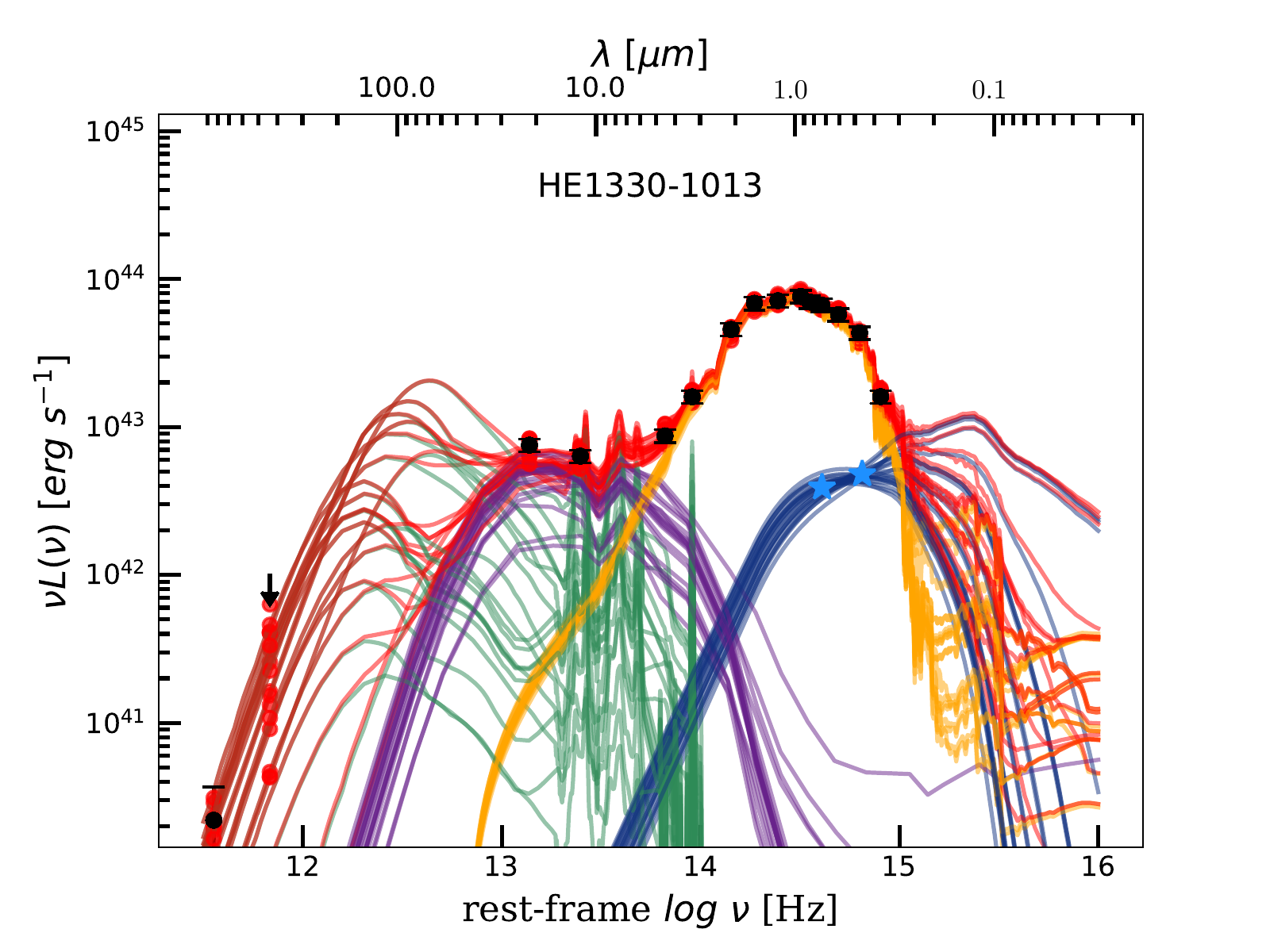}
        \includegraphics[width=0.33\textwidth]{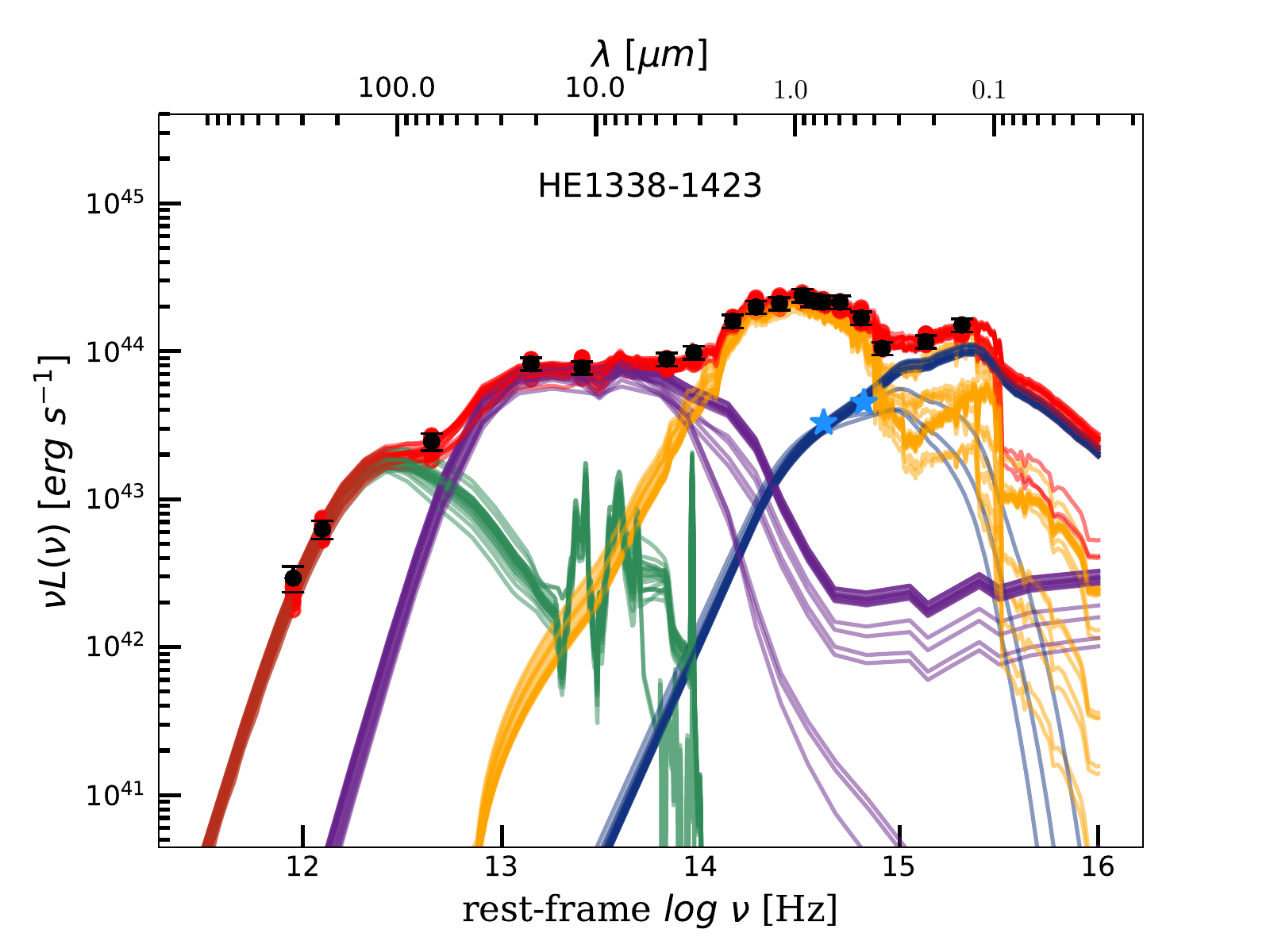}\\
        \includegraphics[width=0.33\textwidth]{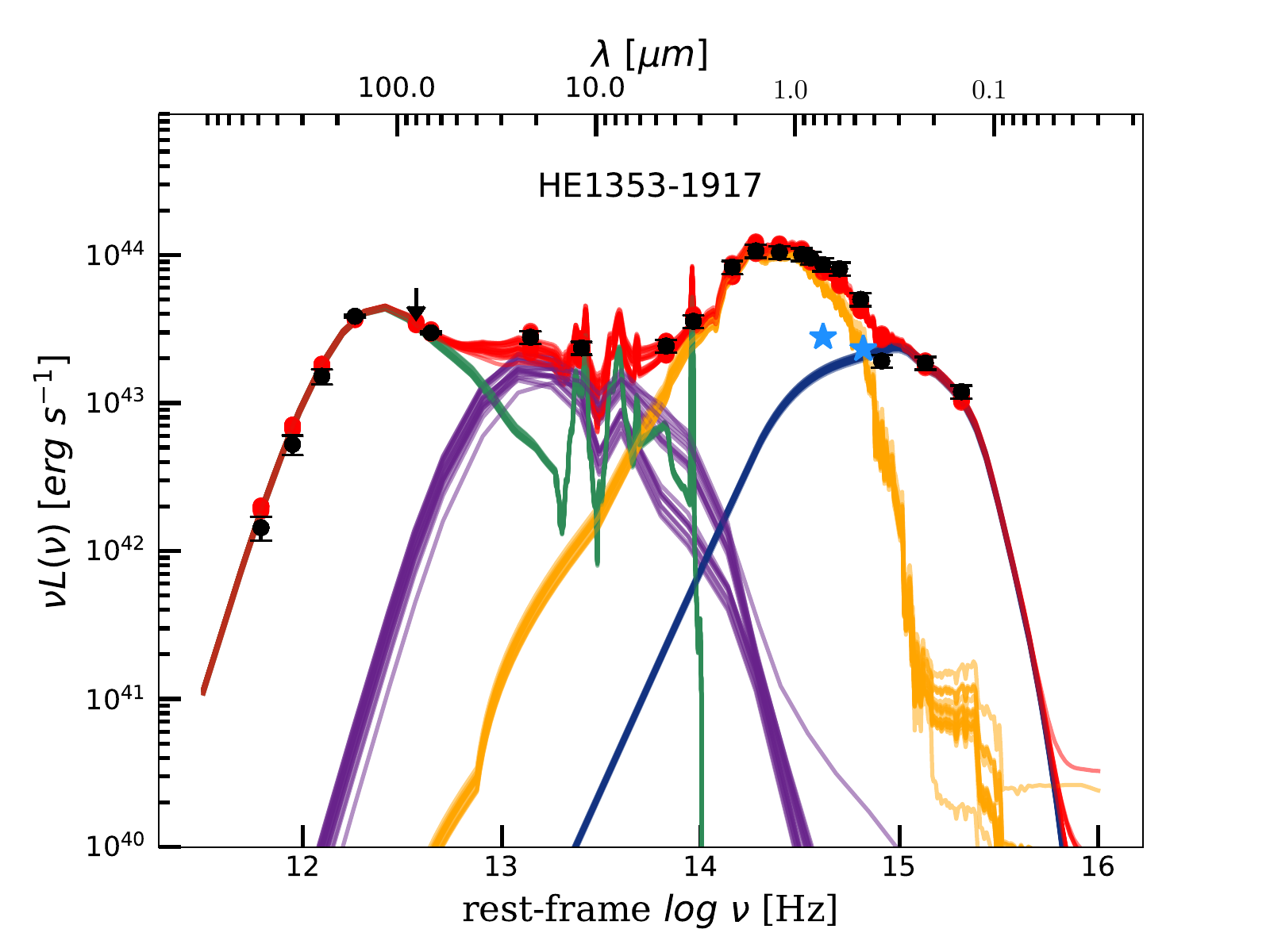}
        \includegraphics[width=0.33\textwidth]{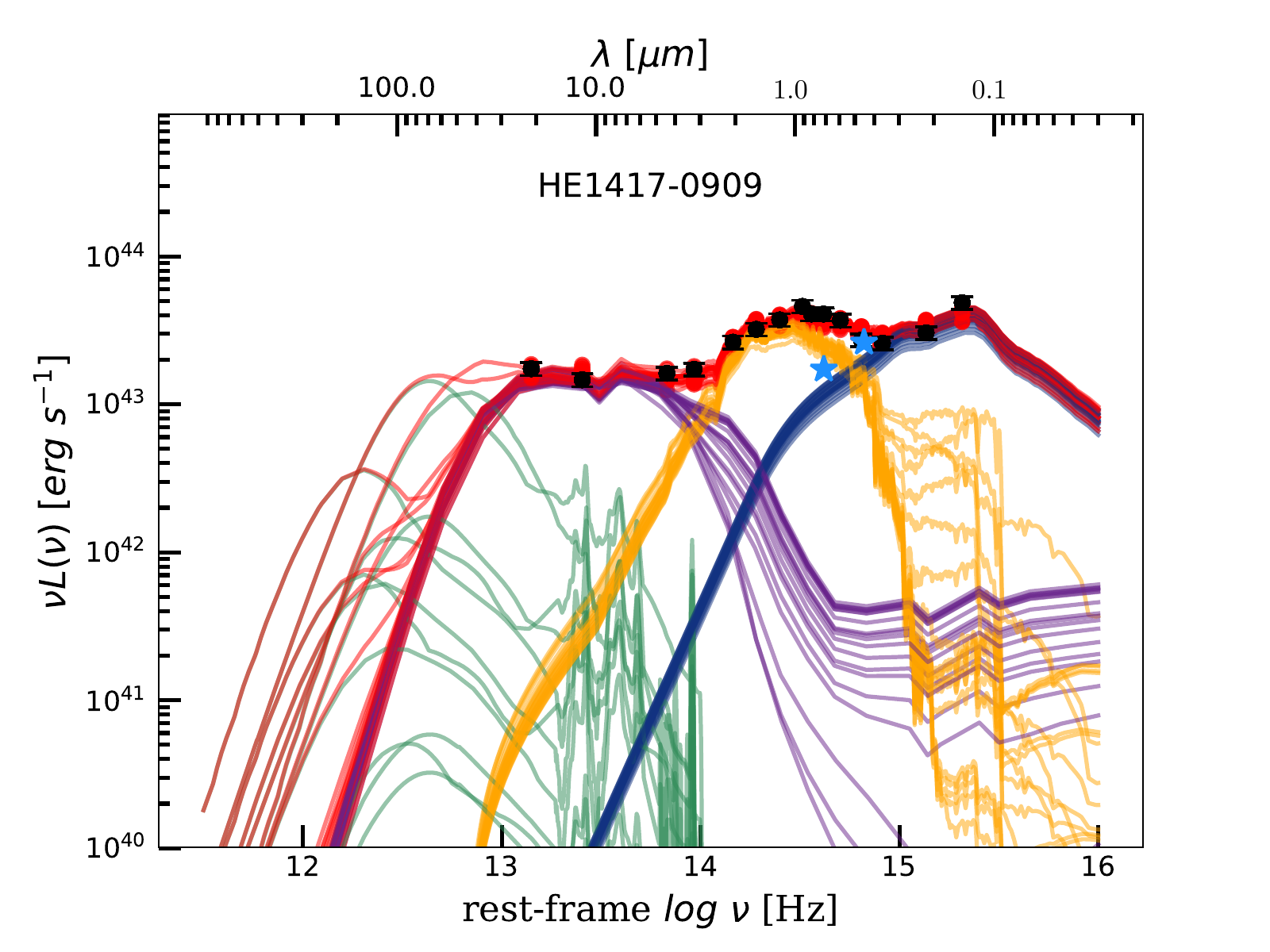}
        \includegraphics[width=0.33\textwidth]{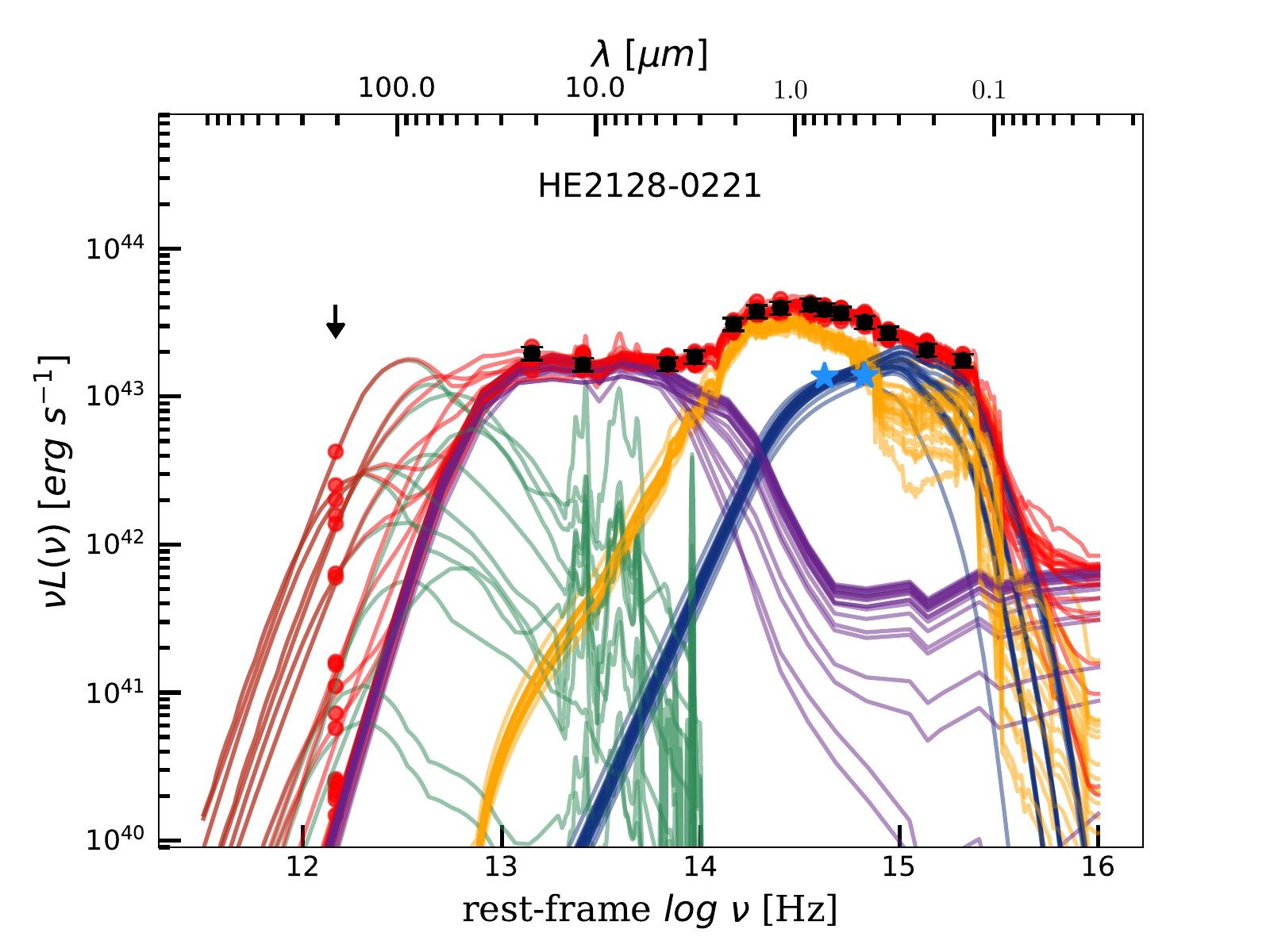}\\
        \includegraphics[width=0.33\textwidth]{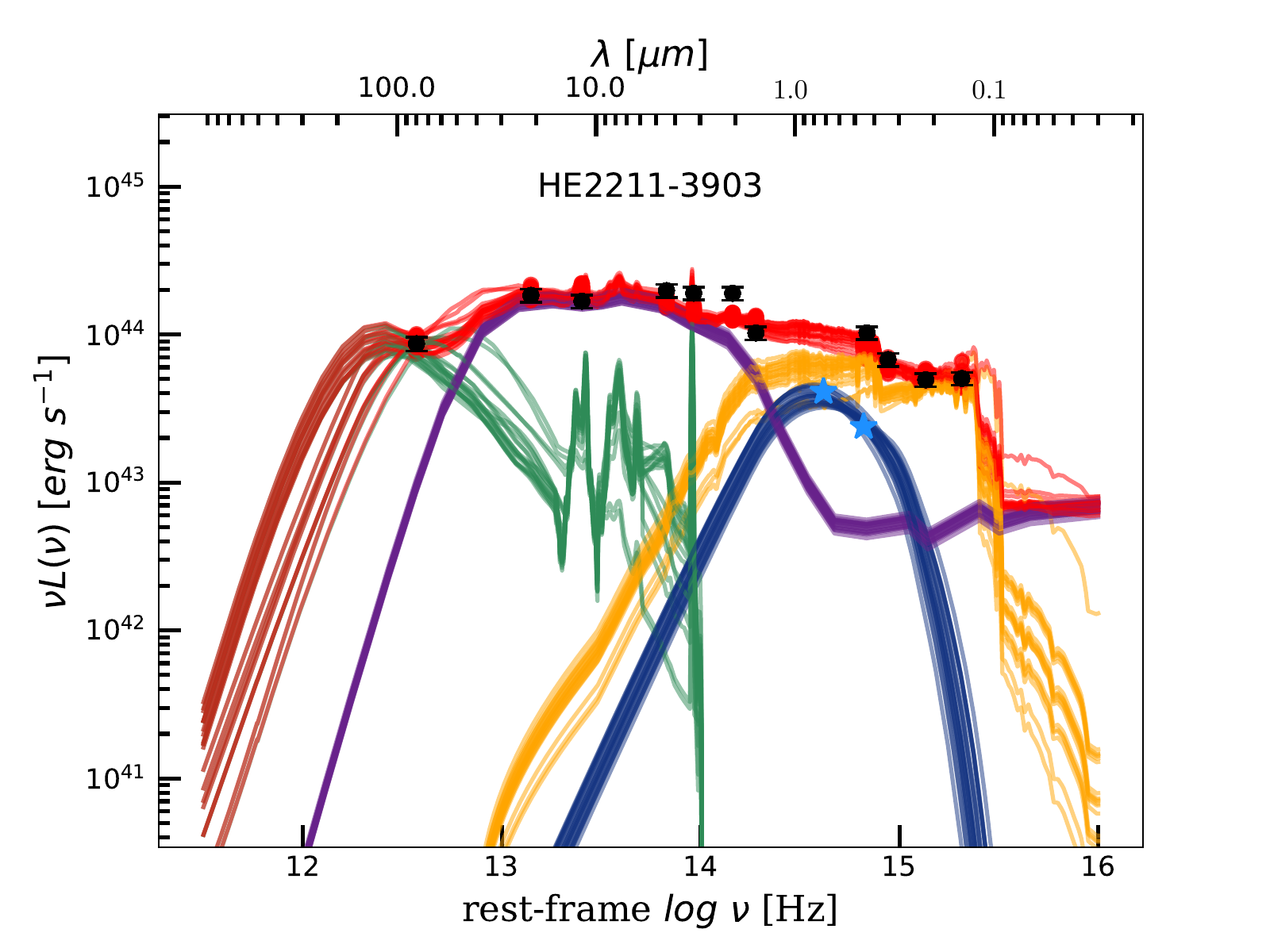}
        \includegraphics[width=0.33\textwidth]{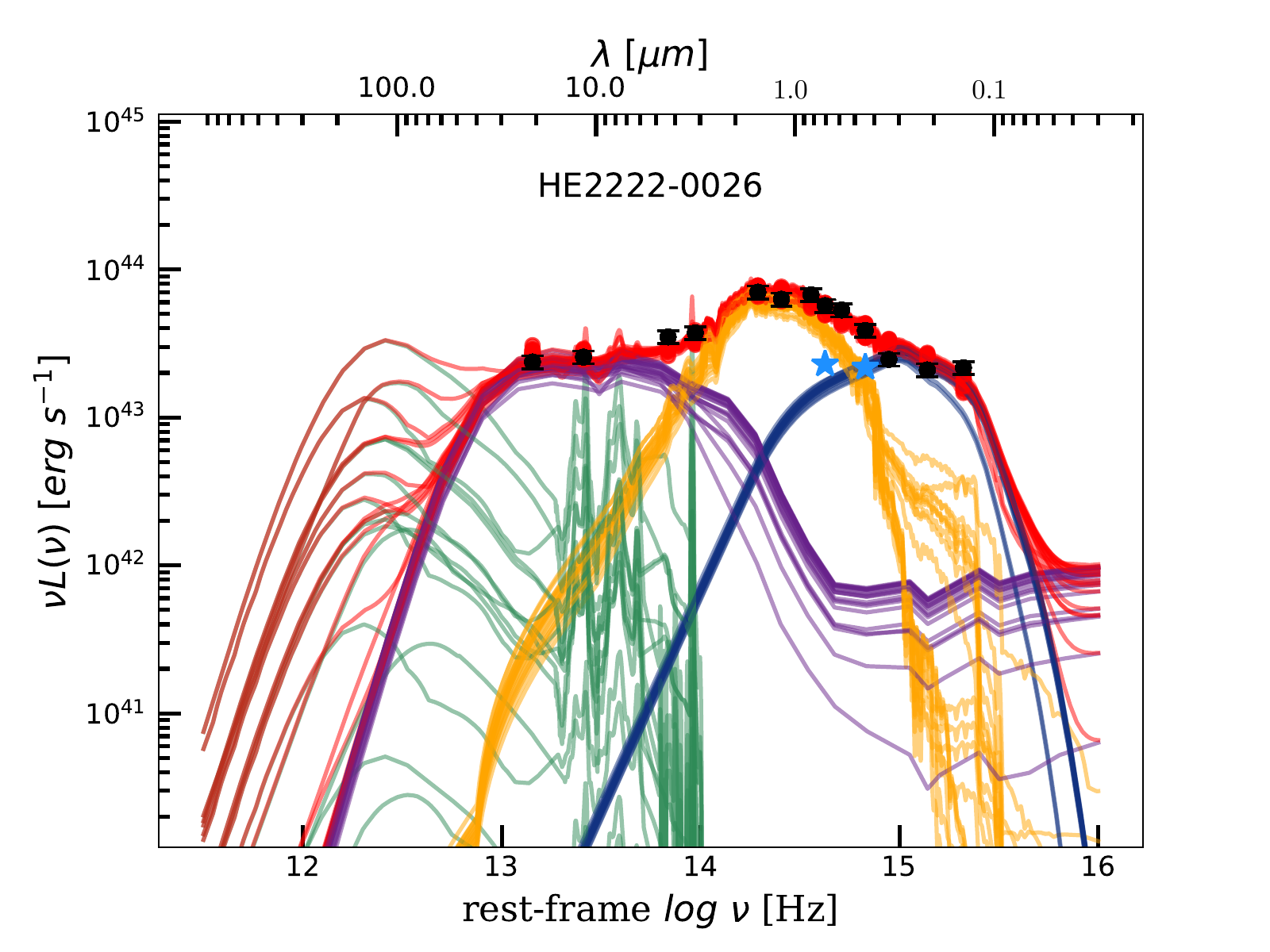}
        \includegraphics[width=0.33\textwidth]{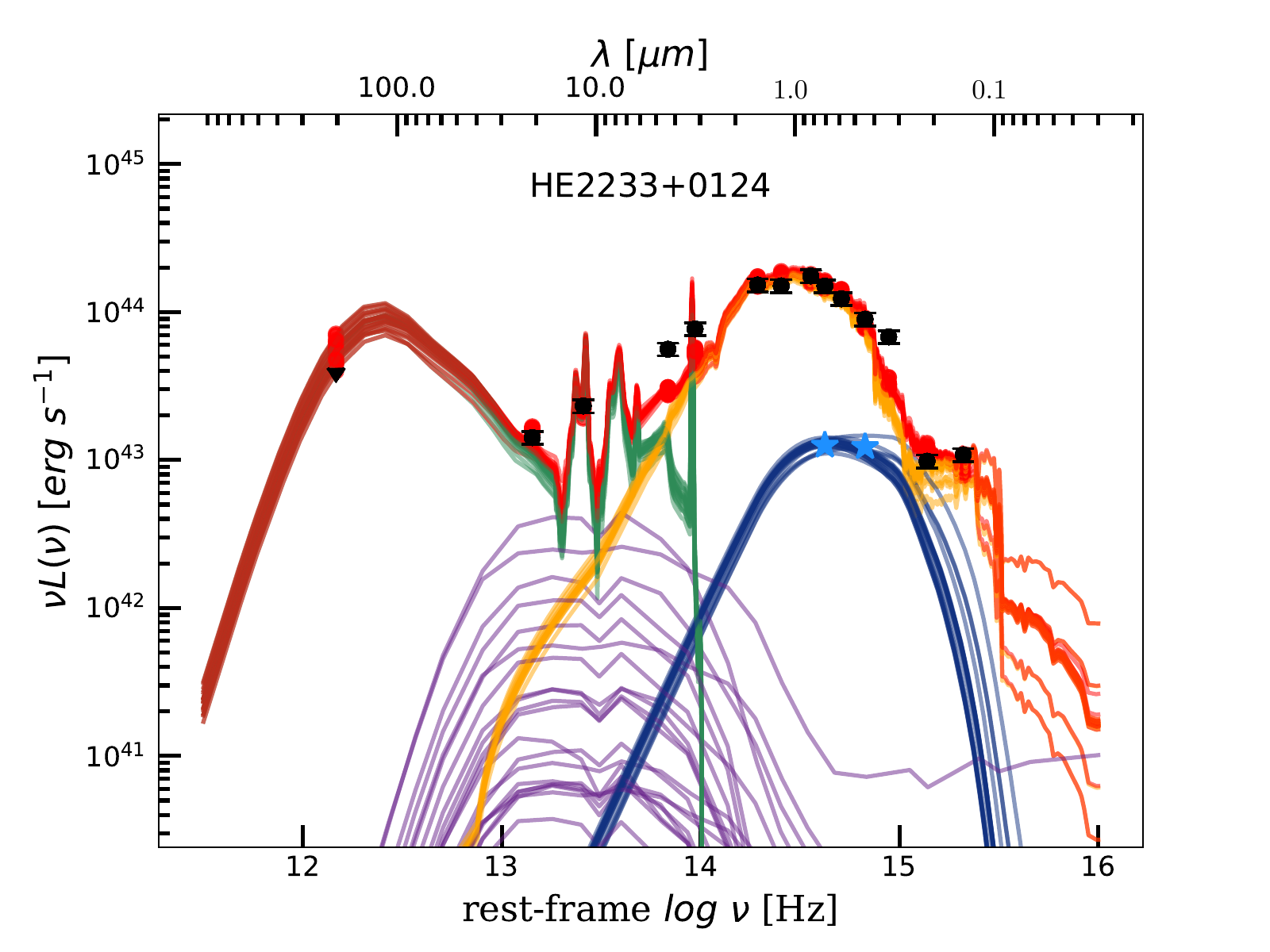}\\
        \includegraphics[width=0.33\textwidth]{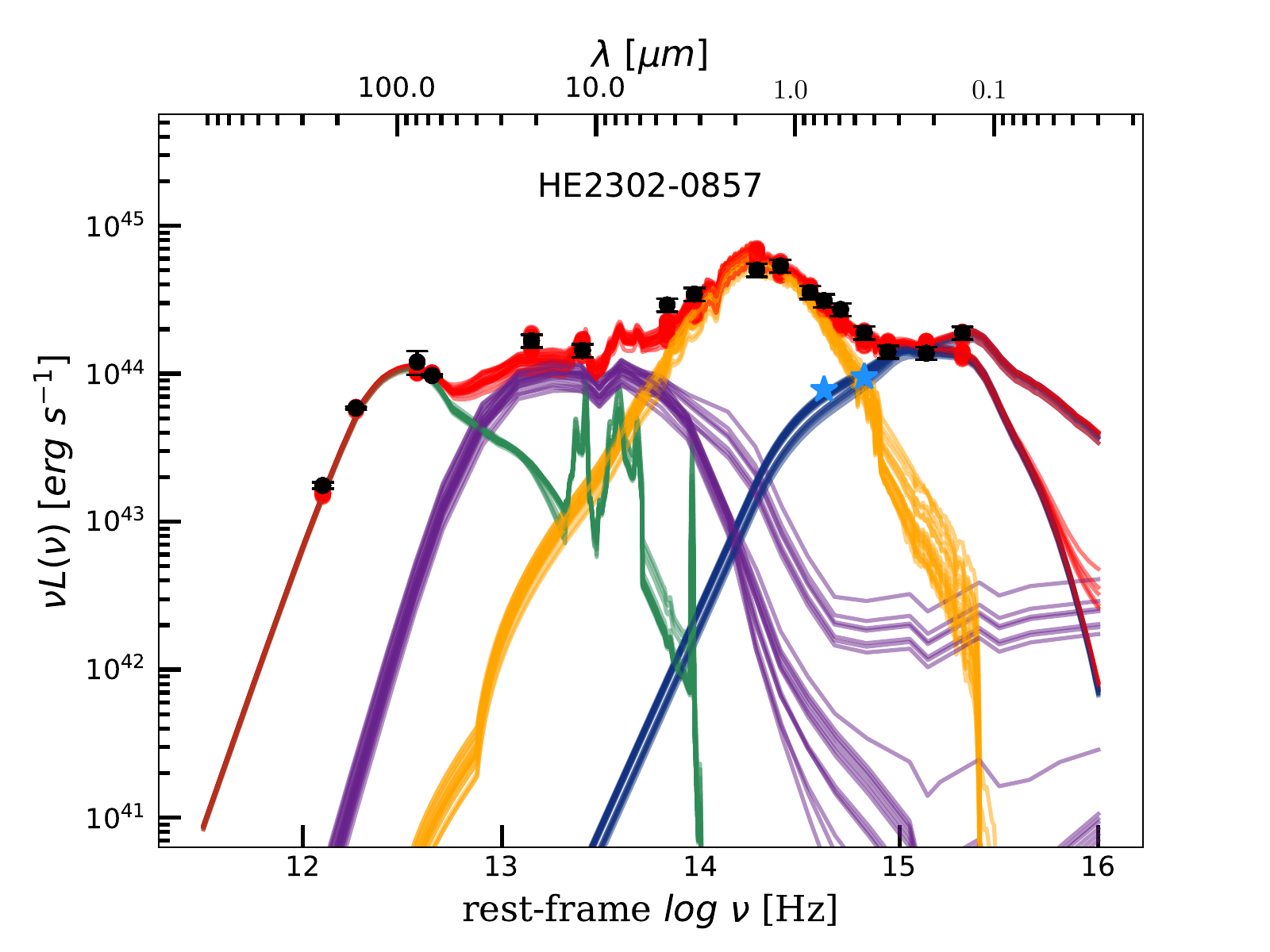}
        \caption{Continued.}
\end{figure*}

 \clearpage
 \section{CARS sample BPT morphology}
 \label{app:bpt_morphology}
 
 \input{Tables/host_properties/BPT_morphology_radii.tex}
 
 \begin{figure*}
 \includegraphics[width=1.\textwidth]{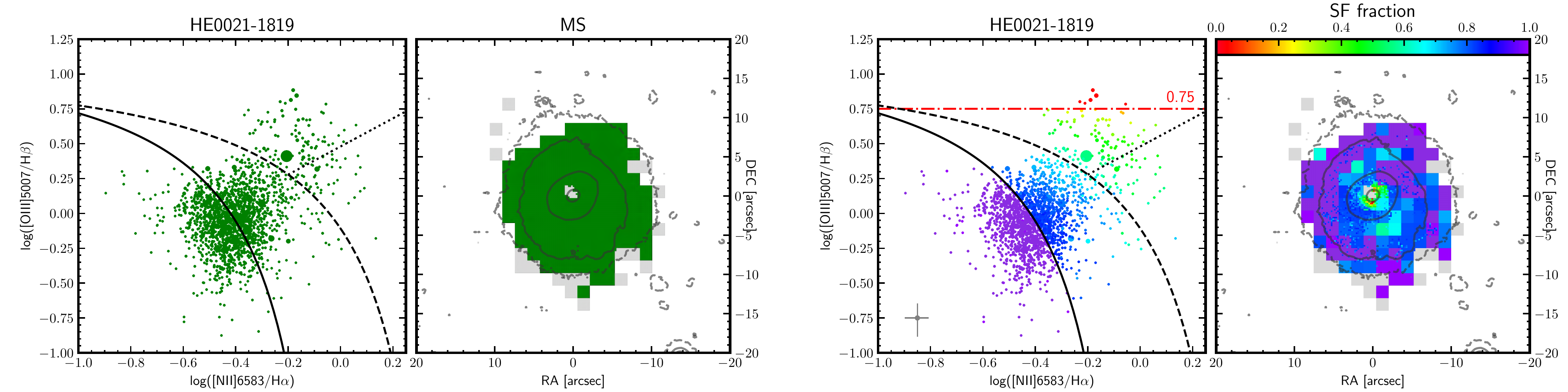}
 \includegraphics[width=1.\textwidth]{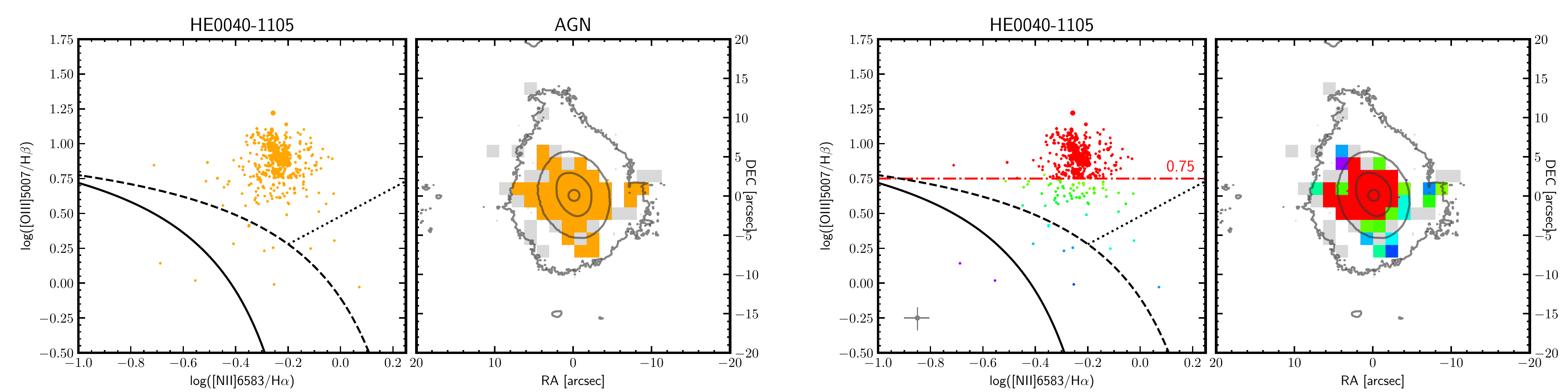}
 \includegraphics[width=1.\textwidth]{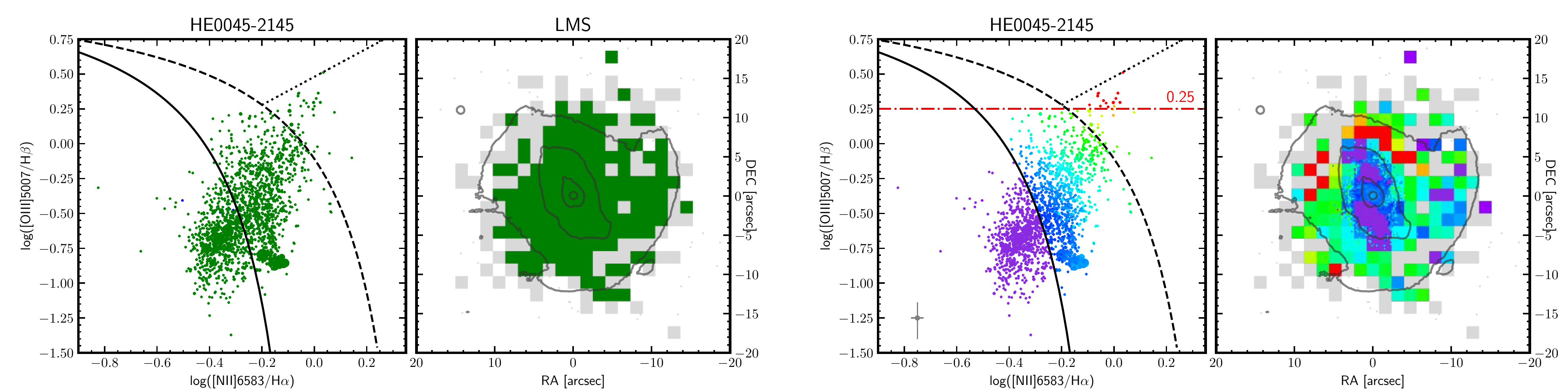}
 \includegraphics[width=1.\textwidth]{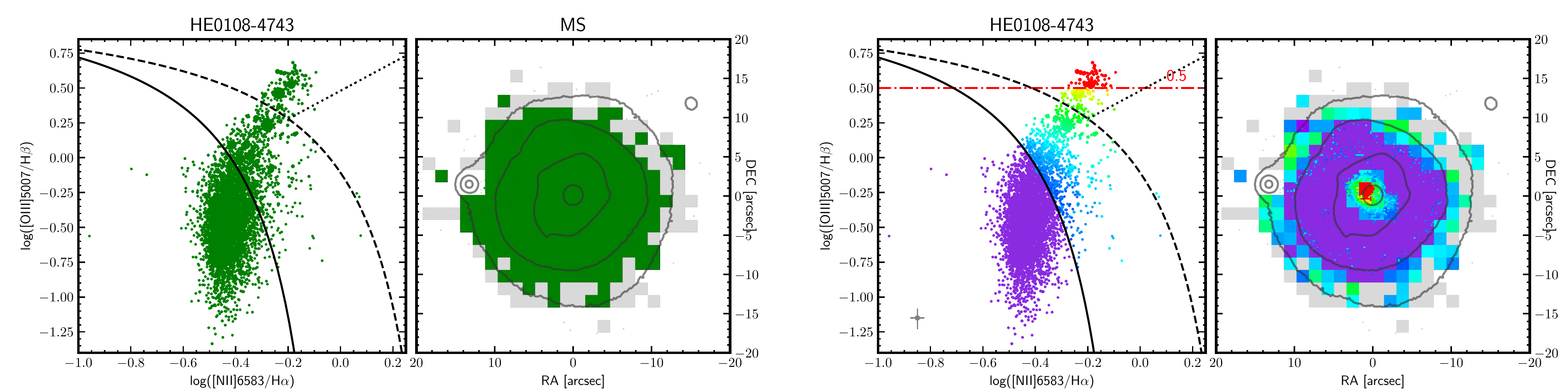}
 \includegraphics[width=1.\textwidth]{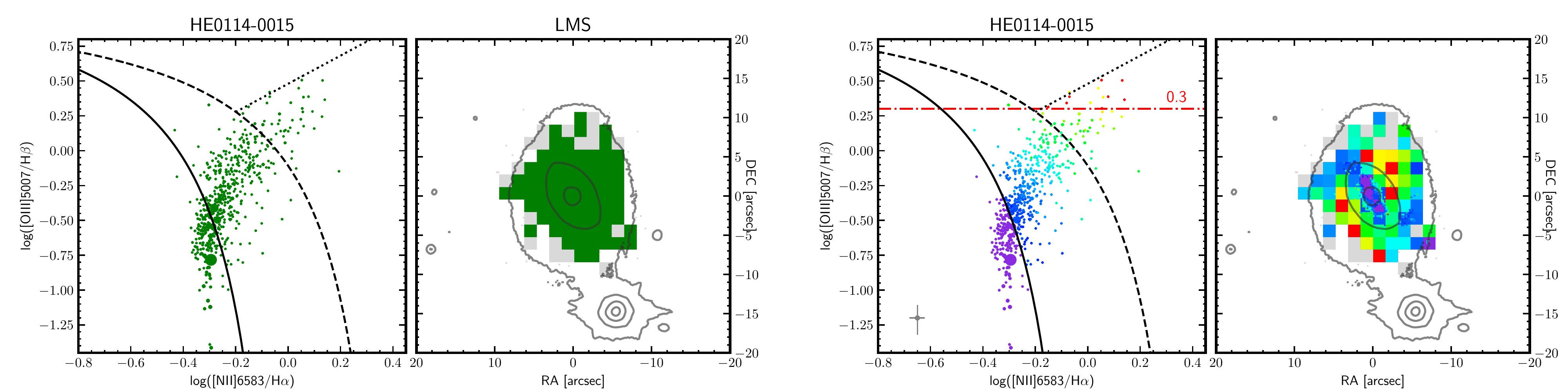}
  \caption{BPT morphology and \textsc{rainbow} results for the sample. The BPT and a spatial map for each galaxy show SF fraction. The red dotted line shows the threshold for the AGN basis spaxels. The contours on the maps represent MUSE whitelight contours. The gray pixels on the maps represent those pixels, where H$\upalpha$ SNR is more than 3$\upsigma$, but the other lines are weaker so the datapoint cannot be shown on a BPT.}\label{fig:bpt_morphology_objects}
 \end{figure*}

\addtocounter{figure}{-1}
\begin{figure*}
 \includegraphics[width=1.\textwidth]{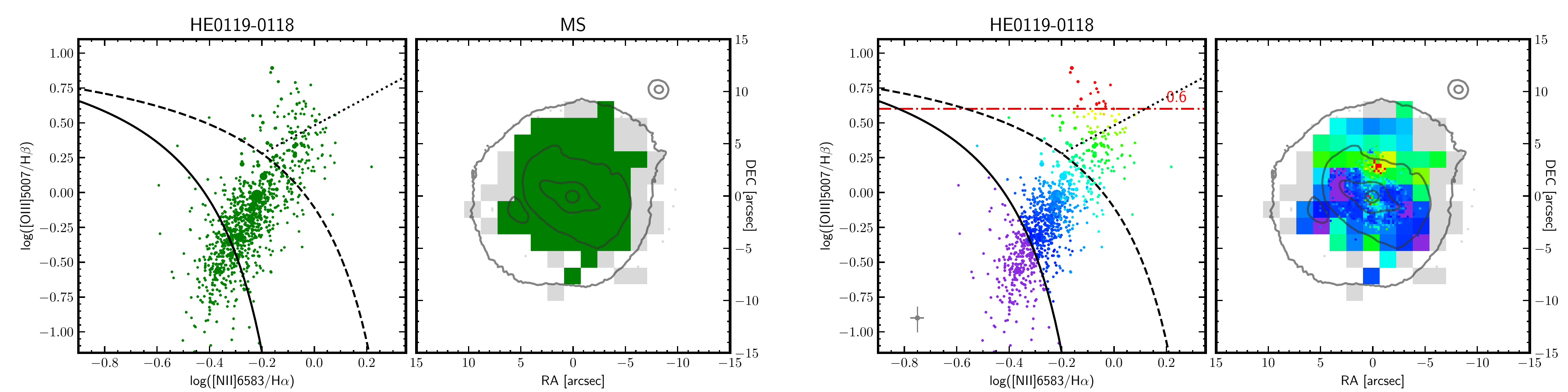}
 \includegraphics[width=1.\textwidth]{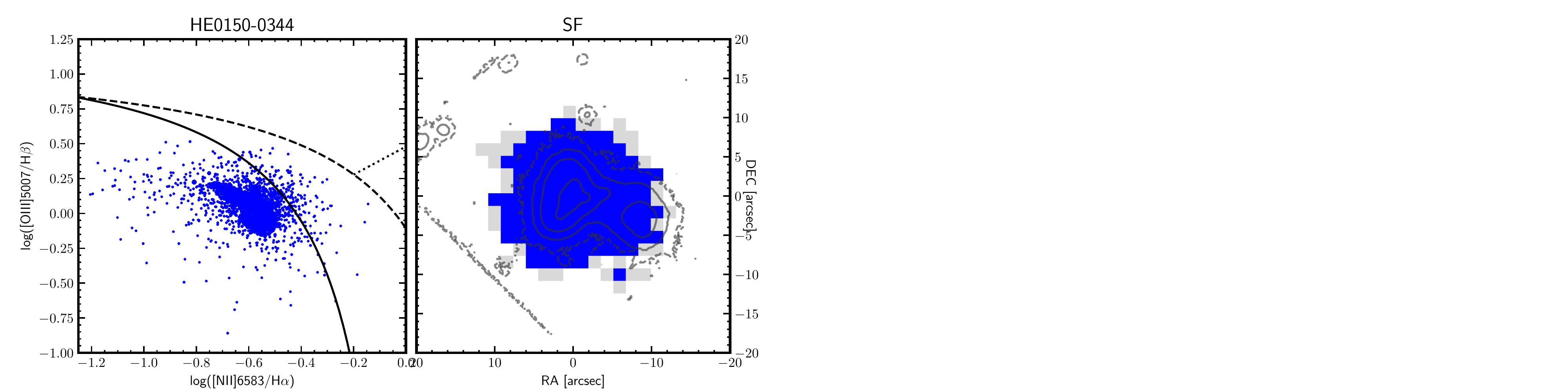}
 \includegraphics[width=1.\textwidth]{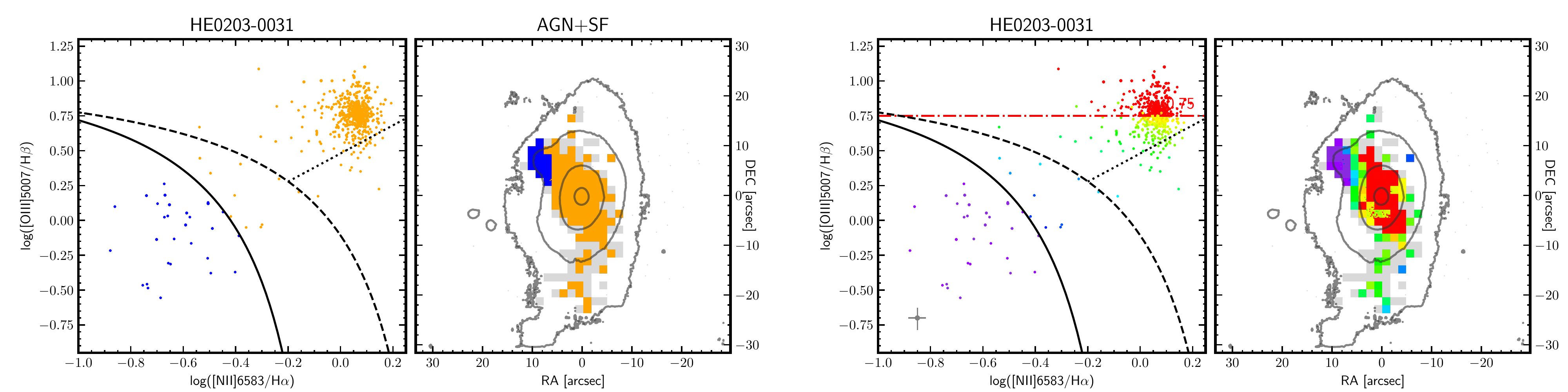}
 \includegraphics[width=1.\textwidth]{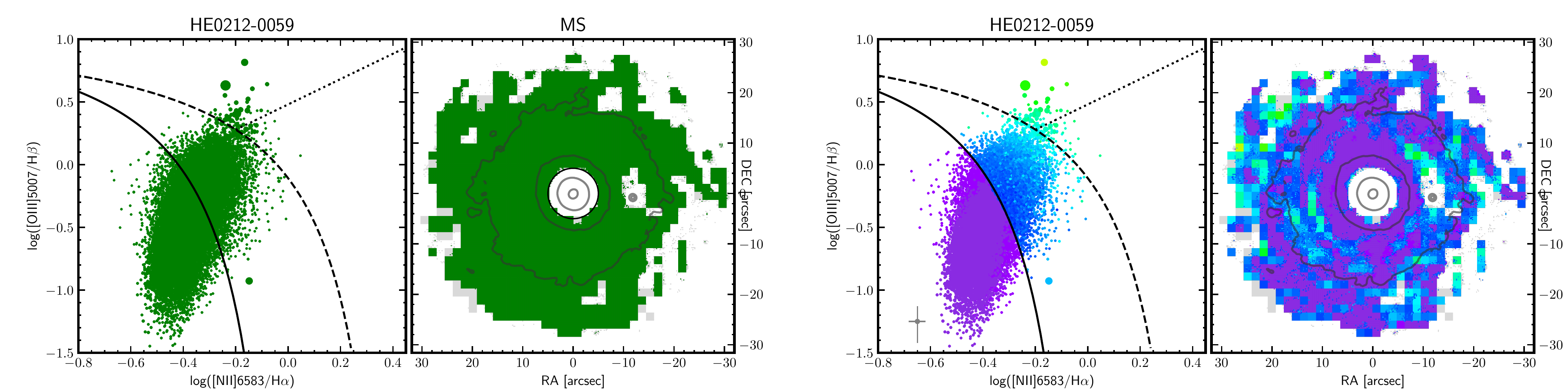}
 \includegraphics[width=1.\textwidth]{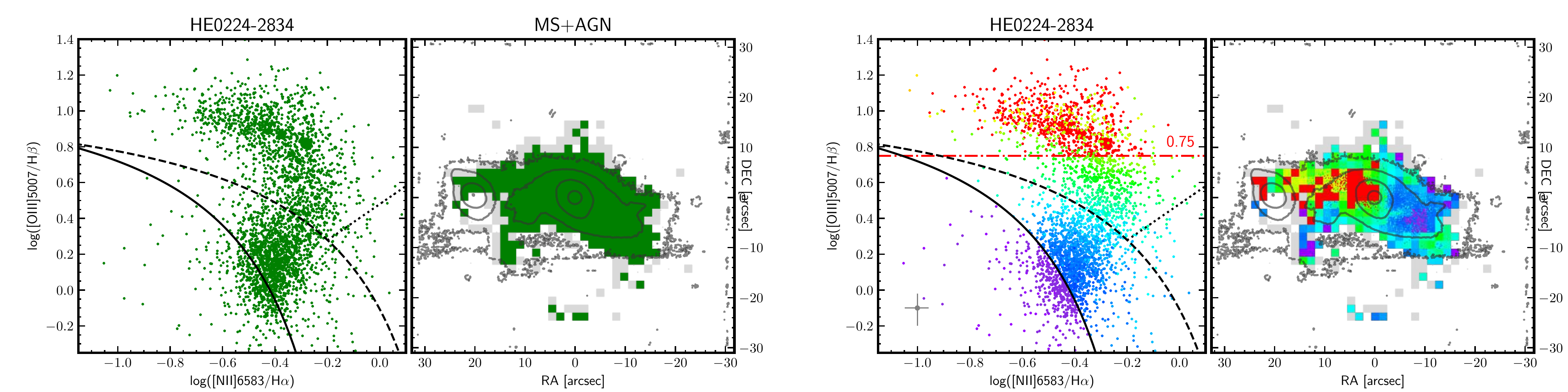}
 \caption{Continued.}
\end{figure*}

\addtocounter{figure}{-1}
\begin{figure*}
 \includegraphics[width=1.\textwidth]{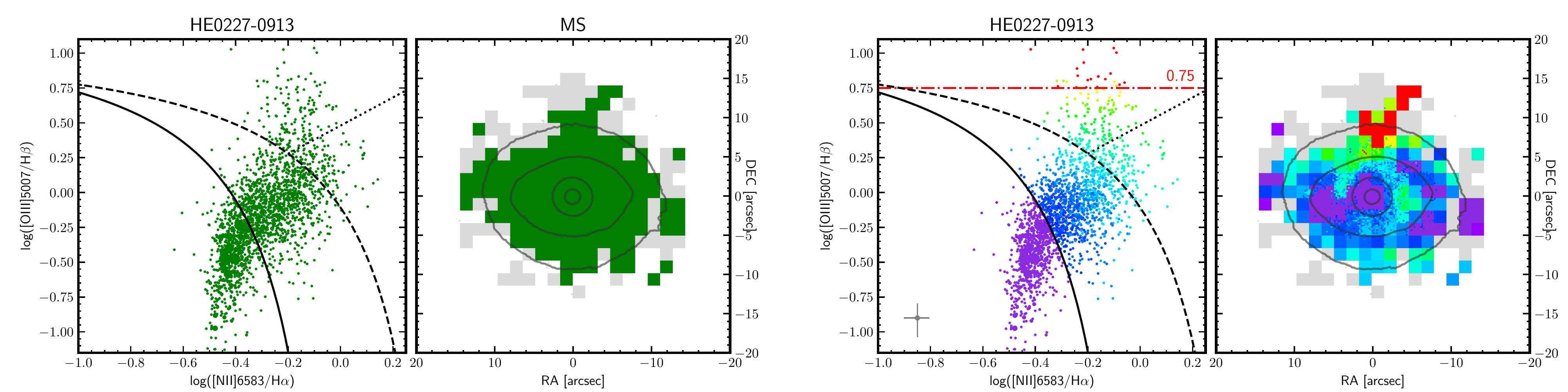}
 \includegraphics[width=1.\textwidth]{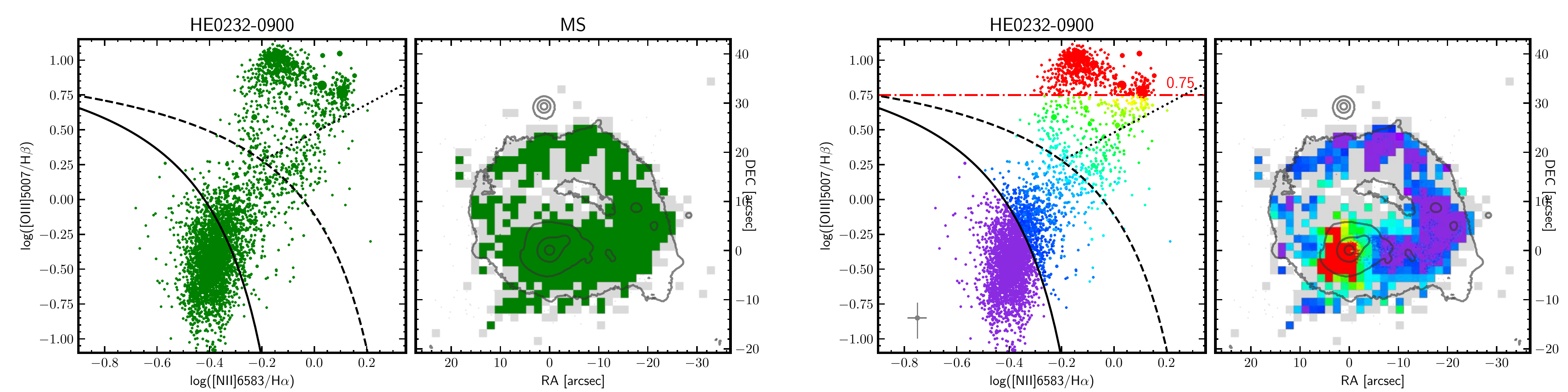}
 \includegraphics[width=1.\textwidth]{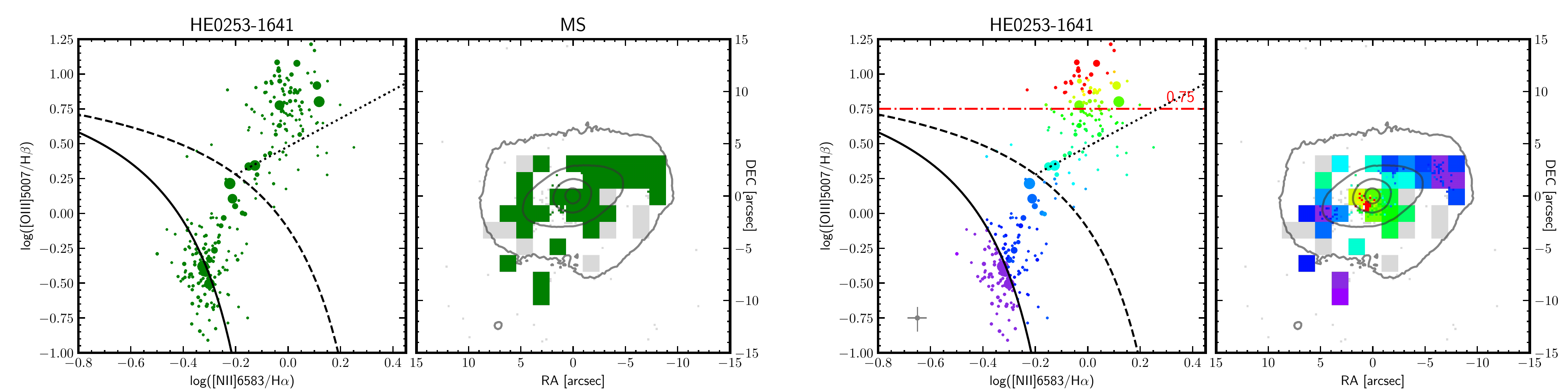}
 \includegraphics[width=1.\textwidth]{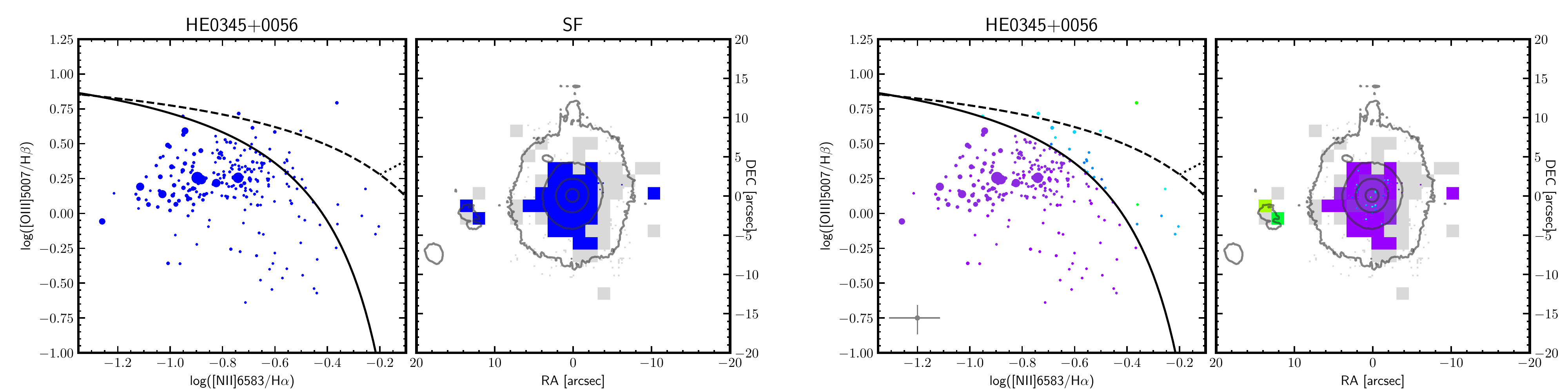}
 \includegraphics[width=1.\textwidth]{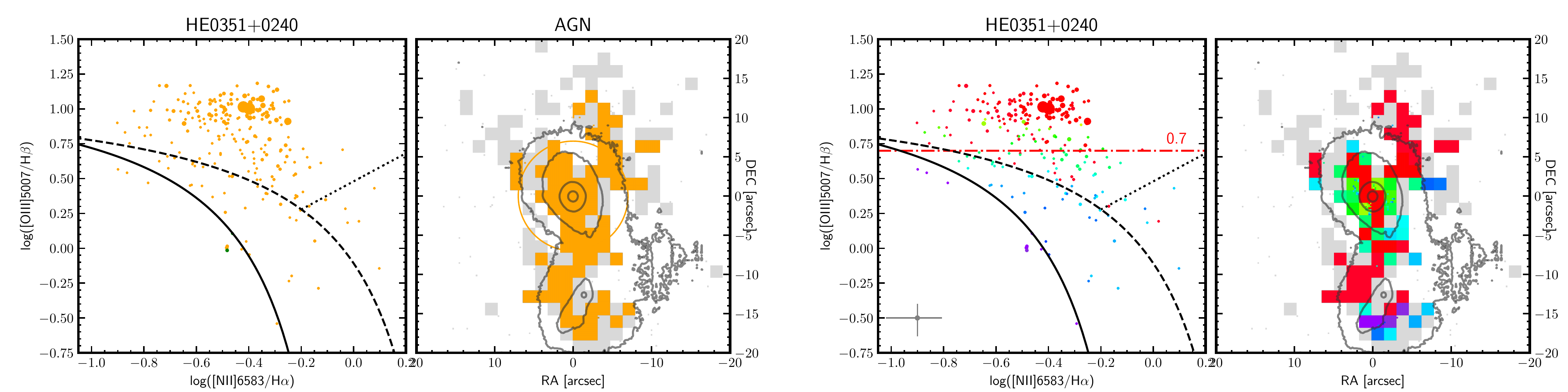}
 \caption{Continued.}
\end{figure*}

\addtocounter{figure}{-1}
\begin{figure*}
 \includegraphics[width=1.\textwidth]{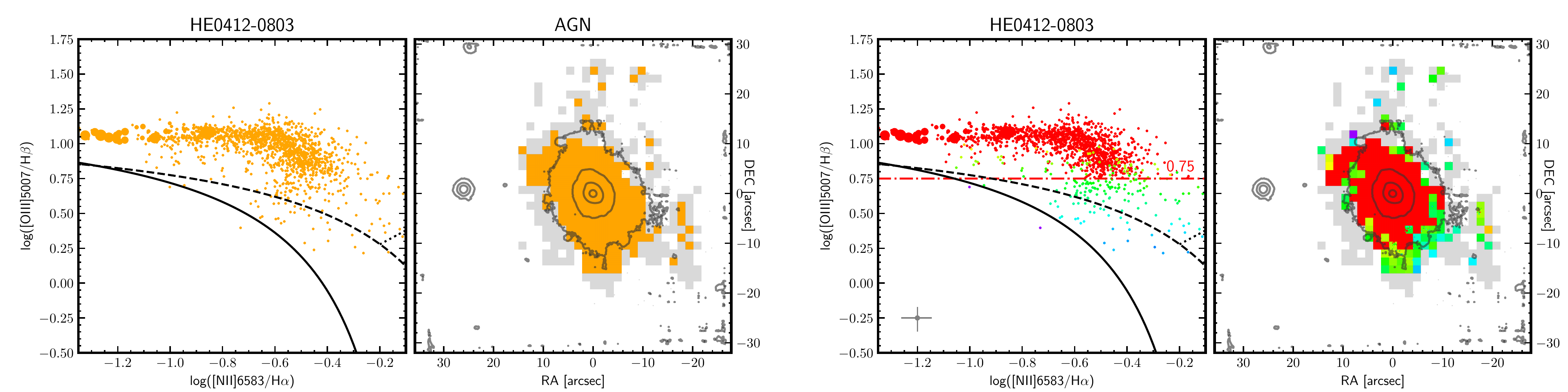}
 \includegraphics[width=1.\textwidth]{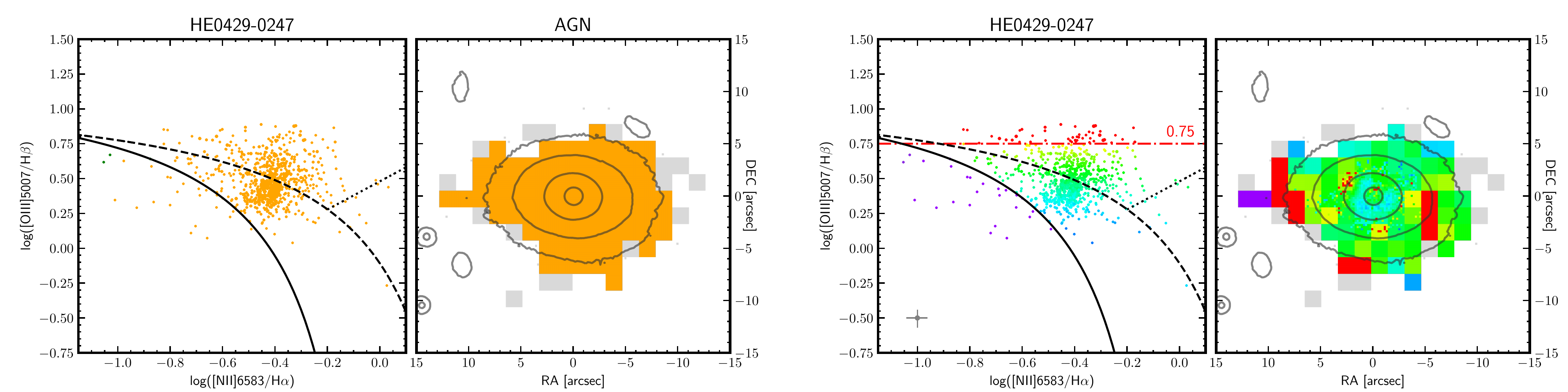}
 \includegraphics[width=1.\textwidth]{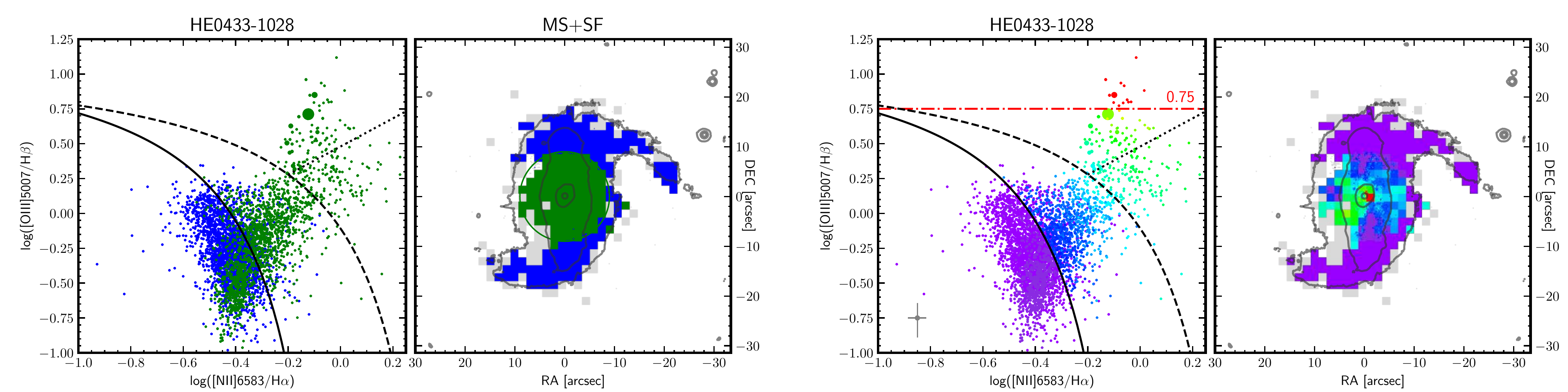}
 \includegraphics[width=1.\textwidth]{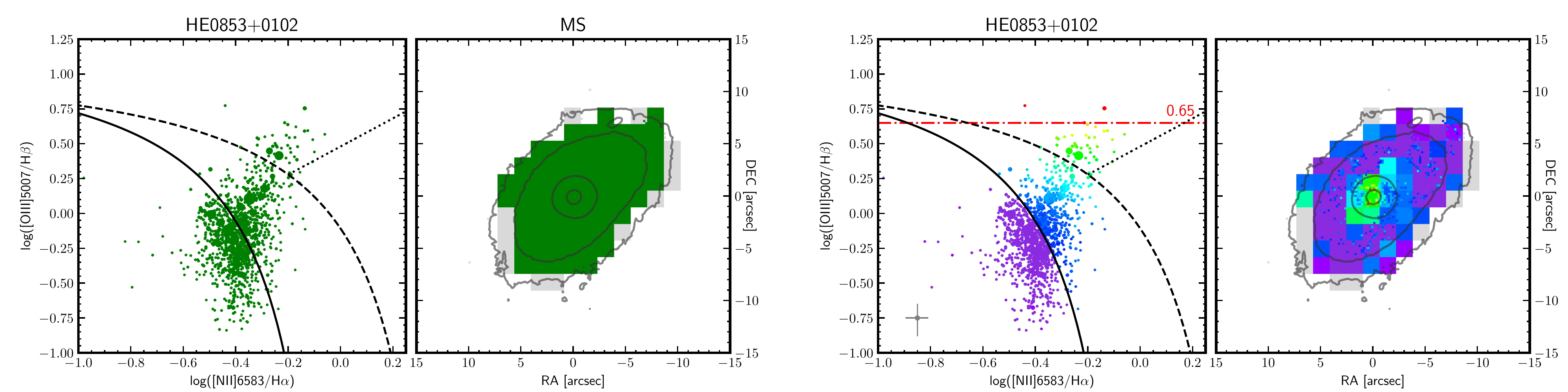}
 \includegraphics[width=1.\textwidth]{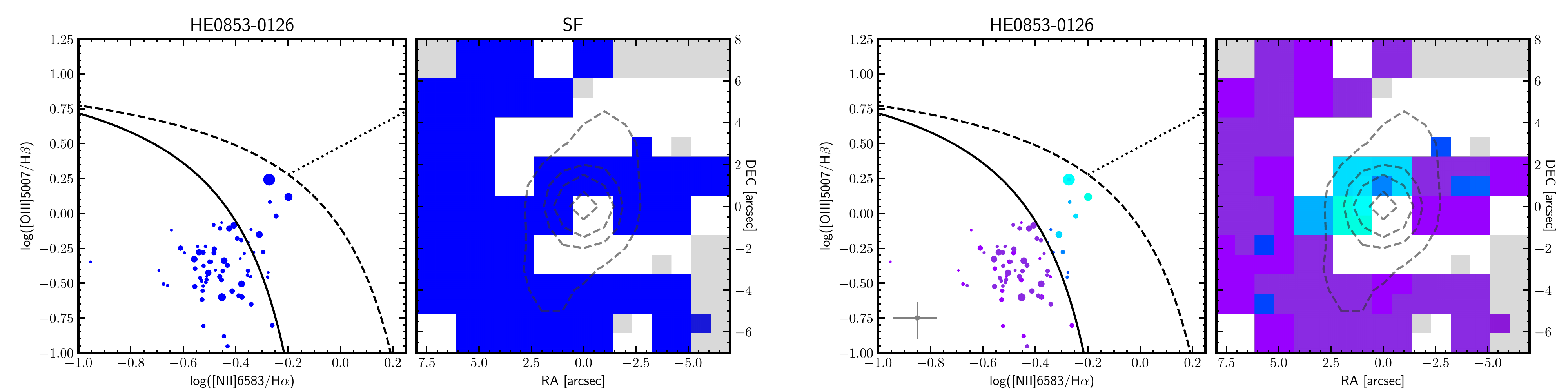}
 \caption{Continued.}
\end{figure*}

\addtocounter{figure}{-1}
\begin{figure*}
 \includegraphics[width=1.\textwidth]{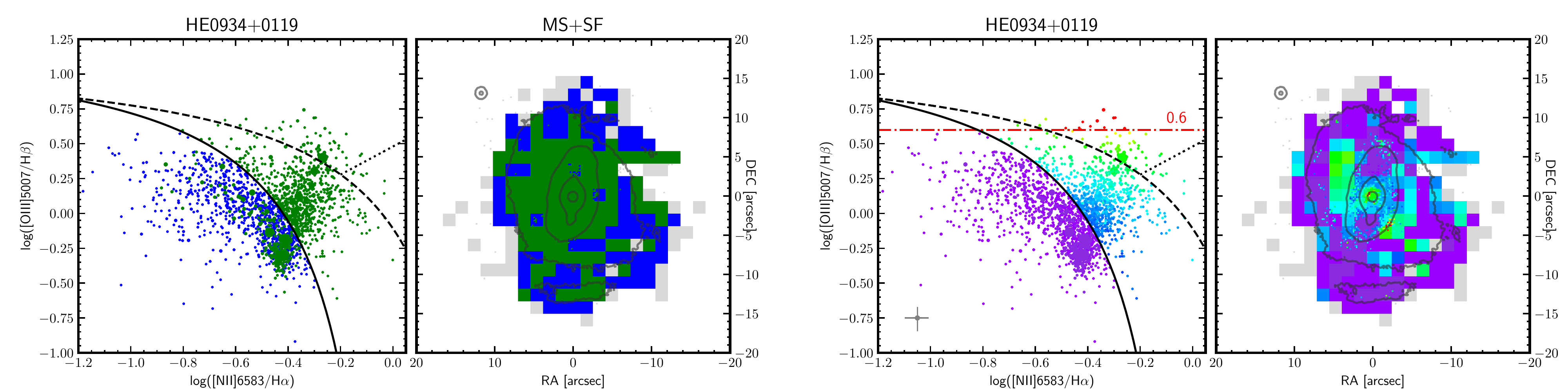}
 \includegraphics[width=1.\textwidth]{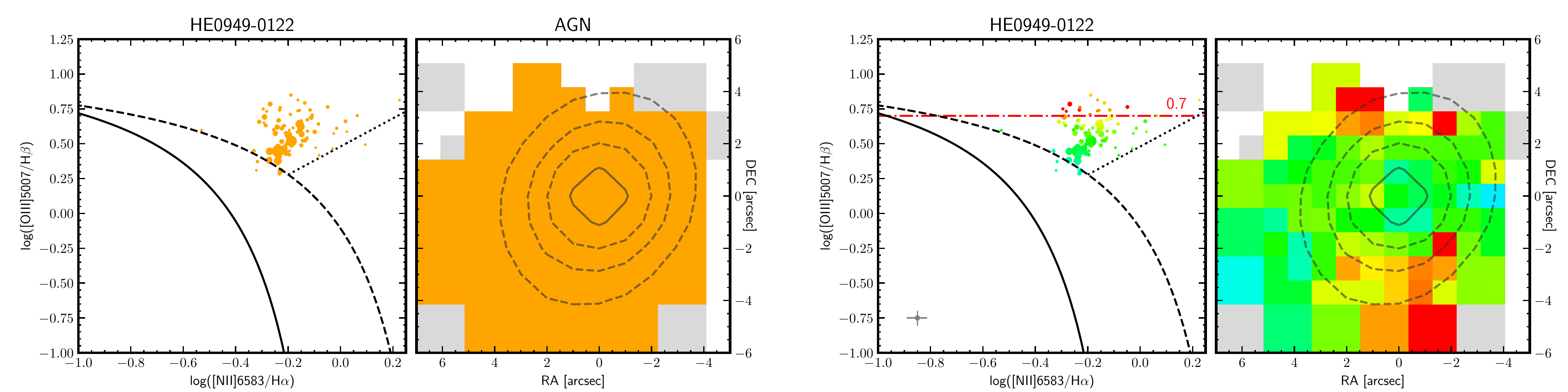}
 \includegraphics[width=1.\textwidth]{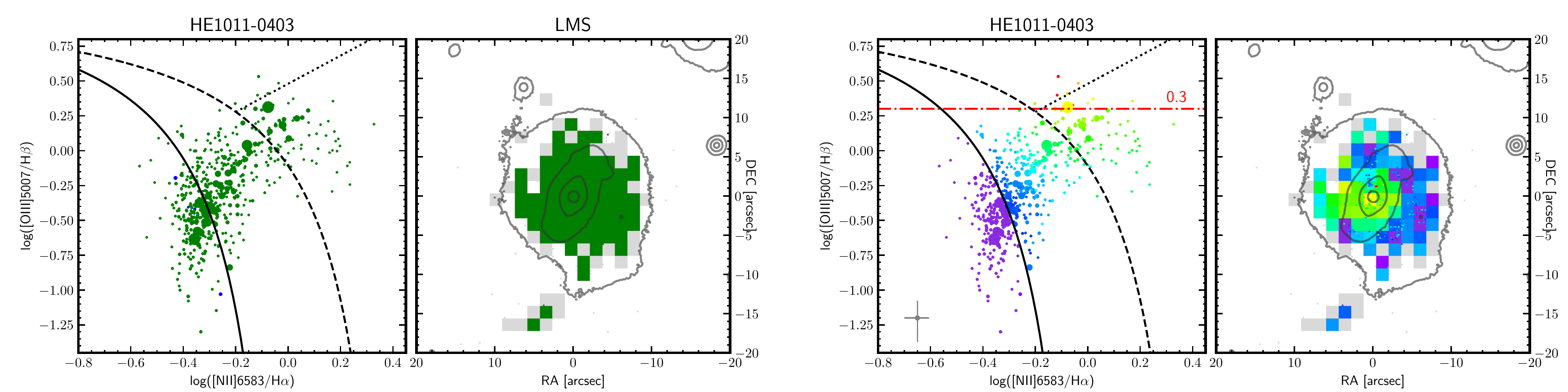}
 \includegraphics[width=1.\textwidth]{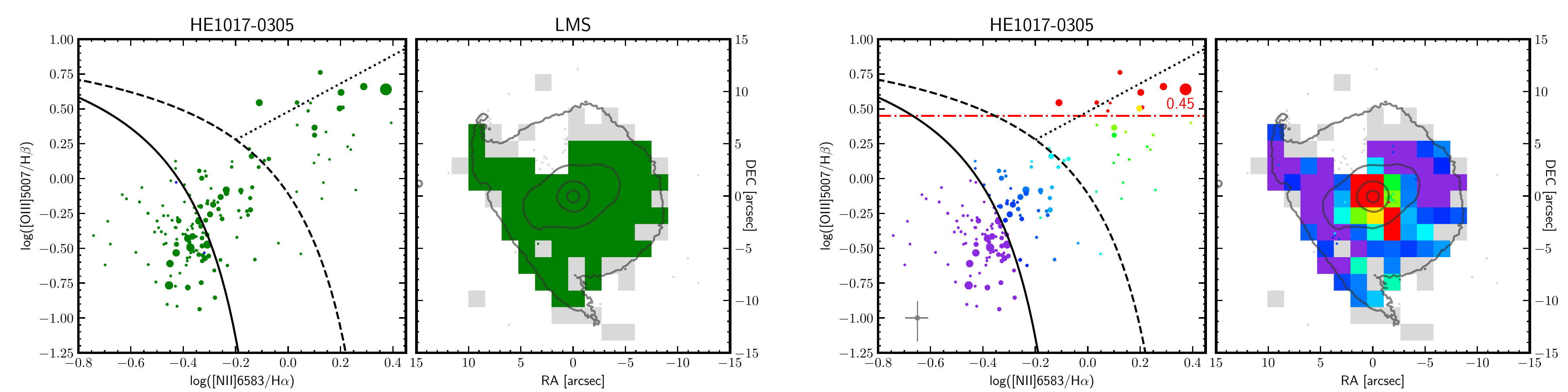}
 \includegraphics[width=1.\textwidth]{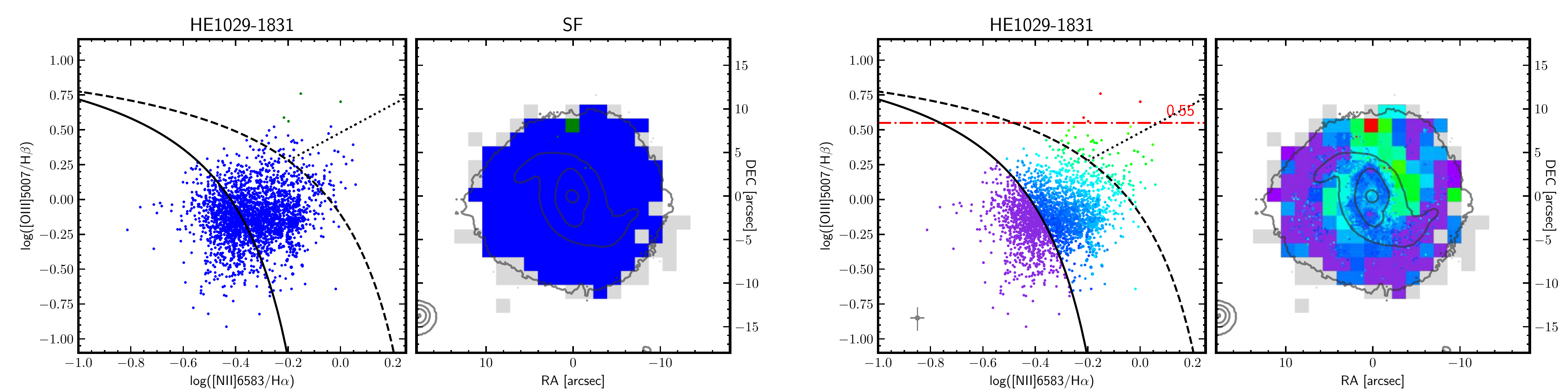}
 \caption{Continued.}
\end{figure*}

\addtocounter{figure}{-1}
\begin{figure*}
 \includegraphics[width=1.\textwidth]{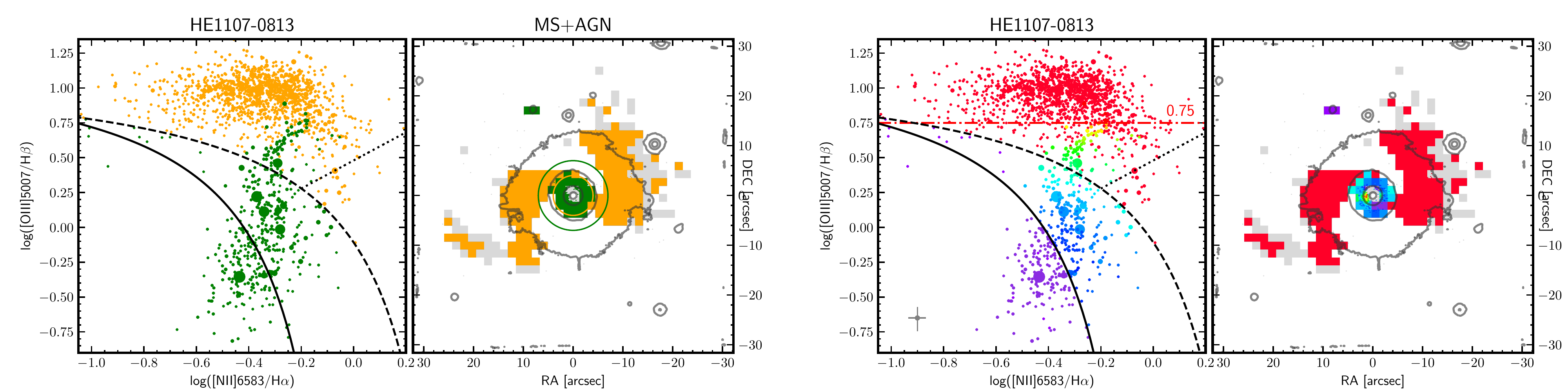}
 \includegraphics[width=1.\textwidth]{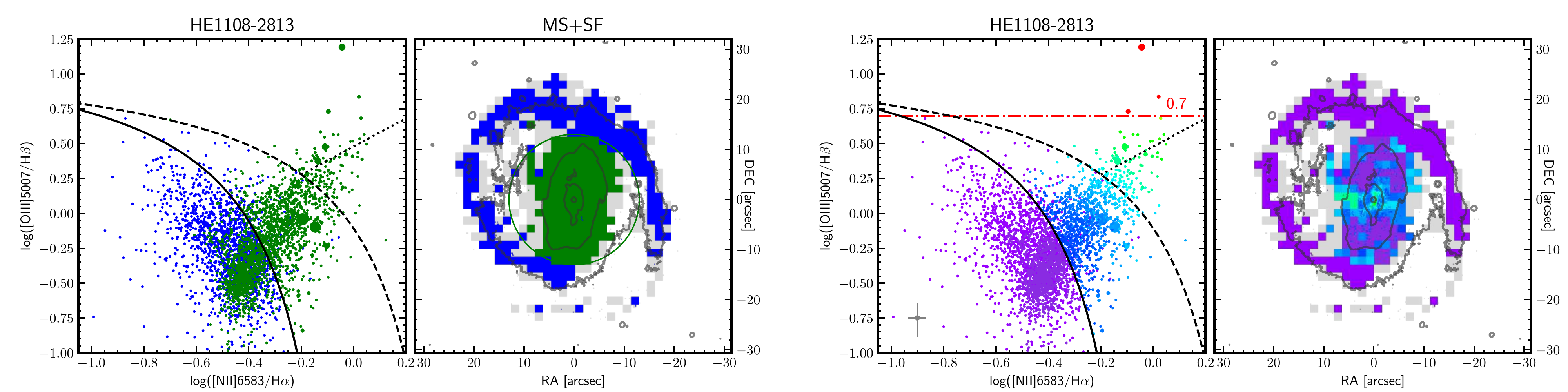}
 \includegraphics[width=1.\textwidth]{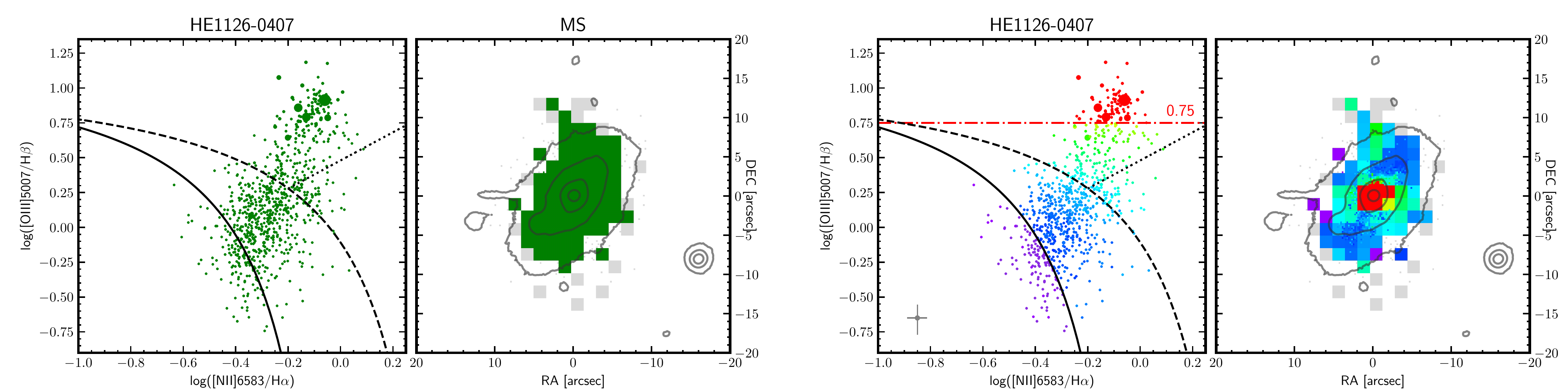}
 \includegraphics[width=1.\textwidth]{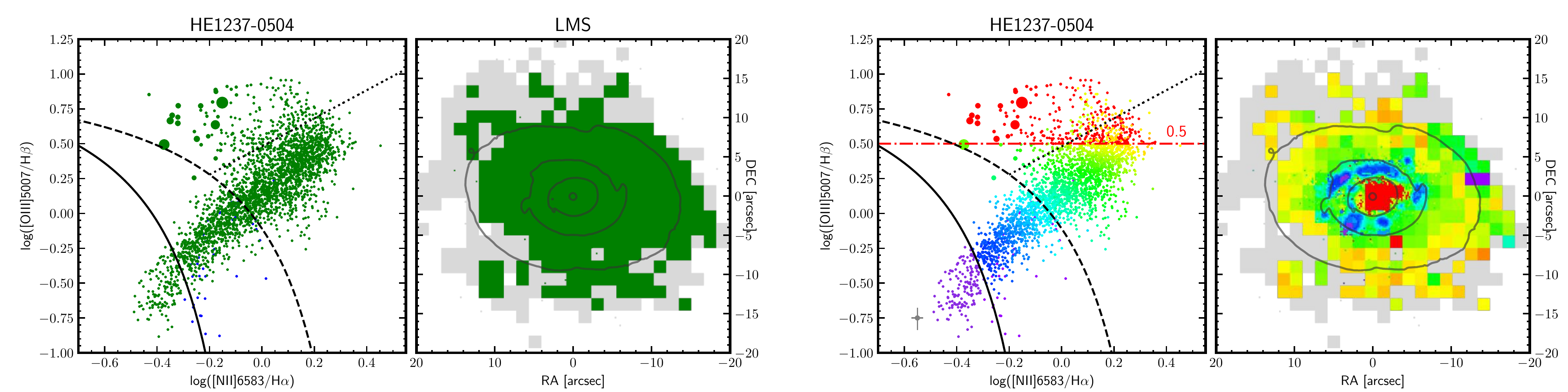}
 \includegraphics[width=1.\textwidth]{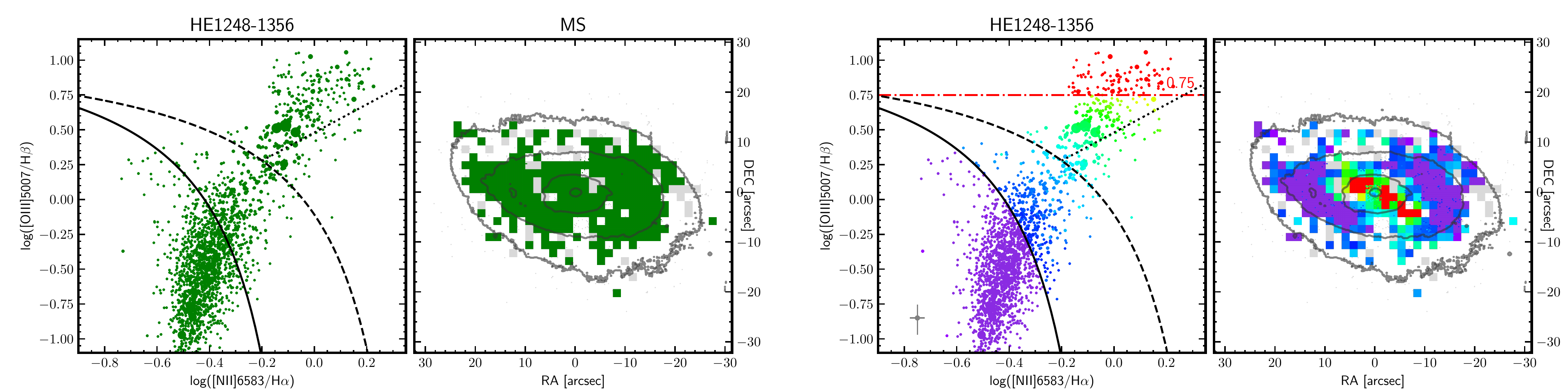}
 \caption{Continued.}
\end{figure*}

\addtocounter{figure}{-1}
\begin{figure*}
 \includegraphics[width=1.\textwidth]{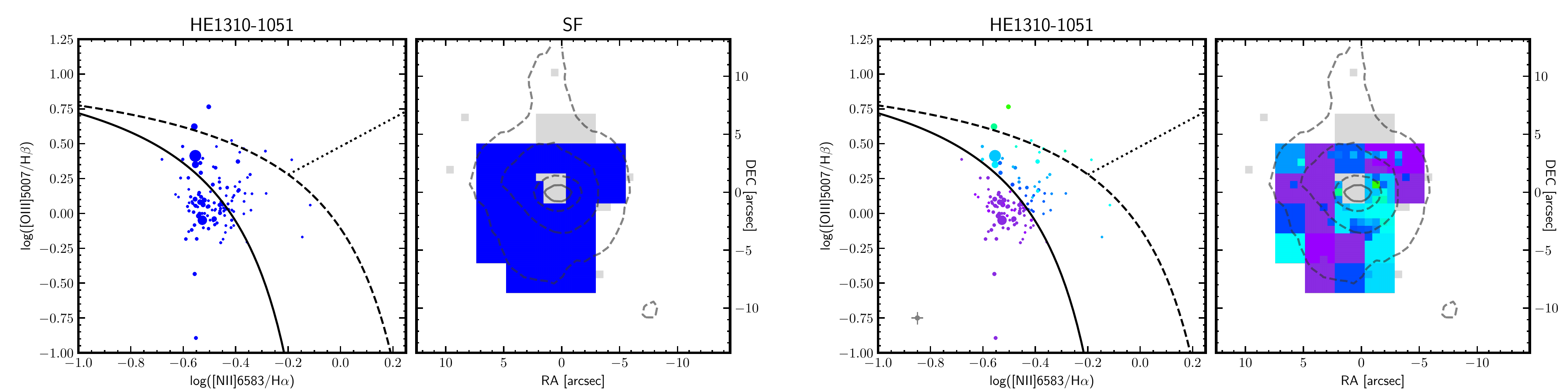}
 \includegraphics[width=1.\textwidth]{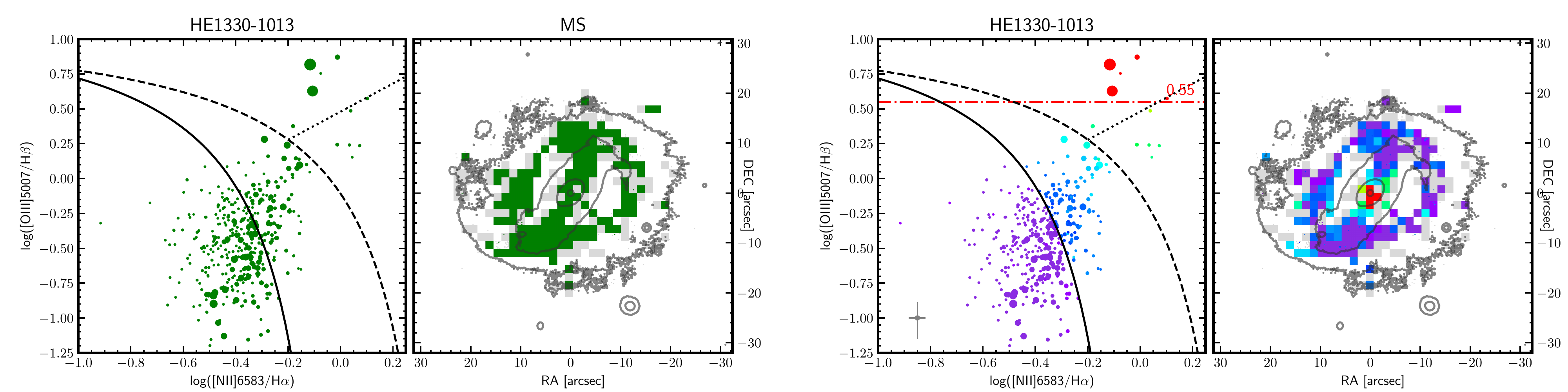}
 \includegraphics[width=1.\textwidth]{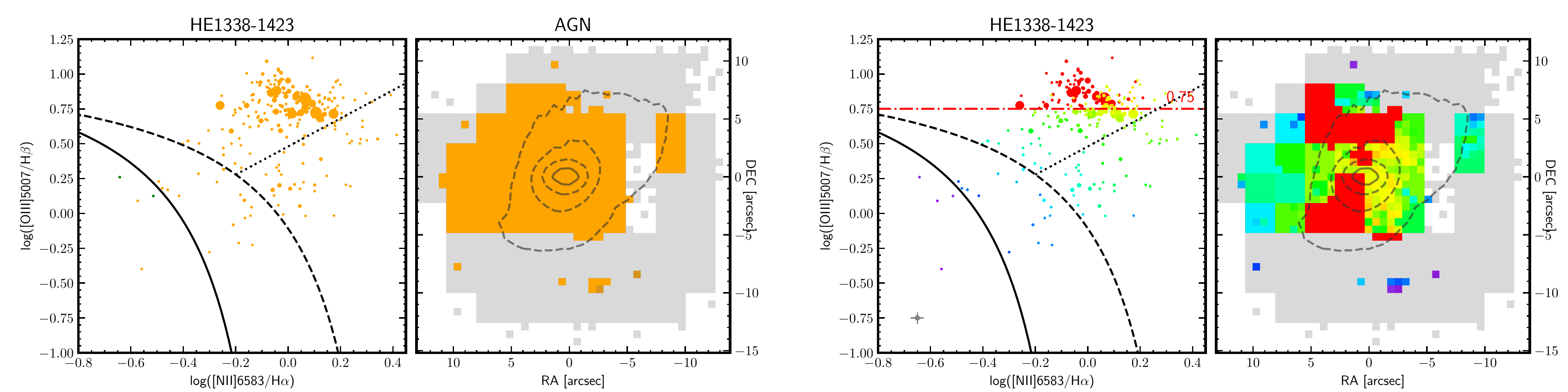}
 \includegraphics[width=1.\textwidth]{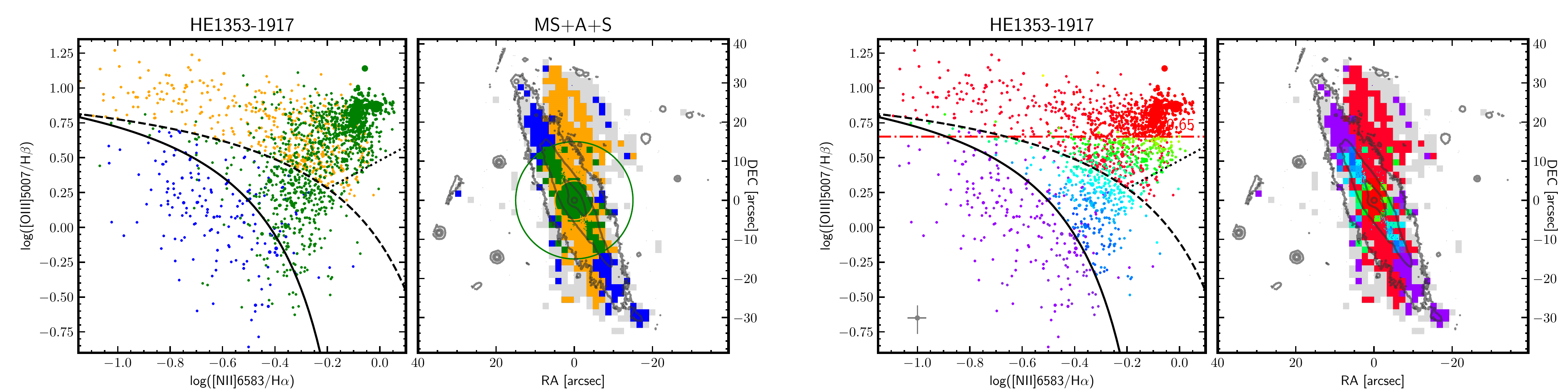}
 \includegraphics[width=1.\textwidth]{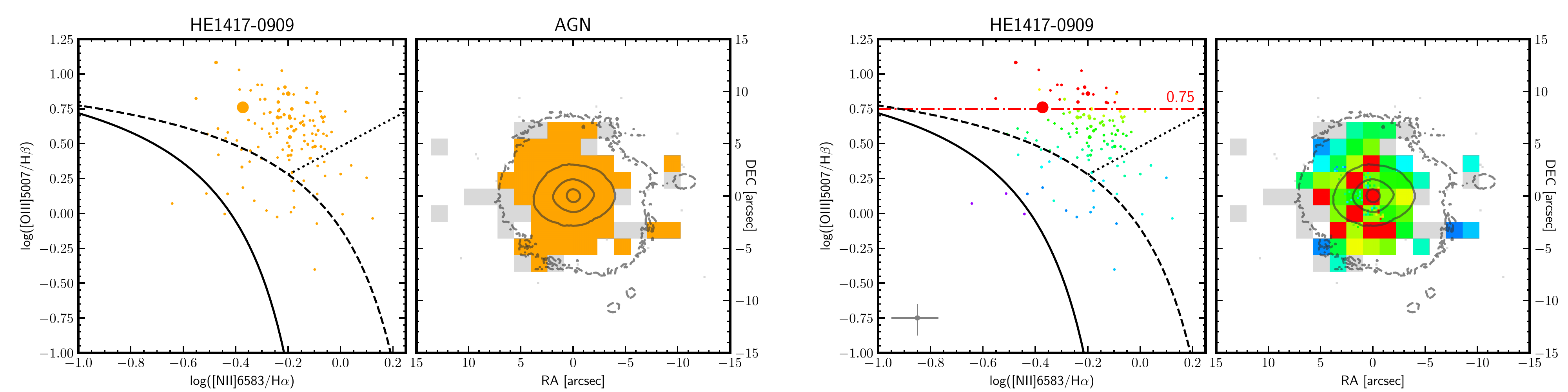}
 \caption{Continued.}
\end{figure*}

\addtocounter{figure}{-1}
\begin{figure*}
 \includegraphics[width=1.\textwidth]{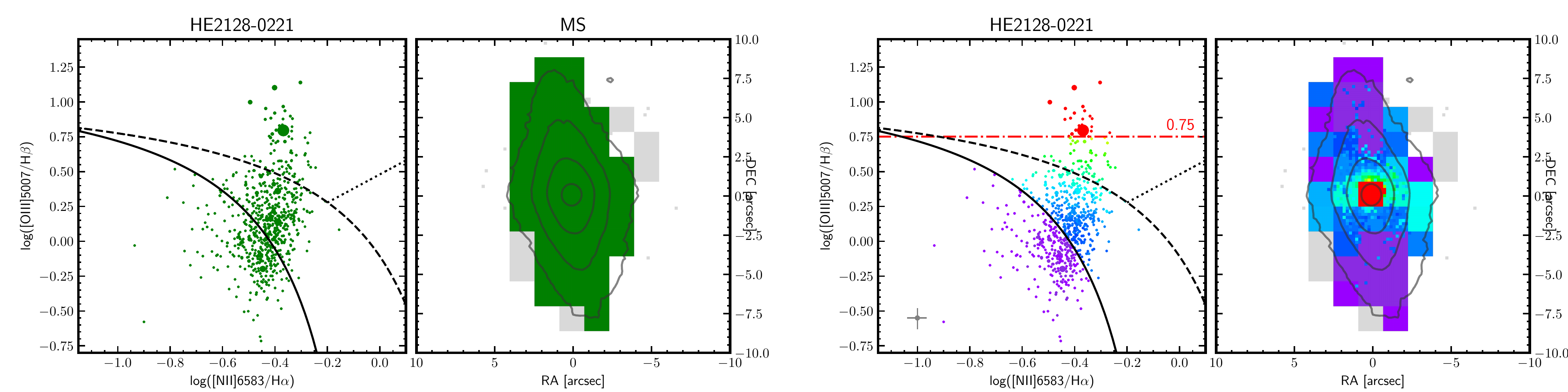}
 \includegraphics[width=1.\textwidth]{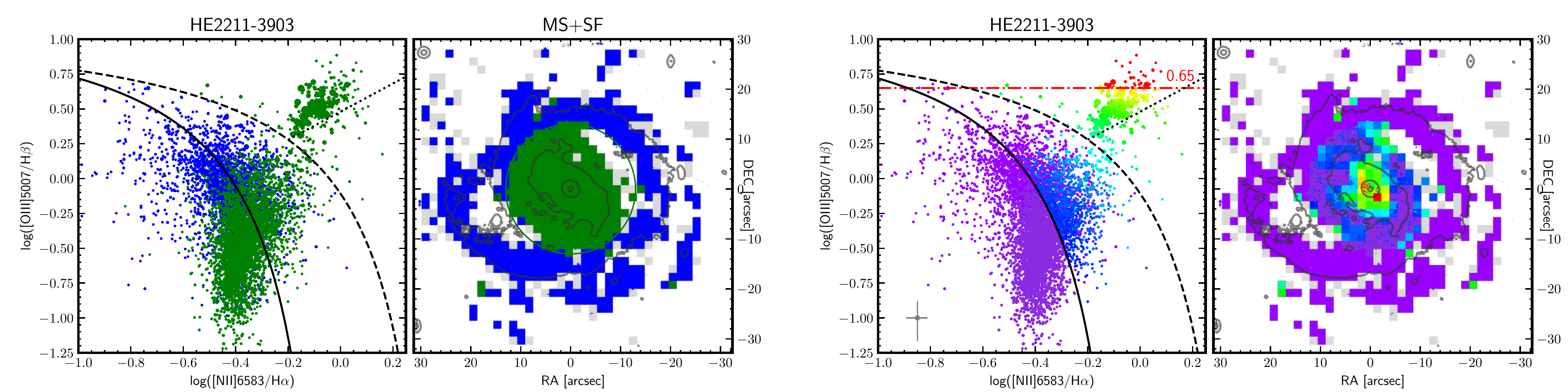}
 \includegraphics[width=1.\textwidth]{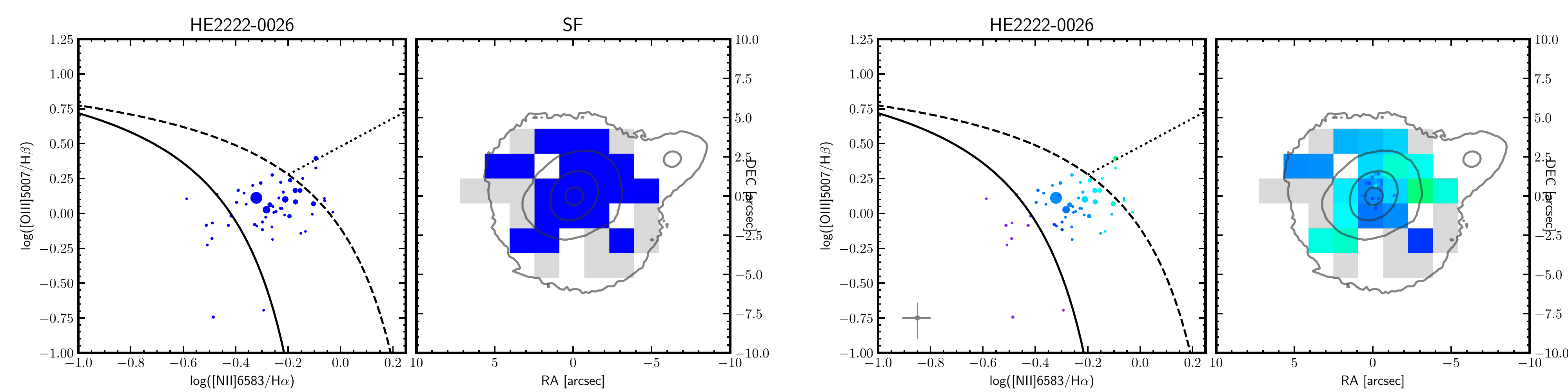}
 \includegraphics[width=1.\textwidth]{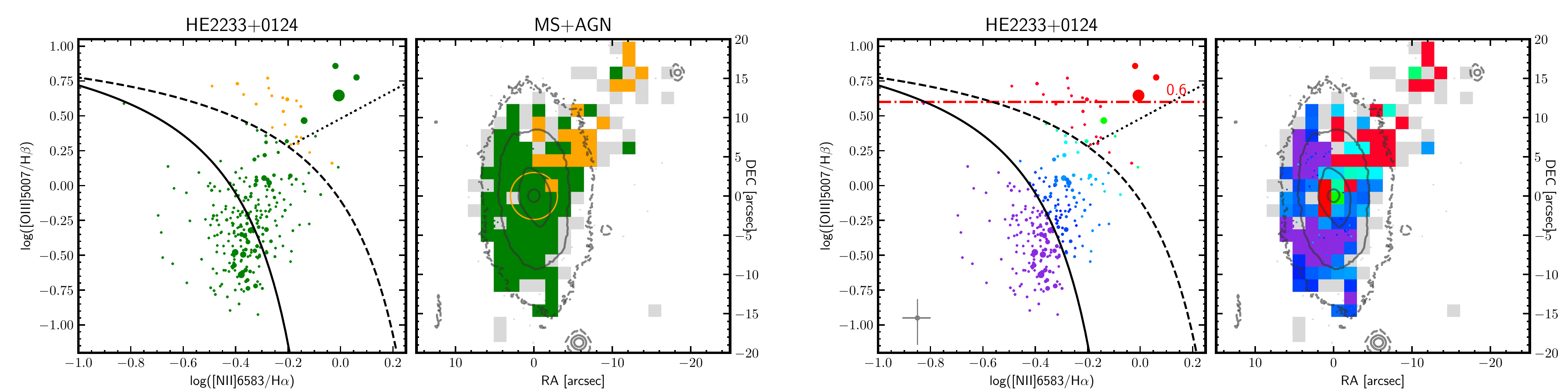}
 \includegraphics[width=1.\textwidth]{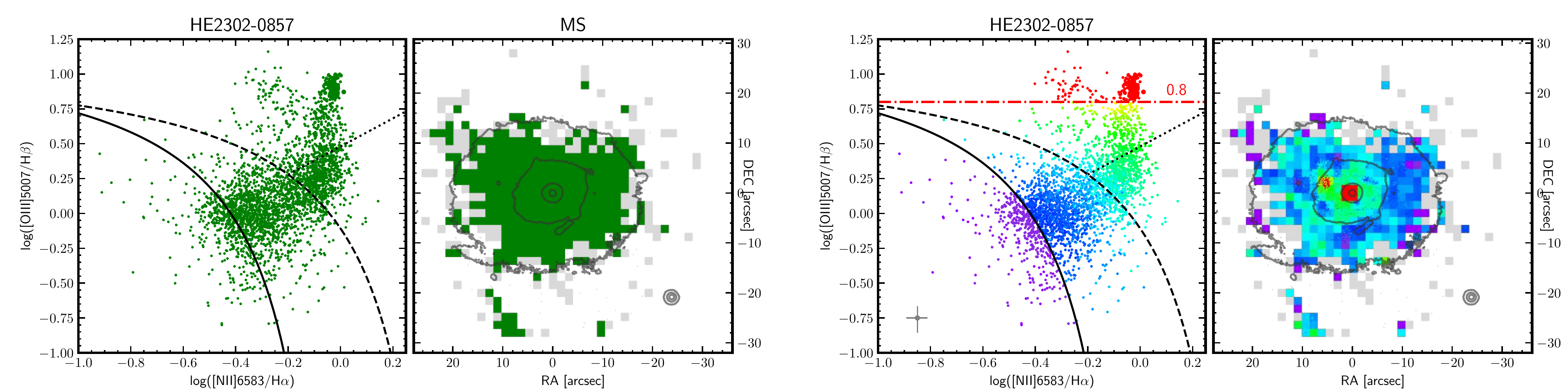}
 \caption{Continued.}
\end{figure*}
\section{Metallicity maps}
 \begin{figure*}
  \includegraphics[width=.5\textwidth]{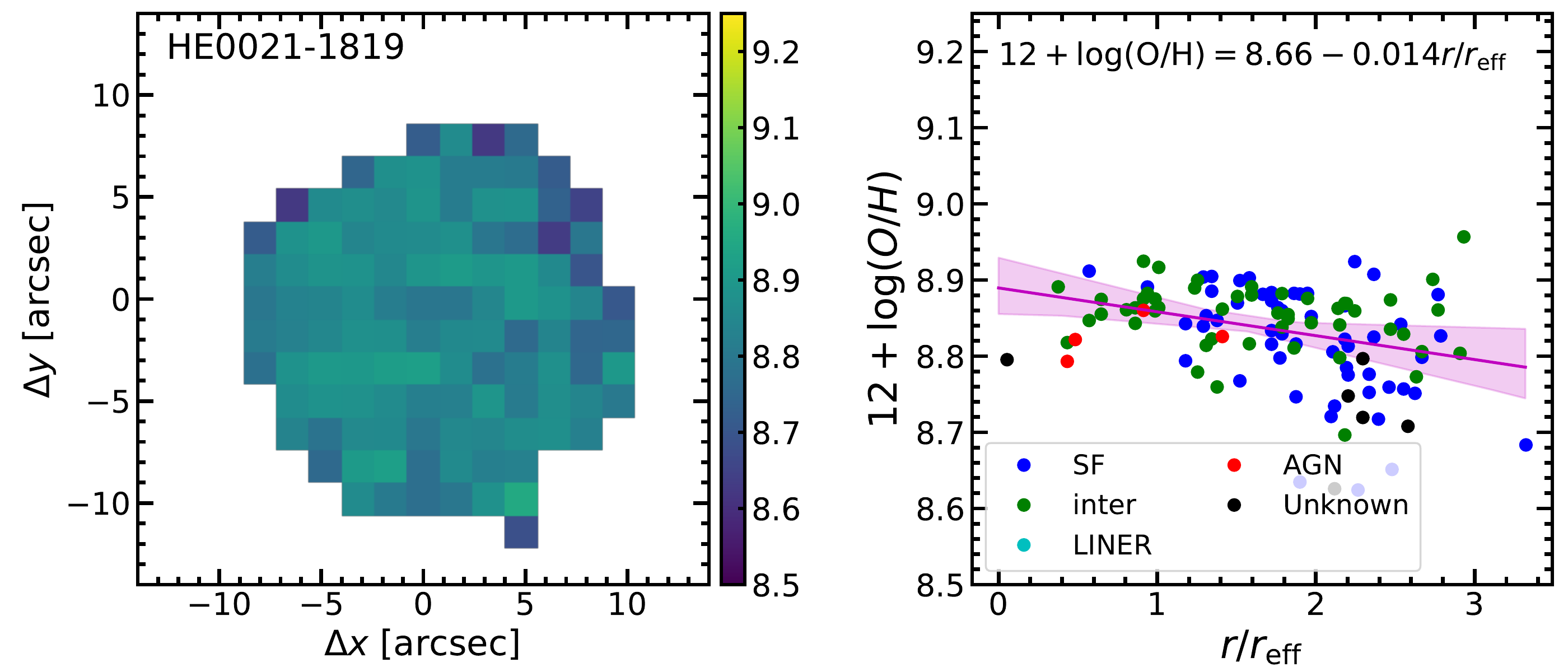}
  \includegraphics[width=.5\textwidth]{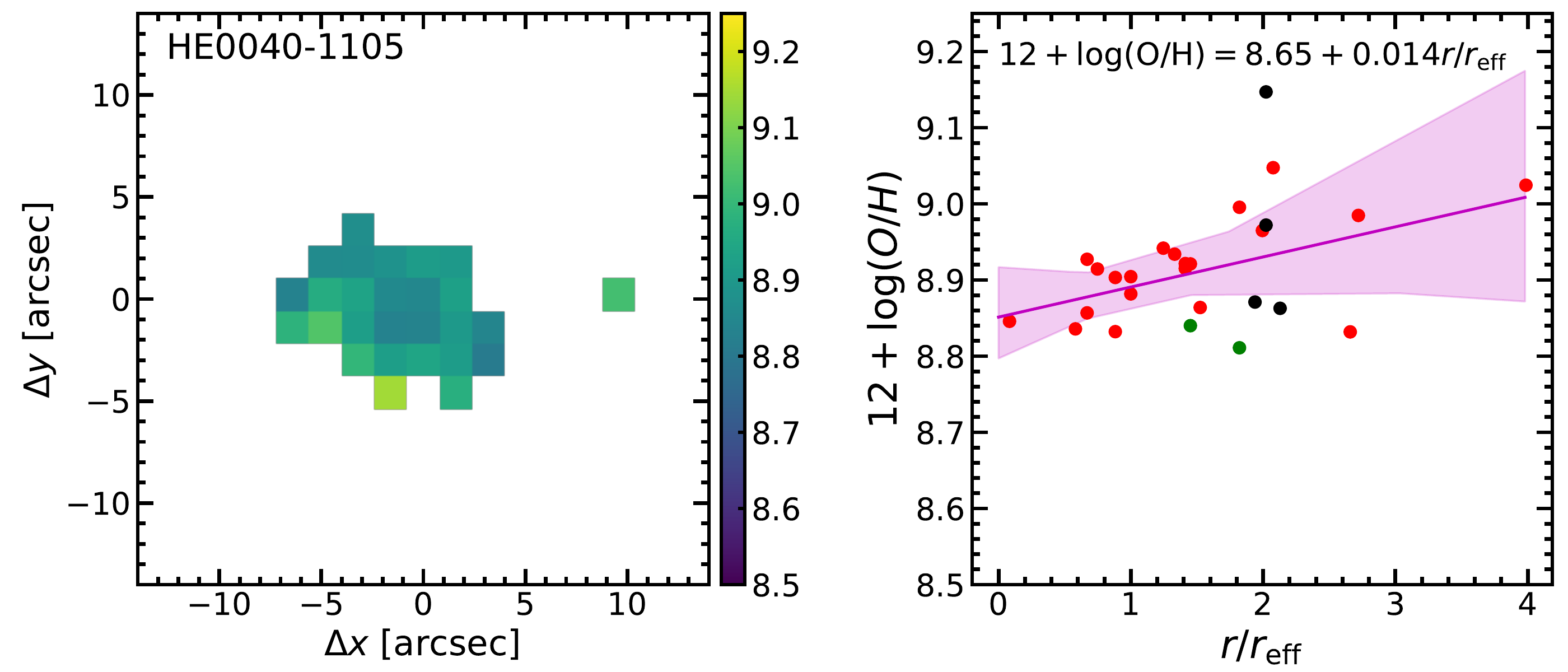}\\
  \includegraphics[width=.5\textwidth]{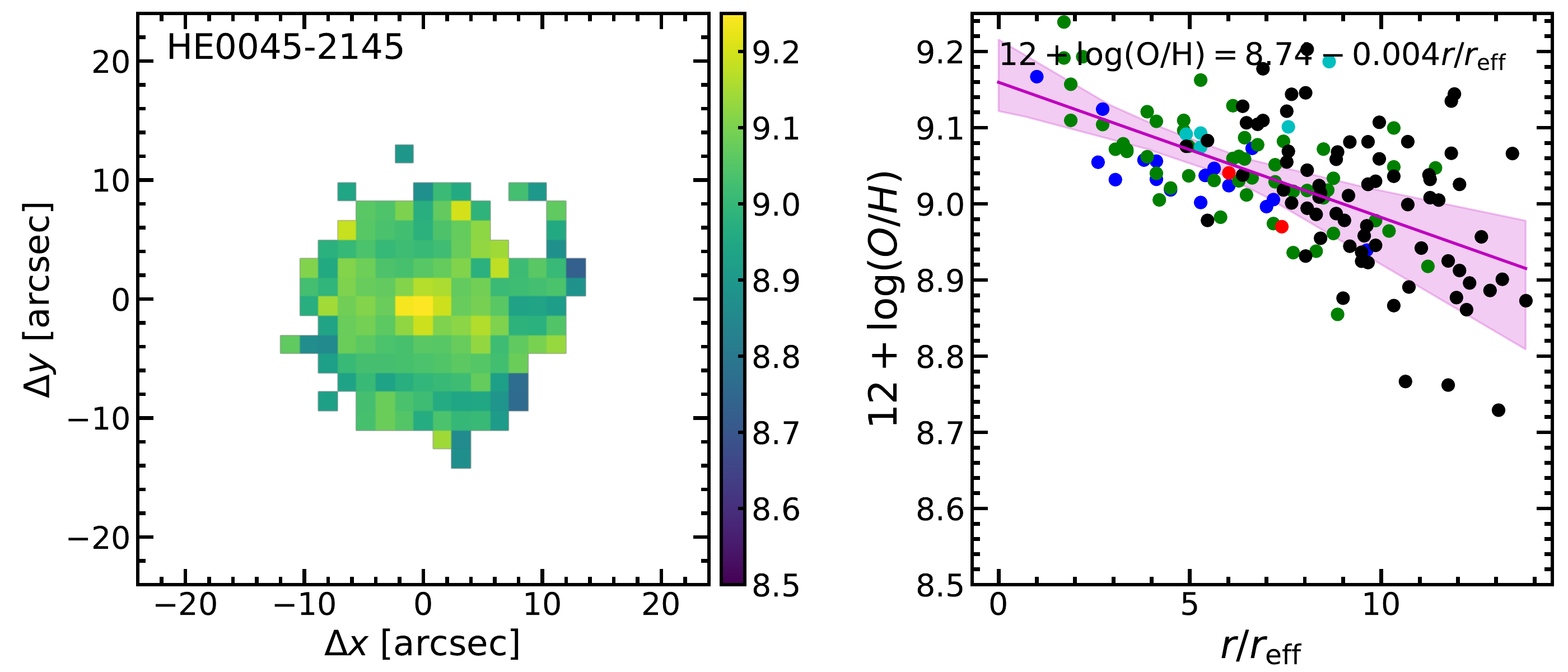}
  \includegraphics[width=.5\textwidth]{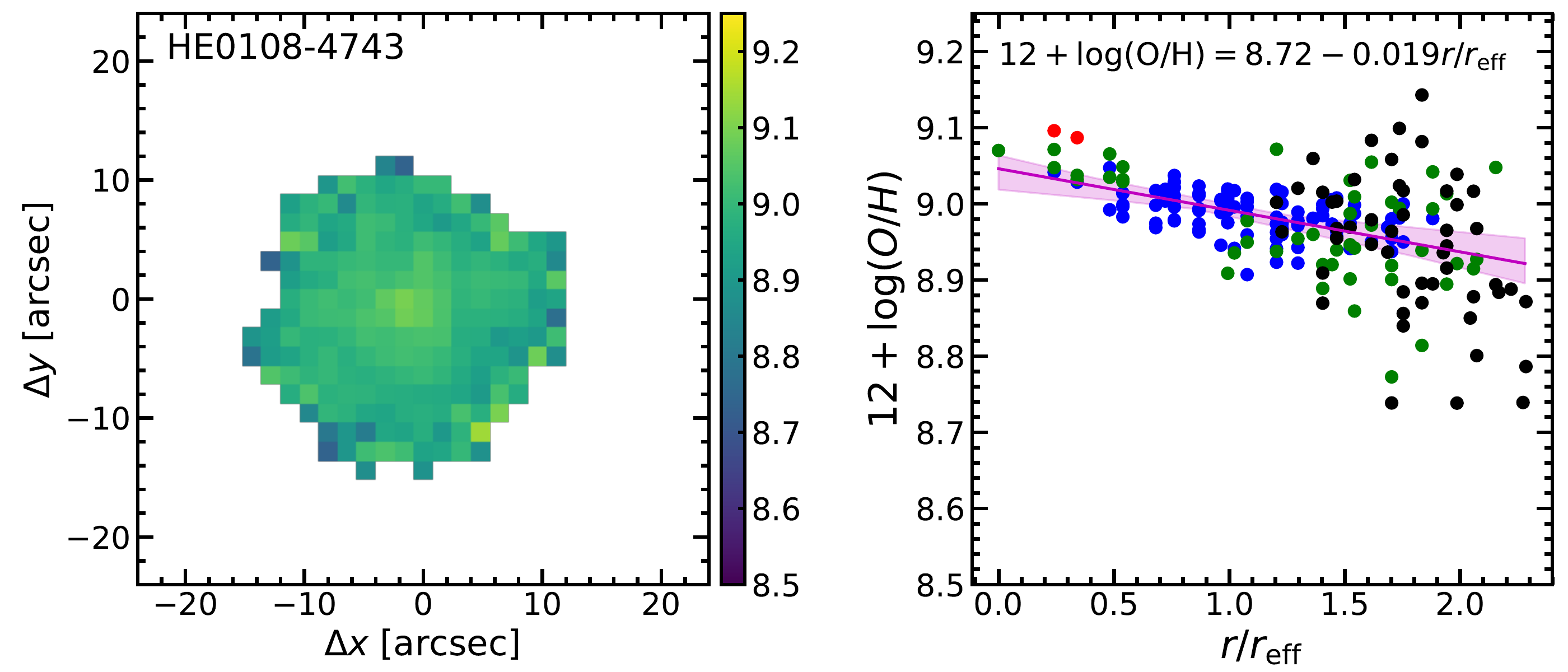}\\
  \includegraphics[width=.5\textwidth]{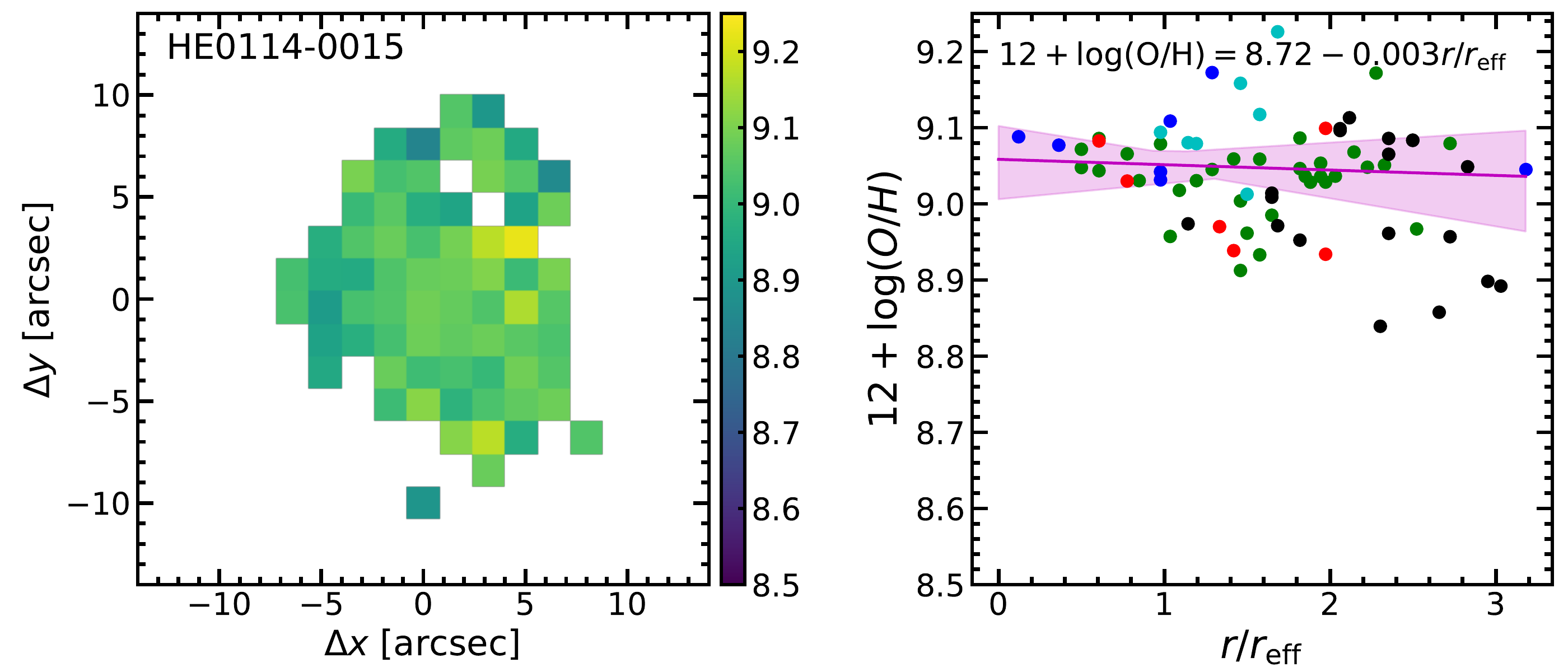}
  \includegraphics[width=.5\textwidth]{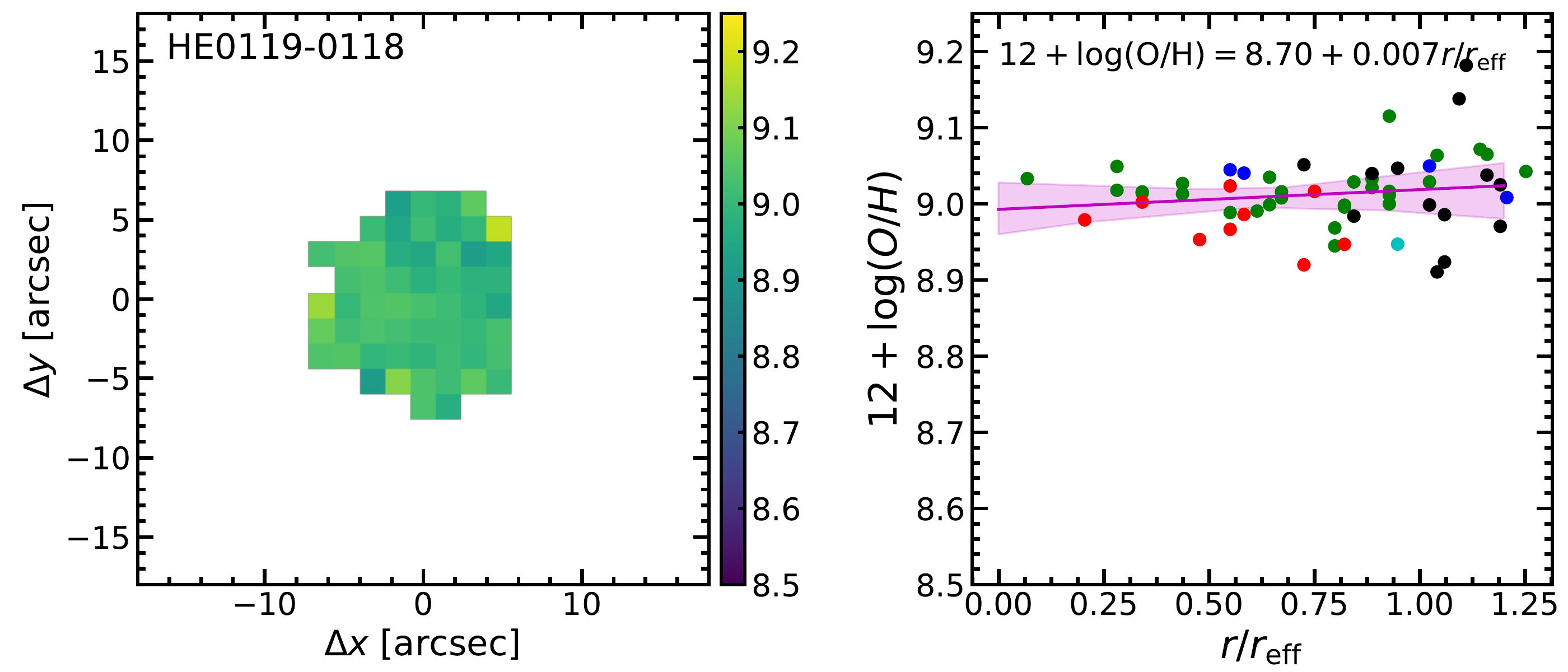}\\
  \includegraphics[width=.5\textwidth]{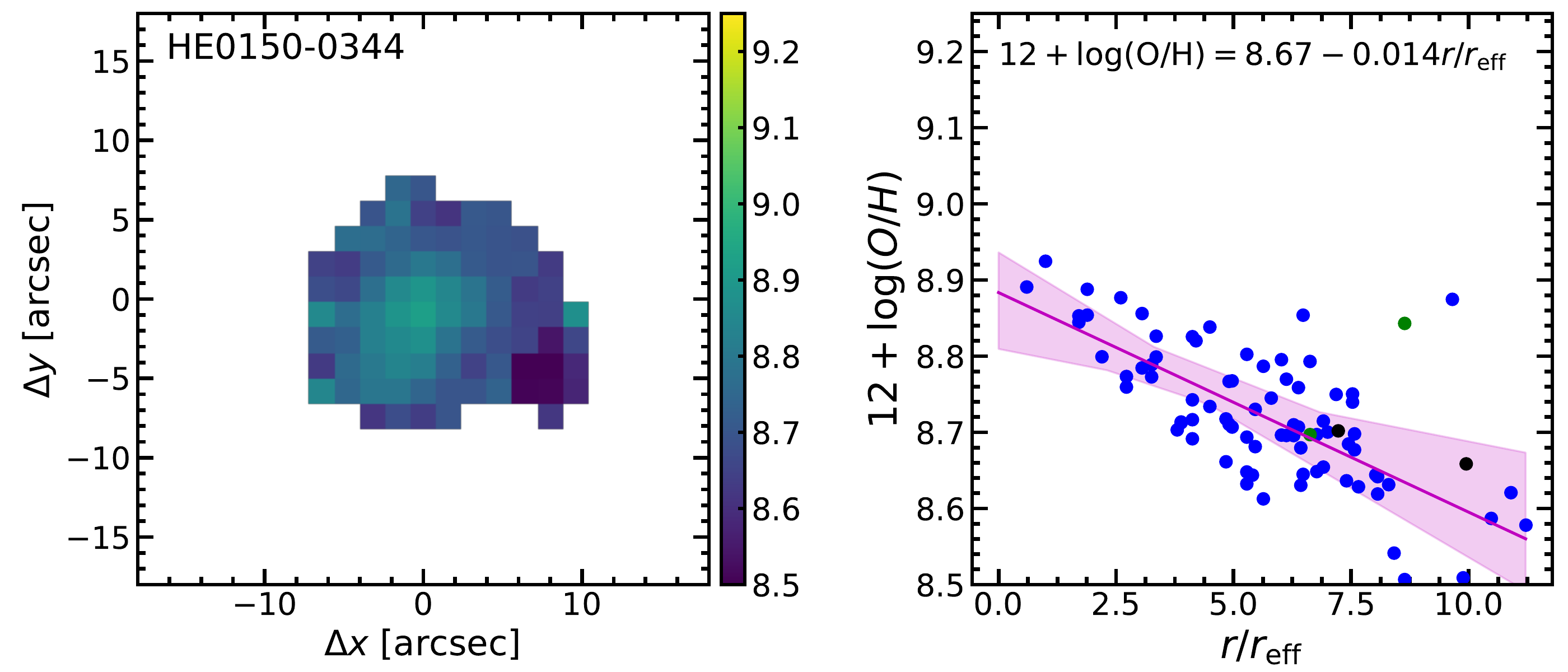}
  \includegraphics[width=.5\textwidth]{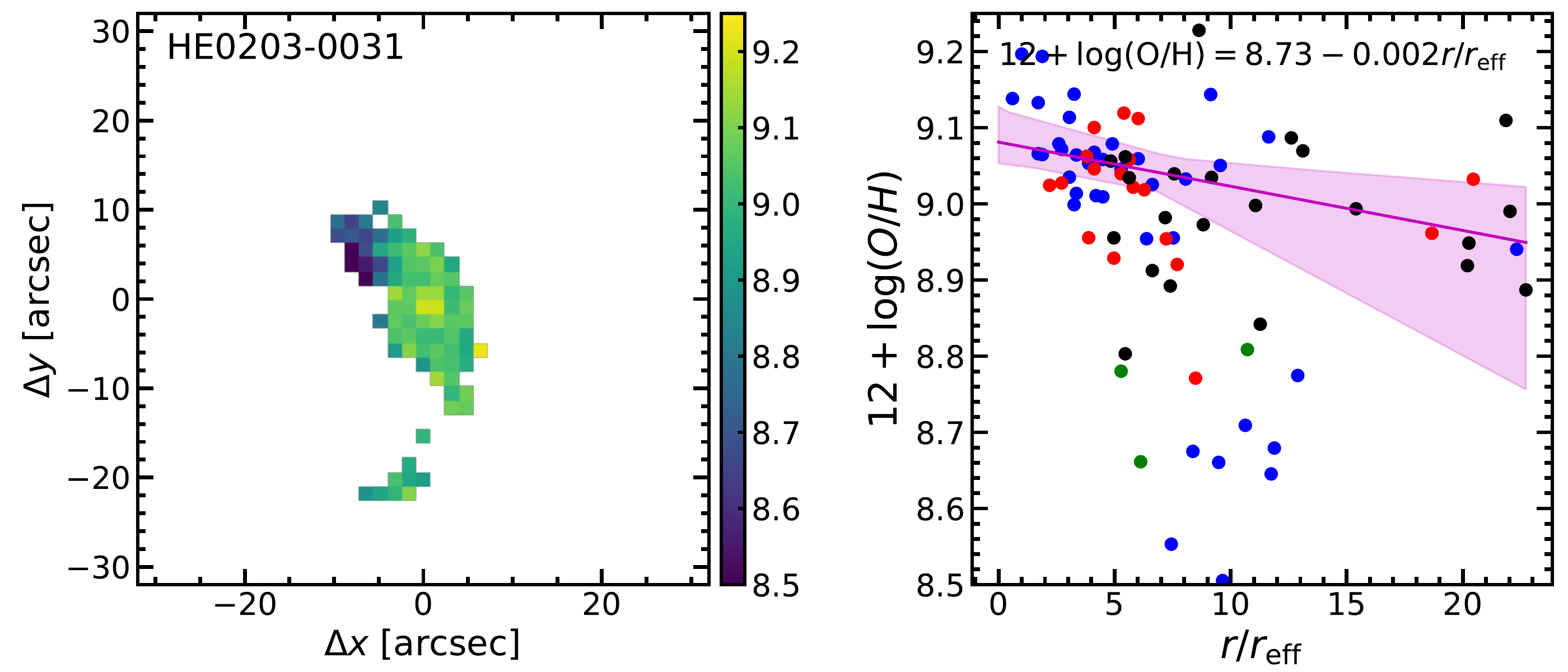}\\
  \includegraphics[width=.5\textwidth]{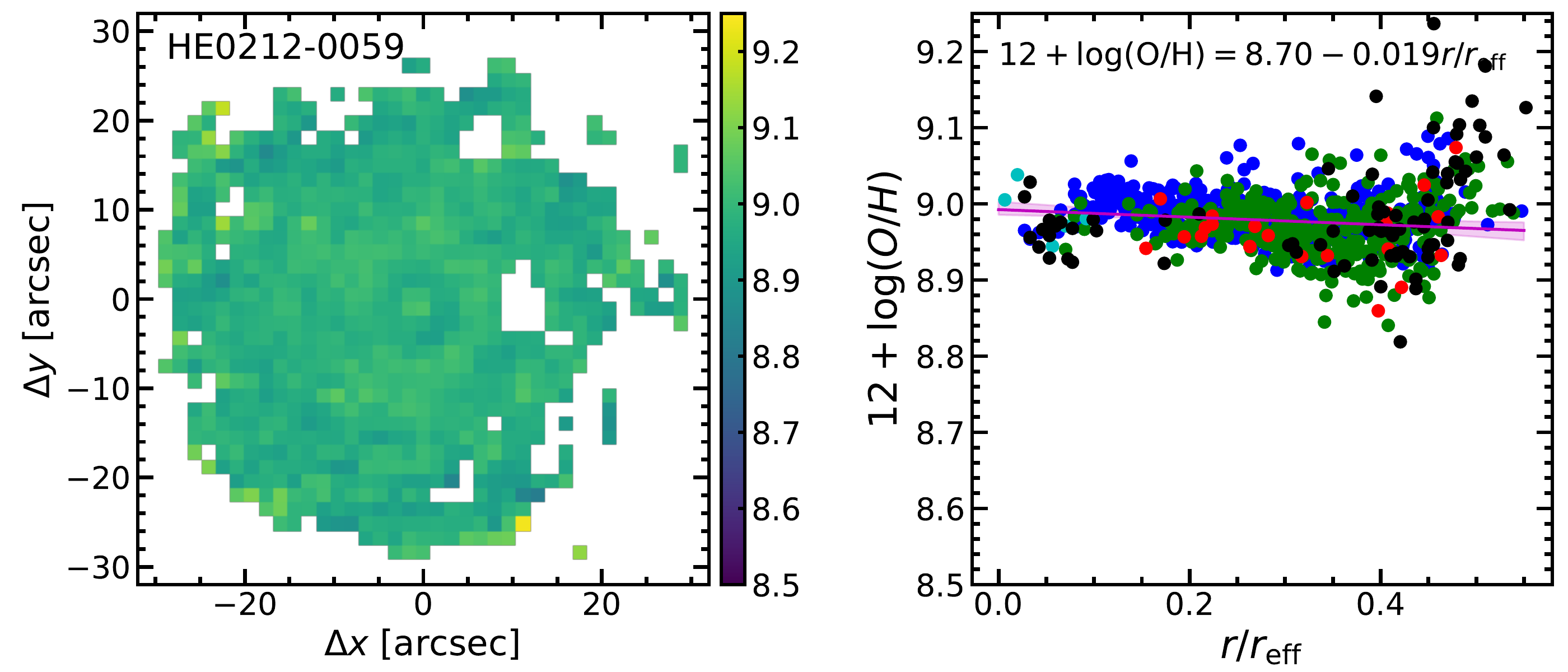}
  \includegraphics[width=.5\textwidth]{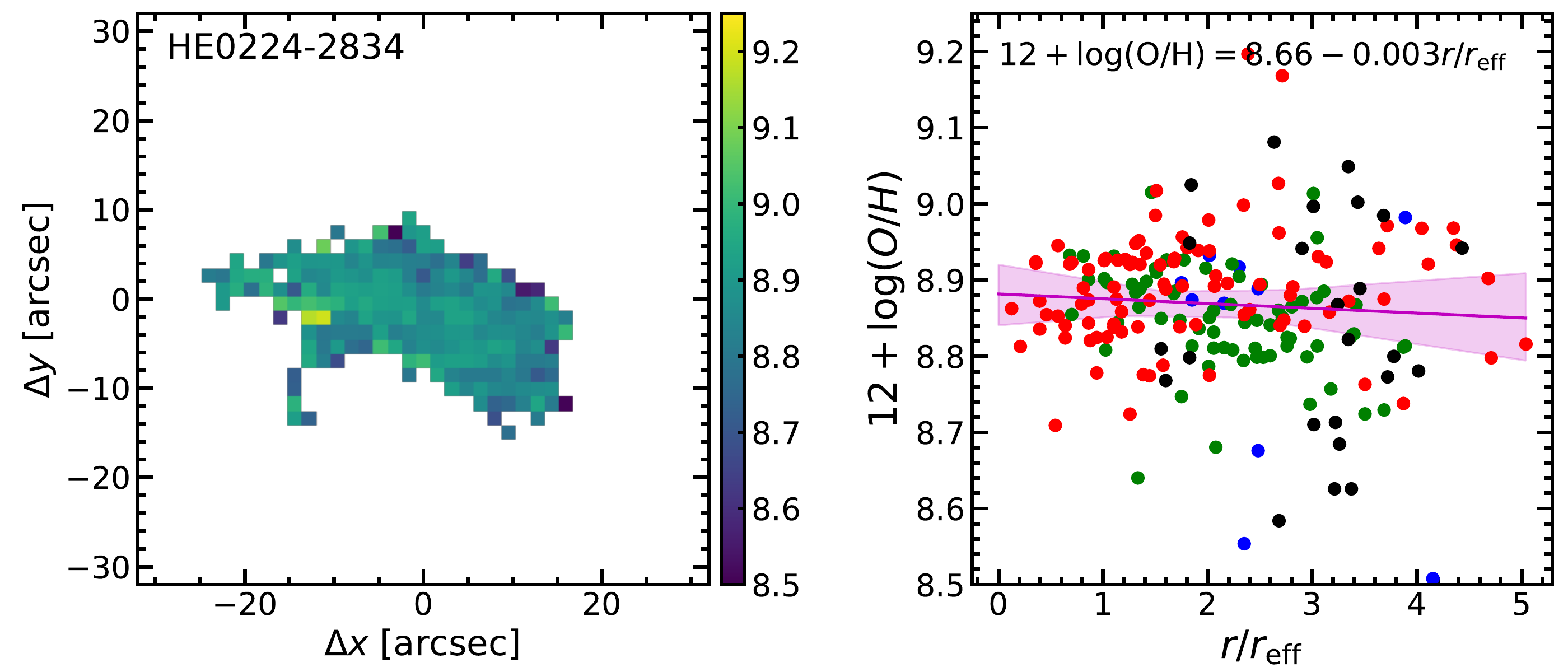}\\
  \caption{} \label{fig:metals_apx}
 \end{figure*}
 
 \begin{figure*}
  \includegraphics[width=.5\textwidth]{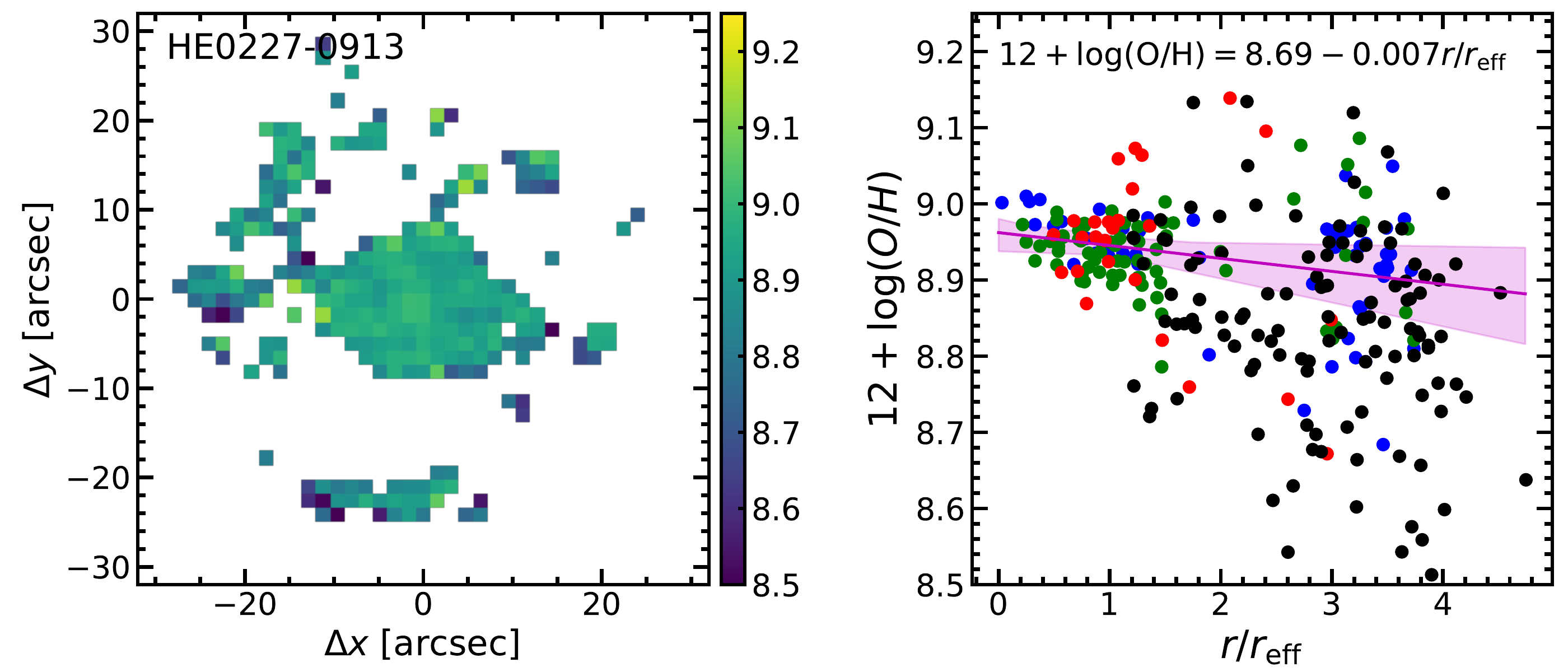}
  \includegraphics[width=.5\textwidth]{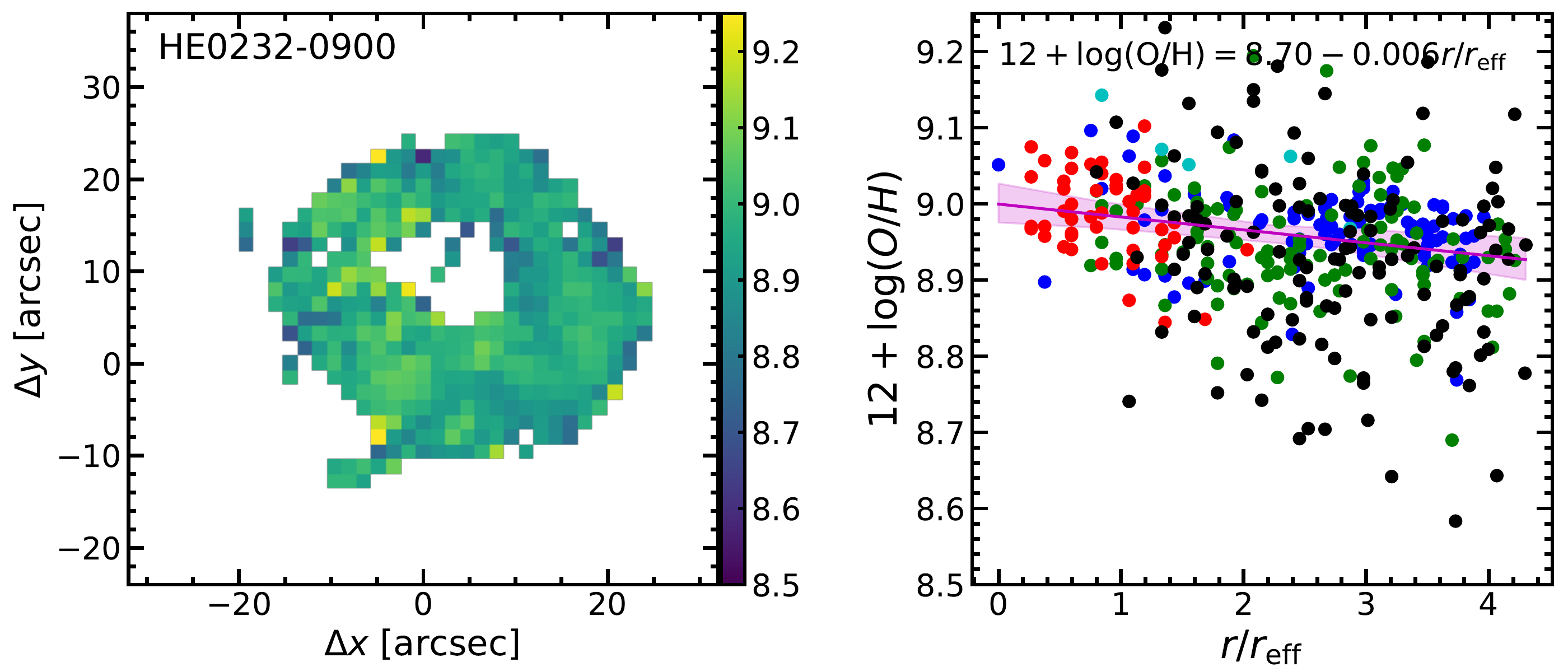}\\
  \includegraphics[width=.5\textwidth]{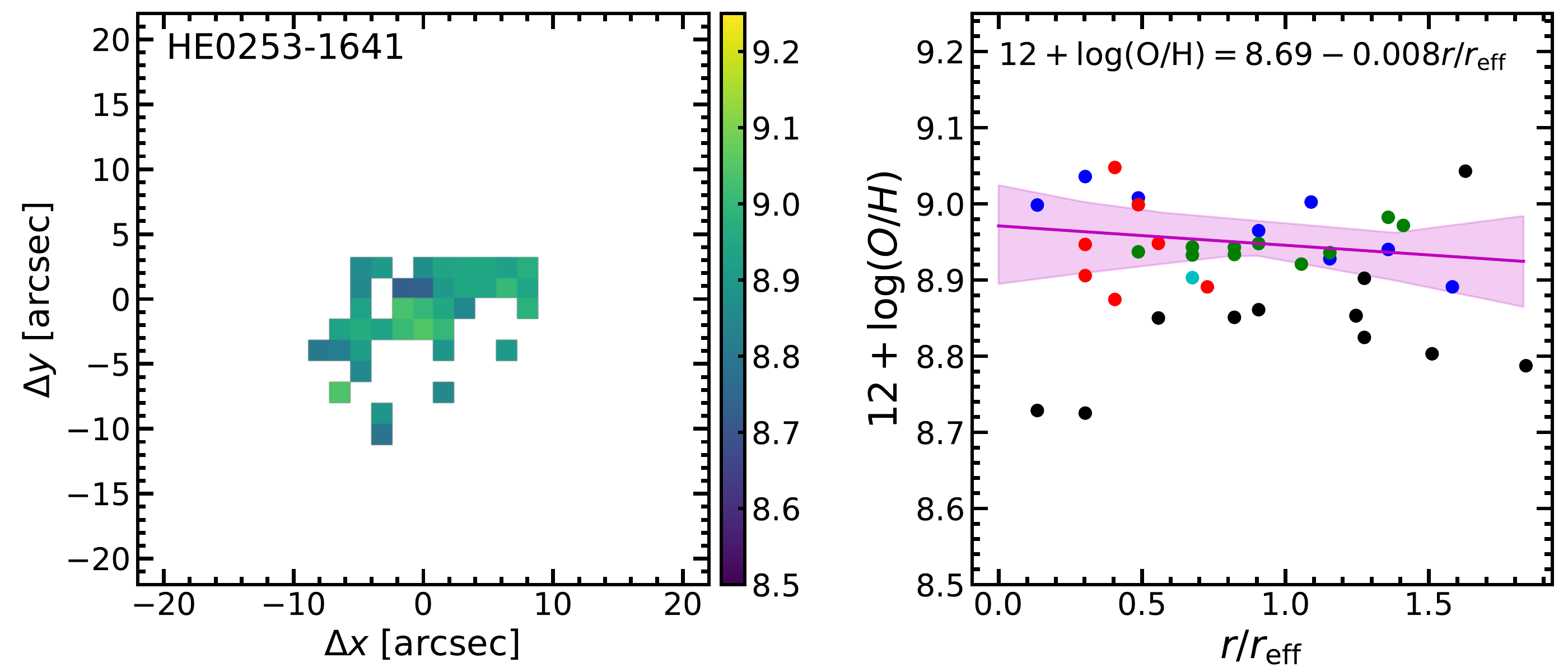}
  \includegraphics[width=.5\textwidth]{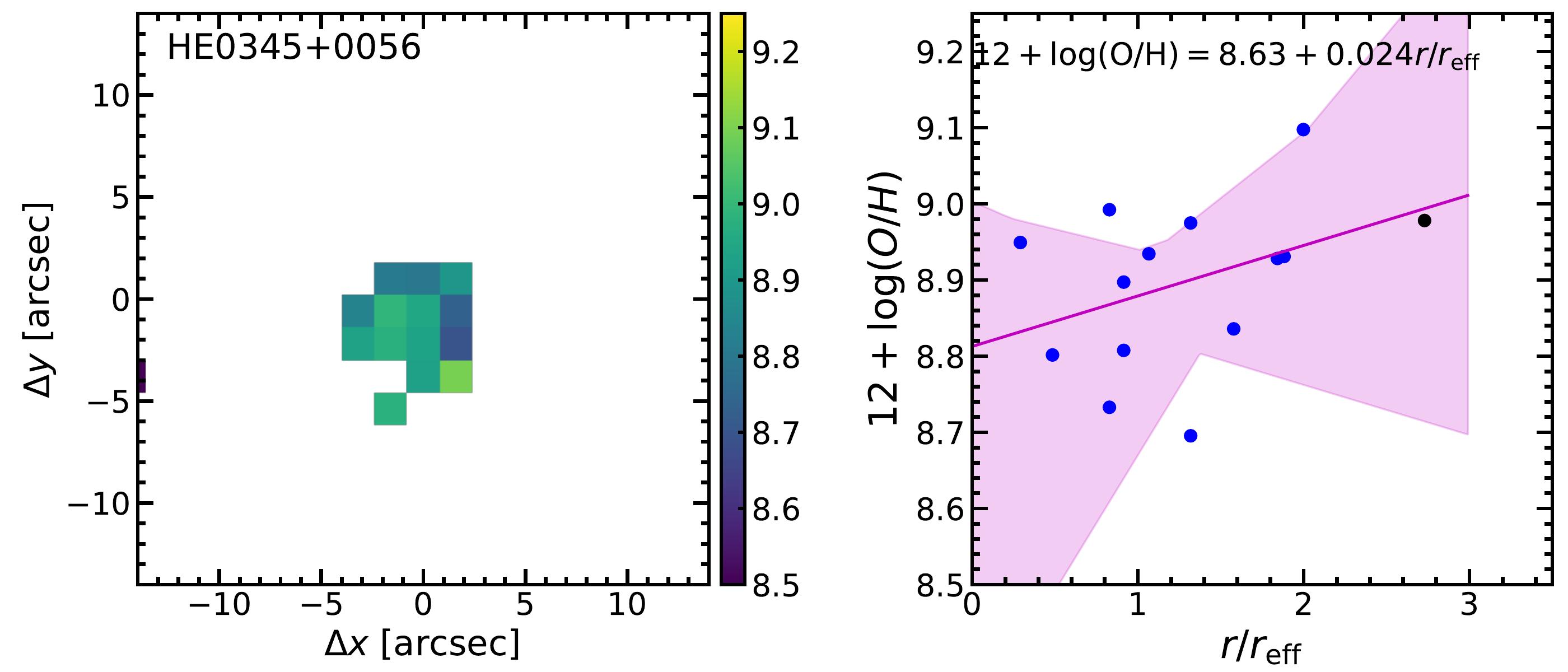}\\
  \includegraphics[width=.5\textwidth]{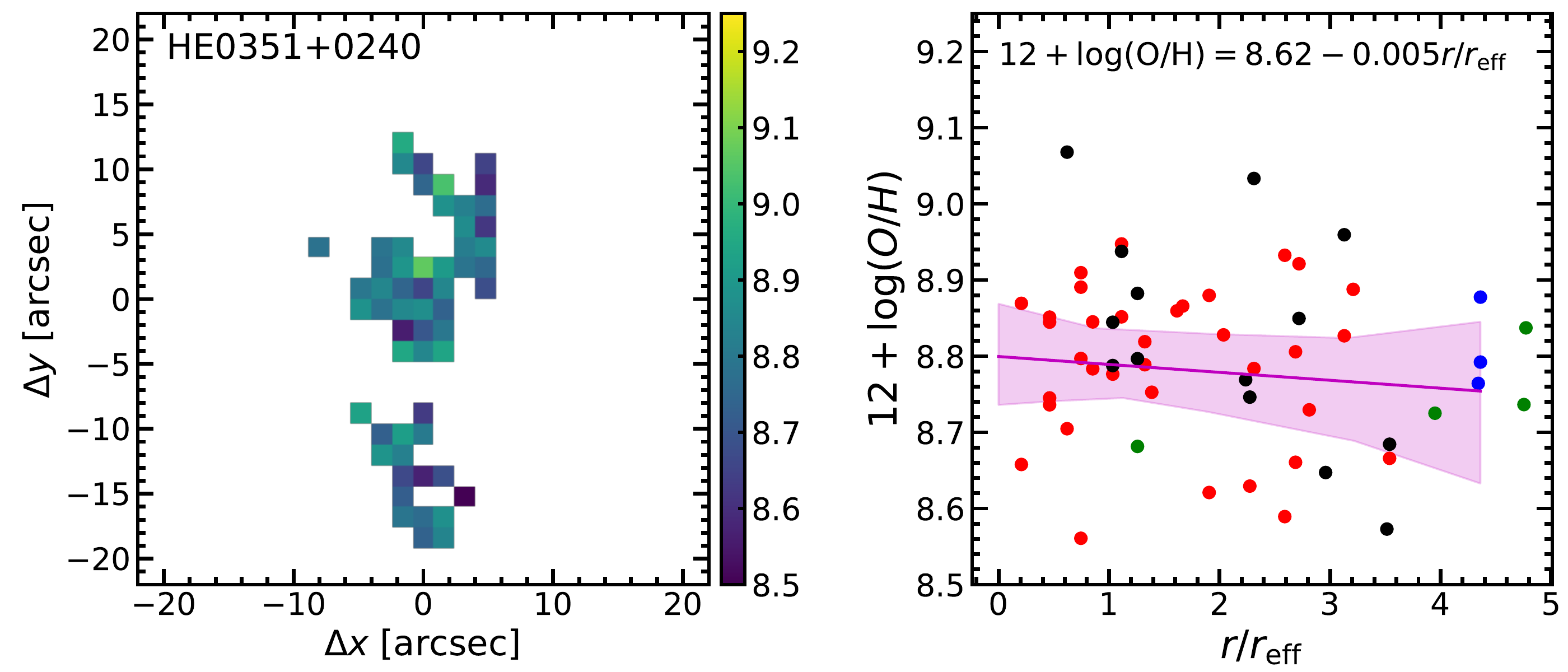}
  \includegraphics[width=.5\textwidth]{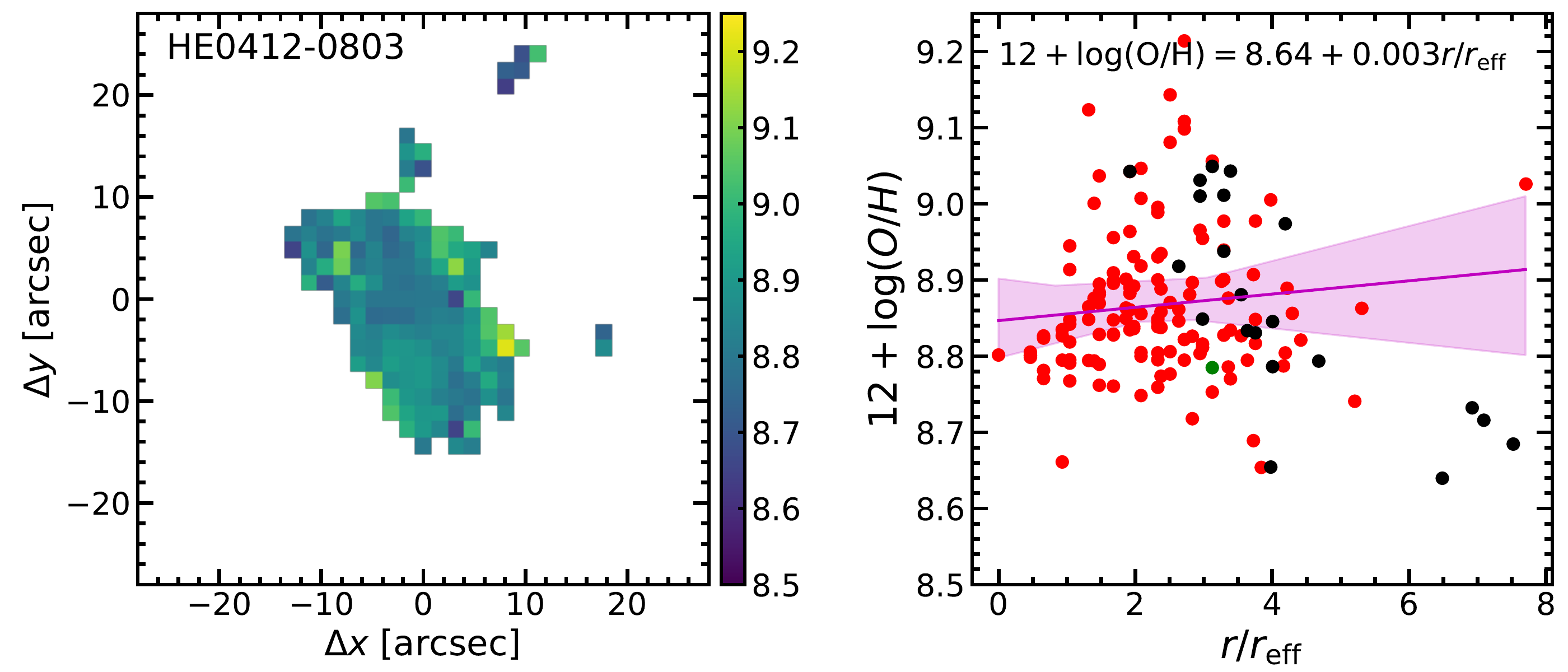}\\
  \includegraphics[width=.5\textwidth]{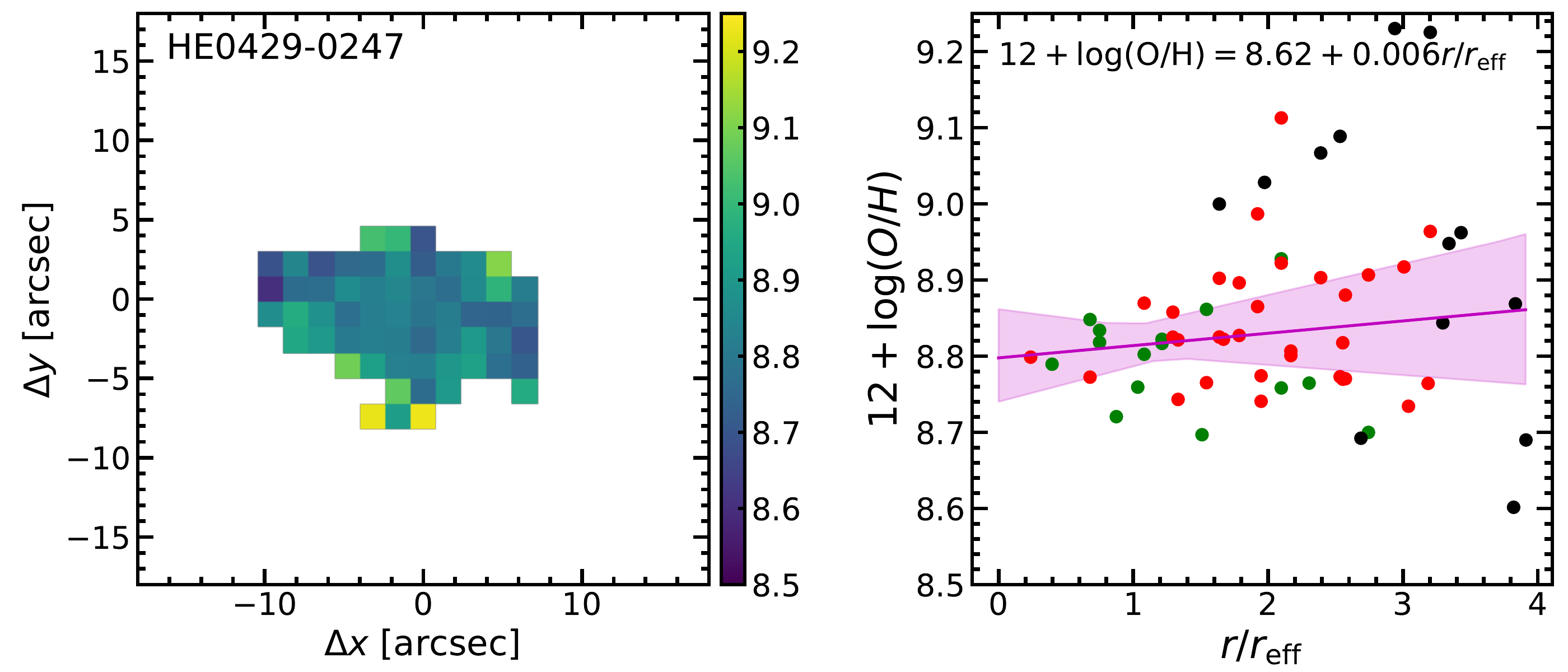}
  \includegraphics[width=.5\textwidth]{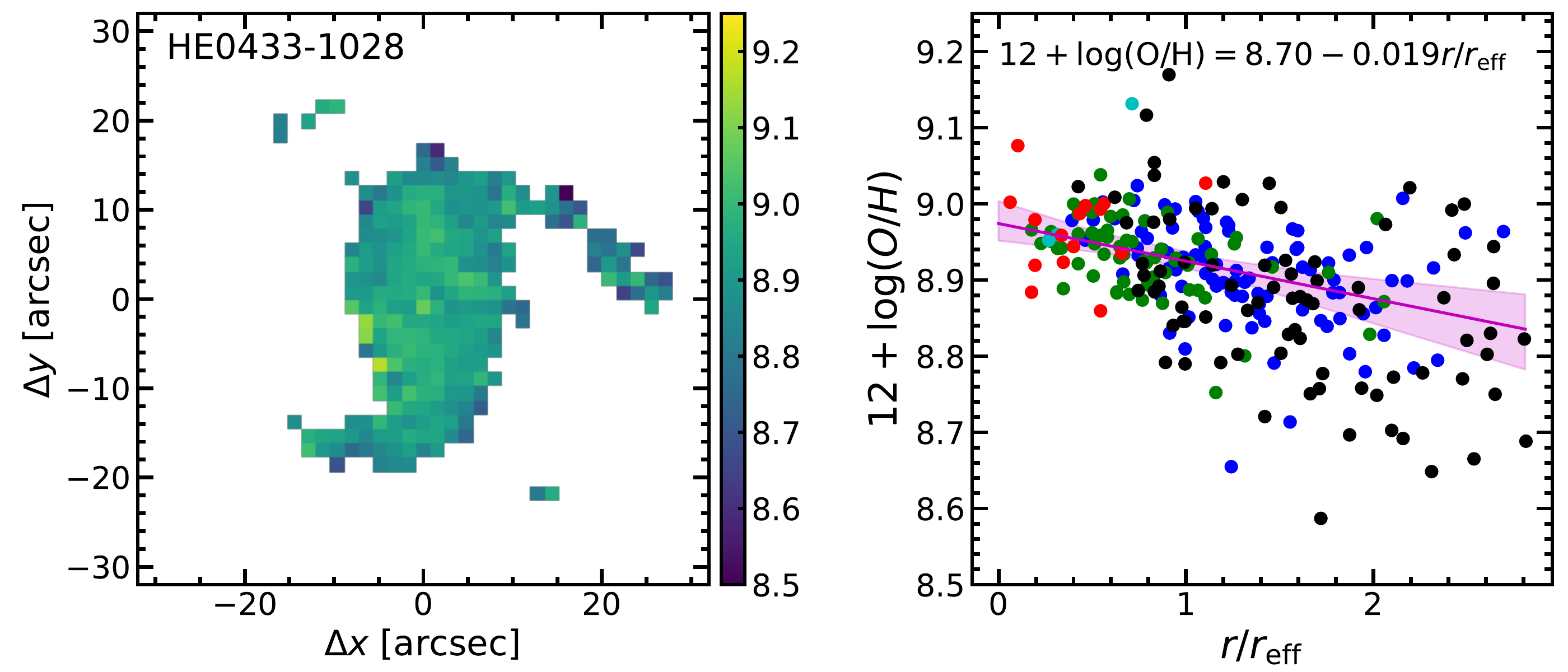}\\
  \includegraphics[width=.5\textwidth]{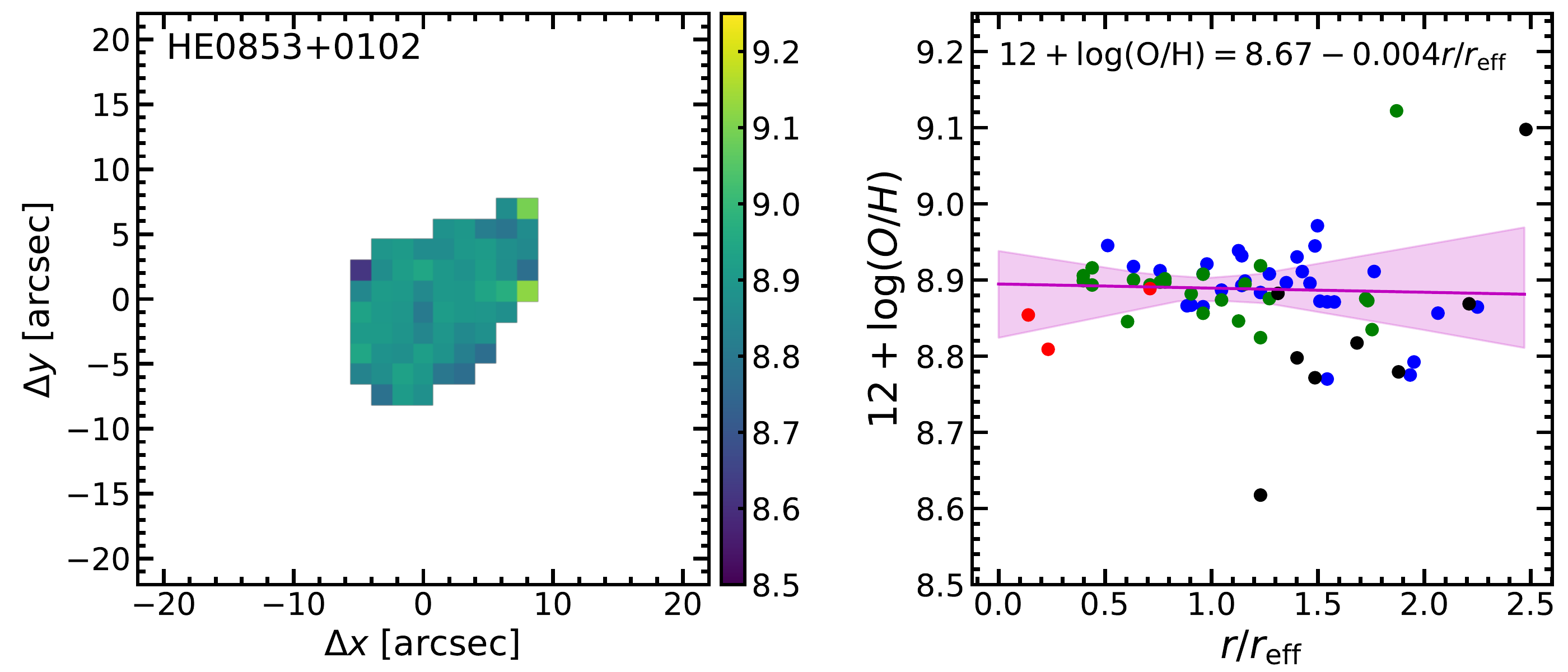}
  \includegraphics[width=.5\textwidth]{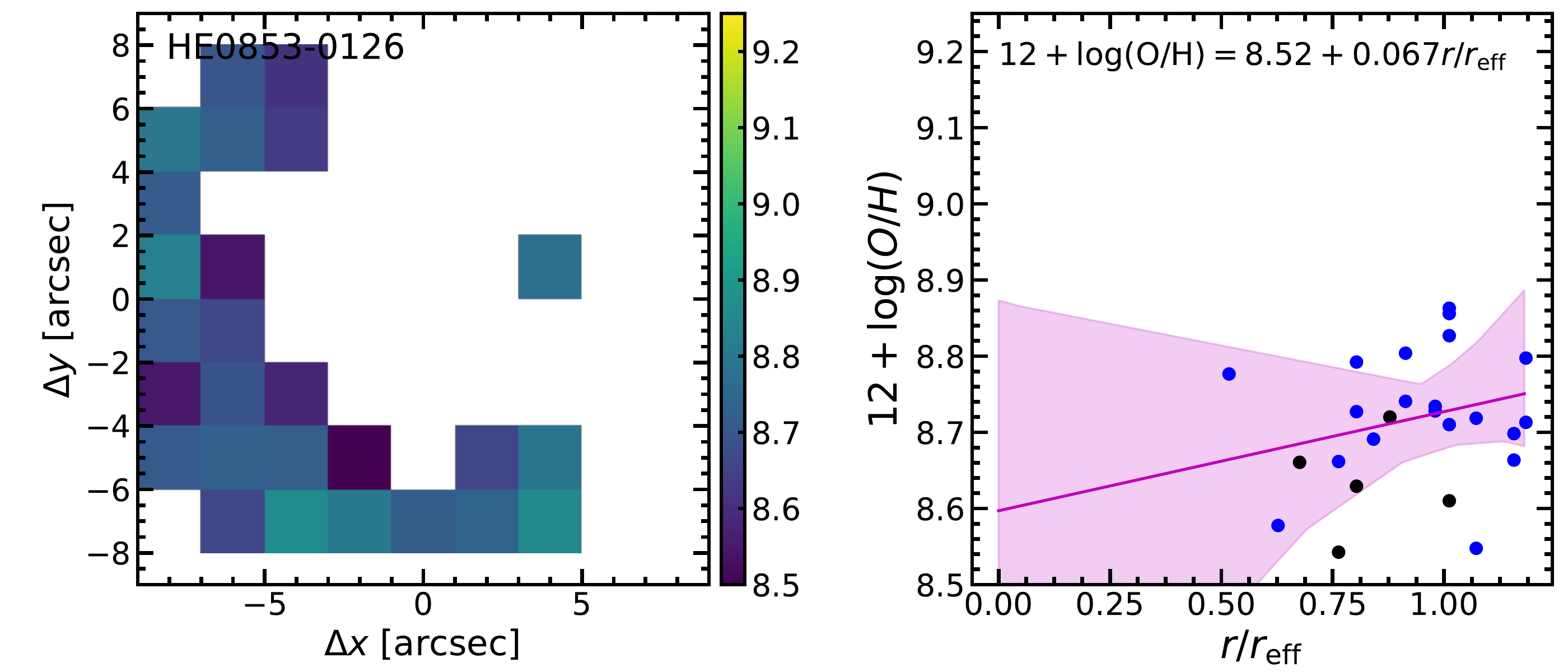}\\
  \includegraphics[width=.5\textwidth]{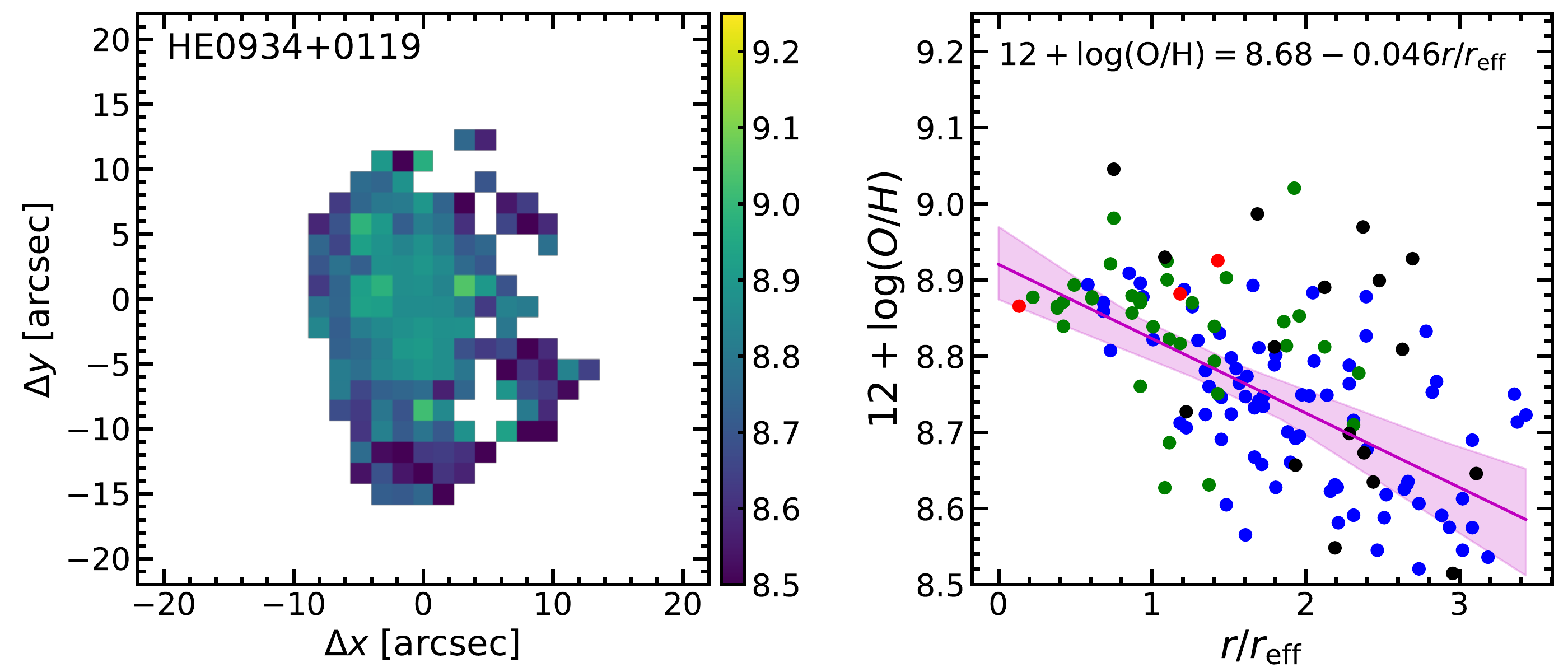}
  \includegraphics[width=.5\textwidth]{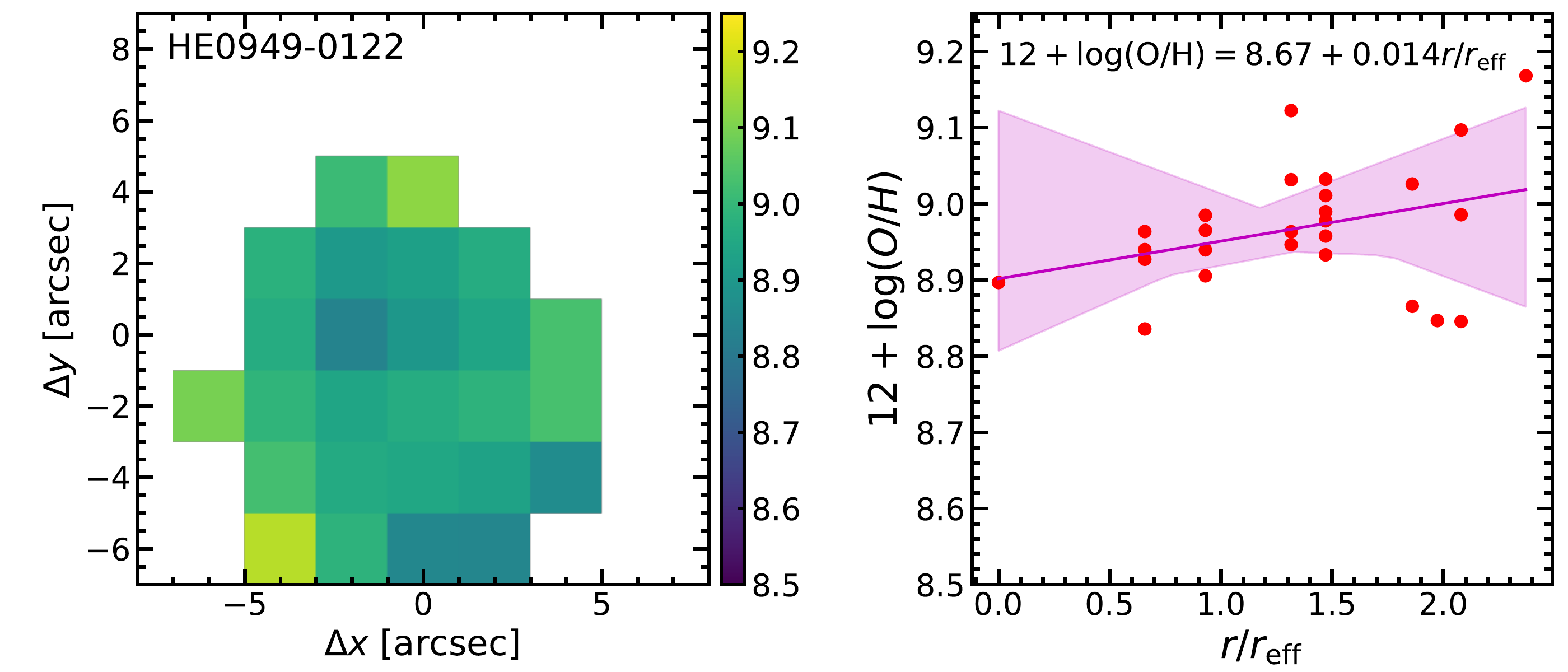}
  \caption{Continued.}

 \end{figure*}
 \begin{figure*}
  \includegraphics[width=.5\textwidth]{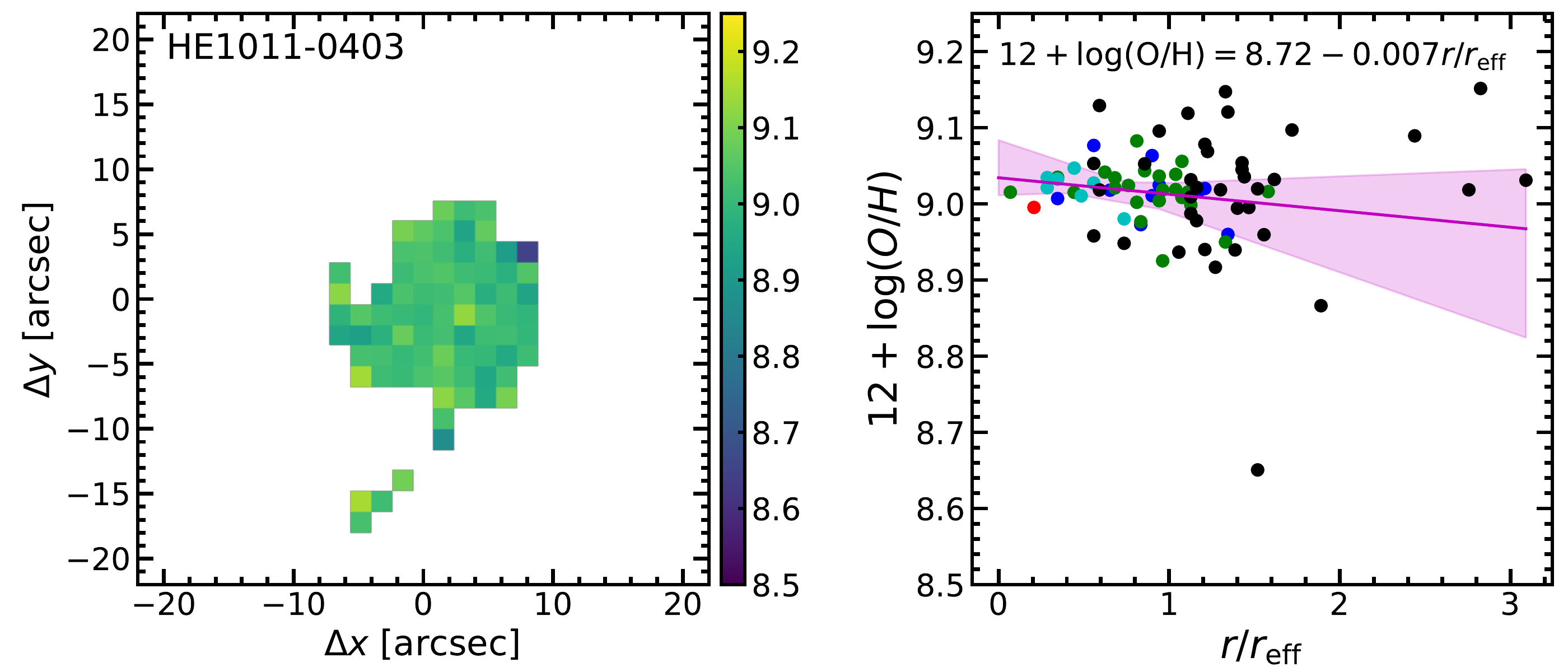}
  \includegraphics[width=.5\textwidth]{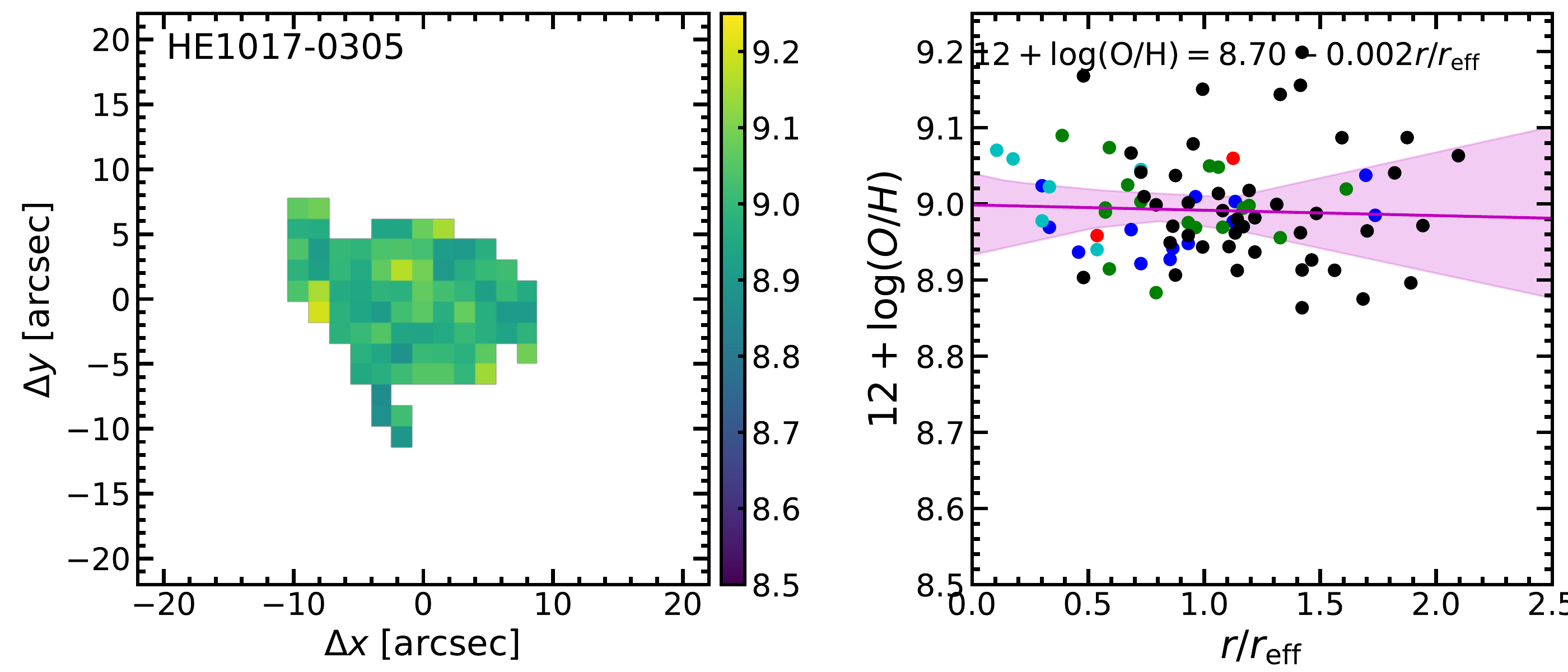}\\
  \includegraphics[width=.5\textwidth]{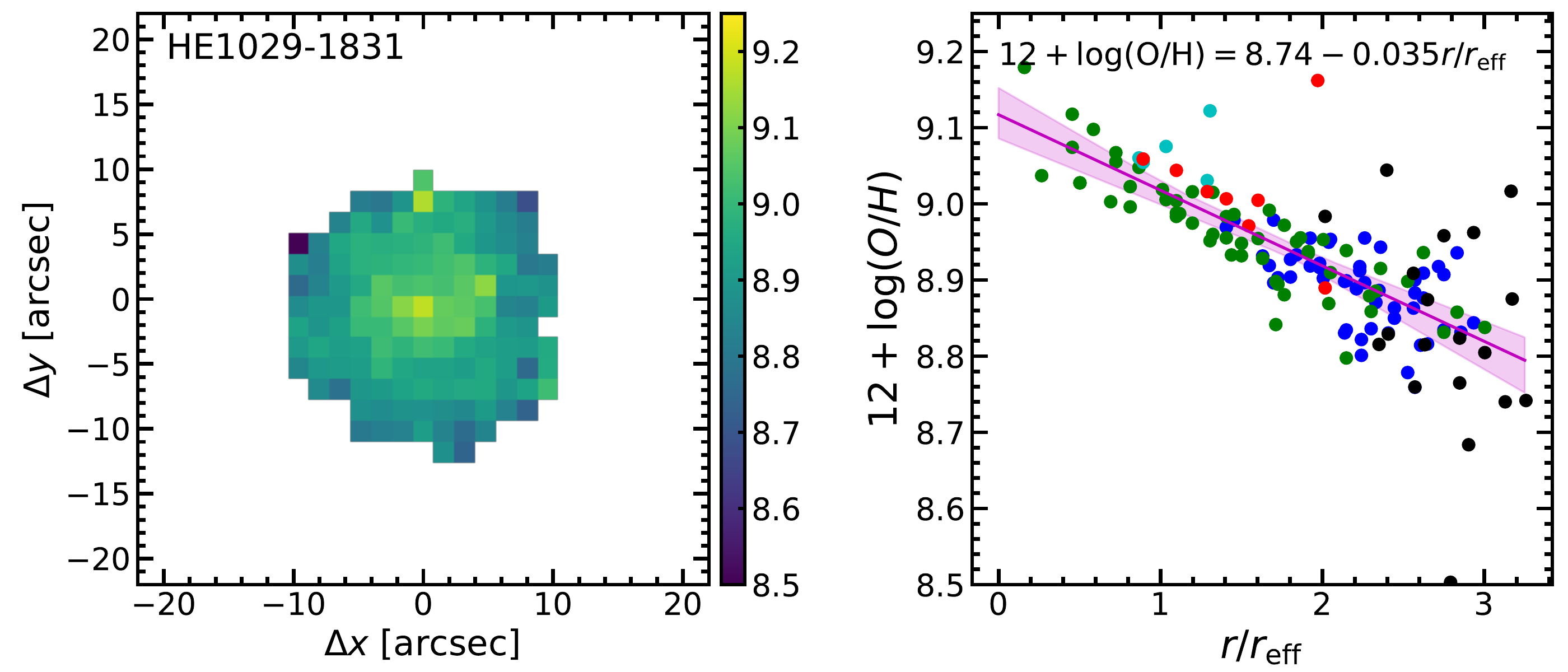}
  \includegraphics[width=.5\textwidth]{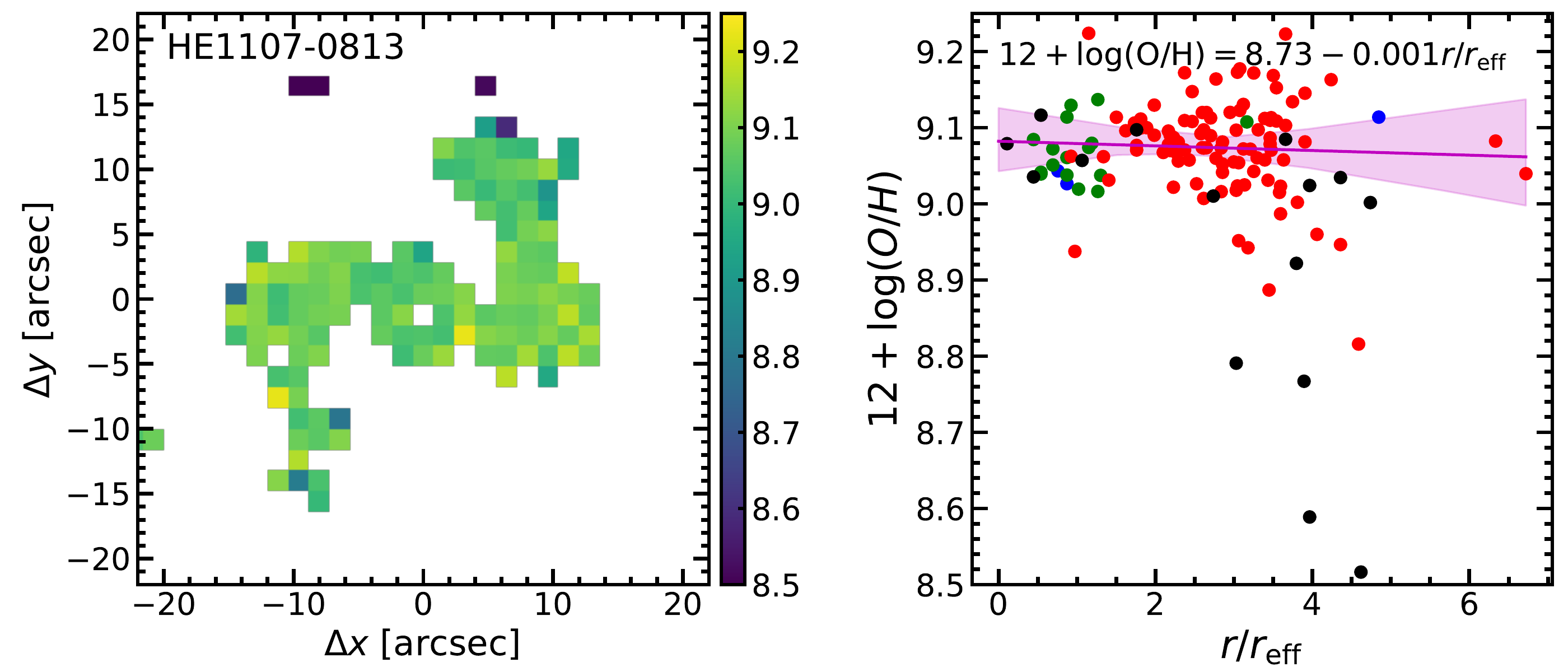}\\
  \includegraphics[width=.5\textwidth]{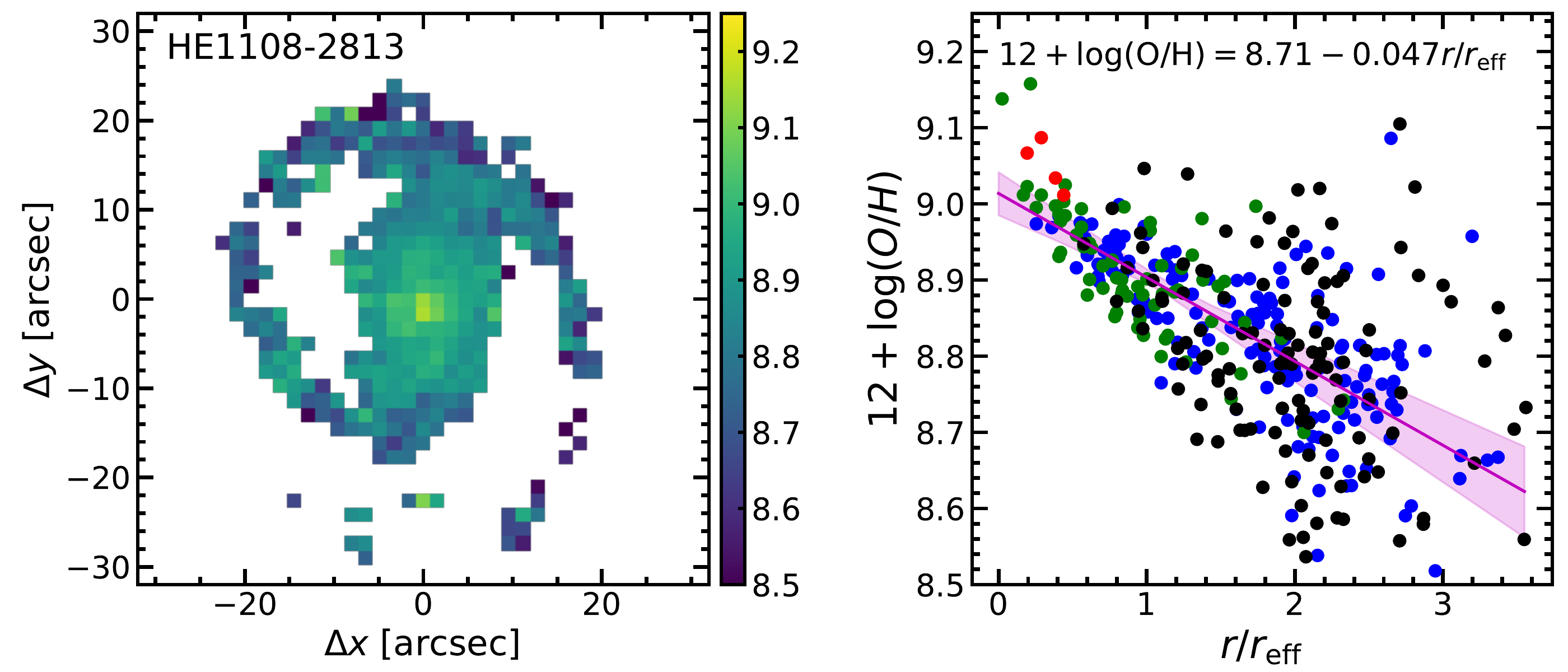}
  \includegraphics[width=.5\textwidth]{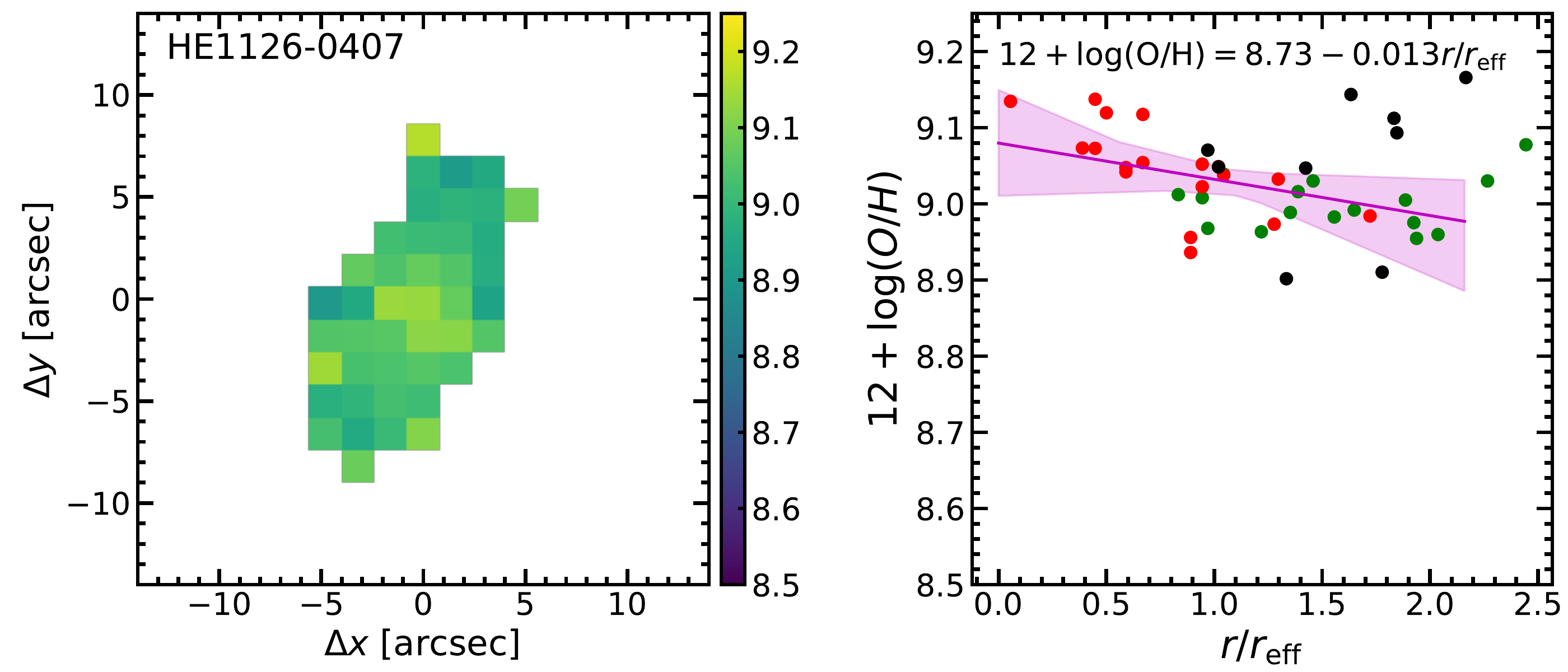}\\
  \includegraphics[width=.5\textwidth]{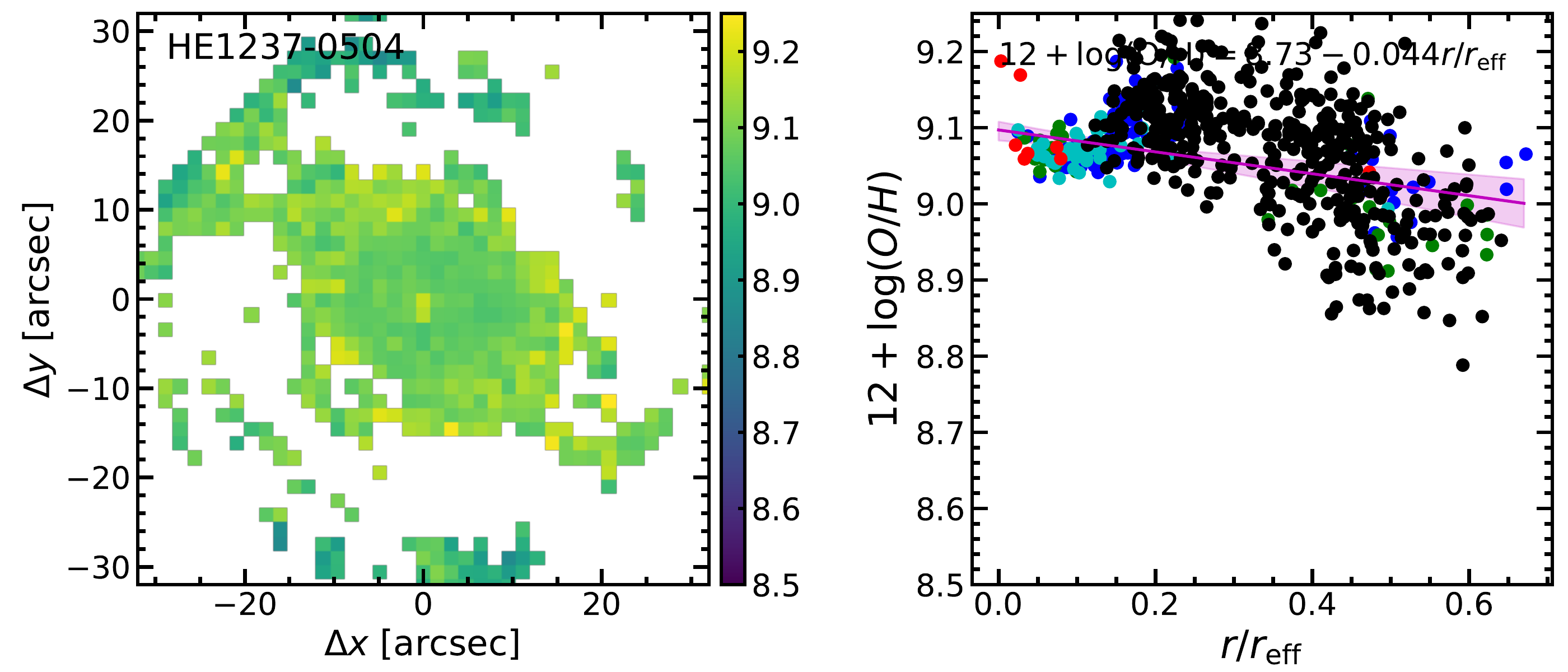}
  \includegraphics[width=.5\textwidth]{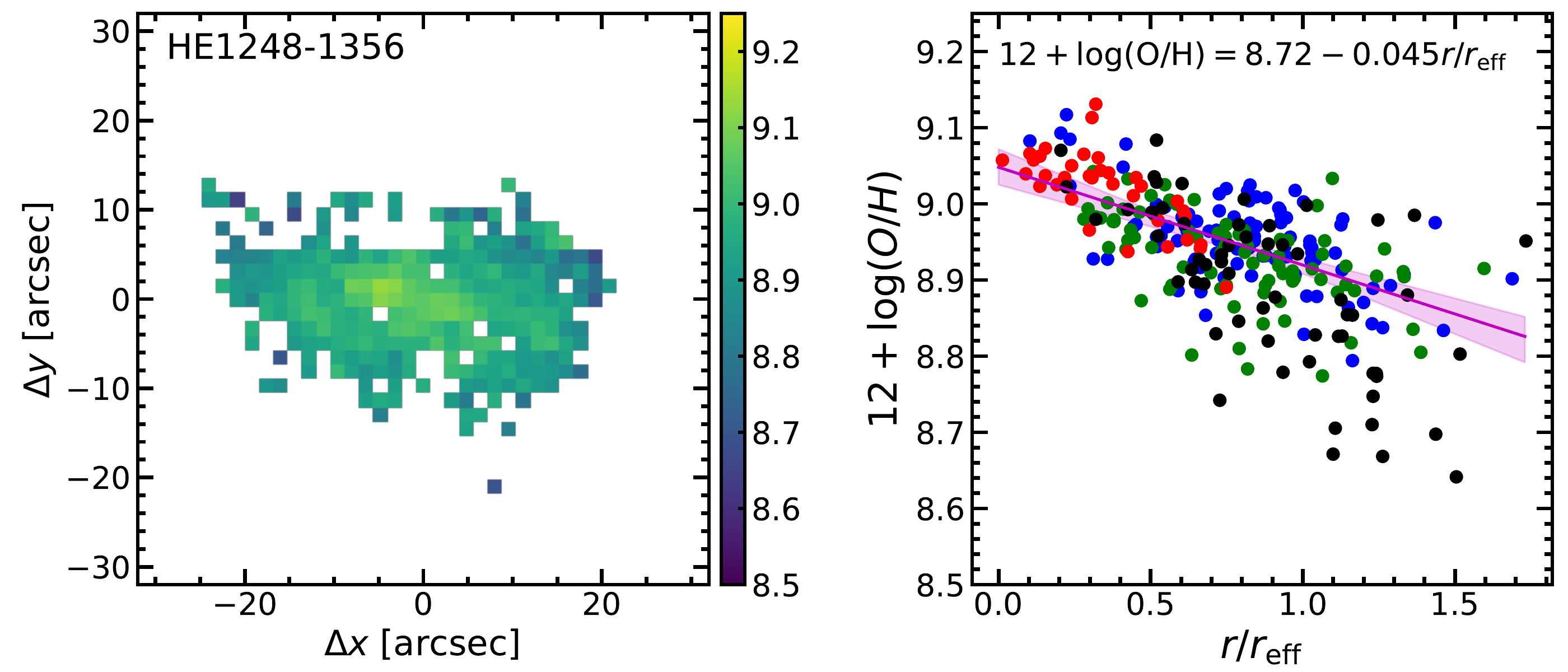}\\
  \includegraphics[width=.5\textwidth]{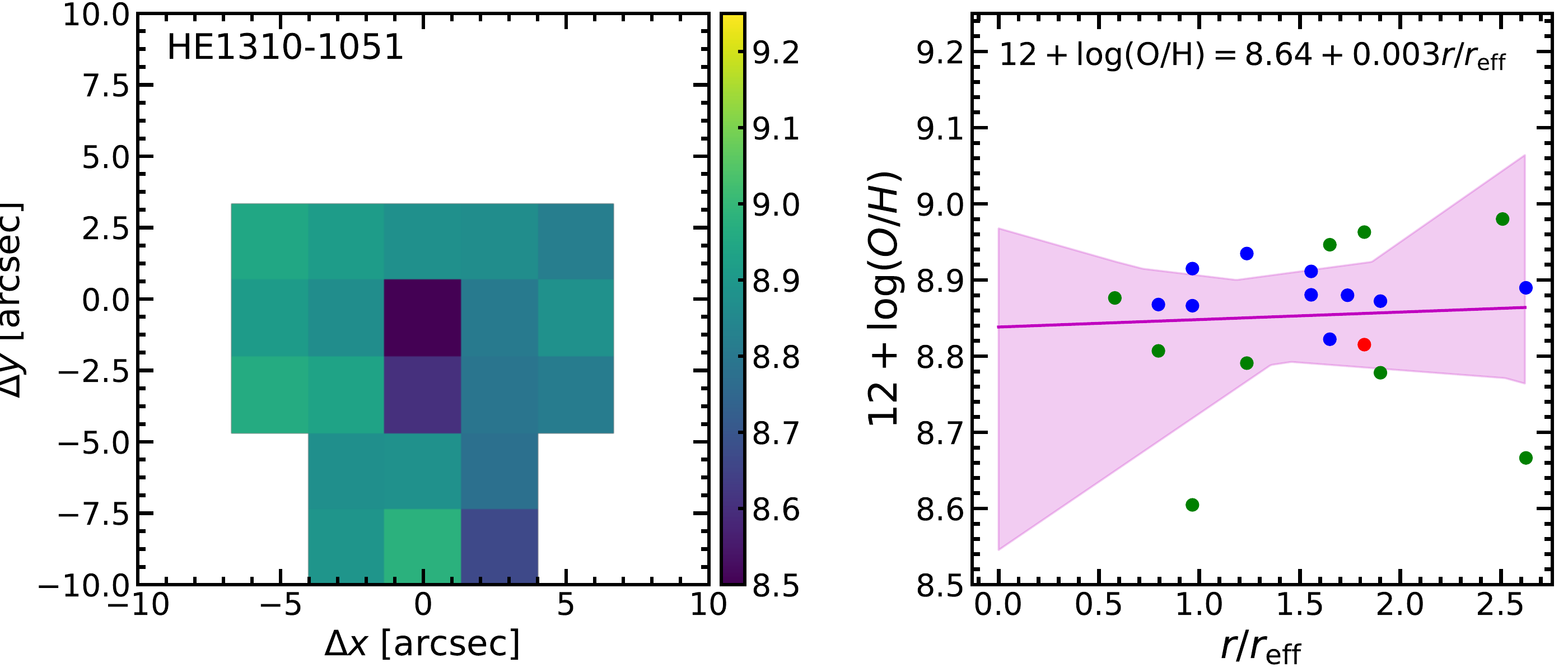}
  \includegraphics[width=.5\textwidth]{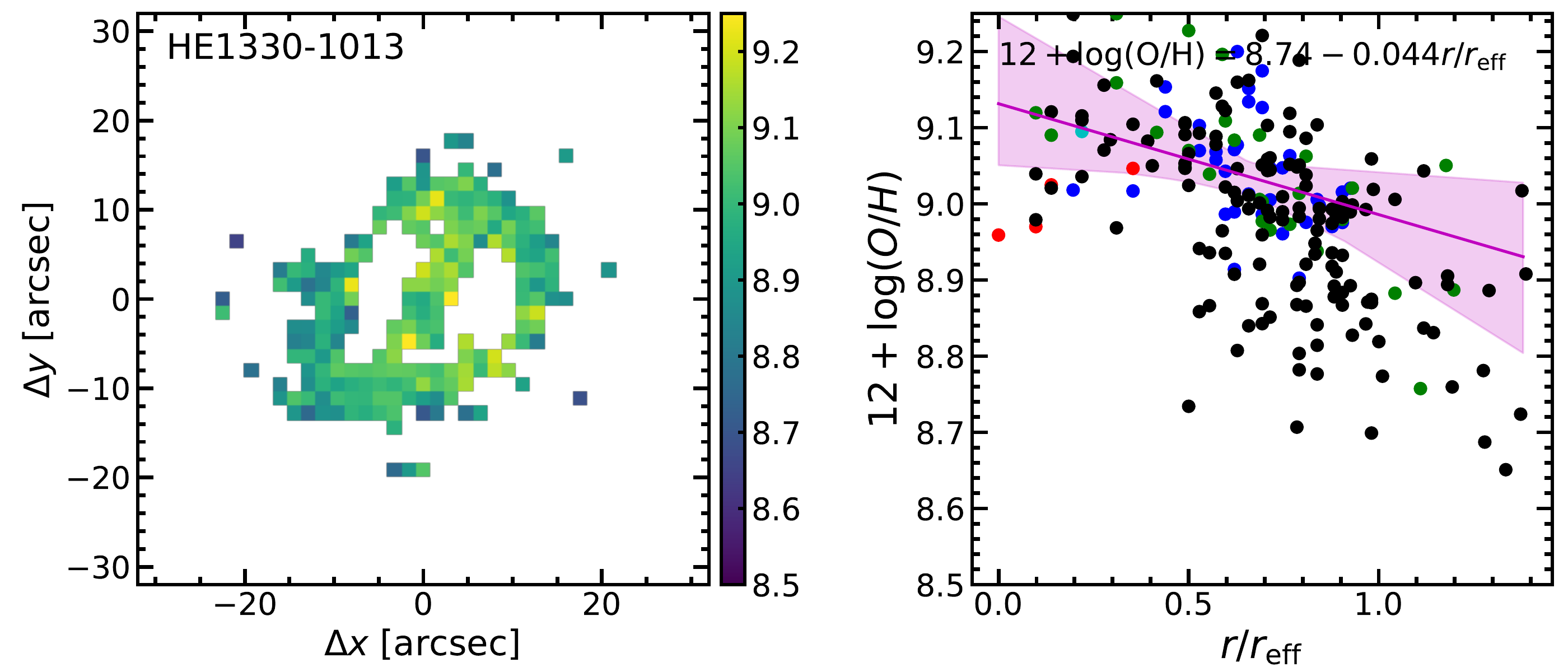}\\
  \includegraphics[width=.5\textwidth]{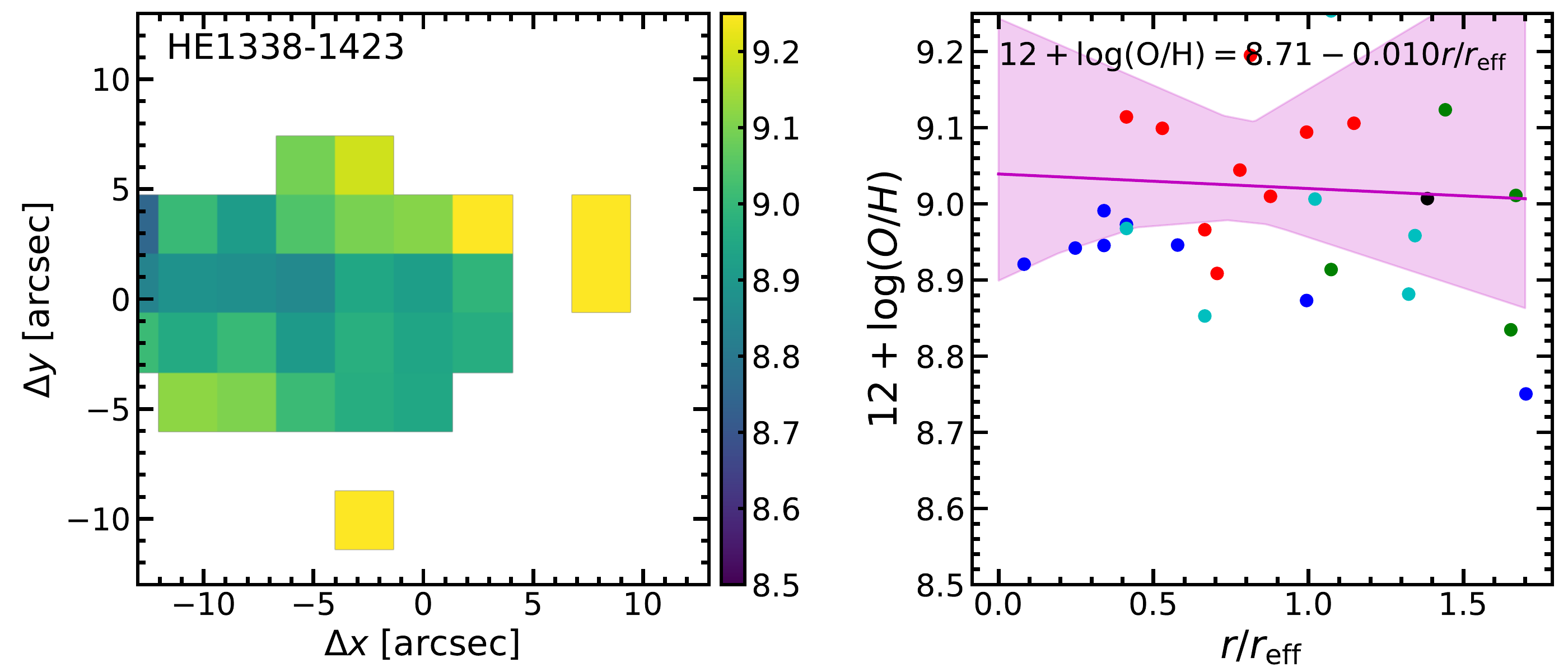}
  \includegraphics[width=.5\textwidth]{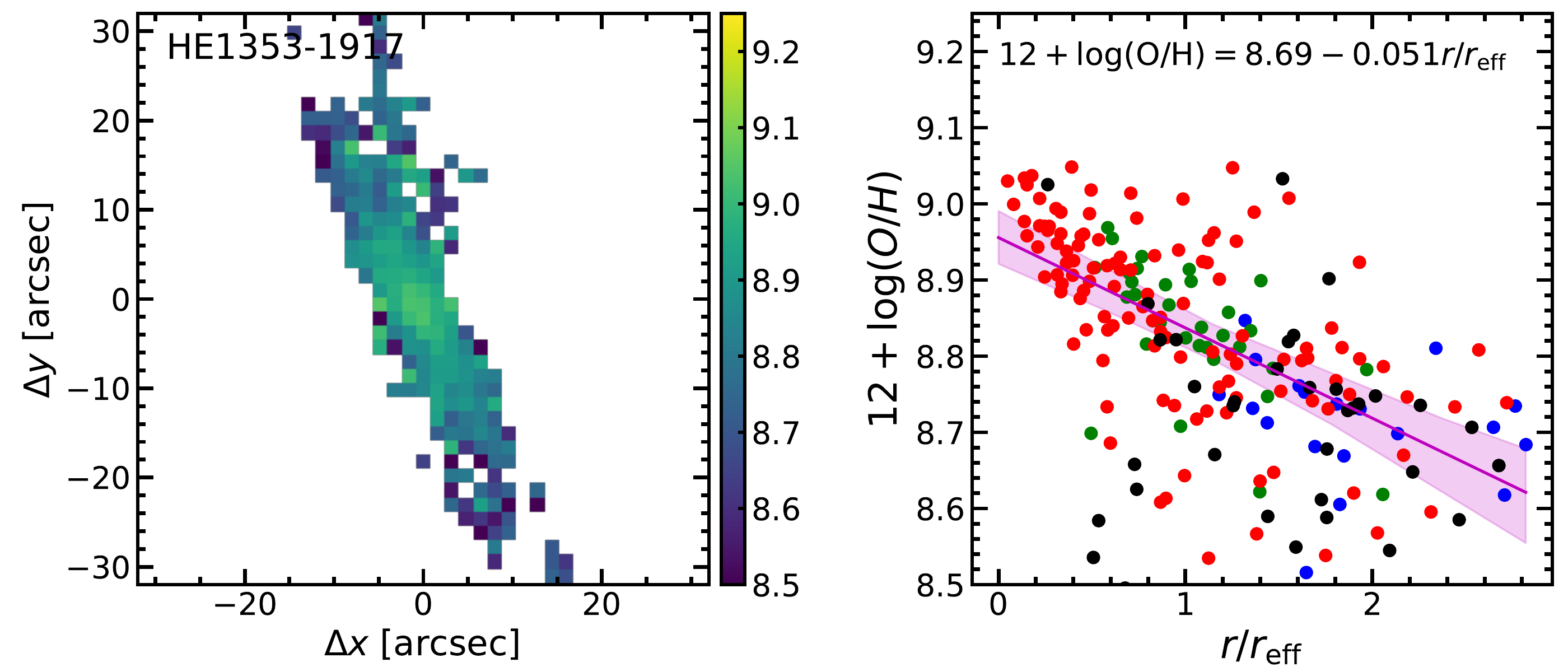}
  \caption{Continued.}

 \end{figure*}
 
 \begin{figure*}
  \includegraphics[width=.5\textwidth]{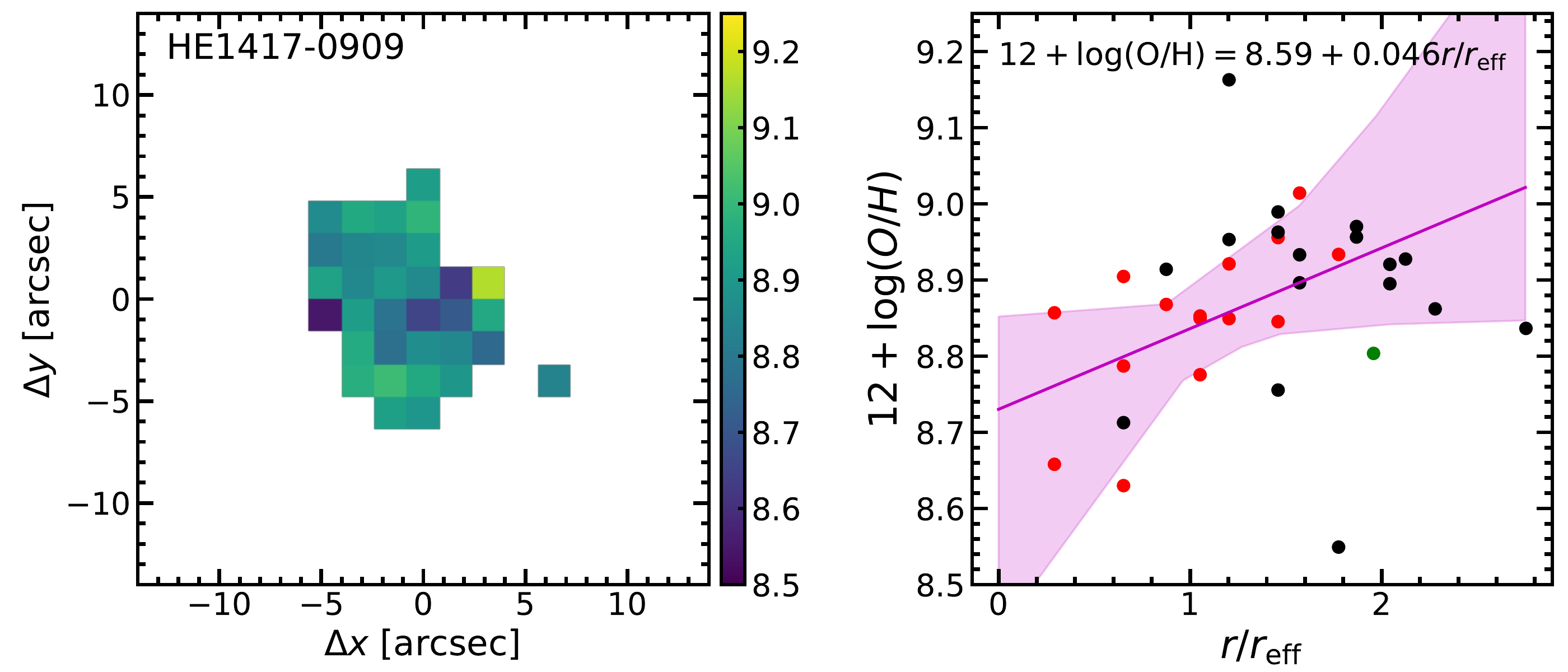}
  \includegraphics[width=.5\textwidth]{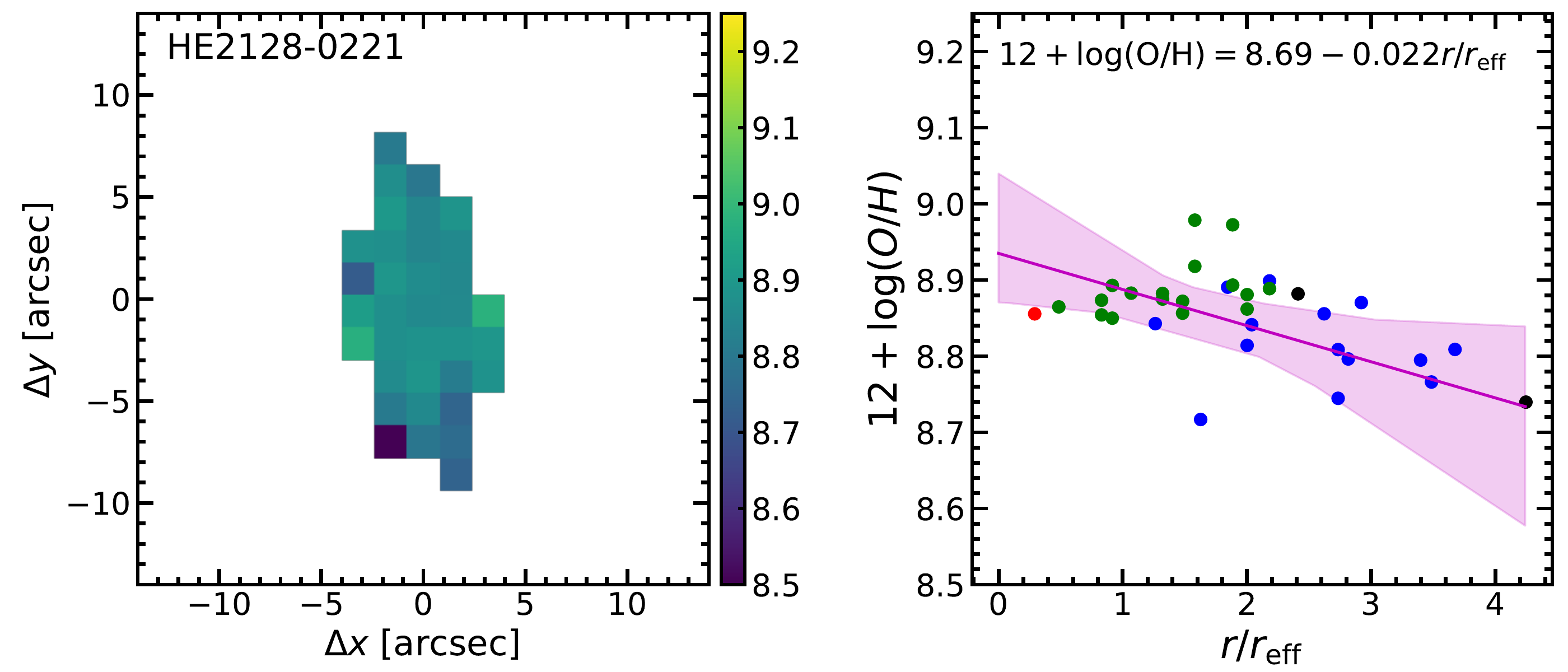}\\
  \includegraphics[width=.5\textwidth]{{Images/metallicity_distribution_N2S2/HE2211-3903.OH_dist}.pdf}
  \includegraphics[width=.5\textwidth]{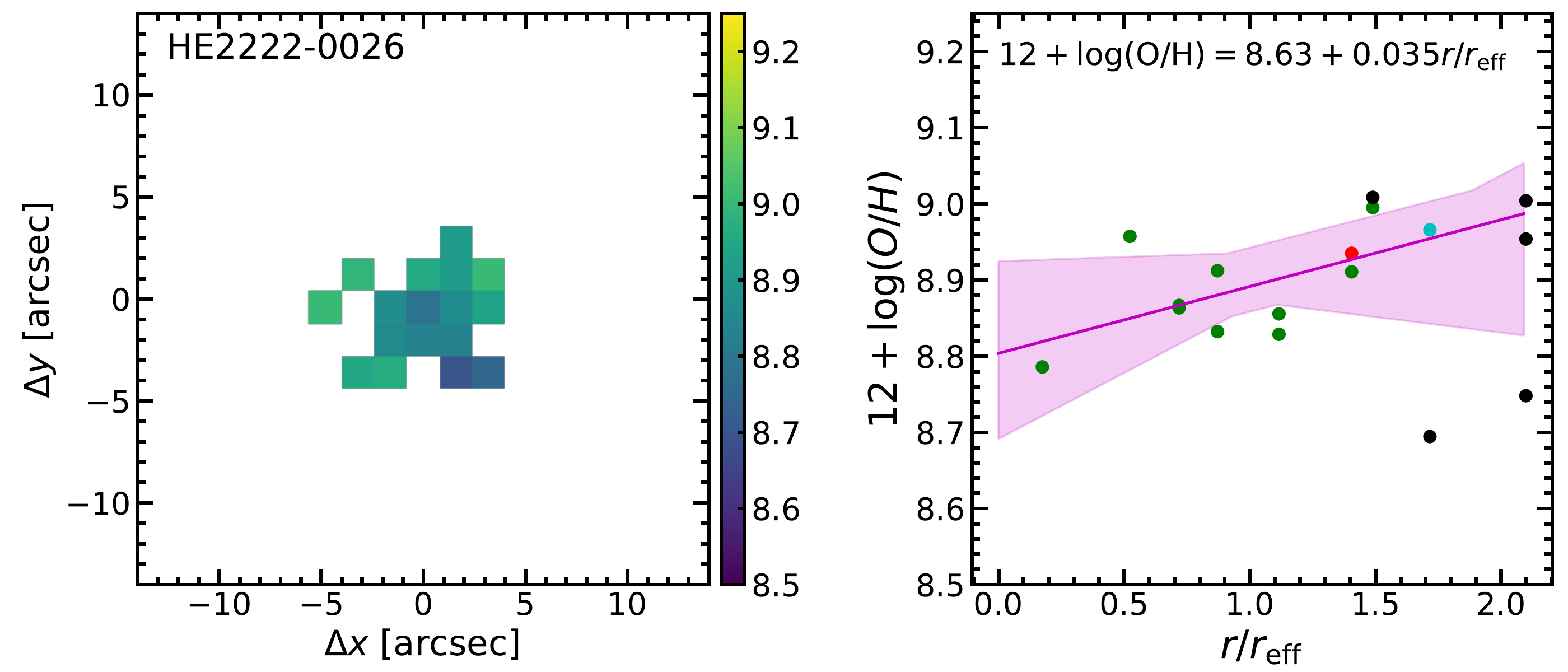}\\
  \includegraphics[width=.5\textwidth]{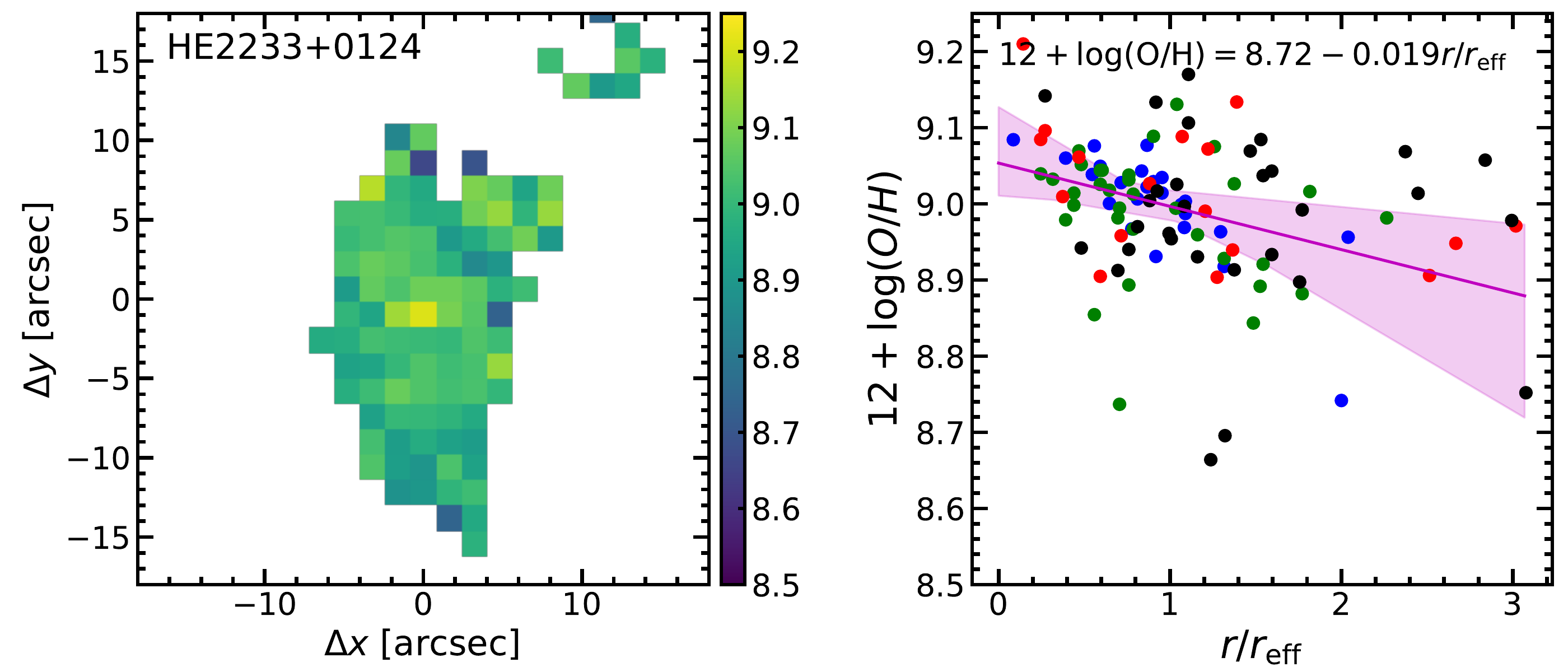}
  \includegraphics[width=.5\textwidth]{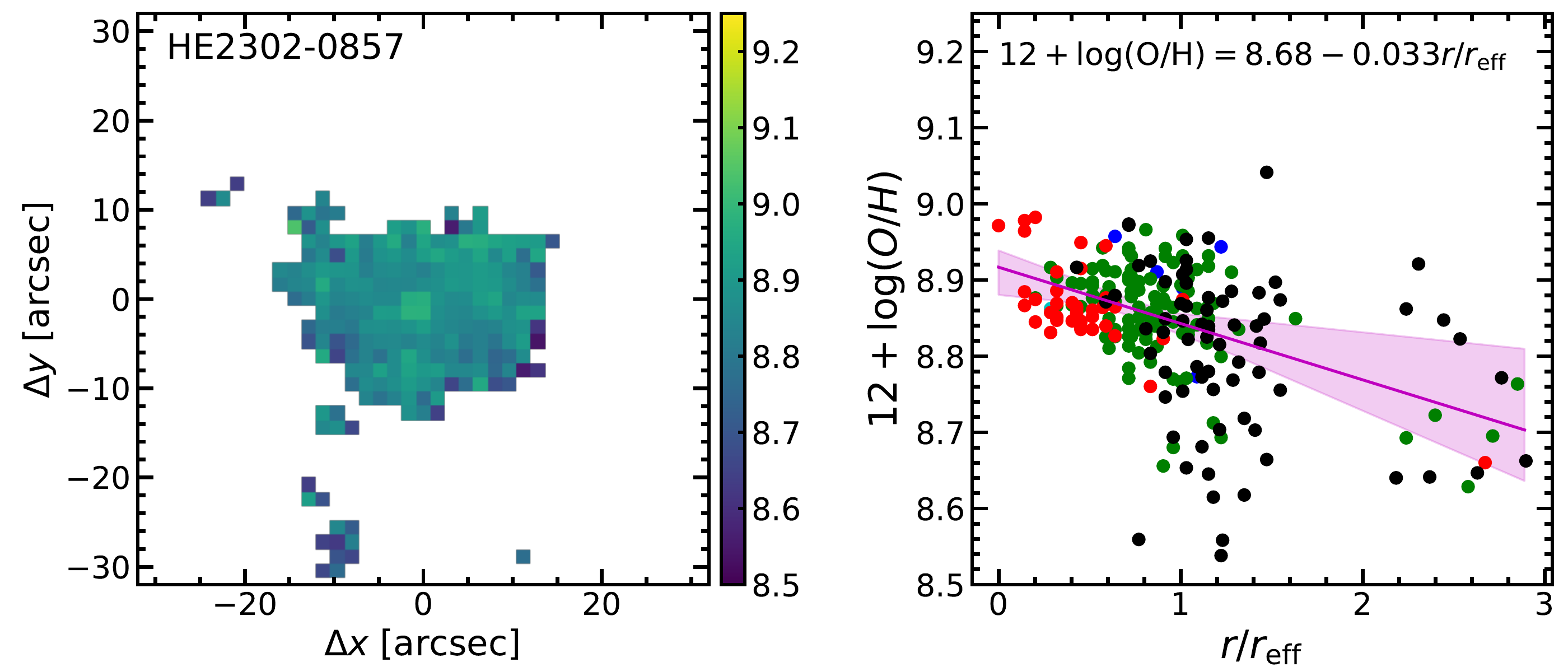}\\
  
  \caption{Continued.}
 \end{figure*}
\end{appendix}

\end{document}

%% file: Tables/photometry_SED/photometry_SED_optical.tex
\begin{tabular}{lccccccccccccc}\hline\hline
& \multicolumn{2}{c}{GALEX} & & \multicolumn{6}{c}{SDSS/DECAM/PanSTARRS1}& &\multicolumn{3}{c}{WFI}\\\cline{2-3}\cline{5-10}\cline{12-14}
Object & $FUV$ & $NUV$ & & $u$ & $g$ & $r$ & $i$ & $z$ & $y$ & & $U$ & $B$ & $V$\\
       & [mJy] & [mJy] & & [mJy] & [mJy] & [mJy] & [mJy] & [mJy] & [mJy] & & [mJy] & [mJy] & [mJy]\\\hline
HE0021-1810 & $0.17\pm0.02$ & $0.21\pm0.02$ & & $...$ & $1.4\pm0.1$ & $2.3\pm0.2$ & $3.2\pm0.3$ & $3.6\pm0.4$ & $4.8\pm0.5$ & & $0.5\pm0.1$ & $...$ & $...$\\
HE0021-1819 & $0.18\pm0.02$ & $0.25\pm0.03$ & & $...$ & $1.0\pm0.1$ & $1.6\pm0.2$ & $2.1\pm0.2$ & $2.3\pm0.2$ & $2.9\pm0.3$ & & $0.3\pm0.0$ & $...$ & $...$\\
HE0040-1105 & $0.09\pm0.01$ & $0.18\pm0.02$ & & $0.62\pm0.06$ & $1.5\pm0.1$ & $2.4\pm0.2$ & $3.4\pm0.3$ & $4.0\pm0.4$ & $...$ & & $...$ & $...$ & $...$\\
HE0045-2145 & $...$ & $...$ & & $...$ & $...$ & $...$ & $...$ & $...$ & $...$ & & $...$ & $...$ & $...$\\
HE0108-4743 & $0.69\pm0.07$ & $1.29\pm0.13$ & & $...$ & $...$ & $...$ & $...$ & $...$ & $...$ & & $1.7\pm0.2$ & $4.9\pm0.5$ & $8.0\pm0.8$\\
HE0114-0015 & $0.21\pm0.02$ & $0.34\pm0.03$ & & $0.68\pm0.07$ & $1.8\pm0.2$ & $3.1\pm0.3$ & $4.3\pm0.4$ & $5.3\pm0.5$ & $...$ & & $...$ & $...$ & $...$\\
HE0119-0118 & $0.84\pm0.02$ & $1.26\pm0.03$ & & $1.66\pm0.17$ & $3.5\pm0.3$ & $5.1\pm0.5$ & $6.9\pm0.7$ & $8.0\pm0.8$ & $...$ & & $...$ & $...$ & $...$\\
HE0150-0344 & $0.32\pm0.03$ & $0.44\pm0.04$ & & $0.65\pm0.07$ & $1.3\pm0.1$ & $1.6\pm0.2$ & $2.1\pm0.2$ & $2.3\pm0.2$ & $...$ & & $...$ & $...$ & $...$\\
HE0203-0031 & $1.78\pm0.18$ & $1.95\pm0.19$ & & $4.61\pm0.46$ & $9.1\pm0.9$ & $15.5\pm1.5$ & $21.6\pm2.2$ & $26.4\pm2.6$ & $...$ & & $...$ & $...$ & $...$\\
HE0212-0059 & $0.81\pm0.08$ & $1.26\pm0.13$ & & $4.66\pm0.47$ & $17.2\pm1.7$ & $32.9\pm3.3$ & $46.3\pm4.6$ & $58.4\pm5.8$ & $...$ & & $...$ & $...$ & $...$\\
HE0224-2834 & $0.19\pm0.02$ & $0.35\pm0.03$ & & $...$ & $2.0\pm0.2$ & $3.0\pm0.3$ & $3.9\pm0.4$ & $4.3\pm0.4$ & $5.2\pm0.5$ & & $0.7\pm0.1$ & $...$ & $...$\\
HE0227-0913 & $3.23\pm0.32$ & $4.25\pm0.43$ & & $5.74\pm0.57$ & $9.4\pm0.9$ & $14.5\pm1.5$ & $17.9\pm1.8$ & $22.1\pm2.2$ & $...$ & & $...$ & $...$ & $...$\\
HE0232-0900 & $4.46\pm0.45$ & $5.22\pm0.52$ & & $7.48\pm0.75$ & $12.0\pm1.2$ & $16.9\pm1.7$ & $23.1\pm2.3$ & $27.4\pm2.7$ & $...$ & & $...$ & $...$ & $...$\\
HE0253-1641 & $0.14\pm0.01$ & $0.31\pm0.03$ & & $1.16\pm0.12$ & $3.1\pm0.3$ & $5.1\pm0.5$ & $6.8\pm0.7$ & $8.3\pm0.8$ & $...$ & & $...$ & $...$ & $...$\\
HE0345+0056 & $2.79\pm0.28$ & $4.32\pm0.43$ & & $6.36\pm0.64$ & $8.3\pm0.8$ & $9.7\pm1.0$ & $10.2\pm1.0$ & $11.5\pm1.2$ & $...$ & & $...$ & $...$ & $...$\\
HE0351+0240 & $0.86\pm0.09$ & $1.14\pm0.11$ & & $...$ & $2.0\pm0.2$ & $3.2\pm0.3$ & $3.8\pm0.4$ & $4.3\pm0.4$ & $5.2\pm0.5$ & & $...$ & $...$ & $...$\\
HE0412-0803 & $0.30\pm0.03$ & $0.45\pm0.05$ & & $...$ & $4.8\pm0.5$ & $7.5\pm0.7$ & $8.2\pm0.8$ & $9.5\pm1.0$ & $11.5\pm1.1$ & & $2.1\pm0.2$ & $...$ & $...$\\
HE0429-0247 & $...$ & $...$ & & $...$ & $1.6\pm0.2$ & $2.4\pm0.2$ & $2.8\pm0.3$ & $3.1\pm0.3$ & $3.6\pm0.4$ & & $...$ & $...$ & $...$\\
HE0433-1028 & $4.72\pm0.47$ & $4.88\pm0.49$ & & $...$ & $10.2\pm1.0$ & $15.1\pm1.5$ & $17.5\pm1.7$ & $20.3\pm2.0$ & $25.3\pm2.5$ & & $5.3\pm0.5$ & $...$ & $...$\\
HE0853-0126 & $0.32\pm0.03$ & $0.53\pm0.05$ & & $1.07\pm0.11$ & $2.1\pm0.2$ & $3.2\pm0.3$ & $4.3\pm0.4$ & $5.1\pm0.5$ & $...$ & & $...$ & $...$ & $...$\\
HE0853+0102 & $0.15\pm0.01$ & $0.20\pm0.02$ & & $0.43\pm0.04$ & $1.2\pm0.1$ & $1.9\pm0.2$ & $2.5\pm0.3$ & $3.1\pm0.3$ & $3.8\pm0.4$ & & $...$ & $...$ & $...$\\
HE0934+0119 & $0.69\pm0.07$ & $0.84\pm0.08$ & & $1.79\pm0.18$ & $2.5\pm0.3$ & $3.3\pm0.3$ & $4.0\pm0.4$ & $3.8\pm0.4$ & $4.6\pm0.5$ & & $...$ & $...$ & $...$\\
HE0949-0122 & $0.45\pm0.05$ & $0.71\pm0.07$ & & $2.14\pm0.21$ & $5.8\pm0.6$ & $10.7\pm1.1$ & $12.6\pm1.3$ & $16.7\pm1.7$ & $...$ & & $...$ & $...$ & $...$\\
HE1011-0403 & $2.39\pm0.24$ & $2.86\pm0.29$ & & $2.39\pm0.24$ & $4.0\pm0.4$ & $5.3\pm0.5$ & $6.5\pm0.6$ & $7.0\pm0.7$ & $8.5\pm0.9$ & & $...$ & $...$ & $...$\\
HE1017-0305 & $0.65\pm0.06$ & $0.84\pm0.08$ & & $1.51\pm0.15$ & $3.1\pm0.3$ & $5.1\pm0.5$ & $7.1\pm0.7$ & $8.6\pm0.9$ & $...$ & & $...$ & $...$ & $...$\\
HE1029-1831 & $0.61\pm0.06$ & $1.13\pm0.11$ & & $1.52\pm0.15$ & $4.0\pm0.4$ & $6.1\pm0.6$ & $7.1\pm0.7$ & $8.2\pm0.8$ & $10.1\pm1.0$ & & $...$ & $...$ & $...$\\
HE1107-0813 & $1.26\pm0.13$ & $2.30\pm0.23$ & & $...$ & $5.7\pm0.6$ & $7.2\pm0.7$ & $7.9\pm0.8$ & $8.2\pm0.8$ & $9.6\pm1.0$ & & $...$ & $...$ & $...$\\
HE1108-2813 & $...$ & $...$ & & $3.80\pm0.38$ & $10.9\pm1.1$ & $16.2\pm1.6$ & $19.6\pm2.0$ & $22.8\pm2.3$ & $26.6\pm2.7$ & & $...$ & $...$ & $...$\\
HE1126-0407 & $1.54\pm0.15$ & $2.77\pm0.28$ & & $4.16\pm0.42$ & $6.7\pm0.7$ & $7.7\pm0.8$ & $10.8\pm1.1$ & $10.5\pm1.1$ & $12.0\pm1.2$ & & $...$ & $...$ & $...$\\
HE1237-0504 & $4.94\pm0.49$ & $9.47\pm0.95$ & & $...$ & $72.1\pm7.2$ & $136.8\pm13.7$ & $189.6\pm19.0$ & $217.0\pm21.7$ & $261.6\pm26.2$ & & $...$ & $...$ & $...$\\
HE1248-1356 & $0.49\pm0.05$ & $0.97\pm0.10$ & & $2.91\pm0.29$ & $12.2\pm1.2$ & $21.5\pm2.2$ & $28.7\pm2.9$ & $34.7\pm3.5$ & $41.2\pm4.1$ & & $...$ & $...$ & $...$\\
HE1310-1051 & $1.19\pm0.12$ & $1.32\pm0.13$ & & $1.42\pm0.14$ & $3.2\pm0.3$ & $4.3\pm0.4$ & $4.7\pm0.5$ & $5.2\pm0.5$ & $5.6\pm0.6$ & & $...$ & $...$ & $...$\\
HE1330-1013 & $...$ & $...$ & & $1.83\pm0.18$ & $6.5\pm0.7$ & $11.2\pm1.1$ & $15.1\pm1.5$ & $18.2\pm1.8$ & $22.0\pm2.2$ & & $...$ & $...$ & $...$\\
HE1338-1423 & $1.86\pm0.19$ & $2.18\pm0.22$ & & $3.26\pm0.33$ & $6.7\pm0.7$ & $10.9\pm1.1$ & $13.2\pm1.3$ & $15.6\pm1.6$ & $18.7\pm1.9$ & & $...$ & $...$ & $...$\\
HE1353-1917 & $0.21\pm0.02$ & $0.50\pm0.05$ & & $0.86\pm0.09$ & $3.0\pm0.3$ & $6.3\pm0.6$ & $7.7\pm0.8$ & $9.7\pm1.0$ & $11.4\pm1.1$ & & $...$ & $...$ & $...$\\
HE1417-0909 & $0.54\pm0.05$ & $0.51\pm0.05$ & & $0.73\pm0.07$ & $1.1\pm0.1$ & $1.9\pm0.2$ & $2.4\pm0.2$ & $2.6\pm0.3$ & $3.2\pm0.3$ & & $...$ & $...$ & $...$\\
HE2128-0221 & $0.14\pm0.01$ & $0.24\pm0.02$ & & $0.49\pm0.05$ & $0.8\pm0.1$ & $1.2\pm0.1$ & $1.6\pm0.2$ & $1.9\pm0.2$ & $...$ & & $...$ & $...$ & $...$\\
HE2211-3903 & $0.69\pm0.07$ & $1.03\pm0.10$ & & $...$ & $...$ & $...$ & $...$ & $...$ & $...$ & & $2.6\pm0.3$ & $4.5\pm0.5$ & $...$\\
HE2222-0026 & $0.13\pm0.01$ & $0.19\pm0.02$ & & $0.35\pm0.04$ & $0.8\pm0.1$ & $1.3\pm0.1$ & $1.9\pm0.2$ & $2.4\pm0.2$ & $...$ & & $...$ & $...$ & $...$\\
HE2233+0124 & $0.07\pm0.01$ & $0.10\pm0.01$ & & $1.07\pm0.11$ & $1.9\pm0.2$ & $3.3\pm0.3$ & $5.2\pm0.5$ & $6.8\pm0.7$ & $...$ & & $...$ & $...$ & $...$\\
HE2302-0857 & $1.85\pm0.18$ & $2.04\pm0.20$ & & $3.25\pm0.33$ & $6.6\pm0.7$ & $11.2\pm1.1$ & $16.5\pm1.7$ & $20.2\pm2.0$ & $...$ & & $...$ & $...$ & $...$\\
\hline
\end{tabular}

%% file: Tables/photometry_SED/photometry_SED_IR.tex
\begin{tabular}{lcccccccccccccccc}\hline\hline
  & \multicolumn{3}{c}{2MASS/PANIC/VISTA} & & \multicolumn{4}{c}{WISE} & & \multicolumn{3}{c}{Herschel/PACS} \\\cline{2-4}\cline{6-9}\cline{11-13}
Object & $J$ & $H$ & $K_s$ &  & $W1$ & $W2$ & $W3$ & $W4$ & & $70\mu$m & $100\mu$m & $160\mu$m\\
       & [mJy] & [mJy] & [mJy] &  & [mJy] & [mJy] & [mJy] & [mJy] & & [mJy] & [mJy] & [mJy]\\\hline
HE0021-1810 & $5.6\pm0.6$ & $7.4\pm0.7$ & $...$ & & $4.6\pm0.5$ & $3.8\pm0.4$ & $5\pm0$ & $8\pm1$ & & $...$ & $...$ & $...$ \\
HE0021-1819 & $3.4\pm0.3$ & $...$ & $3.6\pm0.4$ & & $2.3\pm0.2$ & $2.3\pm0.2$ & $8\pm1$ & $21\pm2$ & & $...$ & $...$ & $...$ \\
HE0040-1105 & $5.4\pm0.5$ & $6.9\pm0.7$ & $...$ & & $5.4\pm0.5$ & $6.4\pm0.6$ & $15\pm2$ & $34\pm3$ & & $...$ & $...$ & $...$ \\
HE0045-2145 & $15.9\pm1.6$ & $...$ & $16.2\pm1.6$ & & $13.0\pm1.3$ & $10.1\pm1.0$ & $99\pm10$ & $516\pm52$ & & $...$ & $...$ & $...$ \\
HE0108-4743 & $18.0\pm1.8$ & $20.9\pm2.1$ & $18.9\pm1.9$ & & $16.8\pm1.7$ & $15.5\pm1.5$ & $63\pm6$ & $121\pm12$ & & $...$ & $...$ & $...$ \\
HE0114-0015 & $6.7\pm0.7$ & $10.7\pm1.1$ & $...$ & & $6.5\pm0.7$ & $6.2\pm0.6$ & $17\pm2$ & $38\pm4$ & & $...$ & $...$ & $...$ \\
HE0119-0118 & $11.8\pm1.2$ & $13.1\pm1.3$ & $...$ & & $14.9\pm1.5$ & $19.2\pm1.9$ & $60\pm6$ & $239\pm24$ & & $...$ & $...$ & $...$ \\
HE0150-0344 & $2.6\pm0.3$ & $3.0\pm0.3$ & $2.8\pm0.3$ & & $2.3\pm0.2$ & $2.0\pm0.2$ & $16\pm2$ & $54\pm5$ & & $...$ & $...$ & $...$ \\
HE0203-0031 & $31.8\pm0.8$ & $38.5\pm1.2$ & $36.3\pm1.6$ & & $31.0\pm3.1$ & $32.1\pm3.2$ & $42\pm4$ & $65\pm6$ & & $77\pm12$ & $...$ & $...$ \\
HE0212-0059 & $73.0\pm7.3$ & $81.0\pm8.1$ & $71.6\pm7.2$ & & $43.3\pm4.3$ & $31.1\pm3.1$ & $92\pm9$ & $242\pm24$ & & $413\pm21$ & $...$ & $1975\pm99$ \\
HE0224-2834 & $5.5\pm0.5$ & $7.9\pm0.8$ & $7.5\pm0.7$ & & $10.2\pm1.0$ & $14.4\pm1.4$ & $35\pm3$ & $89\pm9$ & & $...$ & $...$ & $...$ \\
HE0227-0913 & $27.0\pm2.7$ & $...$ & $40.4\pm4.0$ & & $43.4\pm4.3$ & $50.1\pm5.0$ & $96\pm10$ & $159\pm16$ & & $...$ & $...$ & $...$ \\
HE0232-0900 & $35.4\pm3.5$ & $45.1\pm4.5$ & $...$ & & $44.7\pm4.5$ & $54.8\pm5.5$ & $156\pm16$ & $409\pm41$ & & $1197\pm26$ & $...$ & $1450\pm57$ \\
HE0253-1641 & $11.3\pm1.1$ & $14.5\pm1.4$ & $...$ & & $12.4\pm1.2$ & $15.8\pm1.6$ & $58\pm6$ & $155\pm16$ & & $...$ & $...$ & $...$ \\
HE0345+0056 & $...$ & $21.1\pm2.1$ & $...$ & & $58.9\pm5.9$ & $80.8\pm8.1$ & $181\pm18$ & $315\pm32$ & & $...$ & $...$ & $...$ \\
HE0351+0240 & $...$ & $8.0\pm0.8$ & $...$ & & $9.8\pm1.0$ & $13.8\pm1.4$ & $32\pm3$ & $85\pm9$ & & $186\pm17$ & $...$ & $...$ \\
HE0412-0803 & $...$ & $16.5\pm1.7$ & $...$ & & $43.1\pm4.3$ & $66.0\pm6.6$ & $175\pm17$ & $386\pm39$ & & $587\pm19$ & $...$ & $423\pm51$ \\
HE0429-0247 & $4.5\pm0.4$ & $5.8\pm0.6$ & $...$ & & $8.1\pm0.8$ & $9.9\pm1.0$ & $15\pm1$ & $27\pm3$ & & $...$ & $...$ & $...$ \\
HE0433-1028 & $...$ & $38.6\pm3.9$ & $...$ & & $50.8\pm5.1$ & $77.6\pm7.8$ & $239\pm24$ & $598\pm60$ & & $2474\pm49$ & $...$ & $2334\pm47$ \\
HE0853-0126 & $4.9\pm0.5$ & $...$ & $6.3\pm0.6$ & & $4.9\pm0.5$ & $4.2\pm0.4$ & $11\pm1$ & $19\pm2$ & & $...$ & $...$ & $...$ \\
HE0853+0102 & $5.1\pm0.5$ & $6.8\pm0.7$ & $7.3\pm0.7$ & & $5.8\pm0.6$ & $6.0\pm0.6$ & $9\pm1$ & $14\pm1$ & & $...$ & $...$ & $...$ \\
HE0934+0119 & $6.0\pm0.6$ & $6.9\pm0.7$ & $8.7\pm0.9$ & & $5.9\pm0.6$ & $7.7\pm0.8$ & $25\pm2$ & $59\pm6$ & & $219\pm1$ & $248\pm8$ & $257\pm9$ \\
HE0949-0122 & $28.4\pm0.7$ & $53.3\pm1.4$ & $96.1\pm2.3$ & & $175.1\pm17.5$ & $256.3\pm25.6$ & $490\pm49$ & $921\pm92$ & & $...$ & $...$ & $...$ \\
HE1011-0403 & $...$ & $11.0\pm1.1$ & $...$ & & $10.6\pm1.1$ & $12.7\pm1.3$ & $34\pm3$ & $73\pm7$ & & $165\pm2$ & $209\pm9$ & $234\pm10$ \\
HE1017-0305 & $14.2\pm1.4$ & $...$ & $10.8\pm1.1$ & & $14.0\pm1.4$ & $16.1\pm1.6$ & $36\pm4$ & $71\pm7$ & & $...$ & $...$ & $...$ \\
HE1029-1831 & $11.4\pm1.1$ & $...$ & $14.9\pm1.5$ & & $15.2\pm1.5$ & $16.5\pm1.6$ & $86\pm9$ & $306\pm31$ & & $...$ & $...$ & $...$ \\
HE1107-0813 & $...$ & $20.9\pm2.1$ & $26.6\pm2.7$ & & $21.5\pm2.1$ & $24.7\pm2.5$ & $30\pm3$ & $46\pm5$ & & $...$ & $...$ & $...$ \\
HE1108-2813 & $26.4\pm2.6$ & $...$ & $33.7\pm3.4$ & & $39.7\pm4.0$ & $41.2\pm4.1$ & $127\pm13$ & $471\pm47$ & & $...$ & $...$ & $...$ \\
HE1126-0407 & $11.5\pm1.1$ & $18.9\pm1.9$ & $28.5\pm2.8$ & & $34.7\pm3.5$ & $46.1\pm4.6$ & $98\pm10$ & $262\pm26$ & & $454\pm9$ & $...$ & $354\pm7$ \\
HE1237-0504 & $415.0\pm10.5$ & $494.2\pm13.9$ & $426.5\pm14.6$ & & $241.1\pm24.1$ & $181.5\pm18.2$ & $362\pm36$ & $829\pm83$ & & $...$ & $...$ & $...$ \\
HE1248-1356 & $43.7\pm4.4$ & $48.0\pm4.8$ & $46.2\pm4.6$ & & $28.6\pm2.9$ & $21.5\pm2.1$ & $68\pm7$ & $138\pm14$ & & $...$ & $...$ & $...$ \\
HE1310-1051 & $6.6\pm0.7$ & $7.8\pm0.8$ & $8.8\pm0.9$ & & $10.8\pm1.1$ & $17.2\pm1.7$ & $53\pm5$ & $113\pm11$ & & $95\pm2$ & $115\pm8$ & $116\pm11$ \\
HE1330-1013 & $26.8\pm2.7$ & $33.7\pm3.4$ & $29.4\pm2.9$ & & $16.2\pm1.6$ & $12.1\pm1.2$ & $23\pm2$ & $50\pm5$ & & $...$ & $...$ & $...$ \\
HE1338-1423 & $21.5\pm2.1$ & $26.7\pm2.7$ & $28.0\pm2.8$ & & $27.0\pm2.7$ & $33.3\pm3.3$ & $78\pm8$ & $149\pm15$ & & $141\pm18$ & $...$ & $...$ \\
HE1353-1917 & $15.3\pm1.5$ & $20.5\pm2.1$ & $20.9\pm2.1$ & & $14.1\pm1.4$ & $13.2\pm1.3$ & $34\pm3$ & $73\pm7$ & & $246\pm5$ & $...$ & $768\pm15$ \\
HE1417-0909 & $3.4\pm0.3$ & $3.9\pm0.4$ & $4.2\pm0.4$ & & $4.3\pm0.4$ & $5.5\pm0.6$ & $13\pm1$ & $28\pm3$ & & $...$ & $...$ & $...$ \\
HE2128-0221 & $2.5\pm0.3$ & $3.2\pm0.3$ & $3.4\pm0.3$ & & $3.2\pm0.3$ & $3.9\pm0.4$ & $10\pm1$ & $22\pm2$ & & $...$ & $...$ & $...$ \\
HE2211-3903 & $...$ & $15.3\pm1.5$ & $37.0\pm3.7$ & & $58.2\pm5.8$ & $83.1\pm8.3$ & $188\pm19$ & $371\pm37$ & & $...$ & $...$ & $...$ \\
HE2222-0026 & $3.1\pm0.3$ & $4.6\pm0.5$ & $...$ & & $5.0\pm0.5$ & $6.4\pm0.6$ & $13\pm1$ & $21\pm2$ & & $...$ & $...$ & $...$ \\
HE2233+0124 & $8.2\pm0.8$ & $10.9\pm1.1$ & $...$ & & $11.4\pm1.1$ & $11.3\pm1.1$ & $13\pm1$ & $14\pm1$ & & $...$ & $...$ & $...$ \\
HE2302-0857 & $43.0\pm4.3$ & $53.0\pm5.3$ & $...$ & & $75.1\pm7.5$ & $87.0\pm8.7$ & $114\pm11$ & $239\pm24$ & & $441\pm9$ & $...$ & $641\pm13$ \\
\hline
\end{tabular}

%% file: Tables/photometry_SED/photometry_SED_FIR.tex
\begin{tabular}{lcccccccccccccccc}\hline\hline
 & \multicolumn{4}{c}{AKARI} & & \multicolumn{2}{c}{SOFIA} & & \multicolumn{3}{c}{Herschel/SPIRE} & & \multicolumn{2}{c}{JWST/SCUBA2}\\\cline{2-5}\cline{7-8}\cline{10-12}\cline{14-15}
Object & N60 & WIDE-S & WIDE-L & N160 & & $D$ band  & $E$ band  & & $250\mu$m & $350\mu$m & $500\mu$m & & $450\mu$m & $850\mu$m\\
 & [mJy] & [mJy] & [mJy] & [mJy] & & [mJy]  & [mJy]  & & [mJy] & [mJy] & [mJy] & & [mJy] & [mJy]\\\hline
HE0021-1810 &$...$ & $...$ & $...$ & $...$ & & $<296$ & $...$ & &$...$ & $...$ & $...$ & &$...$ & $...$\\
HE0021-1819 &$...$ & $...$ & $...$ & $...$ & & $228\pm23$ & $...$ & &$...$ & $...$ & $...$ & &$...$ & $...$\\
HE0040-1105 &$...$ & $...$ & $...$ & $...$ & & $<873$ & $...$ & &$...$ & $...$ & $...$ & &$<76$ & $<5$\\
HE0045-2145 &$3130\pm82$ & $3904\pm103$ & $3518\pm391$ & $4134\pm954$ & & $...$ & $...$ & &$...$ & $...$ & $...$ & &$<239$ & $12\pm6$\\
HE0108-4743 &$<450$ & $1006\pm71$ & $3778\pm328$ & $2245\pm979$ & & $...$ & $...$ & &$...$ & $...$ & $...$ & &$...$ & $...$\\
HE0114-0015 &$...$ & $...$ & $...$ & $...$ & & $...$ & $...$ & &$214\pm17$ & $86\pm16$ & $...$ & &$...$ & $...$\\
HE0119-0118 &$...$ & $1411\pm98$ & $1879\pm605$ & $<1917$ & & $...$ & $...$ & &$573\pm23$ & $198\pm20$ & $66\pm22$ & &$...$ & $...$\\
HE0150-0344 &$<330$ & $720\pm65$ & $<1445$ & $<1261$ & & $...$ & $...$ & &$293\pm21$ & $108\pm23$ & $...$ & &$...$ & $...$\\
HE0203-0031 &$...$ & $...$ & $...$ & $...$ & & $...$ & $...$ & &$82\pm16$ & $72\pm16$ & $...$ & &$...$ & $...$\\
HE0212-0059 &$<701$ & $747\pm69$ & $2648\pm929$ & $...$ & & $...$ & $...$ & &$612\pm37$ & $349\pm31$ & $186\pm21$ & &$...$ & $...$\\
HE0224-2834 &$...$ & $...$ & $...$ & $...$ & & $...$ & $...$ & &$...$ & $...$ & $...$ & &$...$ & $...$\\
HE0227-0913 &$<989$ & $455\pm85$ & $<1072$ & $...$ & & $...$ & $407\pm41$ & &$...$ & $...$ & $...$ & &$<270$ & $<12$\\
HE0232-0900 &$...$ & $1291\pm16$ & $2204\pm389$ & $...$ & & $...$ & $...$ & &$699\pm21$ & $356\pm24$ & $144\pm28$ & &$...$ & $...$\\
HE0253-1641 &$...$ & $...$ & $...$ & $...$ & & $661\pm66$ & $...$ & &$...$ & $...$ & $...$ & &$...$ & $...$\\
HE0345+0056 &$...$ & $...$ & $...$ & $...$ & & $...$ & $...$ & &$...$ & $...$ & $...$ & &$...$ & $...$\\
HE0351+0240 &$...$ & $...$ & $...$ & $...$ & & $...$ & $...$ & &$...$ & $...$ & $...$ & &$<129$ & $<9$\\
HE0412-0803 &$...$ & $...$ & $...$ & $...$ & & $...$ & $...$ & &$...$ & $...$ & $...$ & &$<105$ & $<8$\\
HE0429-0247 &$...$ & $...$ & $...$ & $...$ & & $<445$ & $...$ & &$...$ & $...$ & $...$ & &$...$ & $...$\\
HE0433-1028 &$...$ & $2572\pm103$ & $<4505$ & $...$ & & $...$ & $...$ & &$1102\pm20$ & $496\pm26$ & $204\pm18$ & &$...$ & $...$\\
HE0853-0126 &$...$ & $...$ & $...$ & $...$ & & $...$ & $...$ & &$...$ & $...$ & $...$ & &$...$ & $...$\\
HE0853+0102 &$...$ & $...$ & $...$ & $...$ & & $...$ & $...$ & &$100\pm11$ & $...$ & $...$ & &$...$ & $...$\\
HE0934+0119 &$...$ & $...$ & $...$ & $...$ & & $...$ & $...$ & &$...$ & $...$ & $...$ & &$<229$ & $<10$\\
HE0949-0122 &$...$ & $1375\pm164$ & $...$ & $...$ & & $...$ & $...$ & &$...$ & $...$ & $...$ & &$...$ & $...$\\
HE1011-0403 &$...$ & $...$ & $...$ & $...$ & & $...$ & $...$ & &$126\pm12$ & $51\pm12$ & $...$ & &$...$ & $...$\\
HE1017-0305 &$...$ & $...$ & $...$ & $...$ & & $...$ & $...$ & &$...$ & $...$ & $...$ & &$<323$ & $<10$\\
HE1029-1831 &$2134\pm407$ & $2596\pm93$ & $3535\pm984$ & $...$ & & $...$ & $...$ & &$...$ & $...$ & $...$ & &$<331$ & $15\pm7$\\
HE1107-0813 &$...$ & $...$ & $...$ & $...$ & & $...$ & $...$ & &$...$ & $...$ & $...$ & &$...$ & $...$\\
HE1108-2813 &$3203\pm109$ & $3330\pm174$ & $...$ & $...$ & & $...$ & $...$ & &$...$ & $...$ & $...$ & &$191\pm120$ & $14\pm6$\\
HE1126-0407 &$...$ & $484\pm30$ & $...$ & $<1163$ & & $...$ & $...$ & &$198\pm14$ & $80\pm15$ & $...$ & &$...$ & $...$\\
HE1237-0504 &$2659\pm345$ & $2926\pm174$ & $5120\pm423$ & $4676\pm808$ & & $...$ & $...$ & &$...$ & $...$ & $...$ & &$...$ & $...$\\
HE1248-1356 &$...$ & $972\pm100$ & $2409\pm1520$ & $...$ & & $...$ & $...$ & &$...$ & $...$ & $...$ & &$<172$ & $10\pm6$\\
HE1310-1051 &$...$ & $...$ & $...$ & $...$ & & $...$ & $...$ & &$61\pm11$ & $...$ & $...$ & &$...$ & $...$\\
HE1330-1013 &$...$ & $...$ & $...$ & $...$ & & $...$ & $...$ & &$...$ & $...$ & $...$ & &$<137$ & $6\pm4$\\
HE1338-1423 &$...$ & $...$ & $...$ & $...$ & & $...$ & $...$ & &$127\pm18$ & $83\pm16$ & $...$ & &$...$ & $...$\\
HE1353-1917 &$...$ & $<585$ & $...$ & $...$ & & $...$ & $...$ & &$444\pm52$ & $216\pm32$ & $85\pm16$ & &$...$ & $...$\\
HE1417-0909 &$...$ & $...$ & $...$ & $...$ & & $...$ & $...$ & &$...$ & $...$ & $...$ & &$...$ & $...$\\
HE2128-0221 &$...$ & $...$ & $...$ & $...$ & & $<457$ & $...$ & &$...$ & $...$ & $...$ & &$...$ & $...$\\
HE2211-3903 &$...$ & $657\pm70$ & $...$ & $...$ & & $...$ & $...$ & &$...$ & $...$ & $...$ & &$...$ & $...$\\
HE2222-0026 &$...$ & $...$ & $...$ & $...$ & & $...$ & $...$ & &$...$ & $...$ & $...$ & &$...$ & $...$\\
HE2233+0124 &$...$ & $...$ & $...$ & $...$ & & $<527$ & $...$ & &$...$ & $...$ & $...$ & &$...$ & $...$\\
HE2302-0857 &$...$ & $647\pm119$ & $...$ & $...$ & & $...$ & $...$ & &$282\pm14$ & $...$ & $...$ & &$...$ & $...$\\
\hline
\end{tabular}

%% file: Tables/host_properties/host_properties.tex
\begin{tabular}{lcccccccccccc}\hline\hline
Object & z\tablefootmark{a} & $\log\left(\frac{M_\star}{[M_\sun]}\right)$\tablefootmark{b} & $\log\left(\frac{L_\mathrm{tor}}{[\mathrm{erg\,s}^{-1}]}\right)$\tablefootmark{c} & $\log\left(\frac{L_\mathrm{42.5-122.5\,\mu\mathrm{m}}}{[\mathrm{erg\,s}^{-1}]}\right)$\tablefootmark{d} & $\log\left(\frac{L_\mathrm{8-1000\,\mu\mathrm{m}}}{[\mathrm{erg\,s}^{-1}]}\right)$\tablefootmark{e} & $\mathrm{SFR}_\mathrm{IR}$\tablefootmark{f} & $\log\left(\frac{L_\mathrm{CO}}{[\mathrm{K\,km\,s}^{-1}\mathrm{pc}^2]}\right)$\tablefootmark{g} \\\hline
HE0021-1810 & 0.054 & $10.64_{-0.05}^{+0.04}$ & $43.02_{-0.19}^{+0.07}$ &  $<$43.67 & $<$43.98 & $<$3.7 & $<6.8$\smallskip\\
HE0021-1819 & 0.053 & $10.50_{-0.05}^{+0.04}$ & $42.82_{-0.59}^{+0.16}$ &  $43.76_{-0.04}^{+0.05}$ & $44.04_{-0.04}^{+0.05}$ & $4.2_{-0.3}^{+0.6}$ & $7.6\pm2.1$\smallskip\\
HE0040-1105 & 0.042 & $10.16_{-0.10}^{+0.13}$ & $43.50_{-0.04}^{+0.05}$ &  $<$43.82 & $<$44.03 & $<$4.1 & $7.7\pm1.7$\smallskip\\
HE0045-2145 & 0.021 & $9.36_{-0.02}^{+0.03}$ & ... &  $44.18_{-0.01}^{+0.01}$ & $44.45_{-0.01}^{+0.01}$ & $10.9_{-0.2}^{+0.2}$ & $8.1\pm0.4$\smallskip\\
HE0108-4743 & 0.024 & $9.77_{-0.10}^{+0.18}$ & $43.12_{-0.08}^{+0.07}$ &  $43.67_{-0.02}^{+0.02}$ & $44.04_{-0.02}^{+0.02}$ & $4.3_{-0.2}^{+0.2}$ & $7.6\pm0.6$\smallskip\\
HE0114-0015 & 0.046 & $10.47_{-0.19}^{+0.13}$ & $43.26_{-0.09}^{+0.10}$ &  $43.74_{-0.03}^{+0.03}$ & $44.02_{-0.03}^{+0.03}$ & $4.0_{-0.3}^{+0.3}$ & $7.8\pm2.9$\smallskip\\
HE0119-0118 & 0.055 & $10.91_{-0.06}^{+0.02}$ & $43.76_{-0.24}^{+0.10}$ &  $44.65_{-0.02}^{+0.01}$ & $44.89_{-0.02}^{+0.01}$ & $30.4_{-1.1}^{+1.0}$ & $8.4\pm0.8$\smallskip\\
HE0150-0344 & 0.048 & $9.57_{-0.09}^{+0.12}$ & ... &  $44.10_{-0.02}^{+0.02}$ & $44.36_{-0.02}^{+0.02}$ & $9.0_{-0.4}^{+0.4}$ & $7.8\pm3.4$\smallskip\\
HE0203-0031 & 0.043 & $10.88_{-0.02}^{+0.02}$ & $43.90_{-0.07}^{+0.04}$ &  $43.00_{-0.09}^{+0.08}$ & $43.34_{-0.09}^{+0.06}$ & $0.8_{-0.2}^{+0.1}$ & $<7.0$\smallskip\\
HE0212-0059 & 0.026 & $10.59_{-0.01}^{+0.01}$ & ... &  $43.54_{-0.02}^{+0.01}$ & $43.87_{-0.01}^{+0.01}$ & $2.9_{-0.1}^{+0.1}$ & $7.9\pm0.6$\smallskip\\
HE0224-2834 & 0.060 & $10.13_{-0.17}^{+0.19}$ & $44.27_{-0.03}^{+0.04}$ &  $<$44.06 & $<$44.38 & $<$9.3 & $7.9\pm3.2$\smallskip\\
HE0227-0913 & 0.016 & $9.92_{-0.12}^{+0.17}$ & $43.44_{-0.05}^{+0.05}$ &  $42.85_{-0.10}^{+0.11}$ & $43.16_{-0.08}^{+0.10}$ & $0.6_{-0.1}^{+0.1}$ & $7.2\pm1.9$\smallskip\\
HE0232-0900 & 0.043 & $10.88_{-0.12}^{+0.23}$ & $44.45_{-0.04}^{+0.06}$ &  $44.31_{-0.01}^{+0.01}$ & $44.59_{-0.01}^{+0.01}$ & $14.9_{-0.2}^{+0.2}$ & $8.6\pm1.0$\smallskip\\
HE0253-1641 & 0.032 & $10.28_{-0.24}^{+0.11}$ & $42.76_{-1.39}^{+0.54}$ &  $44.17_{-0.05}^{+0.04}$ & $44.41_{-0.05}^{+0.04}$ & $10.0_{-1.2}^{+0.9}$ & $8.0\pm1.1$\smallskip\\
HE0345+0056 & 0.031 & $8.85_{-0.26}^{+0.62}$ & $44.42_{-0.03}^{+0.03}$ &  $<$43.84 & $<$44.13 & $<$5.2 & $<6.7$\smallskip\\
HE0351+0240 & 0.035 & $9.85_{-0.70}^{+0.34}$ & $43.75_{-0.04}^{+0.04}$ &  $43.28_{-0.05}^{+0.04}$ & $43.56_{-0.05}^{+0.05}$ & $1.4_{-0.1}^{+0.2}$ & $<6.9$\smallskip\\
HE0412-0803 & 0.038 & $10.08_{-0.12}^{+0.10}$ & $44.57_{-0.03}^{+0.02}$ &  $43.78_{-0.02}^{+0.02}$ & $44.04_{-0.04}^{+0.02}$ & $4.2_{-0.3}^{+0.2}$ & $<6.8$\smallskip\\
HE0429-0247 & 0.042 & $9.18_{-0.10}^{+0.51}$ & $43.55_{-0.05}^{+0.03}$ &  $<$44.10 & $<$44.43 & $<$10.5 & $<7.3$\smallskip\\
HE0433-1028 & 0.036 & $10.80_{-0.09}^{+0.08}$ & $44.55_{-0.03}^{+0.03}$ &  $44.43_{-0.00}^{+0.00}$ & $44.70_{-0.00}^{+0.00}$ & $19.4_{-0.2}^{+0.2}$ & $8.6\pm0.6$\smallskip\\
HE0853-0126 & 0.060 & $10.32_{-0.08}^{+0.06}$ & $43.54_{-0.24}^{+0.07}$ &  $<$44.05 & $<$44.42 & $<$10.1 & $8.2\pm2.4$\smallskip\\
HE0853+0102 & 0.053 & $10.54_{-0.10}^{+0.04}$ & $43.31_{-0.05}^{+0.08}$ &  $43.16_{-0.30}^{+0.09}$ & $43.51_{-0.20}^{+0.08}$ & $1.3_{-0.5}^{+0.3}$ & $<7.2$\smallskip\\
HE0934+0119 & 0.051 & $10.13_{-0.31}^{+0.25}$ & $43.75_{-0.05}^{+0.05}$ &  $43.71_{-0.01}^{+0.00}$ & $43.98_{-0.01}^{+0.00}$ & $3.7_{-0.0}^{+0.0}$ & $<7.0$\smallskip\\
HE0949-0122 & 0.020 & $10.02_{-0.31}^{+0.09}$ & $44.54_{-0.02}^{+0.02}$ &  $43.45_{-0.10}^{+0.07}$ & $43.74_{-0.09}^{+0.10}$ & $2.1_{-0.4}^{+0.5}$ & $7.2\pm2.3$\smallskip\\
HE1011-0403 & 0.059 & $10.74_{-0.17}^{+0.07}$ & $44.16_{-0.03}^{+0.04}$ &  $43.69_{-0.01}^{+0.01}$ & $43.98_{-0.01}^{+0.01}$ & $3.7_{-0.1}^{+0.1}$ & $8.2\pm1.8$\smallskip\\
HE1017-0305 & 0.049 & $10.93_{-0.15}^{+0.10}$ & $44.06_{-0.05}^{+0.05}$ &  $<$44.10 & $<$44.45 & $<$11.0 & $8.0\pm1.6$\smallskip\\
HE1029-1831 & 0.041 & $10.49_{-0.18}^{+0.06}$ & $43.89_{-0.08}^{+0.05}$ &  $44.57_{-0.02}^{+0.01}$ & $44.84_{-0.02}^{+0.01}$ & $27.1_{-1.3}^{+0.8}$ & $8.4\pm1.0$\smallskip\\
HE1107-0813 & 0.059 & $11.17_{-0.38}^{+0.20}$ & $43.78_{-1.34}^{+0.17}$ &  $<$44.46 & $<$44.87 & $<$28.6 & $8.0\pm2.5$\smallskip\\
HE1108-2813 & 0.024 & $10.29_{-0.05}^{+0.11}$ & $43.74_{-0.09}^{+0.07}$ &  $44.24_{-0.01}^{+0.01}$ & $44.50_{-0.01}^{+0.02}$ & $12.4_{-0.4}^{+0.4}$ & $8.2\pm0.7$\smallskip\\
HE1126-0407 & 0.060 & $10.59_{-0.46}^{+0.37}$ & $44.73_{-0.04}^{+0.03}$ &  $44.15_{-0.01}^{+0.01}$ & $44.42_{-0.01}^{+0.01}$ & $10.2_{-0.1}^{+0.2}$ & $8.2\pm2.0$\smallskip\\
HE1237-0504 & 0.008 & $10.92_{-0.01}^{+0.01}$ & $43.12_{-0.06}^{+0.06}$ &  $43.26_{-0.02}^{+0.03}$ & $43.58_{-0.03}^{+0.02}$ & $1.5_{-0.1}^{+0.1}$ & $7.2\pm1.5$\smallskip\\
HE1248-1356 & 0.015 & $10.31_{-0.01}^{+0.01}$ & $42.65_{-0.16}^{+0.10}$ &  $43.22_{-0.07}^{+0.06}$ & $43.49_{-0.06}^{+0.05}$ & $1.2_{-0.2}^{+0.1}$ & $7.3\pm1.1$\smallskip\\
HE1310-1051 & 0.034 & $10.35_{-0.18}^{+0.06}$ & $43.90_{-0.02}^{+0.02}$ &  $42.87_{-0.02}^{+0.02}$ & $43.17_{-0.02}^{+0.02}$ & $0.6_{-0.0}^{+0.0}$ & $<6.8$\smallskip\\
HE1330-1013 & 0.022 & $10.69_{-0.13}^{+0.03}$ & $42.93_{-0.49}^{+0.10}$ &  $<$43.69 & $<$43.89 & $<$3.0 & $7.5\pm1.8$\smallskip\\
HE1338-1423 & 0.041 & $11.02_{-0.20}^{+0.06}$ & $44.24_{-0.05}^{+0.03}$ &  $43.18_{-0.08}^{+0.06}$ & $43.52_{-0.07}^{+0.05}$ & $1.3_{-0.2}^{+0.2}$ & $<6.9$\smallskip\\
HE1353-1917 & 0.035 & $10.99_{-0.06}^{+0.03}$ & $43.37_{-0.07}^{+0.07}$ &  $43.50_{-0.01}^{+0.01}$ & $43.86_{-0.01}^{+0.01}$ & $2.8_{-0.1}^{+0.1}$ & $8.1\pm1.4$\smallskip\\
HE1417-0909 & 0.044 & $10.23_{-0.17}^{+0.10}$ & $43.54_{-0.05}^{+0.04}$ &  $<$43.39 & $<$43.67 & $<$1.8 & $<6.7$\smallskip\\
HE2128-0221 & 0.053 & $9.93_{-0.67}^{+0.21}$ & $43.58_{-0.05}^{+0.04}$ &  $<$43.75 & $<$44.02 & $<$4.0 & $<6.9$\smallskip\\
HE2211-3903 & 0.040 & $9.83_{-0.21}^{+0.22}$ & $44.64_{-0.03}^{+0.02}$ &  $43.84_{-0.05}^{+0.06}$ & $44.16_{-0.08}^{+0.07}$ & $5.6_{-1.0}^{+0.9}$ & $8.0\pm1.2$\smallskip\\
HE2222-0026 & 0.058 & $10.20_{-0.15}^{+0.09}$ & $43.81_{-0.04}^{+0.02}$ &  $<$44.11 & $<$44.46 & $<$11.2 & $7.4\pm3.9$\smallskip\\
HE2233+0124 & 0.057 & $10.71_{-0.03}^{+0.10}$ & $41.55_{-0.58}^{+0.71}$ &  $43.84_{-0.05}^{+0.03}$ & $44.22_{-0.05}^{+0.03}$ & $6.4_{-0.6}^{+0.4}$ & $8.0\pm1.8$\smallskip\\
HE2302-0857 & 0.047 & $11.20_{-0.13}^{+0.09}$ & $44.31_{-0.06}^{+0.06}$ &  $43.96_{-0.01}^{+0.01}$ & $44.24_{-0.01}^{+0.01}$ & $6.7_{-0.1}^{+0.1}$ & $8.2\pm1.4$\smallskip\\
\hline
\end{tabular}
\tablefoot{
\tablefoottext{a}{Redshift directly measured from the CARS IFU data as reported in Husemann et al. (2021).}
\tablefoottext{b}{Stellar masses and $1\sigma$ uncertainties from stellar template fitting reported by \texttt{AGNFitter}.}
\tablefoottext{c}{Integrated luminosity of the torus template reported by \texttt{AGNFitter}.}
\tablefoottext{d}{Luminosity of the matched star-forming IR template reported by \texttt{AGNFitter} integrated over the 42.5-122.5$\mu$m wavelength range.}
\tablefoottext{e}{Same as (d) but integrated over the full 8-1000$\mu$m wavelength range.}
\tablefoottext{f}{Estimated SFR based on $L_\mathrm{8-1000\,\mu\mathrm{m}}$ using calibration of \citet{murphy2011}, see Eq.~\ref{eq:SFR_IR}. }
\tablefoottext{g}{CO(1-0) line luminosity based on the flux measurements from \citet{bertram2007} and listed in Husemann et al. (2021).}
}

%% file: Tables/gas_properties/gas_properties.tex
\begin{tabular}{lcccccccc}\hline\hline
& \multicolumn{2}{c}{demarcation line approach}  & &  \multicolumn{3}{c}{BPT rainbow method} & & \\\cline{2-3}\cline{5-7}
Object & $f_{\mathrm{H}\upalpha}$\tablefootmark{a} & $\mathrm{SFR}_{\mathrm{H\upalpha}}$\tablefootmark{b} &  & class\tablefootmark{c} &  $f_{\mathrm{H}\upalpha}$\tablefootmark{a} & $\mathrm{SFR}_{\mathrm{H}\upalpha}$\tablefootmark{b} & $12+\log(\mathrm{O/H})$\tablefootmark{d} & $\alpha_{\mathrm{O/H}}\tablefootmark{e}$\\
& [$10^{-16}\,\mathrm{erg\,s}^{-1}\mathrm{cm}^{-2}$] & [$M_\odot\,\mathrm{yr}^{-1}$] & & & [$10^{-16}\,\mathrm{erg\,s}^{-1}\mathrm{cm}^{-2}$] & [$M_\odot\,\mathrm{yr}^{-1}$] &  [$r_\mathrm{eff}^{-1}$] &  [$r_\mathrm{eff}^{-1}$] \\\hline
HE0021-1819 &  $359\pm36$ & $1.3\pm0.1$ &  & MS & $342\pm34$ & $1.2\pm{0.1}$ & $8.89\pm0.02$ & $-0.031\pm0.010$ \smallskip\\
HE0040-1105 &  $16\pm2$ & $0.03\pm0.00$ &  & AGN & $<$41 & $<$0.1 & $8.85\pm0.02$ & $0.040\pm0.016$ \smallskip\\
HE0045-2145 &  $5672\pm567$ & $3.1\pm0.3$ &  & LMS & $4687\pm477$ & $2.5\pm{0.3}$ & $9.16\pm0.02$ & $-0.018\pm0.003$ \smallskip\\
HE0108-4743 &  $8594\pm859$ & $5.9\pm0.6$ &  & MS & $8251\pm825$ & $5.7\pm{0.6}$ & $9.05\pm0.01$ & $-0.055\pm0.009$ \smallskip\\
HE0114-0015 &  $978\pm98$ & $2.5\pm0.3$ &  & LMS & $901\pm90$ & $2.3\pm{0.2}$ & $9.06\pm0.02$ & $-0.007\pm0.011$ \smallskip\\
HE0119-0118 &  $1107\pm111$ & $4.1\pm0.4$ &  & MS & $1006\pm101$ & $3.8\pm{0.4}$ & $8.99\pm0.01$ & $0.026\pm0.020$ \smallskip\\
HE0150-0344 &  $3961\pm396$ & $10.5\pm1.0$ &  & ... & ... & ... & $8.88\pm0.02$ & $-0.029\pm0.005$ \smallskip\\
HE0203-0031 &  $71\pm7$ & $0.16\pm0.02$ &  & AGN+SF & $<$202 & $<$0.5 & $9.08\pm0.02$ & $-0.006\pm0.003$ \smallskip\\
HE0212-0059 &  $4163\pm416$ & $3.5\pm0.4$ &  & MS & $3978\pm398$ & $3.4\pm{0.3}$ & $8.99\pm0.00$ & $-0.050\pm0.013$ \smallskip\\
HE0224-2834 &  $368\pm37$ & $1.7\pm0.2$ &  & MS+AGN & $397\pm40$ & $1.8\pm{0.2}$ & $8.88\pm0.01$ & $-0.006\pm0.007$ \smallskip\\
HE0227-0913 &  $2970\pm297$ & $1.0\pm0.1$ &  & MS & $2933\pm293$ & $0.9\pm{0.1}$ & $8.96\pm0.01$ & $-0.017\pm0.007$ \smallskip\\
HE0232-0900 &  $8226\pm823$ & $19.0\pm1.9$ &  & MS & $8197\pm820$ & $19.0\pm{1.9}$ & $9.00\pm0.01$ & $-0.017\pm0.004$ \smallskip\\
HE0253-1641 &  $197\pm20$ & $0.24\pm0.02$ &  & MS & $232\pm23$ & $0.28\pm{0.03}$ & $8.97\pm0.02$ & $-0.025\pm0.023$ \smallskip\\
HE0345+0056 &  $3474\pm348$ & $4.1\pm0.4$ &  & SF & $3477\pm348$ & $4.1\pm{0.4}$ & $8.81\pm0.09$ & $0.066\pm0.073$ \smallskip\\
HE0351+0240 &  $143\pm14$ & $0.23\pm0.02$ &  & AGN & $<$240 & $<$0.4 & $8.80\pm0.03$ & $-0.010\pm0.013$ \smallskip\\
HE0412-0803 &  $93\pm9$ & $0.17\pm0.02$ &  & AGN & $<$192 & $<$0.3 & $8.85\pm0.02$ & $0.009\pm0.007$ \smallskip\\
HE0429-0247 &  $262\pm26$ & $0.6\pm0.1$ &  & AGN & $<$338 & $<$0.7 & $8.80\pm0.02$ & $0.016\pm0.015$ \smallskip\\
HE0433-1028 &  $5997\pm600$ & $9.3\pm0.9$ &  & MS+SF & $6700\pm671$ & $10.4\pm{1.0}$ & $8.97\pm0.01$ & $-0.049\pm0.009$ \smallskip\\
HE0853+0102 &  $407\pm41$ & $1.4\pm0.1$ &  & MS & $381\pm38$ & $1.3\pm{0.1}$ & $8.89\pm0.02$ & $-0.005\pm0.021$ \smallskip\\
HE0853-0126 &  $142\pm14$ & $0.6\pm0.1$ &  & SF & $130\pm13$ & $0.6\pm{0.1}$ & $8.60\pm0.16$ & $0.130\pm0.154$ \smallskip\\
HE0934+0119 &  $650\pm65$ & $2.1\pm0.2$ &  & MS+SF & $599\pm60$ & $1.9\pm{0.2}$ & $8.92\pm0.02$ & $-0.098\pm0.011$ \smallskip\\
HE0949-0122 &  $26\pm3$ & $0.01\pm0.00$ &  & AGN & $<$214 & $<$nan & $8.90\pm0.04$ & $0.049\pm0.039$ \smallskip\\
HE1011-0403 &  $328\pm33$ & $1.4\pm0.1$ &  & LMS & $294\pm29$ & $1.3\pm{0.1}$ & $9.03\pm0.01$ & $-0.022\pm0.014$ \smallskip\\
HE1017-0305 &  $353\pm35$ & $1.1\pm0.1$ &  & LMS & $340\pm34$ & $1.1\pm{0.1}$ & $9.00\pm0.02$ & $-0.007\pm0.021$ \smallskip\\
HE1029-1831 &  $4689\pm469$ & $9.4\pm0.9$ &  & SF & $3732\pm377$ & $7.5\pm{0.8}$ & $9.12\pm0.01$ & $-0.099\pm0.006$ \smallskip\\
HE1107-0813 &  $121\pm12$ & $0.5\pm0.1$ &  & MS+AGN & $115\pm12$ & $0.49\pm{0.05}$ & $9.08\pm0.02$ & $-0.003\pm0.006$ \smallskip\\
HE1108-2813 &  $6094\pm609$ & $4.2\pm0.4$ &  & MS+SF & $5840\pm587$ & $4.1\pm{0.4}$ & $9.01\pm0.01$ & $-0.110\pm0.008$ \smallskip\\
HE1126-0407 &  $195\pm20$ & $1.0\pm0.1$ &  & MS & $217\pm22$ & $1.1\pm{0.1}$ & $9.08\pm0.03$ & $-0.048\pm0.022$ \smallskip\\
HE1237-0504 &  $1501\pm150$ & $0.14\pm0.01$ &  & LMS & $3249\pm425$ & $0.31\pm{0.04}$ & $9.10\pm0.00$ & $-0.144\pm0.025$ \smallskip\\
HE1248-1356 &  $2766\pm277$ & $0.7\pm0.1$ &  & MS & $3034\pm303$ & $0.8\pm{0.1}$ & $9.05\pm0.01$ & $-0.128\pm0.012$ \smallskip\\
HE1310-1051 &  $1294\pm129$ & $1.9\pm0.2$ &  & SF & $1258\pm127$ & $1.8\pm{0.2}$ & $8.84\pm0.07$ & $0.010\pm0.049$ \smallskip\\
HE1330-1013 &  $763\pm76$ & $0.45\pm0.05$ &  & MS & $716\pm72$ & $0.42\pm{0.04}$ & $9.13\pm0.04$ & $-0.146\pm0.053$ \smallskip\\
HE1338-1423 &  $742\pm74$ & $1.6\pm0.2$ &  & AGN & $1319\pm135$ & $2.9\pm{0.3}$ & $9.04\pm0.06$ & $-0.019\pm0.075$ \smallskip\\
HE1353-1917 &  $609\pm61$ & $0.9\pm0.1$ &  & MS+A+S & $594\pm59$ & $0.9\pm{0.1}$ & $8.96\pm0.02$ & $-0.119\pm0.013$ \smallskip\\
HE1417-0909 &  $6\pm1$ & $0.01\pm0.00$ &  & AGN & $<$23 & $<$0.1 & $8.73\pm0.08$ & $0.106\pm0.066$ \smallskip\\
HE2128-0221 &  $154\pm15$ & $0.5\pm0.1$ &  & MS & $140\pm14$ & $0.49\pm{0.05}$ & $8.93\pm0.04$ & $-0.047\pm0.020$ \smallskip\\
HE2211-3903 &  $2419\pm242$ & $4.7\pm0.5$ &  & MS+SF & $2534\pm253$ & $4.9\pm{0.5}$ & $9.08\pm0.01$ & $-0.145\pm0.010$ \smallskip\\
HE2222-0026 &  $43\pm4$ & $0.19\pm0.02$ &  & SF & $34\pm3$ & $0.15\pm{0.02}$ & $8.80\pm0.05$ & $0.088\pm0.050$ \smallskip\\
HE2233+0124 &  $355\pm36$ & $1.4\pm0.1$ &  & MS+AGN & $339\pm34$ & $1.4\pm{0.1}$ & $9.05\pm0.02$ & $-0.057\pm0.018$ \smallskip\\
HE2302-0857 &  $1339\pm134$ & $3.7\pm0.4$ &  & MS & $1618\pm162$ & $4.4\pm{0.4}$ & $8.92\pm0.01$ & $-0.074\pm0.013$ \smallskip\\
\hline
\end{tabular}
\tablefoot{
\tablefoottext{a}{Integrated extinction-corrected H$\alpha$ flux decontaminated by AGN contribution either using a demarcation line cut or the rainbow method presented here. Given the high S/N of the data the error is dominated by the systematics in the absolute photometric zero-point which is assumed to be 10\%}
\tablefoottext{b}{H$\alpha$-based SFR determined from the asscoicated H$\alpha$ luminosity following the calibration of \citet{calzetti2007}, see Eq.~\ref{eq:SFR_Ha}.}
\tablefoottext{c}{Classification of the BPT in a (LINER) mixing sequence (L)MS, AGN cloud, SF cloud and any combination of the four.}
\tablefoottext{d}{Central oxygen abdundance on the \citet{Tremonti:2004} metallicity scale determined through the N2S2 index following Eq.~\ref{eq:N2S2}. A linear fit to the radial distribution is used to determine the central abdunance as the spatial coverages varies through the sample.}
\tablefoottext{e}{Slope of the linear metallicity gradient normalized to the effective radius of the respective galaxy as reported in Husemann et al. (2021).}
}

%% file: Tables/host_properties/BPT_morphology_radii.tex
\begin{tabular}{lcc}
    \hline\hline
    Object & SF radius & AGN radius\\
     & [arcsec] & [arcsec]\\
    \hline
    HE 0021$-$1810 & 15.0 & ---\\
    HE 0351$+$0240 & --- & 7.0\\
    HE 0433$-$1028 & 9.0 & ---\\
    HE 1107$-$0813 & 7.0 & 4.0\\
    HE 1108$-$2813 & 13.0 & ---\\
    HE 1237$-$0504 & 7.0 & ---\\
    HE 1353$-$1917 & 15.0 & 5.0\\
    HE 2211$-$3903 & 13.0 & ---\\
    HE 2233$+$0124 & --- & 3.0\\
    \hline
\label{tab:spatial_radii}
\end{tabular}
\tablefoot{SF and AGN cloud spaxels are located outside of the corresponding spatial radii.}